\documentclass[10pt]{iopart}
\usepackage{subcaption}
\usepackage{iopams}  
\usepackage{graphicx}
\usepackage{braket}
\usepackage{array}
\newcolumntype{C}[1]{>{\centering\arraybackslash}m{#1}}
\usepackage[numbers,square,sort&compress]{natbib}
\usepackage{upgreek} 
\usepackage{lipsum} 
\usepackage{amsmath} 
\usepackage{multirow} 
\usepackage{pfnote} 
\usepackage[left=2.5cm,right=2.5cm,bottom=2cm,top=3cm]{geometry}
\usepackage[colorlinks=true,linkcolor=blue,urlcolor=blue,citecolor=blue]{hyperref}
\usepackage{feynmf}
\usepackage{enumerate}
\usepackage{threeparttable}

\begin{document}

\title[Methods, Analysis, and the Treatment of Systematic Errors for the eEDM Limit in ThO]{Methods, Analysis, and the Treatment of Systematic Errors for the Electron Electric Dipole Moment Search in Thorium Monoxide}

\author{ACME Collaboration: J~Baron$^1$, W~C~Campbell$^2$, D~DeMille$^3$, J~M~Doyle$^1$, G~Gabrielse$^1$, Y~V~Gurevich$^4$, P~W~Hess$^5$, N~R~Hutzler$^1$, E~Kirilov$^6$, I~Kozyryev$^1$, B~R~O'Leary$^3$, C~D~Panda$^1$, M~F~Parsons$^1$, B~Spaun$^7$, A~C~Vutha$^8$, A~D~West$^3$, E~P~West$^1$}
\address{$^1$Department of Physics, Harvard University, Cambridge, MA 02138, USA, $^2$Department of Physics and Astronomy, University of California, Los Angeles, CA 90095, USA, $^3$Department of Physics, Yale University, New Haven, CT 06511, USA, $^4$ ZAH, Landessternwarte 
K\"{o}nigstuhl 12, D-69117 Heidelberg, Germany, $^5$Joint Quantum Institute, University of Maryland 
College Park, MD 20742, USA, $^6$University of Innsbruck, Technikerstrasse 25, A-6020 Innsbruck, Austria, $^7$Joint Institute for Lab Astrophysics, 440 University Ave, Boulder, CO 80302, USA, $^8$Department of Physics, University of Toronto, Toronto, ON M5S 1A7, Canada.}
\ead{awest@physics.harvard.edu}

\begin{abstract}
We recently set a new limit on the electric dipole moment of the electron (eEDM) (J. Baron et al., ACME collaboration, \emph{Science} \textbf{343} (2014), 269--272), which represented an order-of-magnitude improvement on the previous limit and placed more stringent constraints on many $CP$-violating extensions to the Standard Model. In this paper we discuss the measurement in detail. The experimental method and associated apparatus are described, together with the techniques used to isolate the eEDM signal. In particular, we detail the way experimental switches were used to suppress effects that can mimic the signal of interest. The methods used to search for systematic errors, and models explaining observed systematic errors, are also described. We briefly discuss possible improvements to the experiment.\end{abstract}

\submitto{NJP}

\maketitle

\newcommand{\N}{\mathcal{N}} 
\newcommand{\E}{\mathcal{E}} 
\newcommand{\Elab}{\mathcal{E}_\mathrm{lab}} 
\newcommand{\Eeff}{\mathcal{E}_\mathrm{eff}} 
\newcommand{\vecE}{\vec{\mathcal{E}}} 
\newcommand{\vecElab}{\vec{\mathcal{E}}_\mathrm{lab}} 
\newcommand{\vecEeff}{\vec{\mathcal{E}}_\mathrm{eff}} 
\newcommand{\B}{\mathcal{B}} 
\newcommand{\vecB}{\vec{\mathcal{B}}} 
\newcommand{\Blab}{\mathcal{B}_{\rm lab}} 
\newcommand{\vecBlab}{\vec{\mathcal{B}}_{\rm lab}} 
\newcommand{\de}{d_e} 
\newcommand{\vecde}{\vec{d}_e} 
\newcommand{\expval}[3]{\left\langle #1  \right| #2 \left| #3 \right\rangle}
\newcommand{\wj}[6]{\bpm #1 & #2 & #3 \\ #4 & #5 & #6 \epm} 
\newcommand{\nhat}{\hat{n}} 
\newcommand{\nhatlab}{\hat{n}_{\rm lab}} 
\newcommand{\DHpar}{D_\|} 
\newcommand{\GHpar}{G_\|} 
\newcommand{\Elabhat}{\hat{\mathcal{E}}_{\rm lab}}
\newcommand{\sign}{\textrm{sign}}
\newcommand{\mj}{M} 
\newcommand{\Brot}{B_{v=0}} 
\newcommand{\gj}{g_{J=1}} 
\renewcommand{\dj}{d_{J=1}} 
\newcommand{\dg}{\Delta g}
\newcommand{\gn}{g_{-++}} 
\newcommand{\C}{\mathcal{C}} 
\newcommand{\F}{F} 
\newcommand{\A}{\mathcal{A}} 
\newcommand{\R}{\mathcal{R}} 
\newcommand{\G}{\mathcal{G}} 
\renewcommand{\P}{\mathcal{P}} 
\newcommand{\Ld}{\mathcal{L}} 
\newcommand{\Nsw}{\tilde{\mathcal{N}}} 
\newcommand{\Esw}{\tilde{\mathcal{E}}} 
\newcommand{\Bsw}{\tilde{\mathcal{B}}} 
\newcommand{\Lsw}{\tilde{\mathcal{L}}} 
\newcommand{\Psw}{\tilde{\mathcal{P}}} 
\newcommand{\Gsw}{\tilde{\mathcal{G}}} 
\newcommand{\Rsw}{\tilde{\mathcal{R}}} 
\newcommand{\Thsw}{\tilde{\theta}}
\newcommand{\wNE}{\omega^{\mathcal{NE}}}
\newcommand{\ecm}{e\cdot\mathrm{cm}} 
\newcommand{\Enr}{\E^{\rm{nr}}}
\newcommand{\upperlimit}{$8.7\times 10^{-29}$ $e$ cm}
\newcommand{\result}{$(-2.1 \pm 3.7_\mathrm{stat} \pm 2.5_\mathrm{syst})\times 10^{-29}$ $e$ cm}
\newcommand{\g}{g_1}
\newcommand{\D}{\D_1}
\newcommand{\DW}{\Delta_{\Omega,1}}
\newcommand{\nr}{\rm{nr}}
\newcommand{\kz}{\hat{k}\cdot\hat{z}}
\newcommand{\rabi}{\Omega_{\rm r}}
\newcommand{\rabii}{\Omega_{{\rm r}, i}}
\newcommand{\wNEt}{\omega^{\N\E}_{T}} 
\newcommand{\wLs}[1]{\omega_{\rm{L}, #1}} 
\newcommand{\wL}{\omega_{\rm{L}}} 

\newcommand{\p}{{\rm prep}}
\newcommand{\rX}{X}
\newcommand{\rY}{Y}
\newcommand{\ro}{\rm{read}}

\newcommand{\SP}{S} 

\newcommand{\tdo}{{^3\Delta_1}}
\newcommand{\tdt}{{^3\Delta_2}}
\newcommand{\tpo}{{^3\Pi_1}}
\newcommand{\spo}{{^1\Pi_1}}
\newcommand{\ssz}{{^1\Sigma_0^+}}
\newcommand{\tpz}{{^3\Pi_0}}

\newcommand{\be}{\begin{equation}}
\newcommand{\ee}{\end{equation}}
\newcommand{\bea}{\begin{eqnarray}}
\newcommand{\eea}{\end{eqnarray}}
\newcommand{\bpm}{\begin{pmatrix}}
\newcommand{\epm}{\end{pmatrix}}

\section{Introduction}
Symmetries play a vital role in physics and experimental tests of symmetries have revealed insights into physical theory. Perhaps the most famous early example is the experiment of Michelson and Morley \cite{Michelson1887}, now understood as an early demonstration of Lorentz invariance. Similarly, observed violations of parity ($P$) symmetry \cite{Wu1957} and charge-parity ($CP$) symmetry \cite{Christenson1964} have informed and motivated understanding of the weak and strong forces \cite{Lee1957,Kobayashi1973}. The recent discovery of the Higgs boson \cite{Aad2012short} is a confirmation of a predicted spontaneously broken gauge symmetry \cite{Englert1964}, and the LHC continues to probe physics at high energies, looking for evidence of physics beyond the Standard Model (SM). On a complementary front, precision measurements of charge-parity-time ($CPT$) invariance and Lorentz invariance using low-energy techniques continue to test these fundamental symmetries \cite{Ambrosino2006,Kostelecky2010,Dehmelt1999,DiSciacca2013,Bennett2004,Michimura2013,Hohensee2010,Herrmann2009,Gagnon2004}.

Precision measurements in atomic and molecular systems are well suited to testing fundamental physics, and searches for EDMs of fundamental particles have been at the forefront of such tests \cite{Khriplovich1997,Bernreuther1991}. Measurements of the EDMs of the electron, neutron \cite{Pendlebury2015} and atomic species such as mercury \cite{Graner2016}, are complementary tests of beyond-SM physics and of fundamental symmetries \cite{Pospelov2005}. As discussed in section~\ref{sec:theory}, an EDM of a fundamental particle can only exist if time-reversal ($T$) symmetry is broken, which is equivalent to $CP$ violation for $CPT$-invariant models \cite{Streater2000}. For many theories, intrinsic $CP$ violation is predicted to manifest as eEDMs at an experimentally accessible level \cite{Pospelov2005,Fukuyama2012,Engel2013}. Consequently, discovering an eEDM, or further constraining its value, can inform our understanding of particle physics at high energy and help to shed light on outstanding issues such as the baryon asymmetry problem \cite{Sakharov1967,Gavela1994}. The current best limit on the eEDM was reported by ACME in 2014 \cite{Baron2014}:\footnote{Note that the limit we report here uses an updated value for $\Eeff=78$~GV/cm which is obtained by averaging the results from references \cite{Denis2016,Skripnikov2016}.}
\begin{equation}
d_e\le9.3\times10^{-29}~e\cdot{\rm cm}~(90\%~{\rm conf.~level}).
\end{equation}

Many extensions to the SM predict eEDM values many orders of magnitude higher than the SM prediction of $<10^{-38}~e\cdot{\rm cm}$ \cite{Pospelov1991,Khriplovich1997,Pospelov2005}, meaning measurement of an eEDM at current experimental sensitivity would be a signature of new physics. Supersymmetry is an example of an extension to the SM that predicts a large, potentially measurable eEDM. The current eEDM limit constrains the parameter space associated with supersymmetry such that it is often considered unnatural \cite{Abel2001,Nir1999}.

In most models, the eEDM arises as a radiative correction (Feynman loop diagram) due to $CP$-violating interactions with new particles. An example of such an interaction within generic supersymmetric theory is shown in figure~\ref{fig:susy_feynmann}.
\begin{figure}[!ht]
\centering
\begin{fmffile}{diagram}
\begin{fmfgraph*}(240,150)
\fmfleft{i1}
\fmfright{o1,o2,o3}
\fmftop{o6}
\fmf{fermion,tension=1,label=$e_L$,l.side=right}{i1,v1}
\fmf{wiggly,label=$\tilde{\gamma}$,l.side=right,l.dist=50}{v1,v2}
\fmf{fermion,tension=1,label=$e_R$,l.side=right}{v2,o2}
\fmf{dashes,left,tension=0,label=$\tilde{e}$,l.side=left}{v1,v2}
\fmffreeze
\fmf{wiggly,label=$\gamma$,l.side=left}{o6,o3}
\fmfforce{0.63w,0.67h}{o6}
\fmfforce{0.5w,0.5h}{o1}
\fmfiv{lab=\rotatebox[origin=c]{45}{\scalebox{2}{$\times$}},label.dist=-1.}{(0.438w,0.64h)}
\fmfiv{lab=$e^{i\phi_{\rm CP}}$,label.dist=-1.}{(0.38w,0.75h)}
\end{fmfgraph*}
\end{fmffile}
\vspace{-48pt}
\caption{Example of a supersymmetric 1-loop contribution to the eEDM. The symbols $e_L$ and $e_R$ represent the left and right helicities of the electron, $\tilde{e}$ a selectron, $\tilde{\gamma}$ a photino and $\gamma$ a photon. This generic diagram illustrates how a $CP$-violating phase (represented by the $+$ symbol) can be produced in a straightforward manner by SM extensions. Note that a detailed discussion of associated high-energy theory is beyond the scope of this paper.}
\label{fig:susy_feynmann}
\end{figure}
The $CP$ violation is associated with the presence of non-trivial complex phases in the theory. For a given $CP$-violating phase $\phi_{CP}$, one can make a generic estimate of the mass scale $\Lambda$ of new physics being probed, according to the following formula for an $n$-loop process \cite{Engel2013}:
\begin{equation}
\Lambda^2=e\frac{m_e}{d_e}\left(\frac{\alpha}{4\pi}\right)^n\sin\phi_{CP}
\end{equation}
where $e$ is the electron charge, $m_e$ the electron mass, $\alpha$ is the fine structure constant and $\phi_{CP}$ is a $CP$-violating phase. Assuming that $\sin(\phi_{CP})\sim1$ \cite{Pospelov2005}, we find that our most recent result interrogates energy scales for one-loop processes of around 10~TeV. Similar analysis shows that our result was sensitive to two-loop effects at around the 1~TeV mass scale. While any such estimates are inherently model-dependent, we see that using an apparatus that fits in a room we have been able to probe fundamental physics at energy scales usually associated with the largest particle accelerators. 

\section{Atom and Molecule eEDM Experiments}
\subsection{Theory}
\label{sec:theory}
The eEDM, $\vec{d}_e$, is a vector quantity that is aligned along (or against) the axis of the electron's spin, $\vec{s}$ \cite{Khriplovich1997}. By convention, we write $\vec{d}_e=2d_e\vec{s}$, such that a measurement of any Cartesian component of $\vec{d}_e$ yields a value of $\pm d_e$. (Here and throughout, we set $\hbar=1$.) For an electron moving non-relativistically, the eEDM interacts with an electric field $\vec{\mathcal{E}}$ via the Hamiltonian
\begin{equation}
\mathcal{H}_{\rm EDM}=-\vec{d}_e\cdot\vec{\mathcal{E}}\propto\vec{s}\cdot\vec{\mathcal{E}}.\footnote{A detailed discussion of the sign convention for this Hamiltonian term is provided in section~\ref{sec:sign_conventions}.}
\label{eq:Eedm}
\end{equation}

Under time reversal, $T$, $\vec{s}$ reverses direction, but $\vec{\mathcal{E}}$ is unchanged. Similarly, under space inversion, $P$, $\vec{s}$ is unchanged, but $\vec{\mathcal{E}}$ reverses direction. Hence $\mathcal{H}_{\rm EDM}$ is odd under $P$ and $T$.

To measure the eEDM, one looks for an energy shift due to the interaction in equation~\ref{eq:Eedm}. Since 1964, every improvement in experimental sensitivity to $d_e$ has been obtained by measuring this shift for electrons bound in a neutral atom or molecule \cite{Sandars1964,Stein1967,Weisskopf1968,Player1970,Lamoreaux1987,Murthy1989,Abdullah1990,Commins1994,Regan2002,Hudson2011}. This might seem surprising at first glance, since Schiff's theorem states that there can be no net electric field acting on a non-relativistic point particle bound in a neutral system \cite{Schiff1963}. However, in 1958 Salpeter showed that, when relativistic effects are taken into account, a neutral species can experience an energy shift due to an eEDM when an external electric field $\mathcal{E}_{\rm ext}$ is applied \cite{Salpeter1958}. In 1965 Sandards showed, strikingly, that the size of the resulting energy shifts can be much larger than $d_e\mathcal{E}_{\rm ext}$ \cite{Sandars1965}.

More detailed explanations of this relativistic enhancement can be found elsewhere, e.g.\ \cite{Khriplovich1997,Commins2007,Commins2010}, but we summarise the basic principle here. Taking into account the relativistic length contraction of the eEDM for a moving electron, its interaction with a total electric field $\vec{\mathcal{E}}$ (the sum of an external, applied field and an intra-atomic/molecular field) takes the form
\begin{equation}
\mathcal{H}_{\rm EDM}^{\rm rel}=-\vec{d}_e\cdot\vec{\mathcal{E}}+\frac{\gamma}{1+\gamma}\vec{\beta}\cdot\vec{d}_e\vec{\beta}\cdot\vec{\mathcal{E}},
\end{equation}
where $\vec{\beta}=\vec{v}/c$ is the dimensionless velocity and $\gamma$ is the Lorentz factor \cite{Commins2007}. The first term in this expression is the non-relativistic EDM interaction, whose expectation value vanishes by Schiff's theorem. The second, relativistic term can result in a nonzero net interaction when the electron's velocity and the electric field are non-uniform in space (as, for example, when the electron travels near a charged nucleus in an atom or molecule), and when the atom or molecule is polarised by an external electric field. This interaction can be expressed in terms of an `effective electric field', $\vec{\mathcal{E}}_{\rm eff}$, defined in analogy to equation \ref{eq:Eedm} such that
\begin{equation}
\langle\mathcal{H}_{\rm EDM}^{\rm rel} \rangle \equiv -\langle\vec{d}_e\rangle\cdot\vec{\mathcal{E}}_{\rm eff}.
\label{eq:EeffDef}
\end{equation}
Detailed calculations show that this `effective electric field' within an atom or molecule can be significantly larger in magnitude than the applied external field. The size of $\Eeff$ is maximal for systems where a valence electron has significant wavefunction amplitude near a highly-charged nucleus. In such species with a nucleus of atomic number $Z$, $\Eeff$ scales approximately as \cite{Khriplovich1997}
\begin{equation}
\Eeff\propto P_{\E}Z^3R(Z),
\end{equation}
where $P_{\mathcal{E}}\in[0,1)$ is the degree of electric polarisation of the state and $R$ is a relativistic factor that is roughly constant for $Z\ll\alpha^{-1}$, but grows quickly as $Z$ approaches $\alpha^{-1}$ \cite{Khriplovich1997,Commins2010,DeMille2000,Hinds1997}. For fully polarised systems with $Z\approx90$ (as with our molecule of choice, ThO), the effective electric field can reach values as large as $\Eeff\approx100$~GV/cm. In practice, the maximum polarisation attainable with atoms, even in the highest laboratory static electric fields ($\sim 100$ kV/cm), is $P_{\mathcal{E}}\sim10^{-3}$. Nevertheless, this can lead to values of $\Eeff$ nearly 1,000 times larger than the applied laboratory field (e.g. $\Eeff\sim70$~MV/cm in Tl atoms \cite{Regan2002}). Using this kind of enhancement, the limit on $d_e$ was reduced by six orders of magnitude by the first atom-based eEDM measurement \cite{Sandars1964}. Further improvement is afforded by working with polar molecules, which are much more polarisable than atoms due to having much more closely spaced levels of opposite parity (associated with their rotational motion). In practice, polarisation $P_{\mathcal{E}}\sim1$ is achievable with molecules \cite{Commins2010,DeMille2000,Hinds1997}. In ThO, the effective electric field is $\Eeff\approx 78$~GV/cm \cite{Denis2016,Skripnikov2016}.

To measure the eEDM, the electron spin is prepared in a state oriented perpendicular to $\vec{\mathcal{E}}_{\rm eff}$, i.e. in a superposition of states parallel and antiparallel to $\vec{\mathcal{E}}_{\rm eff}$.  After an interaction time $\tau$, the eEDM energy shift in equation~\ref{eq:Eedm} produces a relative phase accumulation $2\phi_{\rm EDM}=-2d_e\Eeff\tau$ between these states; this is equivalent to a precession of the spin orientation about $\vec{\mathcal{E}}_{\rm eff}$ by an angle $2\phi_{\rm EDM}$.

For a shot-noise-limited measurement, the uncertainty in the eEDM, $\delta d_e$, is given by
\begin{equation}
\label{eq:uncertainty_prop}
\delta d_e=\left(2\tau\Eeff\sqrt{N}\right)^{-1},
\end{equation}
where $N$ is the number of measurements. The large values of $\Eeff$ accessible in many molecules have motivated several recent eEDM searches \cite{Eckel2013,Hudson2011,Leanhardt2011}. This and other advantages associated with the molecule ThO are discussed in the following section.

\subsection{ThO Molecule}
 \label{sec:tho_molecule}
ThO has a number of properties that make it well-suited to an eEDM measurement, both by enhancing statistical sensitivity and by suppressing systematic errors. We performed our measurements in the $H$ electronic state of ThO, which has two valence electrons in a $(\sigma\delta)^3\Delta_1$ state. Such states were first proposed for use in an eEDM measurement by Meyer et al.\ in 2006 \cite{Meyer2006}. The $\sigma$ orbital valence electron wavefunction has a large amplitude near the heavy Th nucleus, facilitating the large $\Eeff$ required for a large eEDM sensitivity, as described in section~\ref{sec:theory}. The $H$ state of ThO has one of the largest calculated values of $\Eeff\approx77.6$~GV/cm \cite{Denis2016,Skripnikov2016}. We note that the value of $\Eeff$ in our experiment with ThO is more than 5 times larger than that attained in experiments using YbF, which set the previous eEDM limit \cite{Mosyagin1998,Sauer2011,Abe2014}, and over 1,000 times larger than that in experiments using Tl atoms \cite{Regan2002}.

All $^3\Delta_1$ states have very small magnetic moments \cite{Herzberg1989} since the $\delta_{3/2}$ orbital valence electron serves to nearly cancel the magnetic moment of the $\sigma_{1/2}$ orbital. The actual magnetic moment of $H$ deviates from zero primarily because of mixing with other states \cite{Petrov2014}. We express ThO molecule states using the basis $\ket{Y,J,M,\Omega}$, where $Y$ is the electronic state, $J$ is the total angular momentum, $M$ is the projection of $J$ onto the laboratory $\hat{z}$-axis, and $\Omega$ is the projection of the electronic angular momentum onto the internuclear axis, $\hat{n}$, which points from the lighter nucleus to the heavier nucleus. We used the $\ket{H,J=1,\Omega=1}$ rotational manifold for our measurement, for which the magnetic moment is $\mu_1=g_1\mu_{\rm B}M$, where $g_1=-0.00440(5)$ is the associated $g$-factor \cite{Petrov2014,Vutha2011} and $\mu_{\rm B}$ is the Bohr magneton. This small $g$-factor, generic to all molecules with this structure, ensures that the $H$ state is particularly insensitive to spurious magnetic fields.

\begin{figure}[!ht]
\centering
\includegraphics[width=0.95\textwidth]{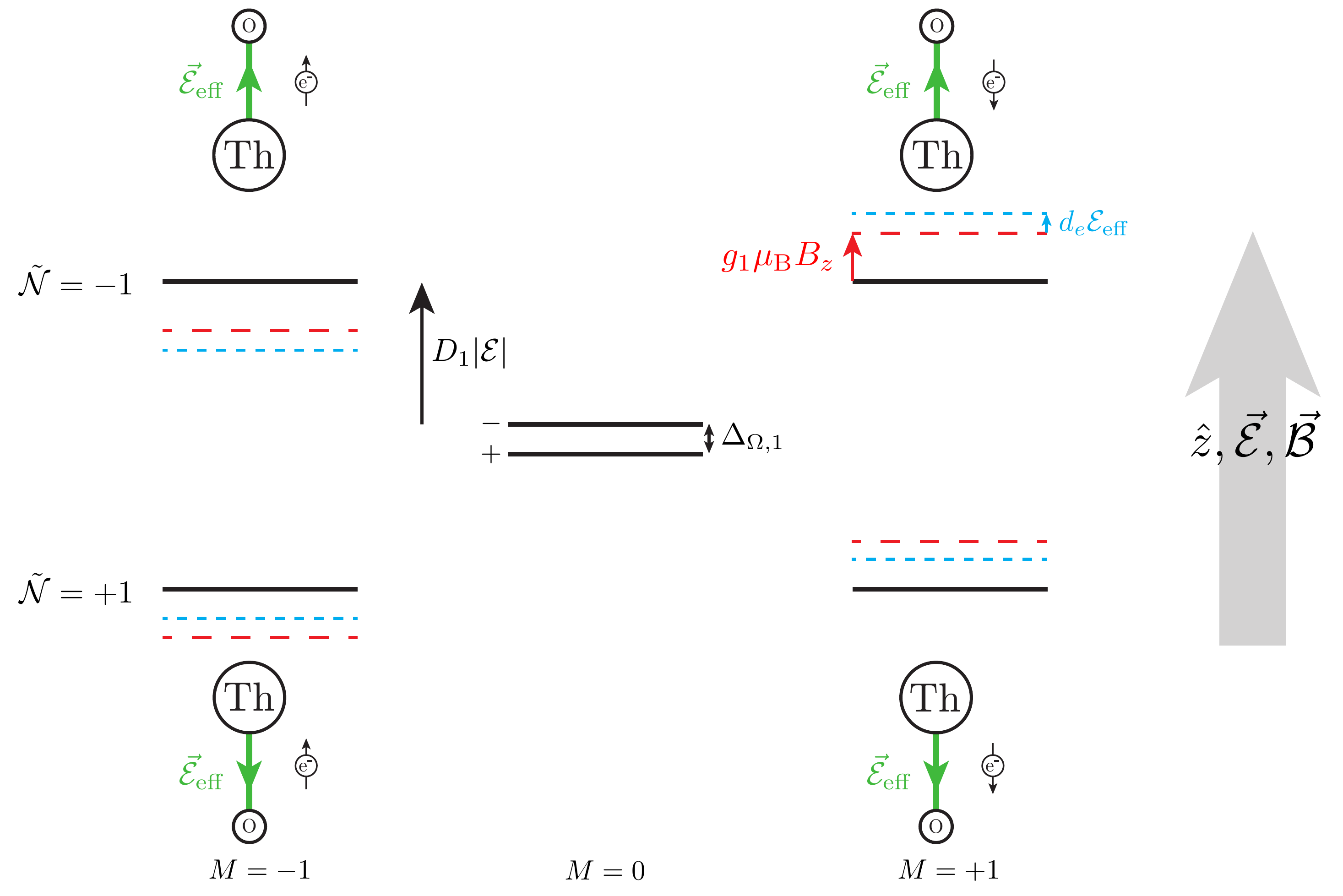}
\caption{Energy level structure of the $|H~^3\Delta_1,J=1, M,\Omega\rangle$ state manifold in ThO in the presence of a polarising electric field. In the absence of applied fields, opposite-parity states $|\pm\rangle\equiv\left(\ket{\Omega=-1}\pm\ket{\Omega=+1}\right)/\sqrt{2}$ are separated by energy $\Delta_{\Omega,1}\approx2\pi\times360$~kHz. The $M=+1$ ($M=-1$) state in $\ket{+}$ is nearly fully mixed with the $M=+1$ ($M=-1$) state in $\ket{-}$ by an electric field $\mathcal{E}{\gtrsim}10$~V/cm. The fully polarised states are denoted by $\ket{\N=\pm1}$. For $\mathcal{E}\gtrsim10$~V/cm, the associated Stark shift is linear and given by $-\Nsw D_1|\mathcal{E}|$, where $D_1\approx2\pi\times1$~MHz/(V/cm) (black arrow/lines) is the expectation value of the molecular electric dipole moment in these states \cite{Vutha2011}. Additionally, a magnetic field $\vec{\B}$ causes a Zeeman shift ${\approx}-M g_1\mu_{\rm B}\B_z$, with $g_1\mu_{\rm B}\approx-2\pi\times6$~kHz/G (red arrows/lines) \cite{Kirilov2013,Petrov2014}. A nonzero eEDM would result in an additional energy shift $\approx-M\Nsw d_e\Eeff$ (blue arrows/lines) where $\Esw=-1$ (+1) when the applied $\E$ field is (is not) reversed. The orientation of $\vec{\mathcal{E}}_{\rm eff}$ (green arrows), the spin of the electron in the $\sigma$ orbital (black arrow next to molecule), the external electric field $\vec{\E}$, and the external magnetic field $\vec{\B}$ are shown relative to the laboratory $\hat{z}$ direction which is oriented upwards on the page. Diagram not to scale.}
\label{fig:H-state}
\end{figure}

States with nonzero $\Omega$ have closely spaced pairs of opposite-parity levels with identical values of $J$ called `$\Omega$-doublets', which are split by energy $\Delta_{\Omega}$ due to the Coriolis effect in the rotating molecule \cite{Herzberg1971,Brown2003,Landau1981}. The application of an electric field $\vec{\mathcal{E}}$ mixes the $M\ne0$ opposite-parity levels via the Stark interaction, $-\vec{D}\cdot\vec{\E}$, where $D$ is the electric dipole operator, and from here on $\vec{\mathcal{E}}$ is the applied (laboratory) field. In the limit $|\langle \vec{D}\rangle\cdot\vec{\E}|\gg\Delta_\Omega$, the molecule is fully polarised, the internuclear axis is nearly aligned or anti-aligned with the applied electric field, and the alignment orientation is described by quantum number $\tilde{\N}\equiv\hat{\mathcal{E}}\cdot\hat{n}=\pm1$. This structure is shown for the $H$ state of ThO in figure~\ref{fig:H-state}.

The use of molecules with $\Omega$-doublet structure for an eEDM measurement, first explored in \cite{DeMille2000,DeMille2001} in the context of experiments using PbO, is of great importance to our experiment. The $\ket{H,J=1}$ manifold has an $\Omega$-doublet splitting $\Delta_{\Omega,1}\approx2\pi\times360$~kHz\footnote{Throughout the paper, we give numerical values of energies (with $\hbar=1$) in terms of angular frequencies by using the notation $2\pi\times f$, where $f$ is a linear frequency in units of Hz.} \cite{HutzlerThesis} and an electric dipole moment $D_1\equiv|\braket{H,J=1,M=\pm1,\Omega|\vec{D}\cdot\hat{z}|H,J=1,M=\pm1,\Omega}|\approx2\pi\times1$~MHz/(V/cm) \cite{Hess2014}; this permits full ($>99~\%$) polarisation of the state in small applied electric fields, $\E\gtrsim10$~V/cm, allowing us to take full advantage of the huge $\Eeff$ in ThO. The $\Omega$-doublet structure is also useful in rejecting systematic errors since it allows for spectroscopic reversal of $\vec{\mathcal{E}}_{\rm eff}\propto-\hat{n}$ by addressing different $\tilde{\N}$ states without reversing the applied electric field \cite{Kawall2004}. This is discussed in greater detail in section~\ref{ssec:E_correlated_phase}.

The $H$ state in ThO is metastable with a lifetime ${\approx}1.8$~ms \cite{Vutha2010}, limiting our measurement time to $\tau\approx1$~ms. We note that this is comparable to previous beam-based eEDM measurements where the atomic/molecular states used had significantly longer lifetimes \cite{Hudson2011,Regan2001,Vutha2010}.

As with many other species, ThO proved nicely compatible with a new approach to creating molecular beams, the hydrodynamically enhanced cryogenic buffer gas beam \cite{Hutzler2012,Patterson2007,Maxwell2005}. This method provides a cold, high-flux and low-divergence beam \cite{Hutzler2011} yielding a large number of molecules in the few lowest-lying quantum states. The molecule beam's forward velocity (${\approx}180$~m/s) was also lower than a typical supersonic beam, which helped minimise the apparatus length for a given coherence time. For more details on the beam source, see section~\ref{sec:beamsource}. 

\section{ACME Experiment}
\subsection{Overview of Measurement Scheme}
 \label{sec:Measurement_scheme}
\subsubsection{Basic Measurement Scheme}
\begin{figure}[!ht]
\centering
\includegraphics[width=0.5\textwidth]{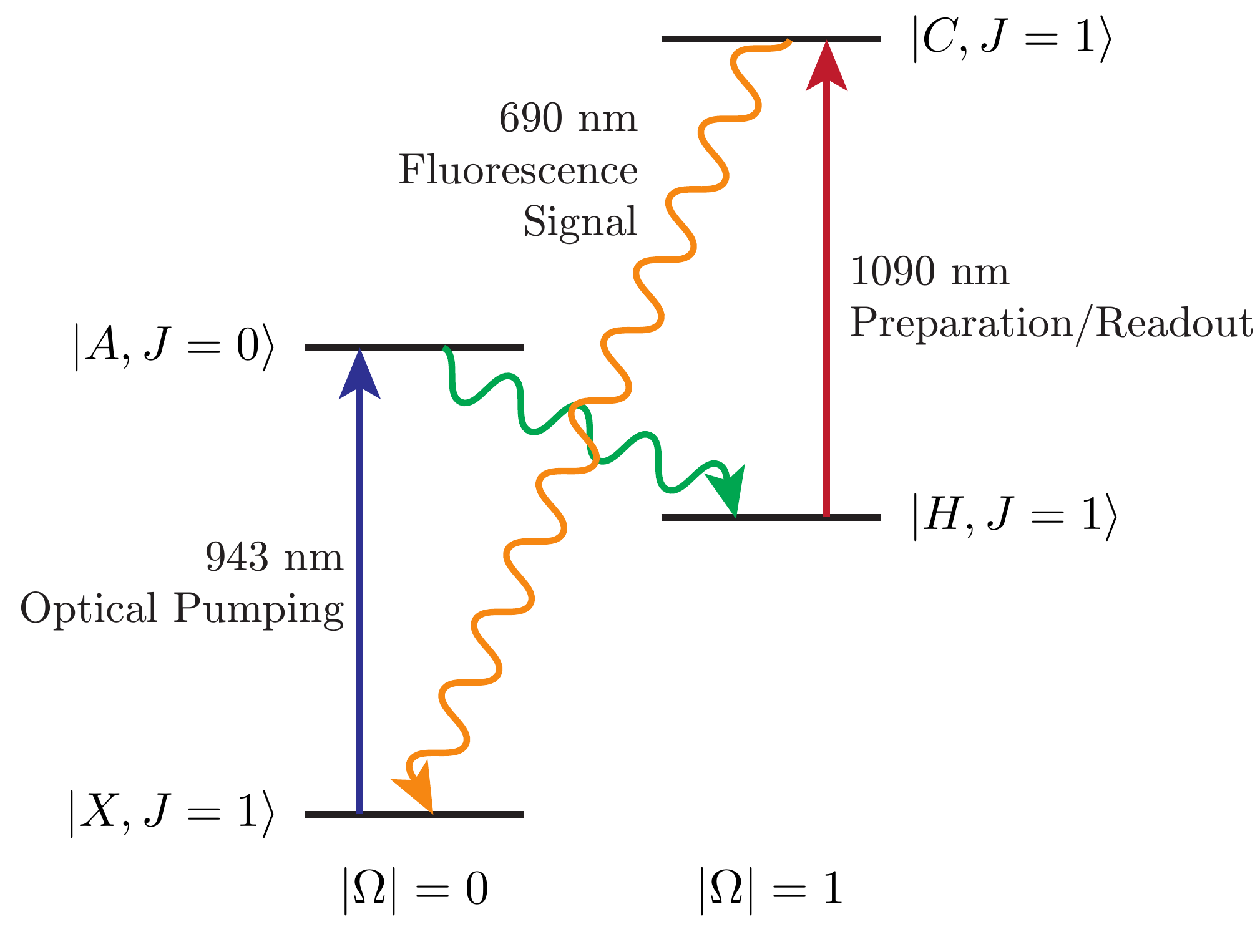}
\caption{Levels and transitions in ThO used in our measurement of the eEDM, based on \cite{Vutha2010,Edvinsson1985,Paulovic2003}. Solid arrows indicate transitions we address with lasers, wavy arrows indicate spontaneous decays of interest. For more details on how these transitions were used, see the main text.}
\label{fig:ThOlevels}
\end{figure}

\begin{figure}[!ht]
\centering
\includegraphics[width=16cm]{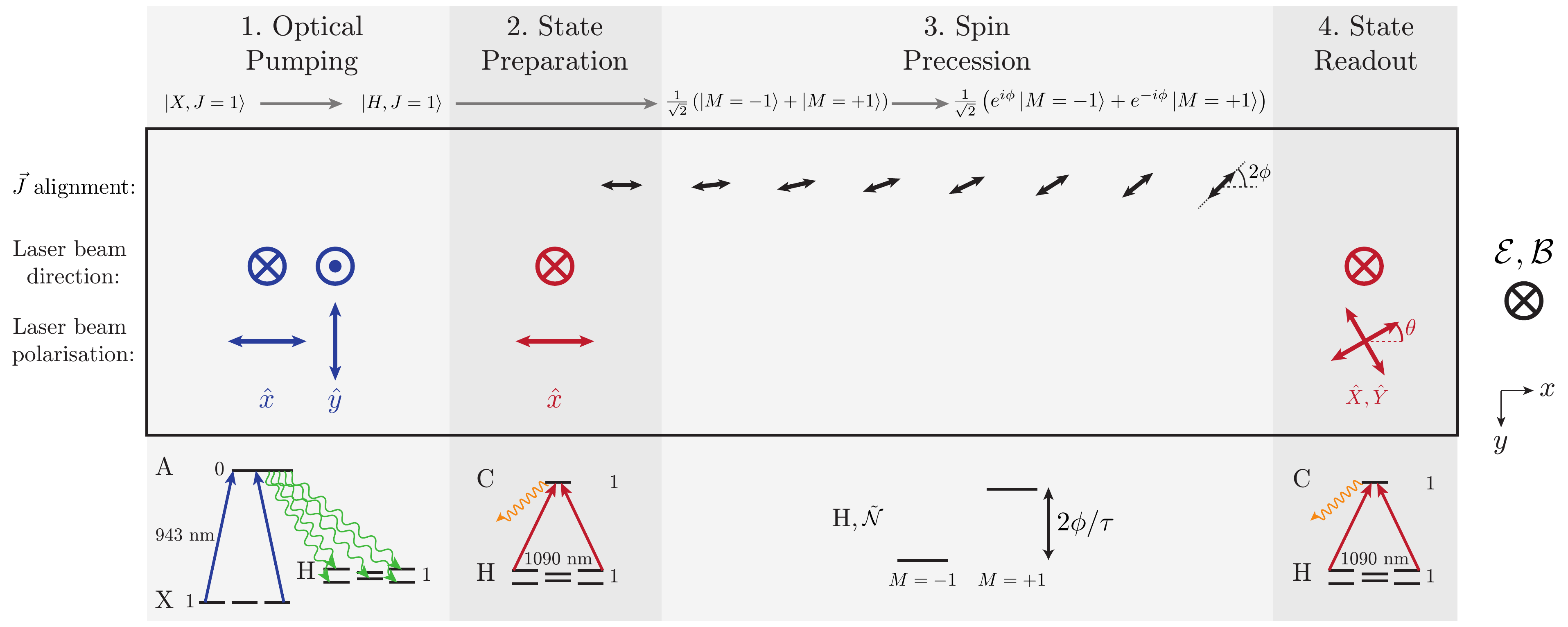}
\caption{Simplified schematic of the measurement scheme; numbers next to energy levels label $J$. \textbf{1.} Molecules in the $\ket{X,J=1}$ state are optically pumped via the $A$ state into $\ket{H,J=1}$ by a retroflected (and offset in $x$) laser beam (blue arrows into/out of page), polarised along $\hat{x}$ and $\hat{y}$ (blue arrows). \textbf{2.} Molecules from one of the $\Nsw$ states are then prepared in a superposition of $M$ sublevels ($M=-1,0,+1$ from left to right) by a linearly polarised laser beam (red) addressing the $H\rightarrow C$ transition. This aligns the molecule's angular momentum, $\vec{J}$, which in turn aligns the spin of the eEDM-sensitive $\sigma$ electron, which is on average aligned with $\vec{J}$. \textbf{3.} The angular momentum (and hence electron spin) then precesses due to the electric and magnetic fields present (into the page) by an angle $\phi$. This precession is dominated by the magnetic interaction but also includes a term linear in $d_e$ (see equation~\ref{eq:simple_phase}). \textbf{4.} The spin state is projected onto orthogonal superpositions of the $M$ sublevels by laser beams polarised along $\hat{X},\hat{Y}$ (red arrows). The resulting fluorescence is determined by the population in each superposition state and hence the precession angle $\phi$.}
\label{fig:meas_scheme_simple}
\end{figure}

\hspace*{\fill} \\
We performed a spin precession measurement, resembling previous beam-based eEDM experiments \cite{Hudson2011,Regan2002,Commins1994}, on $^{232}\rm{Th}^{16}\rm{O}$ molecules in a pulsed molecular beam generated by a cryogenic buffer gas beam source. Figure~\ref{fig:meas_scheme_simple} shows a simplified schematic of the measurement. The molecules fly at velocity $v\approx200$~m/s into a magnetically shielded region with nominally uniform and parallel electric $\vec{\E}$ and magnetic $\vec{\B}$ fields. Molecule population is transfered from $|X^1\Sigma^+,J=1,M=\pm1\rangle$ in the electronic ground state to the metastable $|H,J=1,M=\pm1,\Omega=\tilde{\N}\tilde{\E}M\rangle\equiv|\pm,\tilde{\N}\rangle$ state manifold (in the $\ket{\pm,\Nsw}$ nomenclature we use $\pm$ to refer to $M=\pm1$) by optical pumping through the short-lived $|A^3\Pi_{0^+},J=0,M=0\rangle$ state with a 943~nm laser. This results in an even distribution of population in an incoherent mixture of the four $|\pm,\Nsw\rangle$ states in $H$.\footnote{A glossary of symbols used throughout this paper is provided in section~\ref{sec:glossary}.} Figure~\ref{fig:ThOlevels} shows the electronic states of ThO relevant to the eEDM measurement.

In the absence of any experimental imperfections, we describe our system in terms of coordinate axes $+\hat{z}$ along $+\vec{\E}$ (for a specified sign of applied field that we denote as positive, pointing approximately east to west in the lab) and $+\hat{x}$ along the direction of the molecular beam (which travels approximately south to north) such that $+\hat{y}$ is approximately aligned with gravity (cf.\ figure~\ref{fig:meas_scheme_simple}). Note that when we reverse the direction of the electric field, by construction the laboratory coordinate system does not change and the orientation of the electric field can be described by $\Esw\equiv{\rm sgn}(\hat{z}\cdot\vec{\E})=\pm1$. Analogously, we reverse the direction of the magnetic field between two $\Bsw\equiv{\rm sgn}(\hat{z}\cdot\vec{\B})=\pm1$ states. Since the directions of the fields are encoded by $\Esw$ and $\Bsw$, we define the magnitudes of the fields simply as $\B_z\equiv|\B_z|$ and $\E\equiv|\vec{\E}|$.

A superposition of the $M=\pm1$ sublevels is prepared by optically pumping on the transition at 1090~nm between states $|\pm,\Nsw\rangle$ and $|C^1\Pi_1,J=1,M=0\rangle(|\Omega=+1\rangle-\Psw|\Omega=-1\rangle)/\sqrt{2}\equiv|C,\Psw\rangle$, where $\Psw=\pm1$ is the excited state parity\footnote{In this paper we follow the convention given in \cite{Brown2003}.}, with laser light linearly polarised in the $xy$ plane. The resulting state corresponds to having the total angular momentum of the molecule aligned in the $xy$ plane. Because the $\sigma$ electron's spin is aligned with $\vec{J}$, by the Wigner-Eckart theorem this is equivalent to aligning the spin \cite{Budker2008}, and we use this shorthand from here on. The state preparation laser frequency is tuned to spectroscopically select the molecule alignment $\Nsw$, while the nearly degenerate $M=\pm1$ states remain unresolved. The excited state $C$, which decays at a rate $\gamma_C\approx2\pi\times0.3$ MHz, decays primarily (${\approx}75~\%$ \cite{Hess2014}) to the ground state so that one superposition of the two $|\pm,\Nsw\rangle$ states is optically pumped out of $H$ and the remaining orthogonal superposition, which is `dark' to the preparation laser beam, is the prepared state. The linear polarisation of the state preparation laser beam, $\hat{\epsilon}_{\p}$, sets the relative coupling of each of the two $|\pm,\Nsw\rangle$ states to $\ket{C,\Psw}$ and determines the spin alignment angle of the remaining state in the laboratory frame. The bright superposition $\ket{B(\hat{\epsilon}_{\rm prep})}$
is pumped away, and the orthogonal dark superposition $\ket{D(\hat{\epsilon}_{\rm prep})}$ remains.

For the moment, we consider the specific case $\Psw=+1$ and $\hat{\epsilon}_{\p}=\hat{x}$, (the general case will be discussed in section \ref{sec:Measurement_scheme_more_detail}). In this case, the prepared state
\begin{equation}
\ket{\psi(t=0),\Nsw}=\frac{1}{\sqrt{2}}\left(\ket{+,\Nsw}-\ket{-,\Nsw}\right)
\label{eq:initial_state}
\end{equation}
has the electron spin aligned along the $\hat{y}$ axis. As the molecules traverse the spin precession region of length $L=22$ cm (which takes a time $\tau\approx1$~ms), the electric and magnetic fields exert torques on the electric and magnetic dipole moments, causing the spin to precess in the $xy$ plane by angle $2\phi$; this corresponds to the state
\begin{equation}
|\psi(t=\tau),\Nsw\rangle=\frac{1}{\sqrt{2}}\left(e^{-i\phi}|+,\Nsw\rangle-e^{+i\phi}|-,\Nsw\rangle\right),
\end{equation}
where $\phi$ is given approximately by the sum of the Zeeman and eEDM contributions to the spin precession angles,
\begin{equation}
\phi=-(\Bsw g_1\mu_{\rm B}\B_z+\Nsw\Esw d_e\Eeff)\tau.
\label{eq:simple_phase}
\end{equation}
The sign of the eEDM term, $\Nsw\Esw$, arises from the relative orientation between $\vec{\E}_{\rm eff}$ and the electron spin as illustrated in figure~\ref{fig:H-state}.

At the end of the spin precession region, we measure $\phi$ by optically pumping on the same $H\rightarrow C$ transition with the linearly polarised state readout laser beam. The polarisation alternates rapidly between two orthogonal linear polarisations $\hat{X}$ and $\hat{Y}$, such that each molecule is subject to excitation by both polarisations as it flies through the detection region, and we record the modulated fluorescence signals $F_\rX$ and $F_\rY$ from the decay of $C$ to the ground state at 690 nm. This procedure amounts to a projective measurement of the spin onto $\hat{X}$ and $\hat{Y}$, which are defined such that $\hat{X}$ is at an angle $\theta$ with respect to $\hat{x}$ in the $xy$ plane. To determine $\phi$ we compute the asymmetry,
\begin{equation}
\mathcal{A}\equiv\frac{F_\rX-F_\rY}{F_\rX+F_\rY}\propto\cos{[2(\phi-\theta)]}.
\label{eq:asymmetry}
\end{equation}
We set $\B_z$ and $\theta$ such that $\phi-\theta\approx(\pi/4)(2n+1)$ for integer $n$, so that the asymmetry is linearly proportional to small changes in $\phi$ and maximally sensitive to the eEDM. A simplified schematic of the experimental procedure just described is shown in figure~\ref{fig:meas_scheme_simple}.
\hspace*{\fill} \\

By repeating the measurement of $\phi$ after having reversed any one of the signs $\Nsw$, $\Esw$ or $\Bsw$, we may isolate the eEDM phase from the Zeeman phase. In practice, we repeat the phase measurement under all $2^3$ $(\Nsw,\Esw,\Bsw)$ experiment states to reduce the sensitivity of the eEDM measurement to other spurious phases, and we extract the phase $\phi^{\N\E}=-d_e\Eeff\tau=\phi_{\rm EDM}$. Here, we have introduced the notation $\phi^u$, discussed in detail in the next section, which we use throughout this document to refer to the component of $\phi$ that is odd under the set of switches listed in the superscript $u$, and implicitly even under those which are not listed (see section~\ref{sec:Measurement_scheme_more_detail} and equation~\ref{eq:general_parity} for a rigorous definition). A component which is even under all switches is considered to be `non-reversing' and is given an `nr' superscript.

\subsubsection{Measurement Scheme in Detail}
\label{sec:Measurement_scheme_more_detail}
\hspace*{\fill} \\
To fully describe the method by which we extracted $d_e$ from the data in section \ref{sec:data_analysis}, and to describe the systematic error models in section \ref{sec:systematics}, we must introduce some additional formalism to describe the spin precession measurement to generalize the simple case described in the previous section. 

We work in the regime in which the Stark shift in $H$ is approximately linear, $E_{\rm Stark}\approx-\Nsw D_1\E$, which holds when the Stark interaction energy is large compared to the $\Omega$-doublet energy splitting $\Delta_{\Omega,1}$ but small compared to the rotational energy scale, described by the $H$-state rotational constant $B_H\approx2\pi\times$~9.8 GHz, i.e. $\Delta_{\Omega,1}\ll D_1\E\ll B_H$. In this regime, the molecular alignment is approximately related to $\Omega$ by $\Nsw=\Esw M\Omega$; this relation is assumed throughout this document. This is a good approximation, but it is notable that due to the Stark interaction at first order in perturbation theory, each $|M,\Nsw\rangle$ state is a superposition of all four $|H,J,M,\Omega\rangle$ states with $J=1,2$ and $\Omega=\pm1$. This effect is discussed further in sections~\ref{sssec:correlated_laser_parameters} and \ref{sssec:laser_pointing_and_intensity}.

Let us consider the preparation of a spin-aligned state again. Starting from an incoherent mixture of the four $|\pm,\Nsw\rangle$ states, we perform optical pumping on the electric dipole transition between $\ket{\pm,\Nsw}$ and $\ket{C,\Psw}$, for a specific $\Nsw$, with laser light of polarisation $\hat{\epsilon}_{\p}$ that is nominally linear in the $xy$ plane. This step depletes the bright superposition state (see e.g. \cite{Bickman2009})
\begin{equation}
\ket{B(\hat{\epsilon}_{\p},\Nsw,\Psw)}=(\hat{\epsilon}_{+1}^{*}\cdot\hat{\epsilon}_{\p}^{*})\ket{+,\Nsw}-\Psw(\hat{\epsilon}_{-1}^{*}\cdot\hat{\epsilon}_{\p}^{*})\ket{-,\Nsw},
\label{eq:bright_state}
\end{equation}
where $\hat{\epsilon}_{\pm1}=\mp\left(\hat{x}\pm i\hat{y}\right)/\sqrt{2}$
are unit vectors for circular polarisation. The corresponding dark state (with which the laser does not interact) is the orthogonal superposition
\begin{equation}
\ket{D(\hat{\epsilon}_{\p},\Nsw,\Psw)}=(\hat{\epsilon}_{+1}^{*}\cdot\hat{\epsilon}_{\p})\ket{+,\Nsw}+\Psw(\hat{\epsilon}_{-1}^{*}\cdot\hat{\epsilon}_{\p})\ket{-,\Nsw}.
\label{eq:dark_state}
\end{equation}
This dark state serves as the initial state, $|\psi(0),\Nsw\rangle = |D(\hat{\epsilon}_{\p},\Nsw,\Psw=+1)\rangle$, for the spin-precession experiment, where we fixed the state preparation laser frequency to address the excited state with parity $\Psw=+1$. The state preparation laser polarisation can be parameterised as
\begin{equation}
\hat{\epsilon}_{\p}=-e^{-i\theta_{\p}}\cos\Theta_{\p}\hat{\epsilon}_{+1}+e^{+i\theta_{\p}}\sin\Theta_{\p}\hat{\epsilon}_{-1},
\label{eq:polarization_parametrization}
\end{equation}
where $\Theta_{\p}\approx\pi/4$ defines the ellipticity Stokes parameter $(S_3/I)_{\p}=\cos2\Theta_{\p}\approx0$, and $\theta_{\p}$ defines the linear polarisation angle with respect to $\hat{x}$ in the $xy$ plane. From here on, we refer to the ellipticity Stokes parameter as $S\equiv S_3/I$. There is a one-to-one correspondence between the dark state superposition and the projection of the laser polarisation $\hat{\epsilon}_{\p}$ onto the $xy$ plane. If the laser polarisation does not lie entirely in the $xy$ plane, equations \ref{eq:bright_state} and \ref{eq:dark_state} are still appropriate, but require normalization. Note that if the laser is linearly polarised, switching the excited state parity $\tilde{\mathcal{P}}$ has the same effect on the dark state as rotating the laser polarisation angle by $\pi/2$.

Following the initial state preparation, the molecules traverse the spin-precession region with their forward velocity nominally along $\hat{x}$. In this region there are nominally uniform and parallel electric ($\vec{\E}$) and magnetic ($\vec{\B}$) fields, which produce energy shifts given by
\begin{equation}
E(M,\Nsw) =-|M|D_1\E\Nsw-Mg_1\mu_{\rm{B}}\B_z\tilde{\B}-M\eta\mu_{\rm{B}}\E\B_z\Nsw\tilde{\B}-Md_e\Eeff\Nsw\tilde{\E},
\label{eq:Energy}
\vspace{10pt}
\end{equation}
where $D_1$ is the electric dipole moment of $\ket{H,J=1}$. Here $\eta=0.79(1)$~nm/V accounts for the $\E$-dependent magnetic moment difference between the two sets of $\Nsw$ levels in $\ket{H,J=1}$ \cite{Petrov2014}, as described in section~\ref{sec:compute_phase}. The energy shift terms that depend on the sign of $M$ contribute to the spin precession angle $\phi$, which is given by:
\begin{equation}
\phi=\frac{1}{2}\int_0^L(E(M=+1,\Nsw)-E(M=-1,\Nsw))\frac{{\rm d}x}{v}.
\label{eq:total_phase}
\end{equation}
This phase is dominated by the magnetic (Zeeman) interaction. The Stark shift, proportional to $|M|$, does not contribute. The state then evolves to: 
\begin{equation}
|\psi(\tau),\Nsw\rangle = \left(e^{-i\phi}|+,\Nsw\rangle \langle +,\Nsw| + e^{+i\phi}|-,\Nsw\rangle \langle -,\Nsw|\right)|\psi(0),\Nsw\rangle,
\end{equation}
(recall $\ket{\psi(0),\Nsw}=\ket{D(\hat{\epsilon}_{\p},\Nsw,\Psw=+1)}$ per equation~\ref{eq:dark_state}) and molecules enter a detection region where the state is read out by optically pumping again between the $\ket{H,J=1}$ and $\ket{C,J=1}$ manifolds. This optical pumping is performed alternately by two laser beams with nominally orthogonal linear polarisations 
$\hat{\epsilon}_X$ and $\hat{\epsilon}_Y$.\footnote{For convenience, the notation $\hat{\epsilon}_{\rX}$, $\hat{\epsilon}_{\rY}$ is used interchangeably with the previously used notation $\hat{\rX}$, $\hat{\rY}$.}
These beams excite the projection of $\ket{\psi(\tau),\Nsw}$ onto the bright states
\begin{equation}
\ket{B(\hat{\epsilon}_\rX,\Nsw,\Psw)}\quad{\rm and}\quad\ket{B(\hat{\epsilon}_\rY,\Nsw,\Psw)},
\end{equation}
(with the same $\Nsw$ that was addressed in the state preparation optical pumping step, but with an independent choice of $\Psw$) with probability $P_{X,Y}$ respectively. In the ideal case in which all laser polarisations are exactly linear, this probability is given by
\begin{equation}
\label{eq:xprojection}
P_{\rX,\rY}(\phi,\theta_{\p},\theta_{ \rX,\rY},\Nsw,\Psw)=\left|\braket{B(\hat{\epsilon}_{\rX,\rY},\Nsw,\Psw)|\psi(\tau),\Nsw}\right|^2=\left[1-\Psw\cos(2(\theta_{\p}-\theta_{\rX,\rY}+\phi))\right]/2,
\end{equation}
where $\theta_{X,Y}$ are the linear polarisation angles of the state readout beams, with respect to $\hat{x}$. The result is a signal that varies sinusoidally with the precession angle $\phi$. To measure these probabilities, we observe the associated modulated fluorescence signals, $F_{X,Y}=fN_0P_{X,Y}$, where $N_0$ is the number of molecules in the addressed $\Nsw$ level at the state readout region, and $f$ is the fraction of total fluorescence photons that are detected. 

To distinguish between molecule number fluctuations and phase variations, we normalize with respect to the former by rapidly switching the state readout laser between the two orthogonal polarisations, $\hat{\epsilon}_{X,Y}$, every 5~$\upmu$s. This is significantly quicker than fluctuations in the molecule number and is sufficiently quick that every molecule is interrogated by both polarisations (see section~\ref{sec:data_analysis} or \cite{Kirilov2013} for more details). We then form an asymmetry $\A$, which is immune to molecule number fluctuations, given by
\begin{equation}
\A=\frac{F_\rX-F_\rY}{F_\rX+F_\rY}=\Psw\cos[2(\phi-\theta)],
\label{eq:Asymmetry}
\end{equation}
where we have assumed that the readout polarisations are exactly orthogonal, given by $\theta_\rX=\theta_{\ro}$ and $\theta_\rY=\theta_{\ro}+\pi/2$, and where we have defined $\theta\equiv\theta_{\ro}-\theta_{\p}$.\footnote{Note that this reduces to equation~\ref{eq:asymmetry} for $\theta_{\rm prep}=0$ (i.e. $\hat{\epsilon}_{\rm prep}=\hat{x}$) and $\Psw=+1$.} In this equation and from now on unless otherwise noted, $\Psw$ refers to the excited state parity that is addressed by the state readout laser, not to be confused with the excited state parity addressed by the state preparation laser, which is kept fixed.

The value of $\mathcal{B}_z$ and the state preparation and readout laser beam polarisations are chosen so that $|\phi-\theta|\approx\pi/4$. This corresponds to the linear part of the asymmetry fringe in equation~(\ref{eq:Asymmetry}), where $\A$ is most sensitive to, and linearly proportional to, small changes in $\phi$ (cf.\ figure~\ref{fig:fringe}). A variety of effects including imperfect optical pumping, decay from $C$ back to $H$, elliptical laser polarisation and forward velocity dispersion, reduce the measurement sensitivity by a `contrast' factor
\begin{equation} 
\C\equiv-\frac{1}{2}\frac{\partial\A}{\partial\theta}\approx \frac{1}{2}\frac{\partial\A}{\partial\phi},
\label{eq:Contrast_Definition}
\end{equation}
with $|\C|\le1$. We measure this parameter by dithering $\theta=\theta^{\rm nr} + \Delta\theta\tilde{\theta}$ (where $\theta^{\rm nr}$ is the average or 'non-reversing' polarisation angle)
between states of $\tilde{\theta}=\pm 1$, with amplitude $\Delta\theta=0.05$~rad. 
We found that typically $|\C|\approx0.94$. 
We then extract the measured phase, $\Phi=\mathcal{A}/(2\mathcal{C})+q\pi/4$, by normalising the asymmetry measurements according to the measured contrast --- see section~\ref{sec:data_analysis} for more details on the data analysis methods used to evaluate this quantity. In the ideal case, the measured phase matches closely with the precession phase, $\Phi\approx\phi$. However, a variety effects that are investigated closely in section~\ref{sec:systematics} lead to slight deviations between these two quantities, which can contribute to systematic errors in the measurement. Unless explicitly specified, $\C$ is assumed to be an unsigned quantity from here on. In particular, when averaging over multiple states of the experiment, $|\C|$ is used.

To isolate the eEDM term from other components of the energy shift in equation~(\ref{eq:Energy}), the experiment is repeated under different conditions that are characterised by parameters whose sign is switched regularly during the experiment. The spin precession measurement is repeated for all $2^4$ experiment states defined by the four primary binary switch parameters: $\Nsw$, the molecular orientation relative to the applied electric field (changed every 0.5~s); $\Esw$, the direction of the applied electric field in the laboratory (2~s); $\tilde{\theta}$, the sign of the readout polarisation dither (10~s); and $\tilde{\mathcal{B}}$, the direction of the applied magnetic field in the laboratory (40~s). For each ($\Nsw,\Esw,\tilde{\mathcal{B}}$) state, the asymmetry $\mathcal{A}(\Nsw,\Esw,\tilde{\mathcal{B}})$, contrast $\mathcal{C}(\Nsw,\Esw,\tilde{\mathcal{B}})$, and measured phase $\Phi(\Nsw,\Esw,\tilde{\mathcal{B}})$ are determined as described earlier. The data taken under all $2^4=16$ experimental states derived from these four binary switches constitutes a `block' of data.

We can write the phase $\Phi(\Nsw,\Esw,\tilde{\mathcal{B}})$ in terms of components with particular parity with respect to the experimental switches:
\begin{align}
\Phi(\tilde{\mathcal{N}},\tilde{\mathcal{E}},\tilde{\mathcal{B}})=&\Phi^{\mathrm{nr}}+\Phi^{\mathcal{N}}\tilde{\mathcal{N}}+\Phi^{\mathcal{E}}\tilde{\mathcal{E}}+\Phi^{\mathcal{B}}\tilde{\mathcal{B}}\nonumber\\+&\Phi^{\mathcal{NE}}\tilde{\mathcal{N}}\tilde{\mathcal{E}}+\Phi^{\mathcal{NB}}\tilde{\mathcal{N}}\tilde{\mathcal{B}}+\Phi^{\mathcal{EB}}\tilde{\mathcal{E}}\tilde{\mathcal{B}}+\Phi^{\mathcal{NEB}}\tilde{\mathcal{N}}\tilde{\mathcal{E}}\tilde{\mathcal{B}}.
\label{eq:phase_parity}
\end{align}
We refer to these components as `switch-parity channels'. A channel is said to be odd with respect to some subset of switches (labelled as superscripts) if it changes sign when any of those switches is performed. Thus it will also change sign if an odd number of those switches is performed. It is implicitly even under all other switches. We use this general notation throughout this document to refer to correlations of various measured quantities and experimental parameters with experiment switches. To generalize, if we have $k$ binary experiment switches $(\tilde{\mathcal{S}}_{1},\tilde{\mathcal{S}}_{2},\dots,\tilde{\mathcal{S}}_{k})$ such that $\tilde{\mathcal{S}}_{i}=\pm1$, and we perform a measurement of the parameter $X(\tilde{\mathcal{S}}_{1},\tilde{\mathcal{S}}_{2},\dots,\tilde{\mathcal{S}}_{k})$ for a complete set of the $2^{k}$ switch states, then the component of $X$ that is odd under the product of switches $\left[\tilde{\mathcal{S}}_{a}\tilde{\mathcal{S}}_{b}\dots\right]$ is given by
\begin{equation}
X^{\mathcal{S}_{a}\mathcal{S}_{b}\dots}\equiv \frac{1}{2^{k}}\sum_{\tilde{\mathcal{S}}_{1}\dots\tilde{\mathcal{S}}_{k}=\pm1}
\left[\tilde{\mathcal{S}}_{a}\tilde{\mathcal{S}}_{b}\dots\right]X\left(\tilde{\mathcal{S}}_{1},\tilde{\mathcal{S}}_{2},\dots,
\tilde{\mathcal{S}}_{k}\right).
\label{eq:general_parity}
\end{equation}
The switch parity behavior of a given component is expressed in the superscript which lists the experimental switches with respect to which the component is odd. We order the switch labels in the superscripts such that the fastest switches are listed first and the slowest switches are listed last. Some components give particularly important physical quantities. Most notably, the eEDM precession phase is extracted from the $\tilde{\mathcal{N}}\tilde{\mathcal{E}}$-correlated component of the measured phase: that is, in the ideal case $\Phi^{\mathcal{NE}}=-d_{e}\mathcal{E}_{\mathrm{eff}}\tau$. Additionally, the Zeeman precession phase is nominally given by $\Phi^{\B}=-\mu_{\rm B}\g\mathcal{B}_z\tau$. Recall we label `non-reversing' components with an `nr' superscript. In a few cases, we drop the superscript parity because it is redundant. For example, we drop the superscript on the dominant components of the applied electric and magnetic fields, $\mathcal{E}\equiv\mathcal{E}^{\mathcal{E}}$ and $\mathcal{B}_{z}\equiv\mathcal{B}_{z}^{\mathcal{B}}$.

\begin{figure}[htbp]
\centering
\includegraphics[width=10cm]{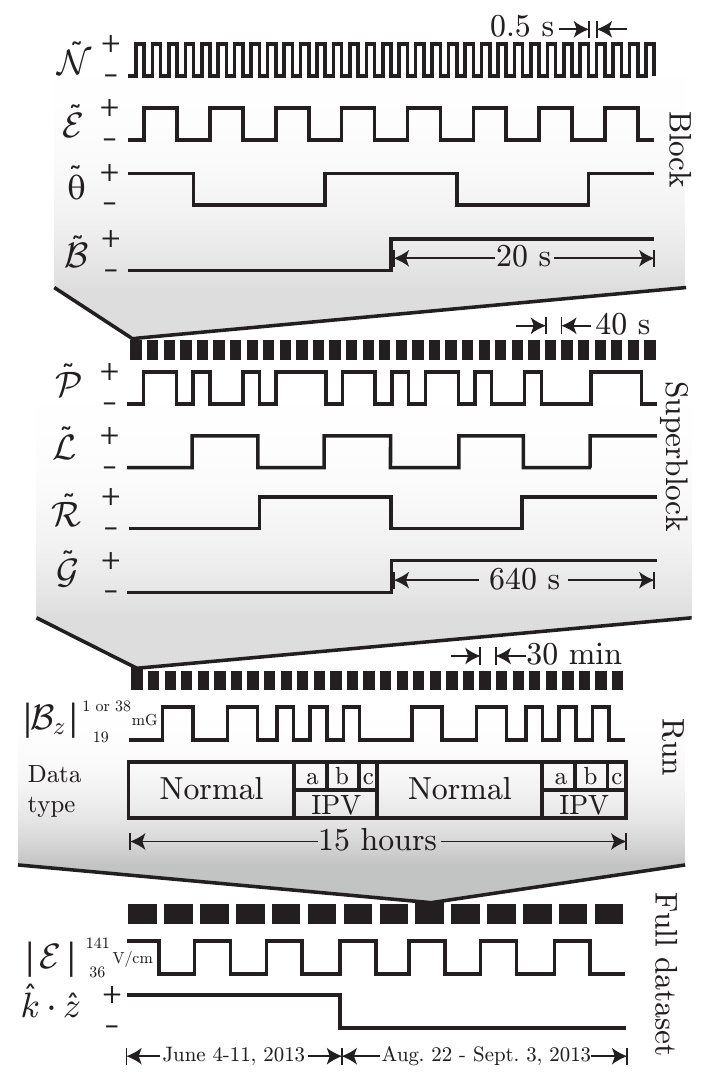}
\caption[Timescales of experimental parameter switches]{A schematic of the switches performed during our experiment and the associated timescales. See the main text for a description of each of the switch parameters and a description of the distinction between the \textit{Normal} and IPV (\emph{Intentional Parameter Variation}) data types. The 15-hour run time and $|\E|$ switching timescale are approximate.}
\label{fig:timing}
\end{figure}

Many other experimental parameters are also varied between blocks of data to suppress and monitor systematic errors (figure~\ref{fig:timing}). These `superblock' switches include: excited-state parity addressed by the state readout laser beams, $\Psw$ (chosen randomly after every block, with equal numbers of $\Psw=\pm1$); simultaneous change of the power supply polarity and interchange of leads connecting the electric field plates to their voltage supply, $\Lsw$ (4~blocks); a rotation of the state readout polarisation basis by $\theta_{\ro}\rightarrow\theta_{\ro}+\pi/2$ to interchange the roles of the $X$ and $Y$ beams, $\Rsw$ (8~blocks); and a global polarisation rotation of both state preparation and readout lasers by $\theta_{\ro}\rightarrow\theta_{\ro}+\pi/2$ and $\theta_{\p}\rightarrow\theta_{\p}+\pi/2$, $\Gsw$ (16~blocks).

Additionally, the magnitude of the magnetic field, $\B_z$, was switched on the timescale of 64--128 blocks (${\sim}1$~hour), and the magnitude of the applied electric field, $\E$, and the laser propagation direction, $\kz$, were changed on timescales of ${\sim}1$~day and ${\sim}1$~week, respectively. 

On these longer timescales, we also alternated between taking eEDM data under \textit{Normal} conditions, for which all experiment parameters were set to their nominally ideal values, and taking data with \textit{Intentional Parameter Variations} (IPVs), during which some experimental parameter was set to deviate from ideal so that we could monitor the size of the known systematic errors described in section \ref{sssec:correlated_laser_parameters}. We took IPV data in which we varied (a) the non-reversing electric field $\E^{\nr}$ and (b) the $\Nsw\Esw$-correlated Rabi frequency, $\rabi^{\N\E}$, to measure the sensitivity of the eEDM measurement to these parameters and we varied (c) the state preparation laser detuning $\Delta_{\p}$ to monitor the size of the residual $\E^{\nr}$. These systematic errors are discussed in more detail in sections~\ref{ssec:efields} and \ref{sssec:correlated_laser_parameters}.

The details of the data analysis required to extract the eEDM-correlated phase $\Phi^{\mathcal{NE}}$ are described in section \ref{sec:data_analysis}. A lower bound on the statistical uncertainty $\delta\Phi^{\mathcal{NE}}$ of the eEDM-correlated phase is given by photoelectron shot noise to be $\delta\Phi^{\mathcal{NE}}=1/(2|\C|\sqrt{N})$ for $N$ detected photoelectrons \cite{Khriplovich1997,VuthaThesis}. In the case where shot noise is the sole contribution, we can express the statistical uncertainty $\delta d_e$ in our measurement of the eEDM as
\begin{equation}
\delta\de=\delta\Phi^{\N\E}\frac{1}{\E_{\rm eff}\tau}=\frac{1}{2|\mathcal{C}|\tau\mathcal{E}_{\rm eff}\sqrt{\dot{N}T}},
\end{equation}
where $\dot{N}\approx f\dot{N_0}$ is the measurement rate (equivalent to the photoelectron detection rate) and $T$ is the integration time (recall $f$ is the fraction of fluorescence photons detected and $N_0$ is the number of molecules in the addressed $\Nsw$ level). Further discussion of the achieved statistical uncertainty is presented in section~\ref{sec:data_analysis}.

\subsection{Apparatus}
 \label{sec:apparatus}
\subsubsection{Overview}
\begin{figure}[!ht]
\centering
\includegraphics[scale=0.46]{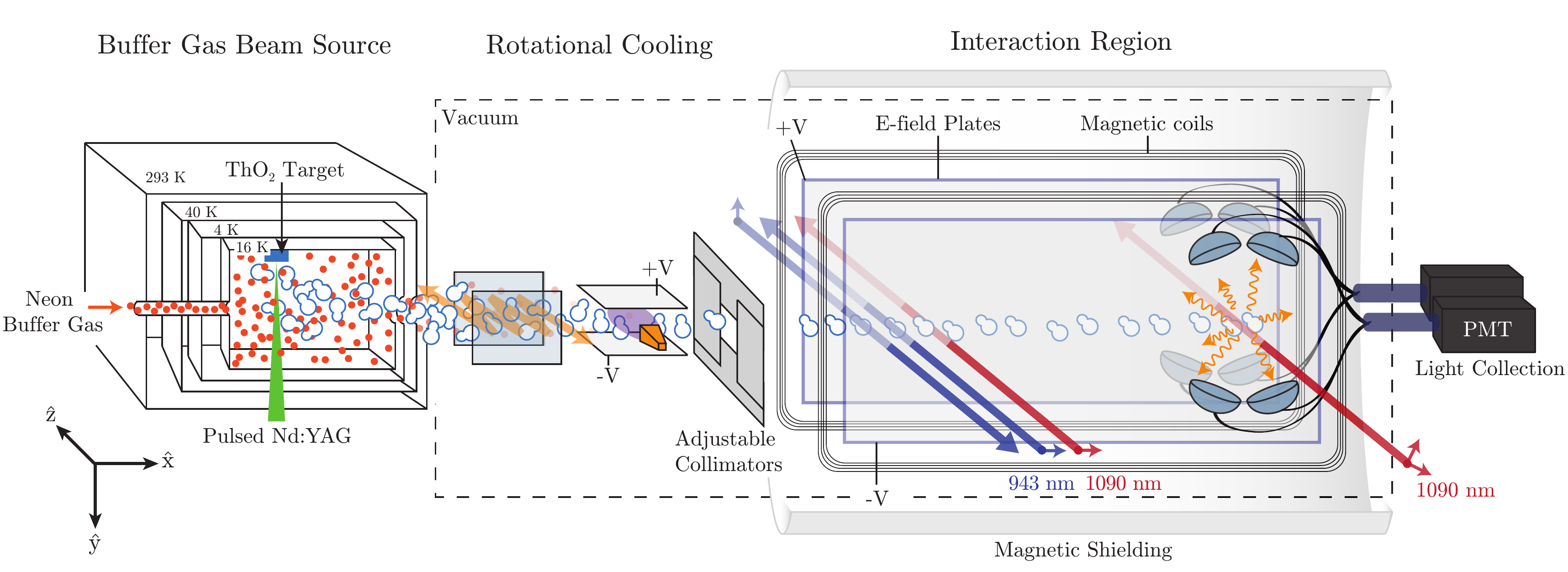}
\caption{A schematic of the overall ACME experimental apparatus. A beam of ThO molecules was produced by a cryogenic buffer-gas-cooled source. After exiting the source, the molecules were rotationally cooled via optical pumping and microwave mixing and then collimated before entering a magnetically shielded spin-precession region where nominally uniform magnetic and electric fields were applied. Using optical pumping, the molecules were transferred into the eEDM-sensitive $H$ state and then a spin superposition state was prepared. The spin precessed for a distance of ${\approx}22$~cm and was then read out via laser-induced fluorescence. The fluorescence photons were collected by lenses and passed out of the chamber for detection by photomultiplier tubes. See main text for further details.\label{fig:apparatus_overview}}
\end{figure}

\hspace*{\fill} \\
In this section we provide an overview of our experimental procedure and the important components of our apparatus. The reader should consult subsequent subsections for further details. A schematic of the experimental apparatus is shown in figure~\ref{fig:apparatus_overview}.

ThO molecules were produced via pulsed laser ablation of a ThO$_2$ ceramic target. This took place in a cryogenic neon buffer gas cell, held at a temperature of ${\approx}16$~K, at a repetition rate of 50~Hz. The resulting molecular beam was collimated and had a forward velocity $v_{\parallel}\approx200$~m/s. In the state readout region the molecular pulses had a temporal (spatial) length of around 2~ms (40~cm). The buffer gas beam source is described in detail in section~\ref{sec:beamsource}.

After leaving the buffer gas source, the molecules had a velocity distribution and rotational level populations consistent with a Maxwell-Boltzmann distribution at a temperature of ${\approx}4$~K. This was lower than the cell temperature due to expansion cooling, which enhanced the number of usable ThO molecules in the relevant rotational state. Further rotational cooling was provided via optical pumping and microwave mixing (see section~\ref{sec:rotcool}). The molecules then passed through adjustable horizontal and vertical collimators consisting of a double layer of razor blades affixed to linear translation vacuum feedthroughs. Under normal running conditions, these collimators were withdrawn so that they did not affect the profile of the molecule beam in the spin-precession region; however, they were used to modify the spatial profile of the molecule beam during systematic checks to investigate the effect of molecule beam position and pointing. Just before the field plates, 126~cm from the beam source, the molecules passed through a 1~cm square collimating aperture, which determined the beam profile in the spin-precession region and prevented particles in the beam from being deposited on the field plates.

As described in section~\ref{sec:Measurement_scheme}, a spin precession measurement was performed where the precession angle provided a measure of the interaction energy of an eEDM with the effective electric field, $\mathcal{E}_{\rm eff}$, in the molecule. A pair of transparent, ITO-coated glass plates provided an electric field that polarised and aligned the molecules. Laser beams passed through these plates to perform state preparation and readout. Around the vacuum chamber were coils that provided a uniform magnetic field in the $+\hat{z}$ direction, and five layers of magnetic shielding which shielded against environmental magnetic fields. The electric and magnetic fields are discussed in detail in sections~\ref{ssec:efields} and \ref{sec:bfields}. The fluorescence induced by the state readout laser beam was collected by a set of eight lenses and transferred out of the spin-precession region using fiber bundles and light pipes (see section \ref{sec:fluorescence_collection}), where it was detected by photo-multiplier tubes\footnote{Hamamatsu R8900U-20.}. 

\subsubsection{Buffer Gas Beam Source}
\begin{figure}[!ht]
\centering
\includegraphics[width=13.5cm]{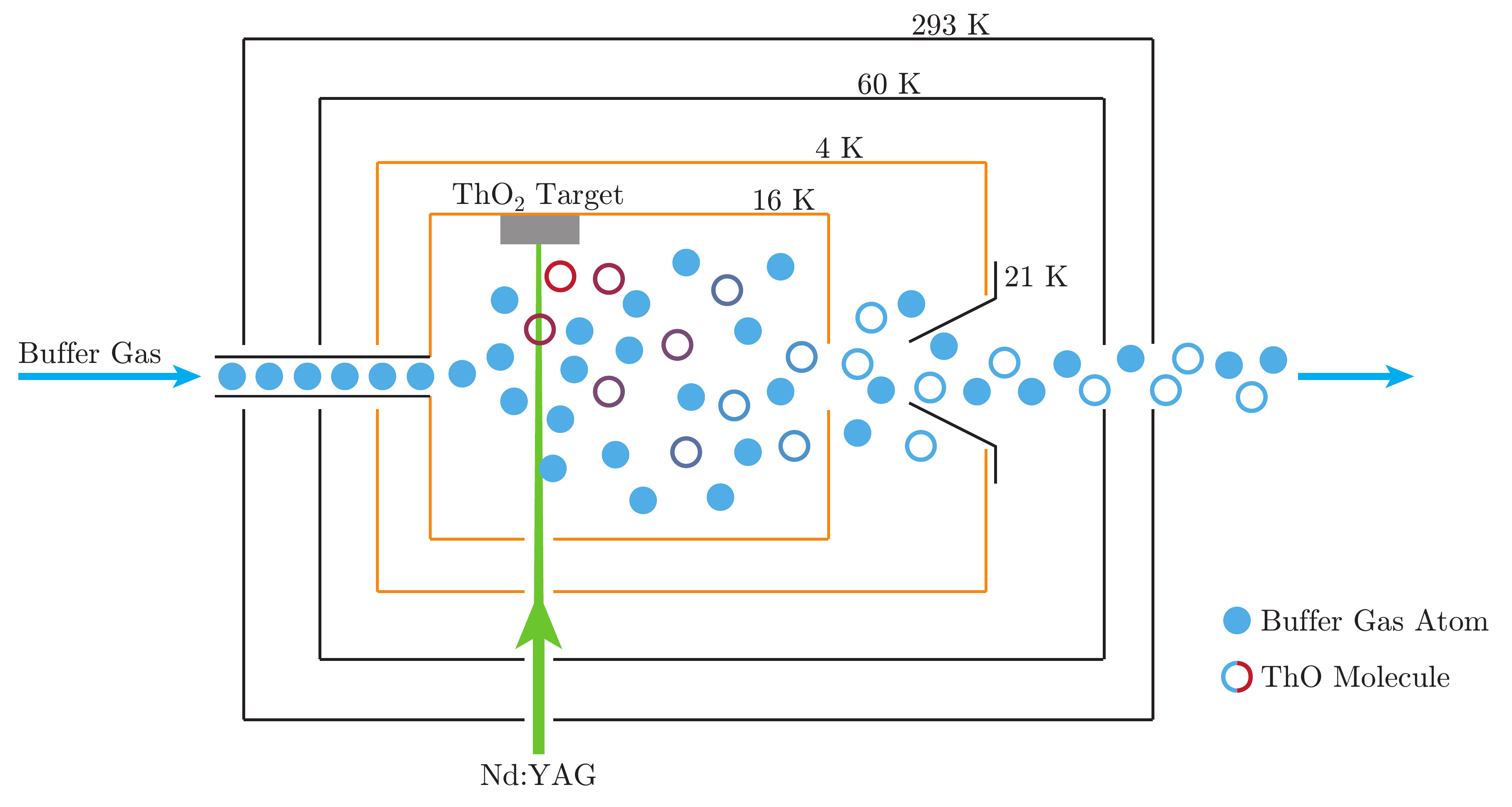}
\caption{A schematic of the buffer gas beam source. Neon buffer gas flowed into a cell at a temperature of 16~K where it served to thermalise the hot ThO molecules produced by laser ablation. The ThO was entrained in the buffer gas flow. The mixture exited the cell and its expansion cooled the ThO to $\approx4$~K. The resulting beam passed through collimating apertures in the 4~K and 50~K radiation shields and exited the beam source into the high vacuum region of the experiment. Solid circles represent buffer gas atoms. Open circles represent ThO molecules being cooled (red to blue transition).}
\label{fig:beam_source_schematic}
\end{figure}

\hspace*{\fill} \\
\label{sec:beamsource}
The basic operation of our beam source \cite{Maxwell2005,Petricka2007,Sushkov2008,Patterson2009,Campbell2009Review,Tu2009,Patterson2010,Hutzler2011,Barry2011,Lu2011,Skoff2011PRA,Skoff2011Thesis,Hutzler2012,HutzlerThesis,Hummon2013,Bulleid2013} is depicted in figure~\ref{fig:beam_source_schematic}.
Neon buffer gas was flowed at a rate of $\approx30$~SCCM (standard cubic centimetres per minute) through a copper cell held at a $T\approx16$~K. The inside of the cell was cylindrical with a diameter of 13~mm and a length of 75~mm. Within the cell ThO was introduced at high temperature via laser ablation: overlapped beams of light with wavelengths 532~nm and 1064~nm emitted by a pulsed Nd:YAG laser\footnote{Litron Nano TRL 80-200.} were focussed onto a 1.9~cm diameter ${\rm ThO}_2$ target fabricated from pressed and sintered powder \cite{Balakrishna1988,KiggansPrivate}. The laser pulses had a duration of a few ns, a pulse energy up to approximately 100~mJ and a repetition rate of $50$~Hz. The resulting hot plume of ejected particles, which contained ThO along with various other ablation byproducts, was cooled by collisions with the neon buffer gas, became entrained, and then exited the cell. The cell temperature was maintained by a combination of a pulse tube refrigerator\footnote{Cryomech PT415.} and a resistive heater.

The cell was surrounded by a 4~K copper shield that protected the cell from black-body radiation and cryopumped most of the neon emerging from the cell. This shield was also partially covered with activated charcoal that acted as a cryopump for residual helium in the neon buffer gas. We observed a background pressure of $10^{-7}$~Torr without any mechanical pumping of the beam source when cold and with no buffer gas flow. The 4~K shield had a stainless steel conical collimator with a circular aperture of diameter 6~mm, located 25~mm from the cell aperture, by which distance the expanding beam was sufficiently diffuse that intra-beam collisions were negligible and most trajectories were ballistic. This collimator thus functioned as a differential pumping aperture without affecting the beam's cooling, acceleration or expansion \cite{Hutzler2011}. The collimator had a thermal standoff relative to the 4~K shield to which it was mounted so that it could be kept at a temperature above the freezing point of neon by a resistive heater to prevent ice buildup on the collimator adversely affecting the beam dynamics. Another layer of shielding surrounded the 4~K copper shield, constructed from aluminium and held at a temperature of 60~K. Both the 4~K and 60~K radiation shields were thermally connected to the pulse tube by heat links made of flexible copper rope.

The aluminium vacuum chamber that housed the buffer gas beam source\footnote{Precision Cryogenic Systems Inc.} had windows on each side, providing optical access for both the ablation laser and spectroscopy lasers, the latter allowing characterisation and monitoring of beam properties. The ThO beam's forward velocity distribution was roughly Gaussian with mean $v_{\parallel}\approx200$~m/s and standard deviation $\sigma_{v_{\parallel}}\approx13$~m/s, corresponding to a temperature of ${\approx}5$~K. The rotational temperature was $T_{\rm rot}\approx4$~K (rotational constant $B_X\approx0.33$~cm$^{-1}$), meaning that ${\approx}90$\% of the population was contained in the levels $J=0$--$3$. Upon exiting the cell, the beam had a FWHM angular spread of $\approx45^{\circ}$. Several stages of collimation were applied before reaching the spin-precession region. The final collimator subtended a solid angle of $\approx6\times10^{-5}~{\rm sr}$, meaning 1 in ${\sim}20,000$ molecules exiting the cell reached the spin-precession region, where the precession measurement was performed (see figure~\ref{fig:apparatus_overview}).

ThO yields from a given ablation spot decreased significantly after ${\sim}10^4--10^5$ YAG pulses (${\sim}10$~mins), at which time the laser spot was moved to an un-depleted region via a motorised mirror to re-optimise the beam flux. Each target was found to provide acceptable levels of molecule flux for around 300~hours of continuous running (${\approx}5\times10^7$ shots) before requiring replacement.

\subsubsection{Rotational Cooling}
\hspace*{\fill} \\
\label{sec:rotcool}
We observed that ${\approx}2$~cm downstream of (further from) the buffer gas beam source cell aperture, $J$-changing collisions were `frozen out' \cite{Hutzler2011}, and the distribution of rotational state populations was fairly well described by a Boltzmann distribution with temperature $T_{\rm rot}\approx4$~K. 
At this temperature the resulting fractions of molecules in the $J=0$--3 levels were estimated to be 0.1, 0.3, 0.3 and 0.2 respectively.

\begin{figure}[!ht]
\centering
\includegraphics[scale=0.55]{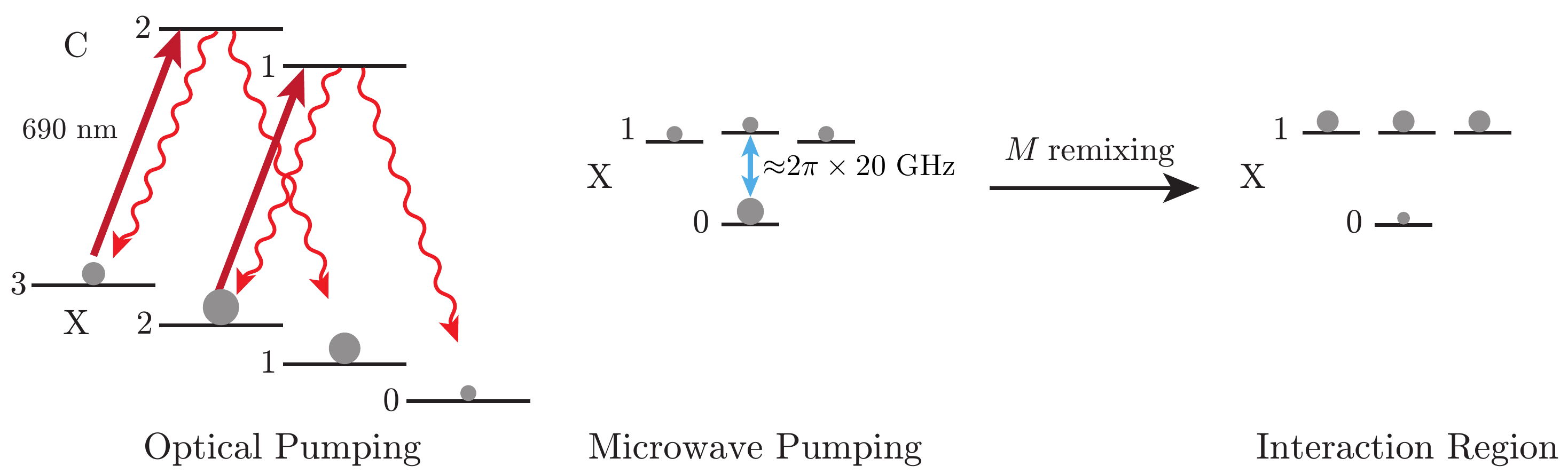}
\caption{Schematic of the rotational cooling process. Numbers label $J$ and $M_y$ (projection of total angular momentum along $y$) sublevels are unlabelled but are $-1$, 0, $+1$ from left to right. Population was first optically pumped out of the $J=2$ and $J=3$ levels ($C$-state $\Omega$-doublet structure and $M_y$ sublevels omitted for clarity) in a nominally field-free region. Next, population was equilibrated between $\ket{J=0}$ and $\ket{J=1,M_y=0}$ via microwave pumping. An electric field of ${\approx}40$~V/cm along $\hat{y}$ was empirically observed to lead to an increased population in $\ket{X,J=1}$. Grey dots represent population before these pumping processes. The schematic on the right represents the populations inside the spin-precession region (after pumping).\label{fig:rotcool}}
\end{figure}

As described in section~\ref{sec:state_prep_read}, we sought to transfer as much of the initial ground state population as possible into $\ket{H,J=1}$ via optical pumping. To enhance the population which was transferred, we accumulated population in a single rotational level of the ground state before state preparation. The scheme used to achieve this, which we refer to as rotational cooling, is illustrated schematically in figure~\ref{fig:rotcool} and discussed in detail in \cite{SpaunThesis}.
The first stage of the process was the optical pumping of molecules out of $\ket{X,J=2}$ ($\ket{X,J=3}$), via $\ket{C,J^{\prime}=1}$ ($\ket{C,J^{\prime}=2}$) into $\ket{X,J=0}$ ($\ket{X,J=1}$) using laser light at 690~nm. The natural linewidth of the $X\rightarrow C$ transition is ${\approx}2\pi\times0.3$~MHz, however the usable molecules had a ${\approx}\pm0.7$~m/s transverse velocity spread, corresponding to a $1\sigma$ Doppler width of ${\approx}2\pi\times1.5$~MHz at 690~nm. Because the lasers used had linewidths of ${\lesssim}1$~MHz, to completely optically pump these molecules we relied on a combination of power broadening and extended interaction time. Optical pumping occured in a magnetically unshielded region where a background field $\B\approx500$~mG was present; however, the magnetic moment of $X$ ($C$) is ${\sim}\mu_{\rm N}$ (${\approx}\mu_{\rm B}/J(J+1)$), the nuclear magneton, which led to a Zeeman shift of ${\sim}2\pi\times400$~Hz (${\lesssim}2\pi\times400$~kHz) such that the $M$ sublevels were not resolved by our lasers. The $\ket{C,J=1}$ state has an $\Omega$-doublet splitting of $\Delta_{\Omega,C,J=1}\approx2\pi\times51$~MHz \cite{Edvinsson1965}. This splitting scales as $\Delta_{\Omega,C,J}\propto J(J+1)$, meaning we could spectroscopically resolve the $\Omega$-doublets for all $\ket{C,J}$. In addition, having no $\E$-field present meant that the $M$ sublevels of $C$ and $X$ remained unresolved and the energy eigenstates remained parity eigenstates. The $X$ state is also insensitive to $\E$-fields due to the lack of $\Omega$-doublet substructure; opposite parity states are separated by ${\sim}10$~GHz and were hence unmixed. Laser beams with linear polarisation alternating between $\hat{x}$ and $\hat{y}$ were used to ensure that all population in $\ket{X,J=2,3}$ was addressed. This was achieved by directing around 10 passes of the beam, offset in $x$, through the vacuum chamber, passing through a quarter-wave plate twice in each pass, over a distance of around 2~cm. 

The laser light for rotational cooling was derived from home-built extended cavity diode lasers (ECDLs). The lasers were frequency-stabilised using a scanning transfer cavity with a computer-controlled servo \cite{YuliaThesis}. Frequency-doubled light at 1064~nm from a frequency-stabilised Nd:YAG laser, locked to a molecular iodine line via modulation transfer spectroscopy \cite{FarkasThesis}, provided the reference for the transfer cavity.

After this first stage of rotational cooling, there was significantly greater population in the $\ket{X,J=0}$ state than in any of the $\ket{X,J=1,M}$ sublevels. We obtained a ${\approx}25~\%$ increase in the $J=1$ population by applying a continuous microwave field, resonant with the $J=0\rightarrow J=1$ transition; a sufficiently high microwave power combined with the inherent velocity dispersion of the molecule beam led to an equilibration of population between the coupled levels \cite{SpaunThesis}. In this second stage of rotational cooling it was empirically observed that applying an electric field to lift the $M_y$ sublevel degeneracy was necessary to obtain the increased population in $\ket{X,J=1}$. A pair of copper electric field plates (spacing $\approx4$~cm) provided a field of ${\approx}40$~V/cm in the $\hat{y}$ (vertical) direction. We applied microwaves resonant with the Stark-shifted $\ket{J=0}\rightarrow\ket{J=1,M_y=0}$ transition at a frequency of $2\pi\times19.904521$~GHz from an \emph{ex vacuo} horn. Between the rotational cooling and spin-precession regions of the experiment (see figure~\ref{fig:apparatus_overview}) there was not a well-defined quantisation axis, and we observe that the populations of the $\ket{J=1,M}$ magnetic sublevels were equalised by the time the molecules reached the state preparation region. 

Overall, we find that rotational cooling provided a factor of between 1.5 and 2.0 increase in the molecule fluorescence signal $F$ in the state readout region. This gain factor was observed to vary slowly over time, possibly due to variations in the rotational temperature of the molecule beam, with significant changes sometimes observed when the ablation target was changed.

\subsubsection{State Preparation and Readout}
\hspace*{\fill} \\
\label{sec:state_prep_read}
Following rotational cooling, the molecular beam passed into the spin-precession region, where the molecules experienced a nominally uniform electric field, $\vec{\E}$, which was nominally collinear with a magnetic field, $\vec{\B}$. Note that since neither of the states $X^1\Sigma^+$ nor $A^3\Pi_{0+}$ have $\Omega$-doublet structure, parity remained a good quantum number for these levels for the small (${\sim}100$~V/cm) electric fields we applied. 

\begin{figure}[!ht]
\centering
\includegraphics[width=10cm]{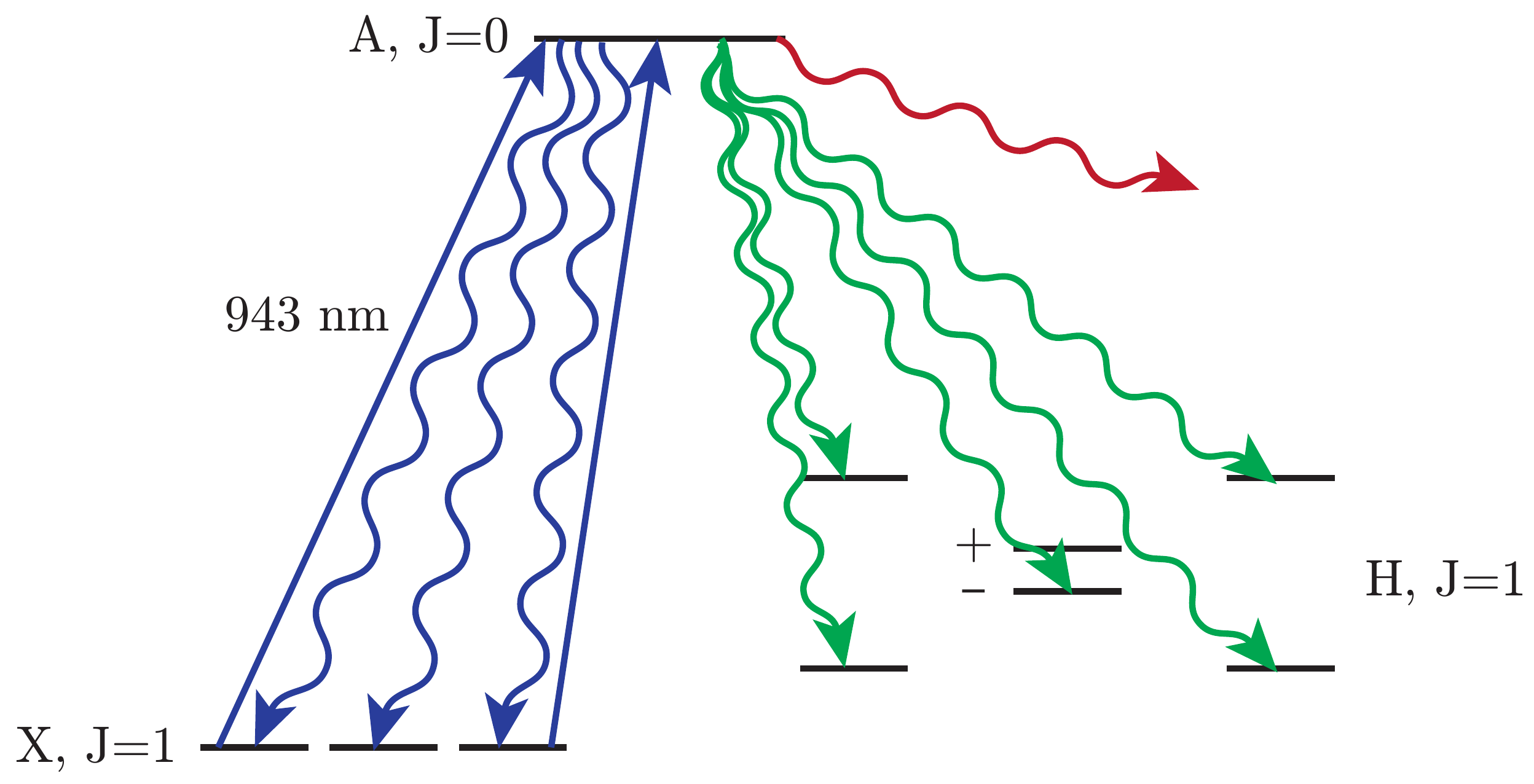}
\caption{Schematic of the optical pumping scheme used to populate the $H$ state. Spontaneous decay to the $H$ state (green arrows) led to an incoherent mixture of all indicated levels. See main text for detailed explanation.}
\label{fig:op_sublevels}
\end{figure}

We transferred the molecules into the $H$ electronic state via optical pumping, as illustrated in figure~\ref{fig:op_sublevels}.
A 943~nm laser beam nominally propagating along $\hat{z}$ excited molecules from the $\ket{X,J=1}$ to $\ket{A,J=0}$. The laser beam passed through a quarter-wave plate, was retroflected and offset in $x$, then passed again through the quarter-wave plate, such that the molecules were pumped by two spatially separated laser beams of orthogonal polarisations, allowing all population in both the $\ket{X,J=1,M=\pm1}$ levels to be excited. After excitation to $A$, the molecules could spontaneously decay into the $\ket{H,J=1}$ manifold of states. We observed a transfer efficiency from $X$ to $H$ of ${\approx}0.3$ \cite{SpaunThesis}. In this decay, five out of the six sublevels were populated; 1/6 of the population decayed to each of $\ket{H,M=\pm1,\Nsw=\pm 1}$ and 1/3 to $\ket{H,\Psw=-1,M=0}$ (see sections~\ref{sec:tho_molecule} and \ref{sec:Measurement_scheme} for definitions of $\Nsw$ and $\Psw$); decay to $\ket{H,\Psw=+1,M=0}$ is forbidden. 
Of these five populated states, only one corresponded to the desired initial state described by equation~\ref{eq:dark_state}, and only 1/6 of the population in the $H$ state was in this desired state.
We estimated a total transfer efficiency from $\ket{X,J=1,M=\pm 1}$ to the state in equation~\ref{eq:dark_state} of $30\%\times1/6=5\%$.

The 943~nm laser light was derived from a commercial ECDL and then amplified by a commercial tapered amplifier\footnote{Toptica DL Pro and BoosTA.}, generating $\approx400$~mW. As with the rotational cooling lasers, we verified that the power was sufficient to drive optical pumping to completion across the entire transverse velocity distribution of the molecular beam. This laser was also stabilised via the previously described (section~\ref{sec:rotcool}) transfer cavity. The frequency of the laser light was monitored every 30--60 mins by scanning across the molecular resonances, allowing for independent fine-tuning and compensation of long-term frequency changes (${\lesssim}2\pi\times100$~kHz per half hour) due to e.g. temperature drifts in the cavity.

\begin{figure}[ht]
\centering
\includegraphics[width=\linewidth]{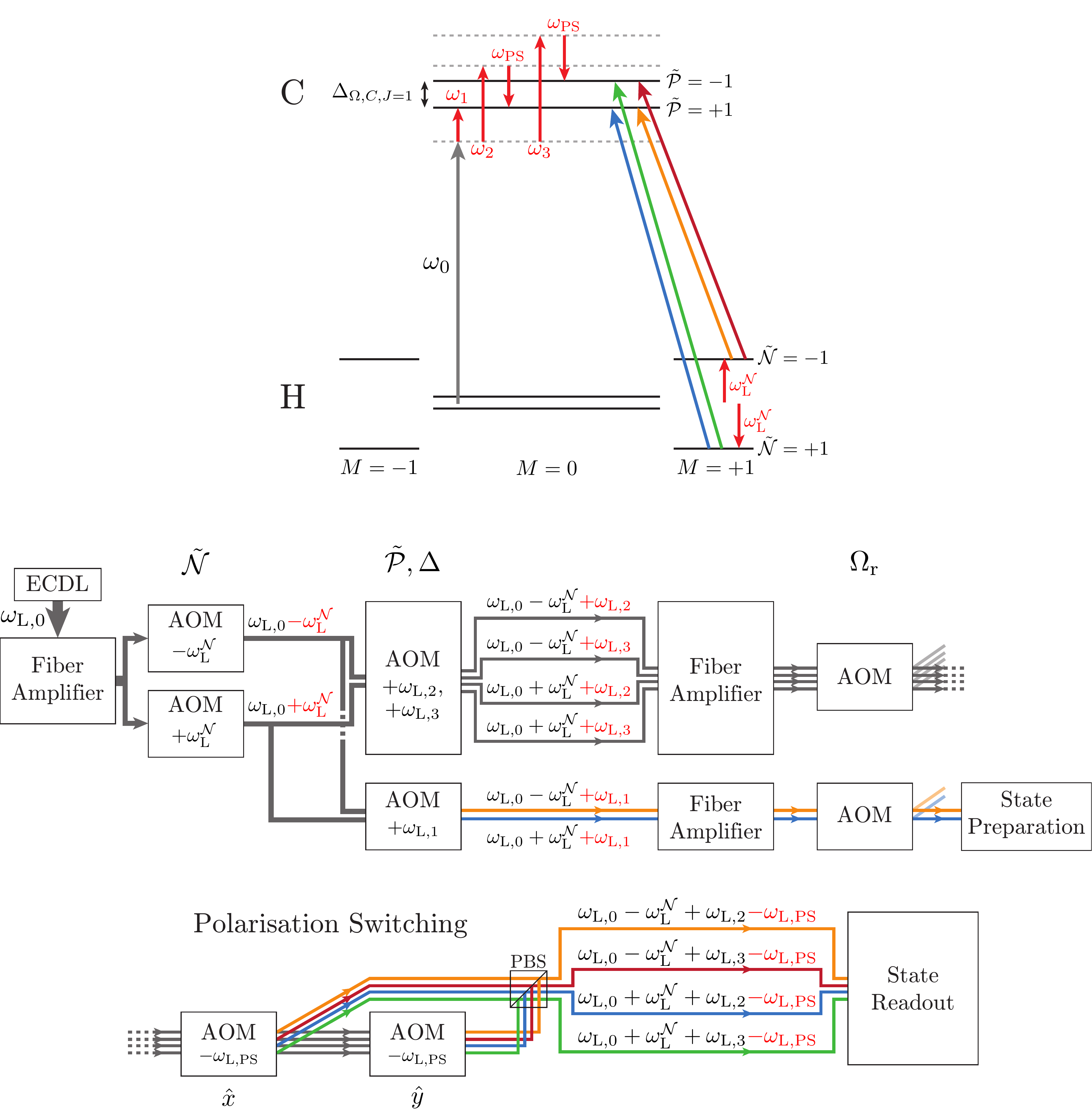}
\caption{Top: transitions addressed during state preparation and readout (not to scale). The grey arrow represents the ECDL output frequency, $\omega_0$, not resonant with any transition and referenced from halfway between the two $H$ state $\Omega$-doublets. Bottom: simplified schematic of how we produced light at the appropriate frequencies. AOM-induced frequency shifts are denoted in the corresponding boxes. Bifurcation of grey lines represents light being split equally. Multiple lines represent different frequencies; only one frequency is used at once. Dashed grey lines represent a continuation of the optical path. AOMs to perform switching between $\Nsw$ states; switching between $\Psw$ states and adding relative detuning $\Delta$; tuning Rabi frequency $\Omega_{\rm r}$; and performing polarisation switching are shown. The setup shown is used with $\E=142$~V/cm and changes slightly if a different value of $\E$ is used. For a full description, consult the main text.}
\label{fig:HC_transitions_setup}
\end{figure}

Around 1~cm downstream of the optical pumping laser beam that transferred population to $H$, we prepared the initial state of $H$ (equation~\ref{eq:dark_state}) by driving the transition between $\ket{H,M=\pm1,\Nsw}$ and $\ket{C,\Psw=+1}$ (see section~\ref{sec:Measurement_scheme} for more details) using laser light at 1090~nm. A distance $L=22$~cm downstream of the preparation laser, a second 1090~nm laser beam was used to read out the molecule state via the same transition (but with the option to excite to either $\Psw$ state). This laser light was also derived from a commercial ECDL. It was then amplified using a fiber amplifier\footnote{Keopsys KPS-BT2-YFA-1083-SLM-PM-05-FA.}, increasing the power to ${\approx}250$~mW. AOMs were then used to split and frequency shift the light to address both $\Nsw$ states in the $H$ state, allowing spectroscopic selection of molecular alignment, and of both $\Psw$ levels in the $C$ state. Switching between these frequencies was achieved with either RF switches\footnote{Mini-Circuits ZYSWA-2-50DR.} or a DDS synthesizer\footnote{Novatech 409B.}. Given the linear Stark shifts $D_1\E\approx2\pi\times146$~MHz ($2\pi\times37$~MHz) in $H$ with an applied electric field strength $|\E|=141$~V/cm (36~V/cm), and the excited state $\Omega$-doublet splitting $\Delta_{\Omega,C,J=1}\approx50$~MHz in $C$, these transitions were spectroscopically well-resolved. We fixed the nominal frequency of the state preparation laser to only address $\Psw=+1$, but periodically switched the state readout laser frequency to address $\Psw=\pm1$ (${\sim}1$~min period). The transition frequencies of the state preparation and state readout laser beams were changed synchronously to always address the same $\Nsw$ level, with a switch between $\Nsw$ levels every 0.5~s. The state preparation and readout laser beams were then independently amplified with a pair of fiber amplifiers\footnote{Nufern PSFA-1084-01-10W-1-3.}, providing ${\sim}3$--4~W of power. Immediately before interrogating the molecules, the polarisation of the state readout laser beam was rapidly (100~kHz) switched between two orthogonal linear polarisations. The scheme for producing the $\Nsw$ and $\Psw$ switches, and this fast polarisation switch, together with the corresponding laser transitions, is shown in figure~\ref{fig:HC_transitions_setup}. We now describe in detail how the appropriate frequency laser light was produced.
\clearpage

Light from the ECDL was amplified and split equally, passing to two AOMs which produced shifts $\pm\omega_{\rm L}^{\mathcal{N}}$ where $\omega_{\rm L}^{\mathcal{N}}$ is half the splitting between the two $\Nsw$ states; these AOMs were switched on and off to perform the $\Nsw$ switch. The two frequency-shifted beams were combined and overlapped. For state preparation (lower branch of diagram), another AOM shifted the light by $+\wLs{1}$, into resonance with the lower $\Omega$-doublet in $C$ ($\Psw=+1$). This light was then amplifed again and passed through an AOM to vary the power (used as a systematic check). For the state readout (upper branch of diagram), a single AOM switched frequency to produce shifts $+\wLs{2,3}$ for the two $\Psw$ states. A relative detuning between state preparation and readout laser beams (not shown) was also implemented with this AOM. (Shifts common to both beams were made by changing $\omega_0$.) The light was then amplified again and passed through an AOM to vary the power. Finally, polarisation switching was achieved with two AOMs switched on and off at 100~kHz, $\pi$ out of phase with each other; light not diffracted (and frequency shifted by $-\wLs{\rm PS}$) by the first AOM was diffracted (and also frequency shifted by $-\wLs{\rm PS}$) by the second AOM. The diffracted light from each path was combined on a polarising beam splitter such that the linear polarisation of the final output beam alternated. 

\begin{figure}[!ht]
\centering
\includegraphics[width=8cm]{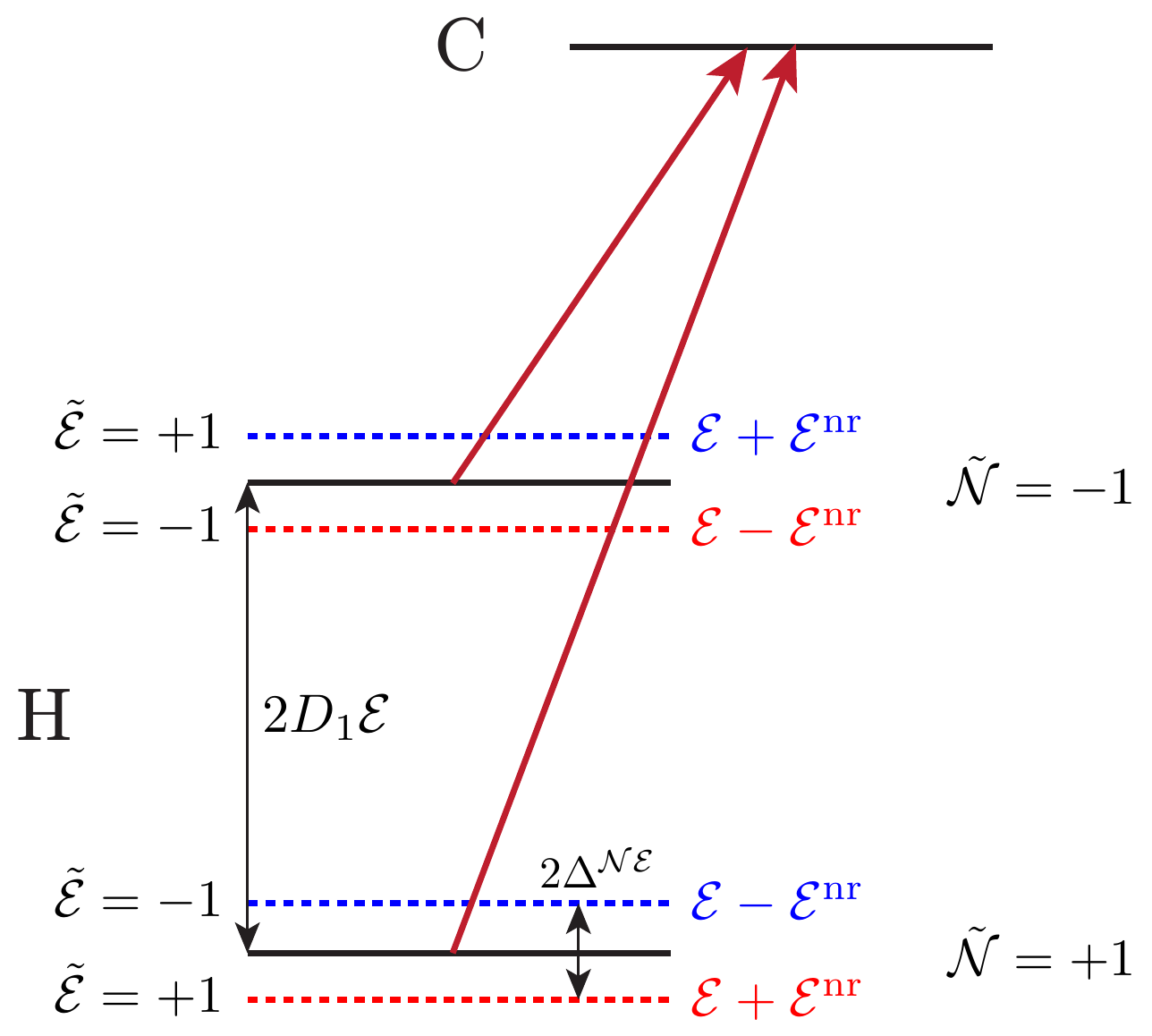}
\caption{Illustration of $\Delta^{\N\E}$ arising from a non-reversing electric field $\Enr$. Dashed lines show energy levels in the presence of $\Enr$. Colours indicate if the the laser shown in dark red is blue- or red-detuned from the transition.}
\label{fig:Enr_wNE}
\end{figure}

Based on the notation above we can now write the components of the frequencies of the state preparation and readout laser beams which do not reverse with any experimental switch as $\omega_{\rm L,prep}^{\nr}=\omega_{\rm L,0}+\omega_{\rm L,1}$ and $\omega_{\rm L,read}^{\nr}=\omega_{\rm L,0}+(\omega_{\rm L,2}+\omega_{\rm L,3})/2-\omega_{\rm L,PS}$, respectively. We can also write the $\Psw$-correlated frequency component of the state readout laser as $\omega_{\rm L,read}^{\P}=(\omega_{\rm L,2}-\omega_{\rm L,3})/2$. We then write the detuning components as $\Delta_i=\omega_{{\rm L},i}-\omega_{HC}$ where $i\in\left\{\mathrm{ prep},X,Y\right\}$ indexes the laser and $\omega_{HC}$ is the transition frequency between the line centres of the $\ket{H,J=1}$ and $\ket{C,J=1}$ manifolds\footnote{Note that this can in principle vary between different laser beams (denoted with the subscript $i$) if there is a relative pointing between them, which produces a relative Doppler shift, but we ignore this effect in our current treatment.}. We can rewrite this overall detuning in terms of various switch parity components:
\begin{align}
\Delta_{i}=&\omega_{{\rm L},i}-\omega_{HC,i}\\
=&\left(\omega_{{\rm L},i}^{\nr}+\Nsw\wL^{\N}+\Psw\wL^{\P}\delta_{i,\left\{\rX,\rY\right\} }\right)-\left(\omega_{HC}^{\nr}+\Nsw D_1\left|\E(x_{i})\tilde{\E}+\E^{\nr}(x_{i})\right|-\frac{1}{2}\Delta_{\Omega,C,J=1}\Psw\delta_{i,\left\{ \rX,\rY\right\} }\right)\\
=&\Delta_{i}^{\nr}+\Nsw\Delta_{i}^{\N}+\Nsw\Esw\Delta_{i}^{\N\E}+\Psw\Delta_{i}^{\P}\delta_{i,\left\{ \rX,\rY\right\}}.
\label{eq:detuningcorrelations}
\end{align}
In the above equations we have defined detuning components of given switch parities --- we shall now explain each component in turn. $\Delta_{i}^{\N}=(\wL^{\N}-D_1\E(x_{i}))$ is the mismatch between the Stark shift $D_1\E(x_{i})$ and the AOM frequency $\wL^{\N}$ used to switch between resonantly addressing the two $\Nsw$ states, where $x_i$ is the $x$ position of laser beam $i$. $\Delta_{i}^{\N\E}=D_1\E^{\nr}(x_{i})$ is a detuning component correlated like an eEDM signal which is due to a non-reversing component of the applied electric field. To understand this relation, consider figure~\ref{fig:Enr_wNE}. Recall that $\Delta_{\Omega,C,J=1}$ is the $\Omega$-doublet splitting of the $C$ state.
For a $\Enr\ne0$, $|\E|$, and hence the splitting between the $\Nsw$ levels in $H$, depends on $\Esw$. If the laser frequency for each $\Nsw$ is set assuming $\Enr=0$, a nonzero $\Enr$ leads to blue or red detuning from resonance, correlated with $\Esw$. Because the sign of the Stark shift is correlated with $\Nsw$, the resulting detuning is also correlated with $\Nsw$.

$\Delta_{i}^{\P}=\wL^{\P}+\Delta_{\Omega,C,J=1}/2$ is the mismatch between the excited state parity splitting and the AOM frequency, $\wL^{\P}=(\omega_{\rm L,3}-\omega_{\rm L,2})/2$, used to switch between the two states ($\delta_{i,\left\{ \rX,\rY\right\}}$ is the Kronecker delta, 1 if $i=X$ or $i=Y$, zero else). We observed that $\Delta^{\N}$ ($\Delta^{\P}$) was typically less than $2\pi\times20$~kHz ($2\pi\times50$~kHz). Although we could measure $\Delta^{\N}$ with ${\sim}2\pi\times1$~kHz precision, fluctuations in the Stark splitting, likely caused by thermally-induced fluctuations of the field plate spacing, limited our ability to zero out this correlated detuning.

We define $\Delta^{\nr}=(\Delta^{\nr}_{\p}+(1/2)(\Delta^{\nr}_{\rX}+\Delta^{\nr}_{\rY}))/2$ as the average non-reversing detuning of the state preparation and readout laser beams; its value typically fluctuated by ${\sim}2\pi\times0.1$~MHz over several hours. Every 30--60 minutes the value of $\Delta^{\rm{nr}}$ was scanned across the molecular resonance in the readout region using the $\Delta$-tuning AOM (see figure~\ref{fig:HC_transitions_setup}), as an auxiliary optimisation. $\Delta^{\rm{nr}}$ was set to the value where the fluorescence signal was maximum. This ensured that the average detuning of the state readout laser beams, $(\Delta_X^{\rm nr}+\Delta_Y^{\rm nr})/2$, was zero, however, if the state preparation and readout laser beams were not exactly parallel, there could be a difference between $\Delta^{\nr}_i$ due to the resulting difference in Doppler shifts. The effect of a detuning difference between the two state readout polarisations $\Delta^{XY}=(\Delta^{\nr}_{\rX}-\Delta^{\nr}_{\rY})/2$ is discussed in section~\ref{ssec:asymmetry_effects}. Additionally, each day we scanned the frequency of the preparation laser across the molecule resonance while monitoring the contrast of our fluorescence signal to ensure $\Delta^{\nr}_{\p}$ was kept below $2\pi\times0.2$~MHz (an example scan is shown in figure~\ref{fig:contrast}). The ways in which detuning components can contribute to systematic errors are discussed in detail in sections~\ref{sssec:AC_stark_shift_phases} and \ref{sssec:correlated_laser_parameters}.

Other polarisation switches of the state preparation and readout laser beams ($\Rsw$ and $\Gsw$) were controlled independently via half-wave plates mounted in high resolution rotation stages\footnote{Newport URS50BCC.}. These switches and their use in the experiment are described in detail in section~\ref{sec:data_analysis}. Both beams were shaped using cylindrical lenses to be extended in $y$ so all molecules in the beam were addressed. The Gaussian standard deviations of the beam intensities were 1.1~mm and 7.5~mm in the $x$ and $y$ directions, respectively \cite{SpaunThesis}. The preparation laser beam was temporally modulated at $50$~Hz with a chopper wheel, synchronous with the molecule beam pulses, to minimise the incident power on the field plates so as to reduce an important systematic error, described in sections~\ref{sssec:AC_stark_shift_phases} and \ref{sssec:polarization_gradients_from_thermal_stress_induced_birefringence}.

\subsubsection{Electric Field}
\hspace*{\fill} \\
\label{ssec:efields}
The applied $\E$-field was generated with a pair of 43~cm~$\times$~23~cm parallel conducting plates composed of ${\approx}1.25$~cm thick Borofloat glass, coated with a ${\sim}200$~nm layer of indium tin oxide on the inner faces\footnote{The plates were fabricated by Custom Scientific, Inc.}. The plates were transparent to the $X\rightarrow A$ optical pumping laser (943~nm), the $H\rightarrow C$ state preparation and readout lasers (1090~nm), and the $C\rightarrow X$ molecule fluorescence (690~nm). The outside faces of the electric field plates were prepared with a broadband anti-reflection coating with a specified \textless1\% reflectivity at normal incidence from 600--1000~nm. The plates were made much larger than the precession region in order to minimise inhomogeneity of the field through which the molecules passed, and to enable large solid angle collection of fluorescence through the plates. One of the field plates was mounted in an aluminium frame fixed to the base of the vacuum chamber. The other field plate was secured a distance of 2.5~cm away in a kinematic aluminium frame. On the inward-facing surfaces, a frame of gold-plated copper clamped each field plate to the aluminium mounts and also functioned as a `guard ring' electrode, suppressing the effect of fringing fields near the edges of the plate. The field plates were protected from impinging molecular beam particles by a $1~\mathrm{cm}\times 1~\mathrm{cm}$ square collimator fixed to the entrance of the assembly.

The applied electric field was controlled by a 20-bit DAC, amplified to produce up to $\pm200$~V\footnote{PA98A Power OpAmp.}. The field plate assembly was referenced to the vacuum chamber ground. Equal and opposite voltages, $\pm V$, were applied to each side of the assembly. The direction of the field (the $\Esw$ switch) was reversed every 1--2~s by reprogramming the output of the DAC channels to reverse their polarity. The configuration of the electrical connections between the amplified voltage and the field plates, denoted by $\Lsw$, was reversed via a pair of mercury-wetted relays every 2.6~minutes\footnote{Note that $\Lsw$ constitutes a reversal of the supply voltages as well as a reversal of the leads connecting the power supply to the field plates, such that $\Esw$ is unchanged.}. Data were also taken with two different values of $\mathcal{E}=36$ and 141~V/cm, varied on a ${\sim}1$~day time scale.

\begin{figure}[!ht]
\centering
\includegraphics[scale=0.4]{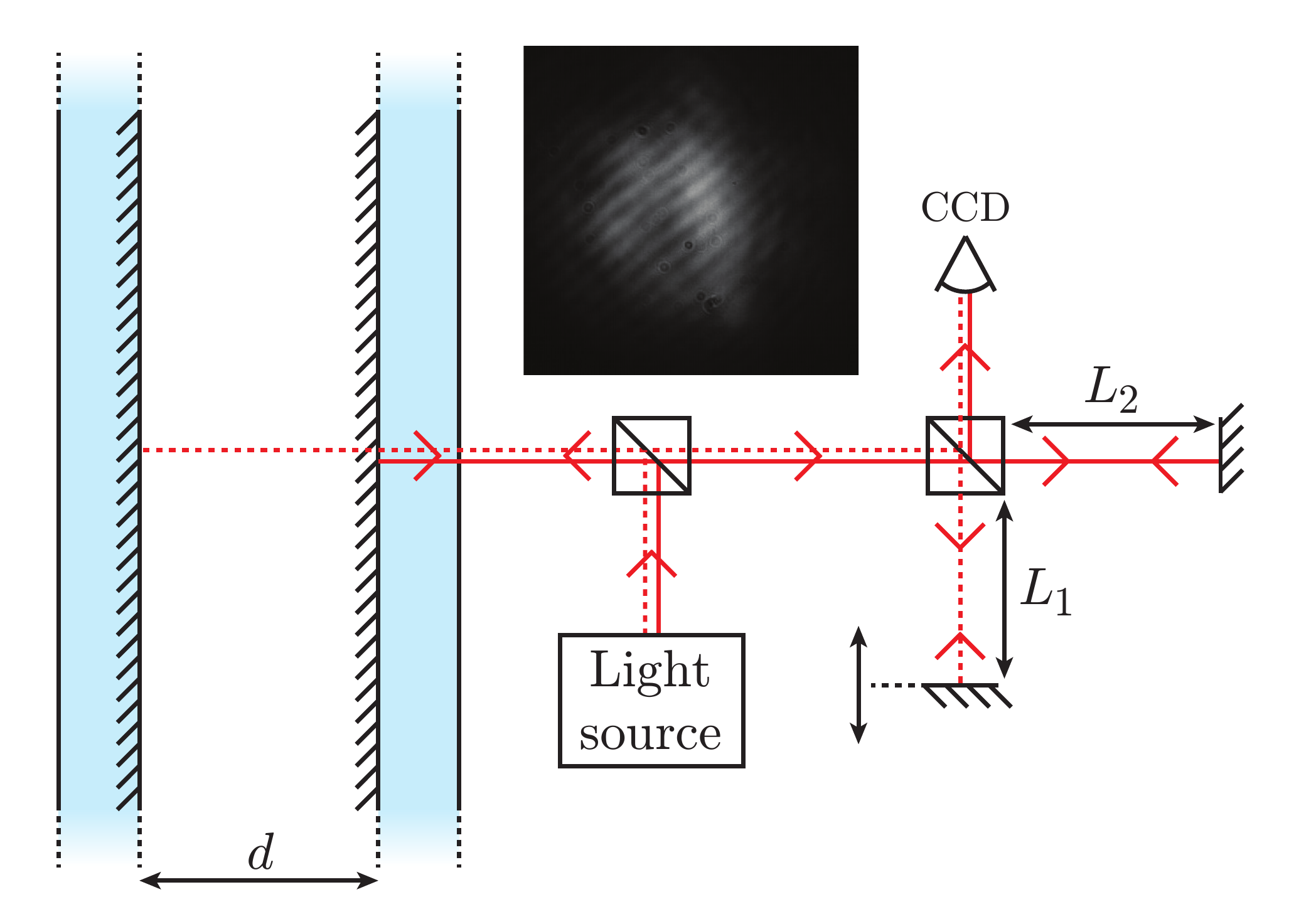}
\caption{\label{fig:Interferometer_setup}Schematic of the apparatus used to perform an interferometric measurement of the electric field plate separation. A spectrally broad light beam is reflected perpendicularly off the field plates and passes into a conventional Michelson interferometer setup with one fixed arm (length $L_2$) and one movable arm (length $L_1$). An example of a pair of beam paths of interest is shown as solid and dashed red lines. If the two paths are slightly tilted relative to each other, a spatial interference pattern (inset) is observed on the CCD detector when the path length difference between the two beams is less than the coherence length, e.g. $L_1+d-L_2<L_{\rm c}$.}
\end{figure}

\begin{figure}[!ht]
\centering
\includegraphics[width=14cm]{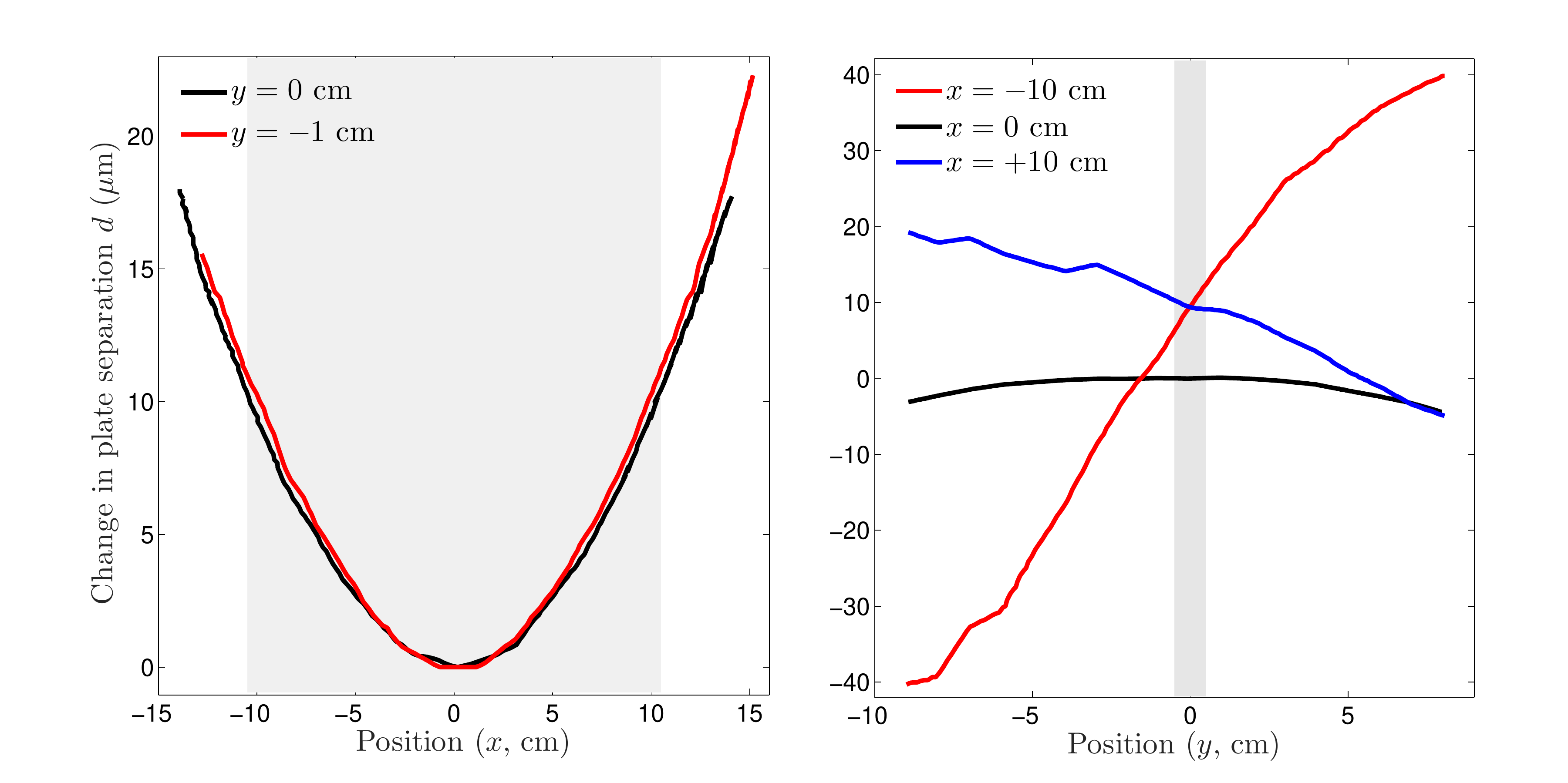}
\caption{Variation in the electric field plate separation as measured by the interferometric method. The left-hand plot shows the variation with $x$, the molecule beam direction, at two different values of $y$. The right-hand plot shows the variation with the $y$ (vertical) position at three different values of $x$. The coordinate origin is at the nominal centre of the plates. The shaded regions indicate the approximate extent of the molecular beam in the spin precession region. The change in separation is quoted relative to a common offset with an estimated error of $\pm0.5~\upmu$m. The mean separation over all $x$ is 25.00~mm.\label{fig:Interferometer_data}}
\end{figure}

We measured the homogoneity of the electric field in a number of ways which we shall describe in turn now. Firstly, an indirect measure was obtained by determining the spatial variation of the field plate separation $d$ using a `white light' Michelson interferometer \cite{Patten1971}. A schematic of the setup is shown in figure~\ref{fig:Interferometer_setup}.
We directed a light beam at normal incidence through the electric field plates. This resulted in multiple reflected beams, but we restrict discussion to the reflections from the conducting surfaces as these are of primary interest and were efficiently experimentally isolated from all others. The reflected beams passed into a Michelson interferometer with one arm of fixed length ($L_2$) and one with length adjustable via a micrometer ($L_1$). Constructive (destructive) interference occured whenever the lengths of two reflected beam paths differed by an integer (odd half-integer) multiple of the wavelength of the light. This condition was restricted further by the use of a broadband superluminescent diode\footnote{QPhotonics QSDM-680-2.} with a short coherence length $L_{\rm c}$ (nominally $L_{\rm c}\approx15~\upmu$m). Thus the interference was only substantial when the two beams differed in length by ${\lesssim}L_{\rm c}$. This occurred when $L_1=L_2$ (for reflections off the same surface) or when $L_1=L_2\pm d$ (for reflections off surfaces spaced by $d$). The case where both beams reflected off the same surface was used as a reference to determine the position $L_1=L_2$. A measure of this interference was achieved by producing a spatial interference pattern (inset figure~\ref{fig:Interferometer_setup}) through a slight tilting of the arms of the interferometer. Analysis of the spatial Fourier components of the resulting interference pattern provided a quantitative measure of the interference fringe contrast; a plot of contrast vs.\ arm position $L_1$ yielded a peak with width $\delta L_1\approx L_{\rm c}$. By performing this analysis while varying the path length $L_1$, the plate separation was deduced. This entire procedure was then performed over a range of transverse ($x,y$) positions on the field plates. The resulting data are shown in figure~\ref{fig:Interferometer_data}.

This measurement clearly showed a bowing of the electric field plates; the plate separation varied approximately quadratically with the position in $x$. This is shown in the left-hand plot of figure~\ref{fig:Interferometer_data}. In the $\hat{x}$ direction we observed a maximum variation in the plate separation of around 20~$\upmu$m. We saw a roughly 80~$\upmu$m variation in the $\hat{y}$ (vertical) direction but note that the collimated molecular beam extended only over $\pm5$~mm in $y$ so the biggest plate spacing variation at a given $x$ was ${\approx}10~\upmu$m. From these measurements and a typical applied voltage of $V=\pm177$~V, we expected $\E$ to vary by around 100~mV/cm in the $\hat{x}$ direction and ${\lesssim}15$~mV/cm in the $y$ direction in the region sampled by the molecules.

The indirect measurements of the spatial variation of the applied electric field provided by interferometric mapping of the field plate separation were later corroborated by direct measurements of $\vec{\E}(x)$. Spatial variation of $\vec{\E}$ could lead to the accumulation of geometric phases during the spin precession measurement \cite{Vutha2009}. There are known mechanisms by which such phases can contribute to eEDM-like systematic errors, as described in section~\ref{ssec:E_correlated_phase}, though simple estimates show that these effects are several orders of magnitude below the sensitivity of this measurement. However, additional $\E$-field imperfections such as non-reversing fields, due to e.g.\ variations in the ITO coating, which could produce patch potentials, are known to contribute to eEDM-like systematic errors and are only revealed by more direct measurements of the electric field, which we will now describe.

We can write the electric field present in the precession region in the following manner:
\begin{equation}
\vec{\E}\cdot\hat{z}=\E\Esw+\Enr+\E^{\Ld}{\Lsw}+\E^{\E\Ld}\tilde{\E}{\Lsw},
\end{equation}
where, as usual, $\Esw={\rm sgn}(\hat{z}\cdot\vec{\E})$ is the direction of the field in the spin-precession region and $\Lsw$ represents the binary state of the physical leads connecting the voltage supply to the field plates. The terms on the right-hand side are: $\E\Esw$, the intentionally applied electric field; $\Enr$, a non-reversing electric field; $\E^{\Ld}$, a non-reversing electric field component from the power supply that can be reversed by switching $\Lsw$; and $\E^{\E\Ld}\tilde{\E}{\Lsw}$, a component of the applied field that is reversed by switching $\Esw$ or $\Lsw$.

\begin{figure}[!ht]
\centering
\includegraphics[width=8cm]{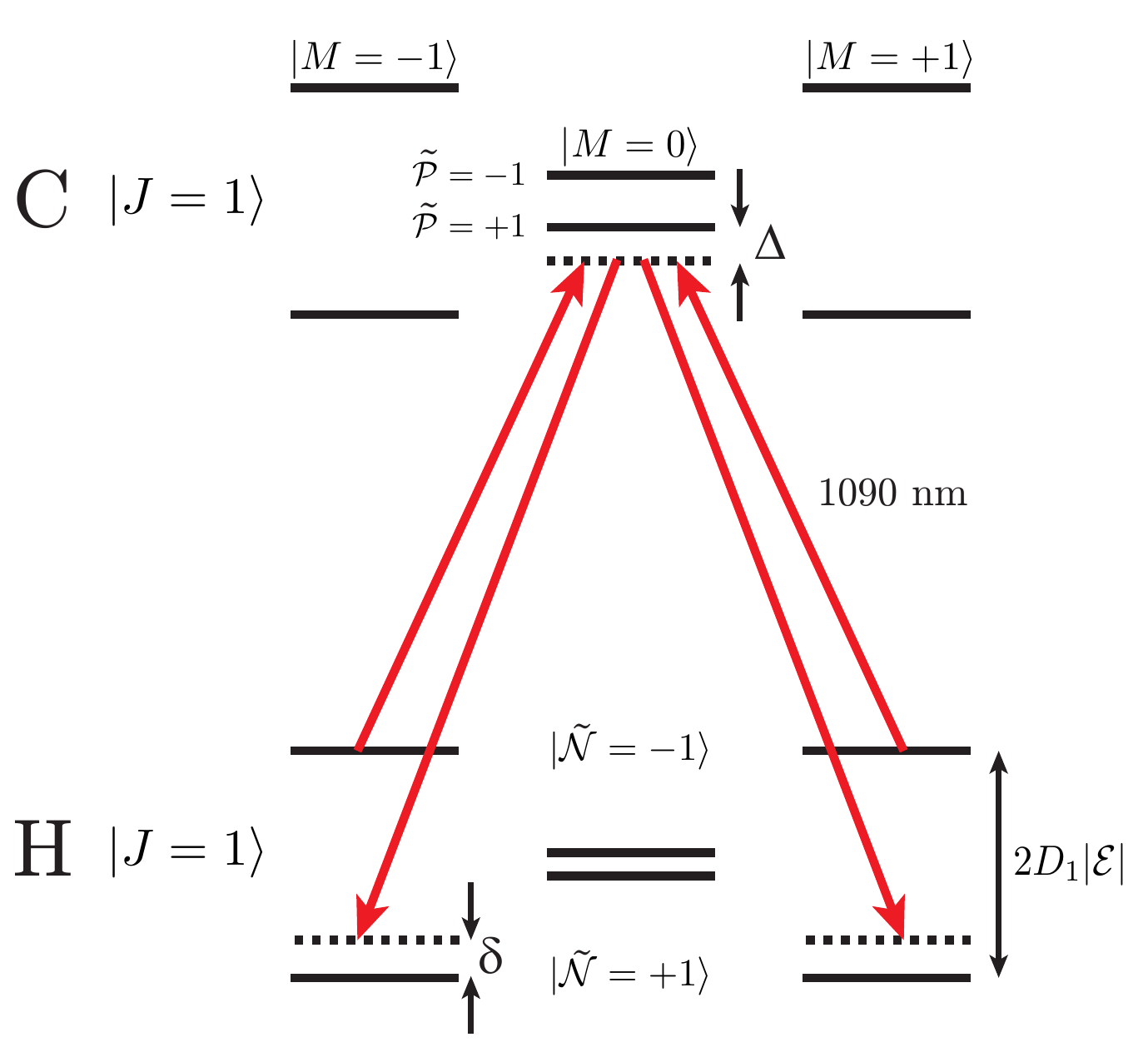}
\includegraphics[width=8cm]{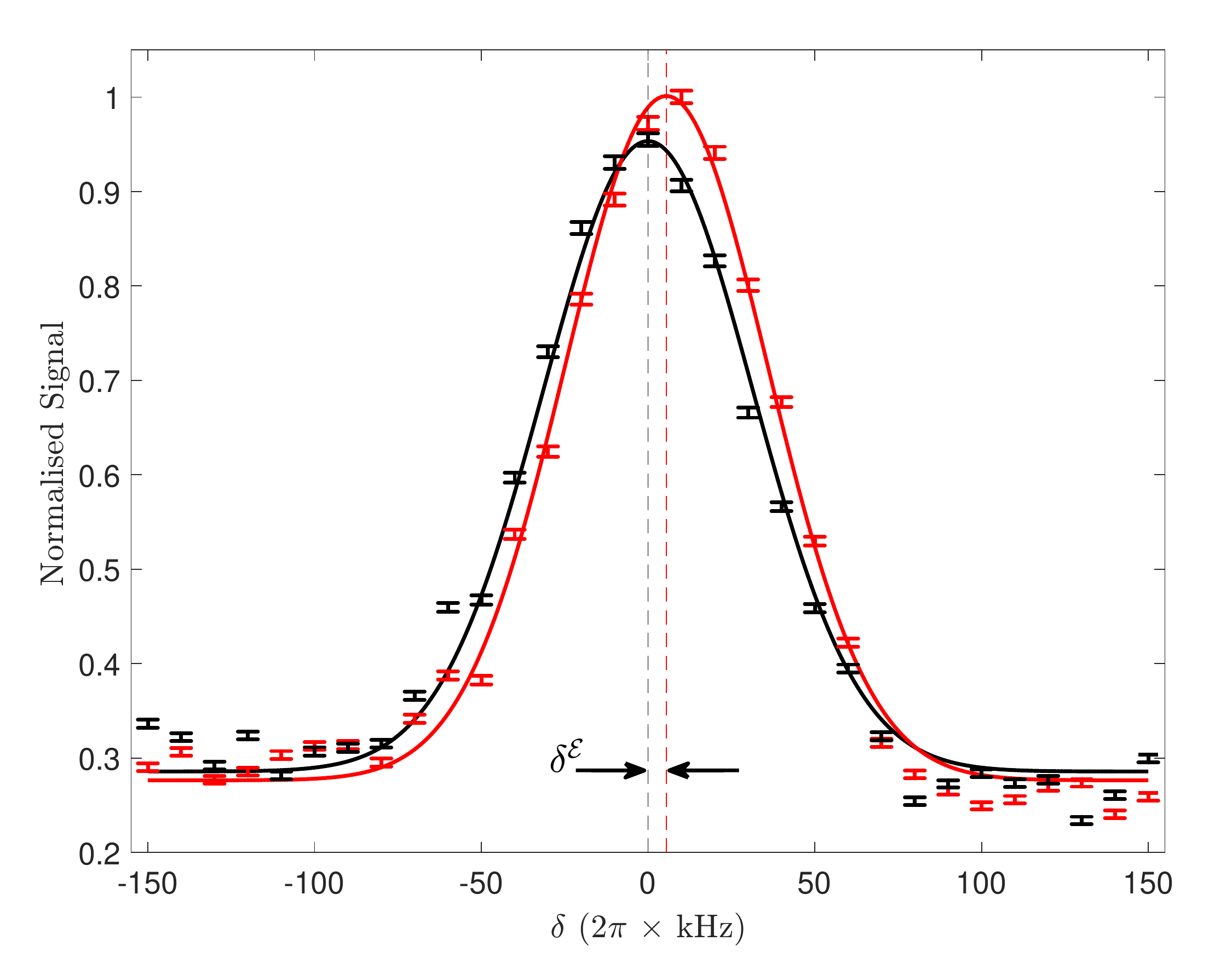}
\caption{Left: Schematic of the Raman-type transition used to perform a measurement of the $\E$-field in the spin-precession region. The pairs of red arrows represent the one-photon transitions driven by linearly polarised light, addressing superpositions of $M=\pm1$. The single-photon detuning is given by $\Delta+\delta/2$ and the two-photon detuning is given by $\delta/2$. $D_1|\mathcal{E}|$ is the magnitude of the Stark shift due to the applied electric field. Right: Example scans for opposite $\Esw$ states obtained by varying the two-photon detuning $\delta/2$ and observing fluorescence, with Gaussian fits to the data.\label{fig:Raman_transition}}
\end{figure}

We directly measured the components of $\E$ using the molecules themselves, in three different ways. The first method used Raman spectroscopy, driving a two-photon Lambda-type transition between $\Nsw$ levels in $\ket{H,J=1}$ as shown in figure~\ref{fig:Raman_transition}. The Raman transfer was performed at positions between, but close to, the state preparation and readout laser beams, where there was sufficient optical access. The procedure was as follows: first, an $\hat{x}$-polarised state preparation laser beam depleted a superposition $\ket{B(\hat{x},\Nsw=+1,\Psw=+1)}$ (recall $\ket{B}$ is the bright state as defined in section~\ref{sec:Measurement_scheme}) by exciting it to the $C$ state. Next, at a point downstream, two co-propagating, $\hat{x}$-polarised Raman beams were used to repopulate this depleted superposition by driving population from the other $\Nsw$ state, via the transition $\ket{B(\hat{x},\Nsw=-1,\Psw=+1)}\rightarrow \ket{C,\Psw=+1}\rightarrow\ket{B(\hat{x},\Nsw=+1,\Psw=+1)}$. The frequencies of the two Raman beams were tuned with a pair of AOMs. The state readout laser then addressed the same transition as the preparation laser and excited the repopulated superposition to the $C$-state from which it spontaneously decayed back to $X$ and fluoresced at 690~nm.

Efficient transfer of population between the two $\Nsw$ states occurred for zero two-photon detuning ($\delta/2$ in figure~\ref{fig:Raman_transition}). This condition was indicated by a peak in fluorescence, giving a measure of the Stark shifted energy, and hence the absolute size of the applied field, $|\E|$. This procedure was repeated for different positions of the Raman laser beams along the $\hat{x}$ direction. The non-reversing component of the electric field was found by repeating the measurement after reversing the applied voltages. An example of such a pair of scans is shown on the right of figure~\ref{fig:Raman_transition}.

Using this method we measured the electric field at $x$ positions where there was sufficient optical access, i.e.\ near the state preparation and readout laser beams. The $\Esw$-correlated two-photon detuning $\delta^{\E}=2\pi\times13$~kHz ($2\pi\times11$~kHz) allowed us to extract a value of the non-reversing electric field component, $\Enr=\delta^{\E}/2D_1=-6.5\pm0.3$~mV/cm ($-5.5\pm0.3$~mV/cm), in the state preparation (readout) region. We did not observe any significant variation within the individual regions. We also observed that this non-reversing component did not vary with the size of the reversing electric field.

\begin{figure}[!ht]
\centering
\includegraphics[scale=0.5]{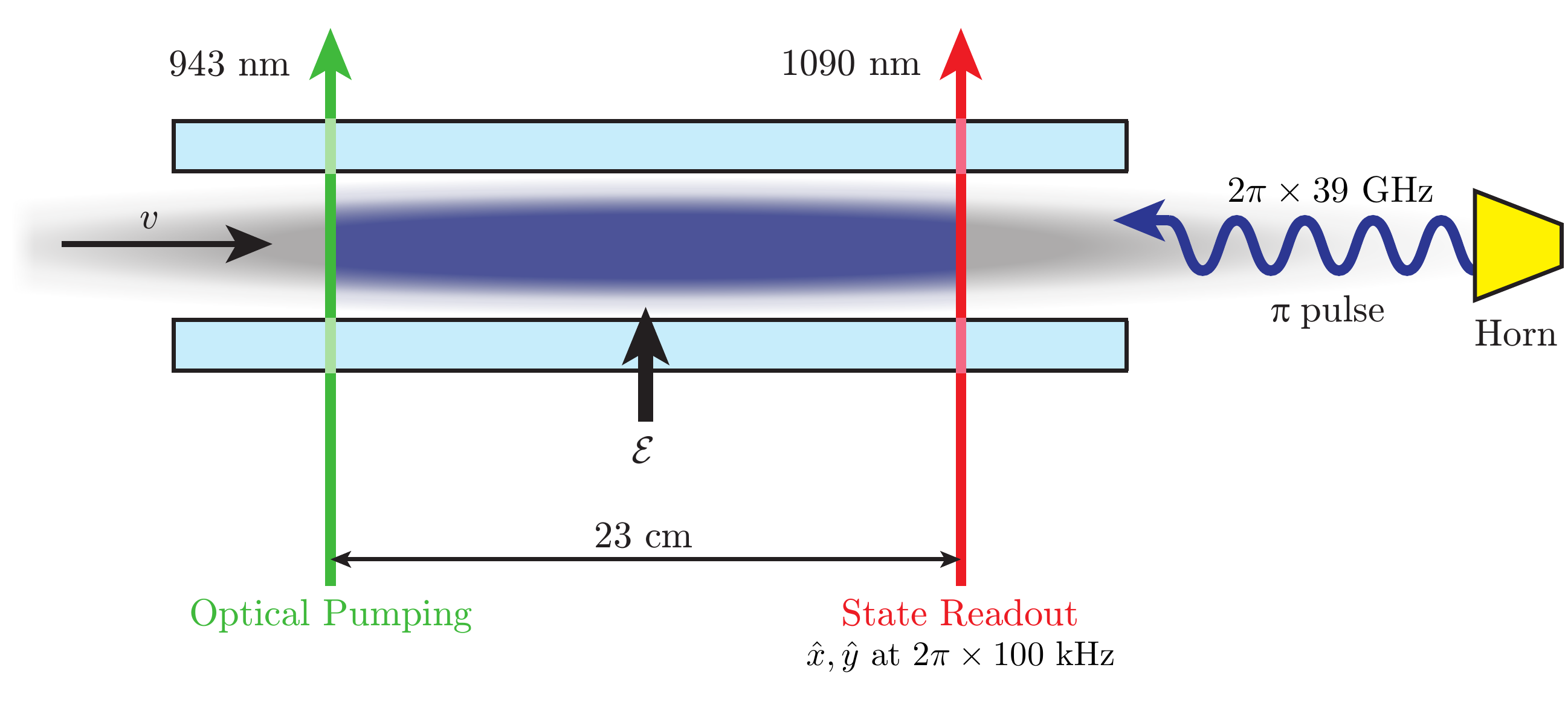}
\caption{Experimental setup for spatial measurement of $\E$ via microwave spectroscopy. A molecular pulse (grey cloud) passed between the electric field plates (light blue). The optical pumping laser beam transferred population from $\ket{X,J=1}$ to an incoherent mixture of states in $\ket{H,J=1}$ as described in section~\ref{sec:state_prep_read}. When the pulse was centred in the spin-precession region, a microwave $\pi$-pulse was applied, driving population in $H$ from $J=1$ to $J=2$ when resonant (dark blue region). The depletion efficiency out of $J=1$ was subsequently read out by laser induced fluorescence as per the normal measurement scheme described in section~\ref{sec:Measurement_scheme}. The time of arrival of the molecules in the state readout region encoded the position where they absorbed the microwaves.
\label{fig:microwave_setup}}
\end{figure}

The second method used to measure the electric field had the greatest utility because it allowed for spatially resolved measurements along $x$ in the spin precession region with comparable precision to the Raman method without perturbing the experimental apparatus. This was achieved via microwave spectroscopy. A schematic of the experimental setup is shown in figure~\ref{fig:microwave_setup}. 
\begin{figure}[!ht]
\centering
\includegraphics[scale=0.65]{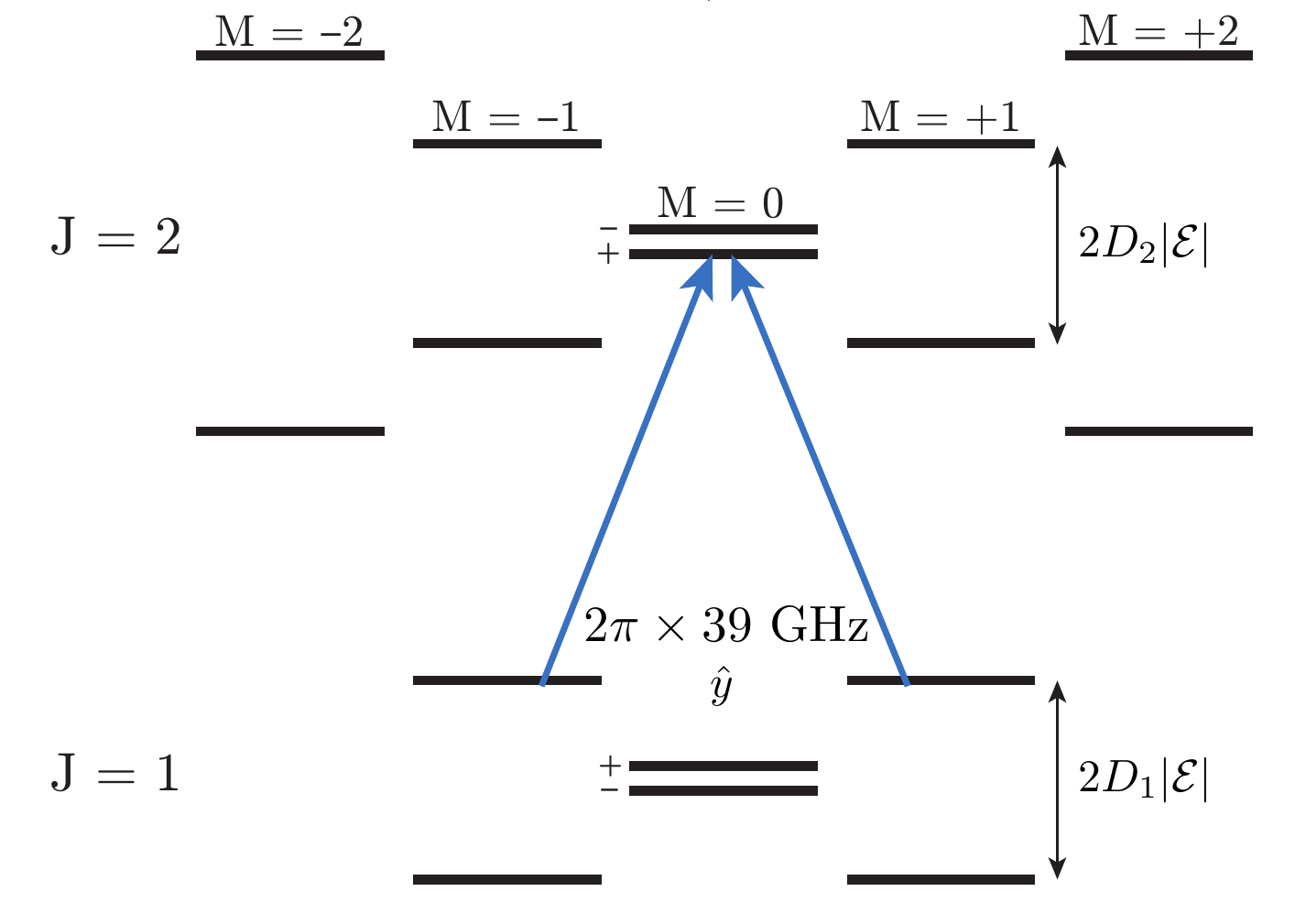}
\caption{The transition driven by microwaves during a measurement of the electric field. We used $\hat{y}$-polarised microwaves of frequency $2\pi\times39$~GHz to drive a rotational transition between $\ket{H,J=1}$ and $\ket{H,J=2}$. The $M=0$ levels are labelled with their parity. We applied a moderate $\E$-field such that $\Delta_{\Omega}\ll D|\E|\ll B_H$ where $B_H=0.33~{\rm cm}^{-1}$ is the rotational constant. The electric dipole moment of the $J=1$ state $D_1\approx2\pi\times1~{\rm MHz/(V/cm)}\approx 3D_2$.\label{fig:microwave_transitions}}
\end{figure}

The measurement procedure began with optical pumping of molecules into the $H$-state. The molecules travelled through the spin-precession region until it was entirely occupied by the molecule pulse. At this time, a $\pi$-pulse of microwaves at $2\pi\times39$~GHz with nominal $\hat{y}$ polarisation was applied counter-propagating to the molecule beam. When on resonance, this transferred population from $\ket{B(\hat{y},\Nsw,\Psw)}$ to $\ket{H,J=2,M=0,\Psw}$ (excitation to (from) either $\Psw$ ($\Nsw$) state was permitted) as shown in figure~\ref{fig:microwave_transitions}. State readout was performed as usual (see section~\ref{sec:Measurement_scheme}) by optically pumping with alternating polarisations $\hat{x}$ and $\hat{y}$. The measured asymmetry (as defined in equation~\ref{eq:asymmetry}) served as a measure of the microwave transfer efficiency. The $x$ position of the molecules at the time of the microwave pulse was mapped onto their arrival time in the detection region and, with knowledge of the longitudinal molecular beam velocity, $v_{\parallel}$, could be extracted. Thus, the spatial dependence of the resonant frequency, $\omega_{\rm MW}(x)$, was provided by the time-dependence of the asymmetry, $\mathcal{A}(t)$. Due to the DC Stark shift, $\omega_{\rm MW}$ was linearly proportional to the electric field magnitude and $|\E(x)|$ could be directly extracted.

\begin{figure}[!ht]
\centering
\includegraphics[width=15cm]{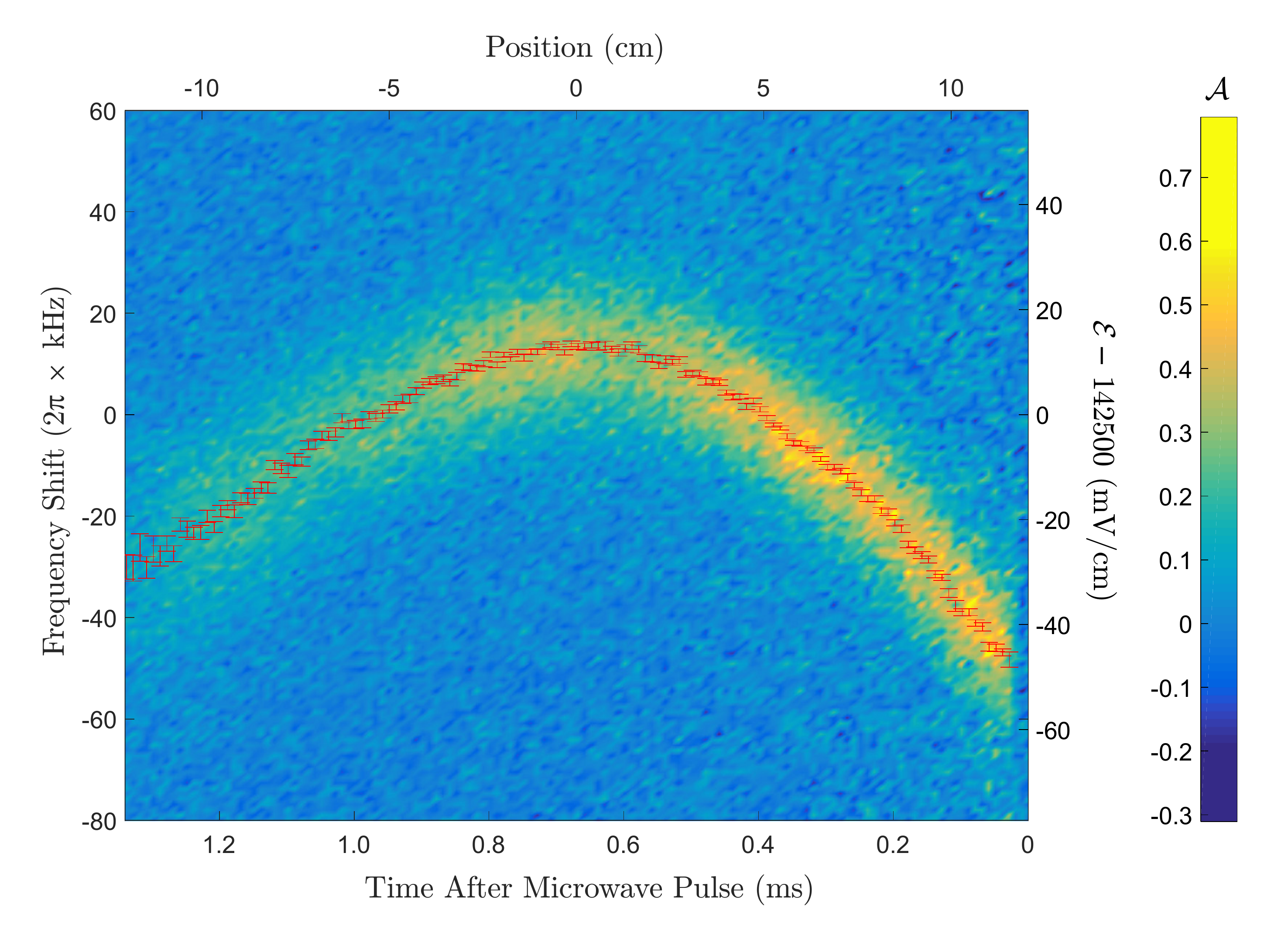}
\caption{Colourmap: Plot of the asymmetry $\A$ induced by a microwave pulse as the frequency of the microwaves was scanned. Red data points: Plot of the corresponding reversing component of the electric field obtained by extracting the centre of the resonance signal. The position is relative to the centre of the spin-precession region.\label{fig:asym_map}}
\end{figure}

We observed a resonance linewidth of ${\approx}2\pi\times25~{\rm kHz}\approx2\pi/T$ which was limited by the microwave $\pi$-pulse duration of $T=40~\upmu\rm{s}$. With our signal-to-noise, we were able to fit the resonance centre to a precision of ${\sim}2\pi\times1$~kHz, typically using ${\sim}50$ detuning values and averaging over ${\sim}50$ molecule pulses per detuning value. Example data obtained via this method are shown in figure~\ref{fig:asym_map}.

In these data, it is evident that the resonant frequency of the microwaves varied across the molecule pulse by around $2\pi\times60$~kHz. The position $x$ of the molecules at the time of the microwave pulse was assumed to be linearly related to the molecule arrival time in the state readout region. The observed spatial variation of $\mathcal{E}$ was roughly consistent with expectations based on the measured variation of the plate spacing described above.

\begin{figure}[!ht]
\centering
\includegraphics[scale=0.35]{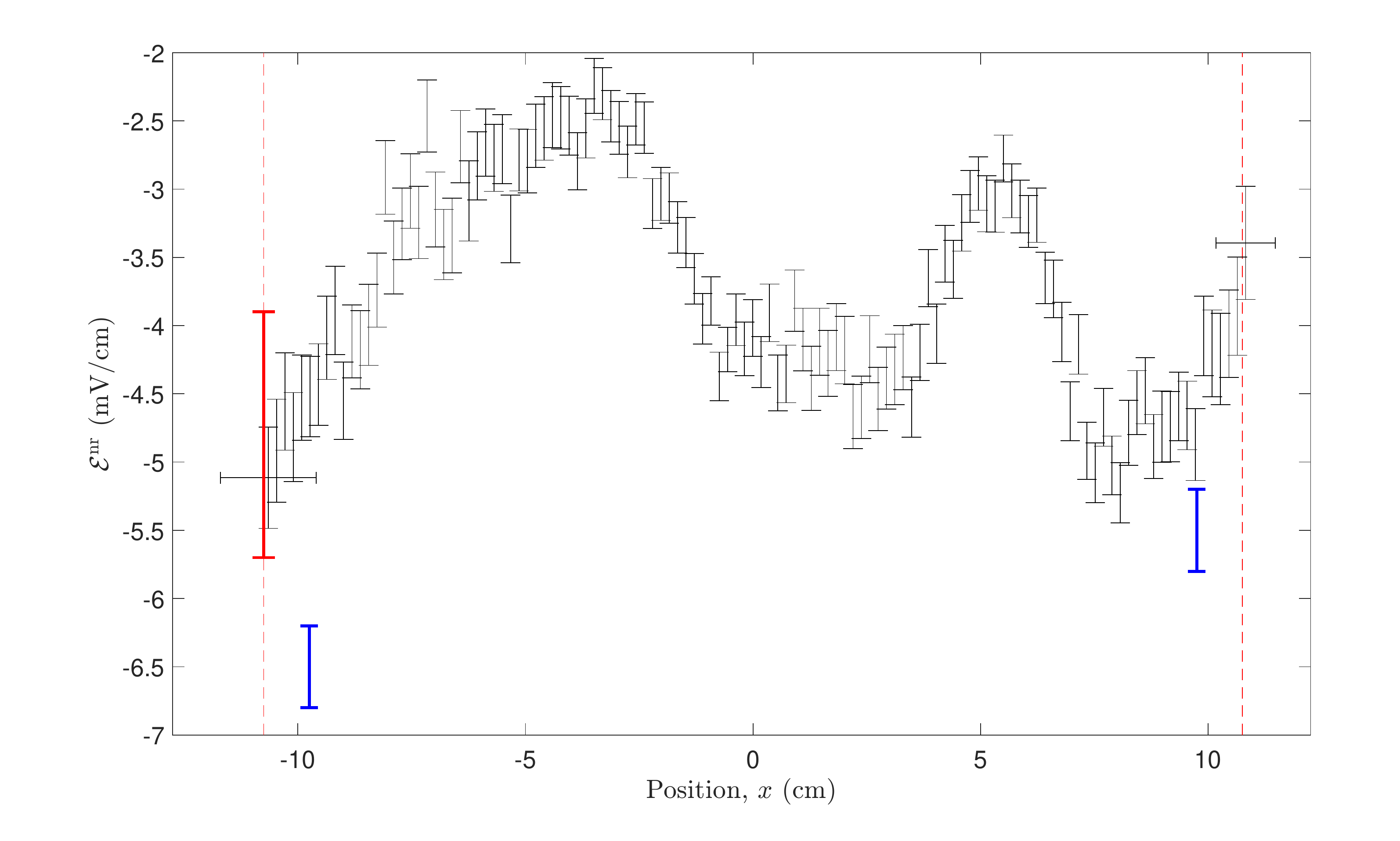}
\caption{A plot of the spatial variation of $\Enr$. The black points are data obtained via microwave spectroscopy. The blue points are data obtained via Raman spectroscopy. The red data point was obtained by examining the variation of contrast with $\Delta_{\rm prep}$. The approximate position of the state preparation (state readout) laser beam is shown as a red dotted line on the left (right) of the figure. For the microwave spectroscopy data the uncertainty/averaging range of the position is around 21~mm at the left-hand side of the plot and decreases to around 13~mm at the right-hand side --- see main text for details.\label{fig:enr}}
\end{figure}

By switching $\Nsw$ and $\Esw$ between measurements of the $\E$-field we were able to extract $\Enr$ from the $\Nsw\Esw$-correlated component of $\omega_{\rm MW}$. These measurements, shown in figure~\ref{fig:enr}, were used to evaluate the corresponding systematic error in equation~\ref{eq:Enr_systematic_error_value}. 

We clearly saw a non-uniform $\Enr$ across the spin precession region. The spatial variation shown in figure~\ref{fig:enr} was reproducible for the period of several weeks over which these measurements of the electric field were taken. We are unsure as to the origin of the $\Enr$ but believe it may have been caused by patch potentials \cite{Robertson2006} present on the electric field plates. We observed unexplained disagreement between the two measurement methods (Raman spectroscopy vs.\ microwave spectroscopy), but note that both report non-reversing fields of a few mV/cm with the same sign.

The mapping between arrival time in the detection region and $x$ position during the microwave pulse was approximate, suffering from spatial averaging due to a variety of effects. For example, velocity dispersion led to averaging of $dx\times\sigma_{v_{\parallel}}/v_{\parallel}$, where $\sigma_{v_{\parallel}}$ is the longitudinal velocity spread of the molecular beam and $dx$ is the distance between microwave interrogation and state readout. This averaging distance was largest, ${\approx}1.6~{\rm cm}$, at the state preparation region. Spatial averaging also occurred across the ${\approx}0.7$~cm distance traversed during the $T=40~\upmu$s microwave pulse. Finally, there was averaging of the spatial position of the molecules due to the finite size of the state readout laser beam and the polarisation switching; molecules were optically pumped (with varying probability) throughout the ${\approx}0.5$~cm wide laser beam.

In addition to spatial averaging, uncertainty in the mean longitudinal velocity also contributed an uncertainty in position. Changes of ${\approx}10$~m/s between molecule pulses were quite typical over the course of the $\mathcal{E}$-field measurement, giving an estimated position uncertainty of ${\lesssim}1$~cm.


By adding the above contributions in quadrature we concluded that the range of positions from which the microwave-induced signals could have originated increased from around ${\approx}1.3$~cm at the state readout beam to ${\approx}2.1$~cm at the optical pumping beam. These ranges are shown as horizontal error bars at the extrema of position in figure~\ref{fig:enr}.


We used a third method to measure $\mathcal{E}$ and $\mathcal{E}^{\mathrm{nr}}$ \emph{in situ} throughout the eEDM dataset by performing `intentional parameter variation' tests with large $\Delta_{\p}$ (denoted by `c' in figure~\ref{fig:timing}). Detuning the state preparation laser resulted in a reduction in the measured contrast $|\mathcal{C}|$ as shown in figure~\ref{fig:contrast}~(B). Setting $\Delta_{\p}\approx\pm2^{\:}\mathrm{MHz}$ gives $\left|\mathcal{C}\right|\approx0.5$, and the contrast was then approximately linearly proportional to $\Delta_{\rm prep}$ with a sensitivity
of about $1/\gamma_{C}\approx1/(2\pi\times2~{\rm MHz})$. Any variation in the electric field would change the Stark shift, and thus also $\Delta_{\rm prep}$, resulting in a change in contrast. Thus, using the previously described spin precession scheme, we indirectly measured parity components of the electric field from the appropriate parity components of the contrast:
\begin{align}
D_1\mathcal{E}^{\mathrm{nr}}(x_{\mathrm{prep}})\approx&\frac{\partial\Delta_{\mathrm{prep}}}{\partial\mathcal{C}}\mathcal{C}^{\mathcal{NE}}\\
D_1\mathcal{E}(x_{\mathrm{prep}})\approx&\omega_{\rm L}^{\mathcal{N}}+\frac{\partial\Delta_{\mathrm{prep}}}{\partial\mathcal{C}}\mathcal{C}^{\mathcal{N}}.
\end{align}

We looked for variation of $\mathcal{E}$ or $\mathcal{E}^{\mathrm{nr}}$ every 3--4 hours. Measurements of $\mathcal{E}^{\mathrm{nr}}$ were consistent with the microwave measurements, with a constant value $\mathcal{E}^{\mathrm{nr}}(x_{\mathrm{prep}})=-4.8\pm0.9^{\:}\mathrm{mV/cm}$. However, the mismatch $\Delta^{\mathcal{N}}=D_1\mathcal{E}-\omega_{\rm L}^{\mathcal{N}}$ between the Stark shift $D_1\mathcal{E}$ and the $\Nsw$-correlated laser frequency shift, $\omega_{L}^{\mathcal{N}}$, was found to drift significantly on the scale of around $2\pi\times20^{\:}\mathrm{kHz}/\mathrm{day}$. This drift of $\Delta^{\N}$ was servoed by tuning $\omega^{\N}_{\rm L}$ after each measurement, ensuring $\left|\Delta^{\mathcal{N}}\right|<2\pi\times30$~kHz at all times \cite{SpaunThesis}, see sections \ref{sssec:correlated_laser_parameters} and \ref{ssec:laser_imperfections} for more details. 

\subsubsection{Magnetic Fields}
\hspace*{\fill} \\
\label{sec:bfields}
Our experimental scheme did not require the application of a magnetic field. This was not the case with some previous eEDM experiments, where the magnetic field was used to define a quantization axis \cite{Commins1994,Regan2002}, or to cause the precession of spin to a direction associated with maximum sensitivity \cite{Hudson2011,Kara2012}. Instead we used the electric field to define a quantization axis, and we used the relative polarisations of the state preparation and readout lasers to define the basis in which we read out the electron's spin precession with maximal sensitivity.

However, we regularly applied a magnetic field $\B$ in order to perform searches for systematic errors. The phase accumulation induced by an eEDM $\delta d_{e}\approx5\times10^{-29}$~$\ecm$ would have the same size as a Zeeman phase produced by a magnetic field of $\mathcal{B}\approx0.2~\upmu{\rm G}$, which is small compared to some of the magnetic field imperfections in the experiment. However, phases associated with magnetic-field-induced precession were distinguished from eEDM-induced precession by the use of the switches at our disposal (e.g.~electric field reversal). Nevertheless, it was important to investigate, quantify and minimize the effects of such magnetic fields, as they could have coupled with other experimental imperfections to give eEDM-like phases.

\begin{figure}[!ht]
\centering
\includegraphics[scale=0.25]{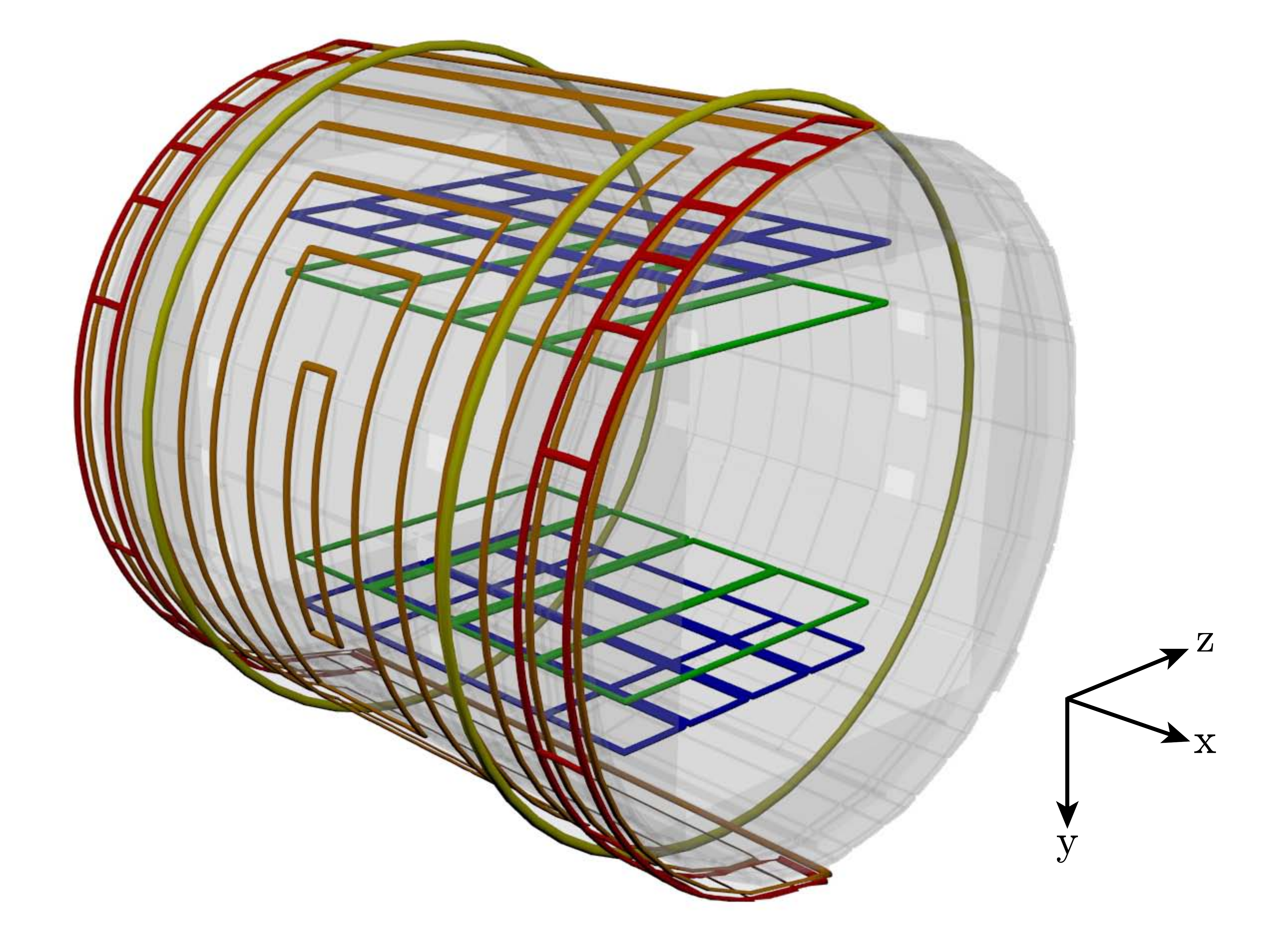}
\caption{A schematic of the magnetic field coils used. The main coils consisted of rectangular cosine coils (orange) wound on the surface of a cylindrical plastic frame together with additional end coils (red) to correct for the low aspect ratio (length/diameter) in our system; a second set of these coils, mirrored in the $xy$ plane, is not shown. Also wrapped around this frame are a pair of circular auxiliary coils shown in yellow. The other auxiliary coils are shown in blue and green and consist of rectangular coils above and below the vacuum chamber. See the main text for descriptions of the functions of all of the coils.\label{fig:coil_schematic}}
\end{figure}

Under normal operating conditions we ran the experiment at three different magnetic field magnitudes, corresponding to a relative precession phase of $\phi^{\mathcal{B}}\approx q\frac{\pi}{4}$ for $q=0,1,2$. The required $z$-component of the field was then $\mathcal{B}_z=q\mathcal{B}_0\tilde{\mathcal{B}}$, where $\B_0=\frac{\pi}{4}\frac{1}{g_1\mu_{\rm B}\tau}\approx 20~{\rm mG}$. We also had the ability to apply transverse magnetic field components along $\hat{x}$ and $\hat{y}$, and all five linearly independent first-order gradients. The various coils that we used are illustrated in figure~\ref{fig:coil_schematic}.

The primary magnetic field, $\mathcal{B}_z$, was produced by two sets of rectangular coils, shown in orange in figure~\ref{fig:coil_schematic}. These were wound on the surface of two hemicylindrical plastic shells, on the $\pm z$ sides of the spin-precession region. The coils were designed to maximize field uniformity and minimize distortion due to the boundary conditions imposed by the magnetic shielding. It was also possible to apply a $\partial\mathcal{B}_z/\partial z$ gradient with these coils. Two end coils (red in figure~\ref{fig:coil_schematic}), located on the $\pm x$ ends of the spin-precession region, enhanced the uniformity of the $\B$-field along $x$ and enabled application of a $\partial\B_z/\partial x$. The main coils were powered by two separate commercial power supplies\footnote{Krohn-Hite 521/522}, and the end coils were powered by custom power supplies. The current flowing through these coils was continuously monitored throughout the course of the experiment by measuring with a digital multimeter the voltage dropped across precision resistors.

We used three sets of auxiliary magnetic field coils in systematic error searches. A pair of circular Helmholtz coils (yellow in figure~\ref{fig:coil_schematic}) were wrapped around the same frame used for the main coils and were formed from ribbon cable. They provided a magnetic field in the $\pm\hat{x}$ directions and could also provide a $\partial\B_x/\partial x$. Above and below the spin-precession region chamber ($\pm y$) there were four sets of rectangular coils (blue and green in figure~\ref{fig:coil_schematic}). These allowed us to produce a field in the $\pm\hat{y}$ directions as well as all three associated first-order gradients. Note that the three first-order magnetic field gradients that we could not apply could be inferred from Maxwell's equations. A summary of the fields that we could apply is given in table~\ref{tab:coils_table}.
\begin{centering}
\begin{table}
\caption{A summary of the magnetic fields and magnetic field gradients that we could produce. The coil colours refer to figure~\ref{fig:coil_schematic}.\label{tab:coils_table}}
\begin{tabular}{ccc}
\br
Coil colour & Fields produced & Field gradients produced\\
\mr
Orange & $\B_z$ & $\partial \B_z/\partial z$\\
Red & $\B_z$ & $\partial \B_z/\partial x$, $\partial \B_z/\partial z$\\
Yellow & $\B_x$ & $\partial \B_x/\partial x$\\
Blue & $\B_y$ & $\partial \B_y/\partial y$, $\partial \B_y/\partial z$\\
Green & $\B_y$ & $\partial \B_y/\partial x$, $\partial \B_y/\partial y$\\
\br
\end{tabular}
\end{table}
\end{centering}

Several measures were taken to minimize stray magnetic fields affecting the molecules. The simplest was to ensure no magnetized objects were placed within the spin-precession region. To ensure this, all components were fabricated from non-magnetic materials (e.g.\ no stainless steel). The magnetization of all objects was also checked before installation by passing them across an AC-coupled magnetometer sensitive to 0.1~mG field variations.

The ambient $\B$-field in the laboratory was dominated by that from the Earth's core (${\sim}500~$mG approximately along $\hat{x}+\hat{y}$). To suppress this and other DC/low-frequency fields, the spin-precession region was surrounded by a set of five concentric cylindrical magnetic shields constructed from ${\approx}1.6$~mm thick mu-metal\footnote{Amuneal Inc.}. Each layer of shielding should have provided around a factor of 10 reduction in the DC magnetic field \cite{Vutha2010}; however, residual magnetisation of the mu-metal was found to limit the field components to $\gtrsim20~\upmu$G for $\mathcal{B}_x$ and $\mathcal{B}_y$, and $\gtrsim500~\upmu$G for $\mathcal{B}_y$. \footnote{We later found that the residual $\mathcal{B}_y$ could be reduced to a level comparable to $\mathcal{B}_x$ and $\mathcal{B}_z$ by performing degaussing with a higher current.} Each shielding layer was divided into two half-cylinders and two end caps. The outermost (innermost) shield was 132~cm (86~cm) long and had a diameter of 107~cm (76~cm). These shields had holes to allow lasers to pass through in the $z$ direction, and to accommodate the molecule beam. There were also holes for the light pipes to extract molecule fluorescence, and some electric connections, in the $x$ direction. Measurements and simulations showed that these holes had a negligible impact on the shielding efficiency. The shielding factor remained approximately constant up to an AC frequency $\sim2\pi\times3$~GHz for which the wavelength becomes comparable to the size of any apertures in the shields, ${\sim}10$~cm, and the magnetic field noise starts to penetrate the shields. However, our measurement was only sensitive to magnetic field noise at frequencies up to roughly the inverse of the spin precession time $1/\tau\approx2\pi\times1$~kHz \cite{VuthaThesis}. The aluminium vacuum chamber also shielded AC magnetic noise above a frequency ${\sim}1/\pi\sigma t^2\mu\approx2\pi\times100$~Hz, where $\sigma\approx3.5\times10^7$~S/m is the electrical conductivity, $t\approx1$~cm is the thickness and $\mu$ is the permeability ${\approx}\mu_0$, the vacuum permeability \cite{Mager1969,Sumner1987}.

The relatively large ($\B\sim10$~mG) fields applied by the $\B_z$ coils caused the inner magnetic shields to become slightly magnetized, inducing a non-reversing magnetic field, $\B^{\nr}\approx30~\upmu{\rm G}$. In order to suppress this remanent field we performed a degaussing procedure on the magnetic shields by passing a $200$~Hz sinusoidal current through sets of loosely wound ribbon cable coils which wrapped axially (in the $xy$ plane at $z=0$) between the shield layers. The maximum current amplitude was 1~A, sufficient to drive the mu-metal to saturation, and the amplitude was decreased with an exponential envelope over a period of 1~s. To fully degauss all layers of the magnetic shielding takes around 4~s. There was also a 1~s period of `dead time' during which the main magnetic field was turned back on and allowed to settle. This degaussing procedure was repeated every time the applied magnetic field was changed, which occured approximately every 40~s.

Variations in the magnetic fields present were continuously measured throughout the experimental procedure. This was achieved using a set of four three-axis fluxgate magnetometers\footnote{Bartington Mag-03.}, which were mounted in a tetrahedral configuration outside the spin-precession region vacuum chamber (but inside the magnetic shielding). We also used an additional fluxgate magnetometer which was positioned at a distance of around 1~m from the apparatus and outside of the magnetic shielding. By continuously recording the measurements provided by these magnetometers we were able to search for correlations of our data with the magnetic field present. In particular, we checked for the presence of a magnetic field correlated with the electric field, $\mathcal{B}^{\E}$, which would have been characteristic of a leakage current flowing between the electric field plates --- an effect known to contribute a significant systematic error in previous eEDM experiments \cite{Regan2001,Kara2012}.

\begin{figure}[!ht]
\centering
\includegraphics[width=0.8\textwidth]{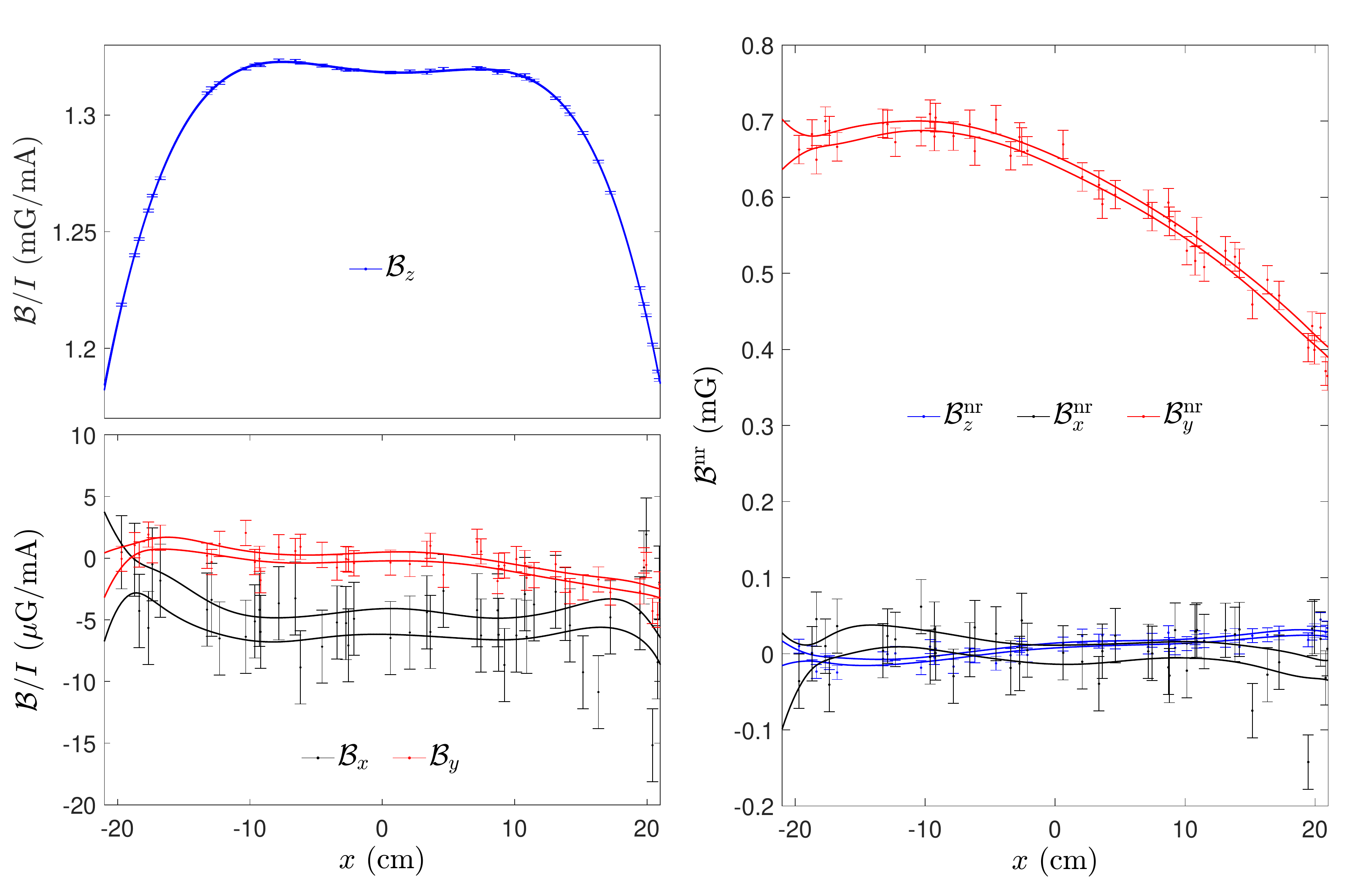}
\caption{Magnetic field data taken with a flux-gate magnetometer passed along the molecular beam line. The left-hand plot shows the reversing components of field whilst a nominal $\mathcal{B}_z$ was applied. The right-hand plot shows the corresponding non-reversing components. The data are fit by polynomial curves.\label{fig:probulator_data}}
\end{figure}
Additional measurement of the magnetic fields was carried out by opening the vacuum system and passing a rotatable flux-gate magnetometer into the chamber. This allowed for measurement of the fields directly along the beam line. The freedom to rotate the magnetometers was crucial to distinguish between electronic offsets and $\vec{\B}^{\rm nr}$ for fields ${\lesssim}1$~mG. From these measurements we were able to directly characterise most of the magnetic fields and first-order field gradients, including non-reversing components. Example data obtained from these measurements are shown in Figure~\ref{fig:probulator_data}. We saw that the applied fields were all flat to within 1~mG, and the non-reversing components, with the exception of $\B^{\rm nr}_y$, were less than 50~$\upmu$G. Systematic uncertainty due to these fields is discussed in section~\ref{ssec:magnetic_field_imperfections}.

\subsubsection{Fluorescence Collection and Detection}
\begin{figure}[!ht]
\centering
\includegraphics[width=16cm]{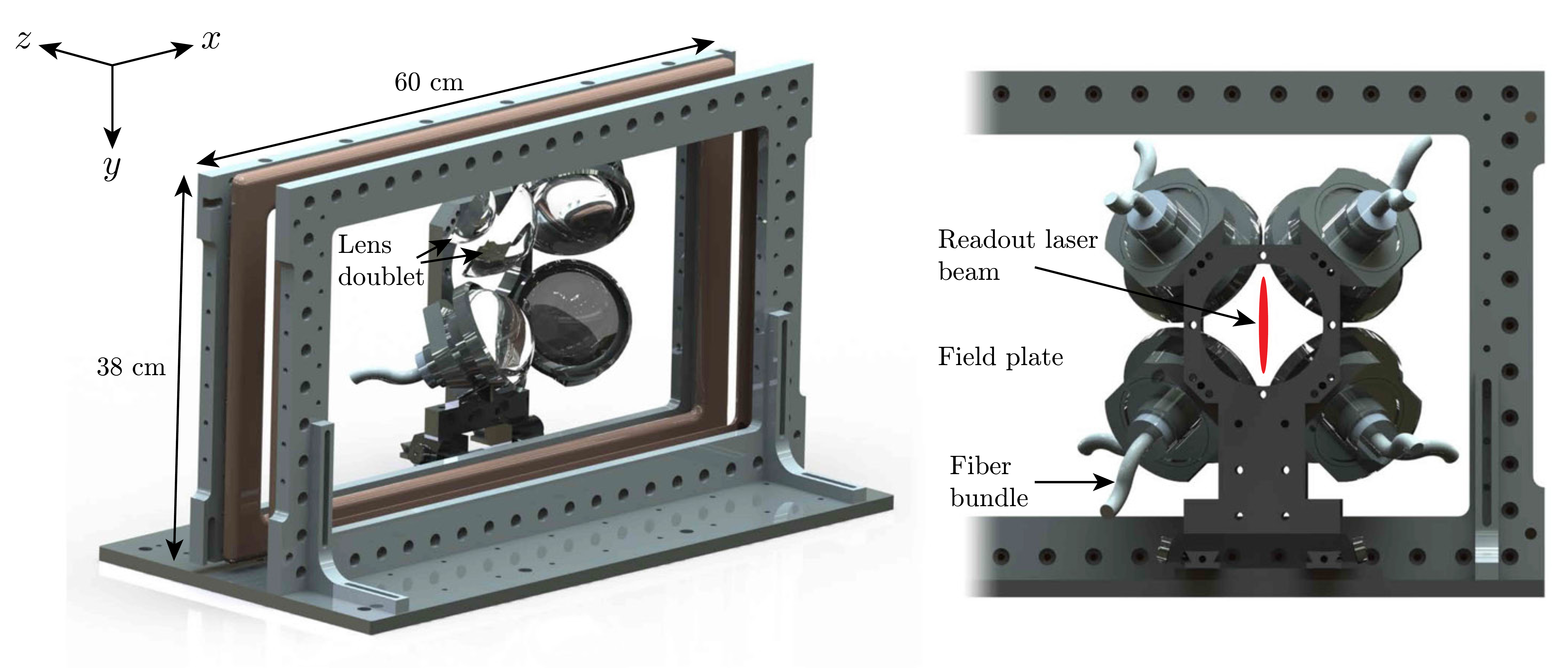}
\caption{Fluorescence collection apparatus. Left: The mounted electric field plates are shown together with one of the two sets of four lens doublets. The mounting for the top-left doublet has been removed to show the lenses. The fiber bundles are shown schematically, fastened into the lens tubes behind the doublets. The lens assembly was mounted on rails and the entire assembly sat on a breadboard which was fastened to the vacuum chamber. The view on the right also shows the approximate position of the state readout laser beam as it passes through the apparatus.}
\label{fig:collection_optics}
\end{figure}

\hspace*{\fill} \\
\label{sec:fluorescence_collection}
As previously described, our experimental data consisted of laser-induced molecule fluorescence, emitted in all directions (with a well-defined angular distribution \cite{Kirilov2013}) when the molecules were interrogated by the state readout laser beam. The apparatus for collecting this light is illustrated in figure~\ref{fig:collection_optics}.
The fluorescence light passed through the transparent electric field plates, whose inner (outer) faces are ITO (anti-reflection [AR]) coated. Behind each field plate was a set of four AR-coated lens doublets, which collimated and then focussed the light. The optical axes of the doublets intersected a ray path from the centre of the fluorescing molecule region, accounting for refraction through the electric field plates. 
The first (second) lens of each doublet was a 75~mm\footnote{CVI Melles Griot LAG-75.0-50.0-C-SLMF-400-700.} (50~mm\footnote{CVI Melles Griot LAG-50.0-35.0-C-SLMF-400-700.}) diameter spherical lens of focal length 50~mm (35~mm). On each side ($\pm z$), each of the four lens doublets focussed light onto one of four sections of a `quadfurcated' fiber bundle\footnote{Fiberoptic Systems.} whose input ends were 9~mm in diameter and fastened in lens tubes. The output of the fiber bundle was connected to a 19~mm diameter fused quartz light pipe with optical couplant gel\footnote{Corning Q2-3067.} in between. The light pipe passed out of the spin-precession region vacuum chamber and magnetic shields and directed the light onto a PMT\footnote{Hamamatsu R8900U-20.}. Bandpass filters\footnote{Semrock FF01-689/23-25-D.} were used to suppress backgrounds from e.g.\ scattered light. Detailed tests of the light collection were carried out \cite{SpaunThesis} which estimated that ${\approx}14$~\% of the fluorescence photons were collected. The major contributions to this efficiency were the finite solid angle subtended by the collection lenses (${\approx}50~$\%), finite coupling efficiency into fiber bundles (${\approx}60~$\%) and finite coupling efficiency between the fiber bundles and the light pipes (${\approx}50~$\%). In addition, the quantum efficiency of the PMT's was specified to be ${\approx}10$~\%, which further reduced the signal obtained.

\subsubsection{Data Acquisition}
\hspace*{\fill} \\
\label{sec:data_acquisition}
The data acqusition system performed the following three functions: 
\begin{enumerate}
\item Digital modulation of the experimental parameters necessary for acquiring the complete set of phase and contrast measurements required to extract the eEDM, as described in section~\ref{sec:Measurement_scheme_more_detail}.
\item Rapid (5~MSa/s) acquisition and storage of high-bandwidth fluorescence waveforms for the spin precession measurement.
\item Monitoring and logging of experimental parameters useful for checking the experimental state and for searching for systematic errors (e.g. magnetic fields, beam source temperatures).
\end{enumerate}
All functions were coordinated with a LabVIEW-based software system.

Data acquisition timing was controlled by a digital delay generator.\footnote{SRS DG645.} Every 20~ms, a TTL signal was produced which triggered the ablation laser Q-switch, in turn creating a pulse of molecules. Molecule fluorescence signals, measured as a PMT photocurrent, were captured on a 20-bit digital oscilloscope\footnote{National Instruments PXI-5922.}. The oscilloscope was triggered 6--7~ms after the ablation pulse, depending on the current molecule beam forward velocity, and recorded a 9~ms window of signal containing the entire molecule signal (1--2~ms) and several ms of background. The 100~kHz square wave that drove the fast polarisation switching of the state readout laser was synchronised with the 50~Hz Q-switch trigger so that the relative phase was fixed. The 5~MSa/s data rate of the oscilloscope enabled resolution of the time-dependent structure within each 5~$\upmu$s polarisation bin; this structure could vary on timescales as short as the C-state lifetime $1/\gamma_C\approx500$~ns \cite{Hess2014}.

Signal waveforms, $S(t)$, were captured from two PMTs --- note that we were not counting individual photoelectrons, but instead amplified and read out a voltage proportional to the count rate. These waveforms were then transferred to the control PC where they were digitally averaged over 25 pulses to form one `trace'. The traces were then written to a hard drive. A file containing auxiliary measurements was recorded synchronously with each fluorescence trace. This file included the states of the experimental switches and other auxiliary measurements such as $\E$-field voltages, $\B$-field currents, laser power and polarisation, magnetic field measurements, molecular beam buffer gas flow rate, buffer gas cell temperature, and the temperature, pressure and humidity in our lab. This data proved useful in searching for systematic errors as described in section~\ref{sec:systematics}.

\section{Data Analysis}
 \label{sec:data_analysis}
In this section we describe the data analysis routine used to extract the eEDM value, and other quantities, from our dataset of nearly $10^6$ PMT fluorescence traces. The entire analysis was implemented with a `blind' offset on the eEDM channel such that the channel's mean value was not known until 
after all the data had been acquired and the systematic error in the measurement had been determined. No analysis changes were made after the blind was revealed.
Several data cuts were applied (before removal of the blind) to ensure that the resulting eEDM measurements would be nearly normally distributed and to filter data that was not taken under normal operating conditions.

\subsection{Signal Asymmetry}
\label{sec:signal_asymmetry}
As described in section~\ref{sec:Measurement_scheme}, the accumulated phase $\Phi$ was read out by resonantly addressing the $H \rightarrow C$ transition with linearly polarised light and monitoring the resulting fluorescence. The state readout laser was switched between orthogonal polarisations, $\hat{X}$ and $\hat{Y}$, at $100$~kHz (with $1.2~\upmu$s of dead time between polarisations) in order to normalize against molecular flux variations. By switching at a rate fast enough that each molecule experienced both polarisations, we achieved nearly photon-shot-noise-limited phase measurements \cite{Kirilov2013}. With a sufficiently wide laser beam, all molecules were completely optically pumped by both laser polarisations during their $\sim$20~$\upmu$s fly-through time. We induced approximately one fluorescence photon from each molecule by projecting the molecule state onto the two orthogonal spin states excited by laser beams with orthogonal polarisations.

\begin{figure}[!htbp]
\centering
\includegraphics[width=12.7cm]{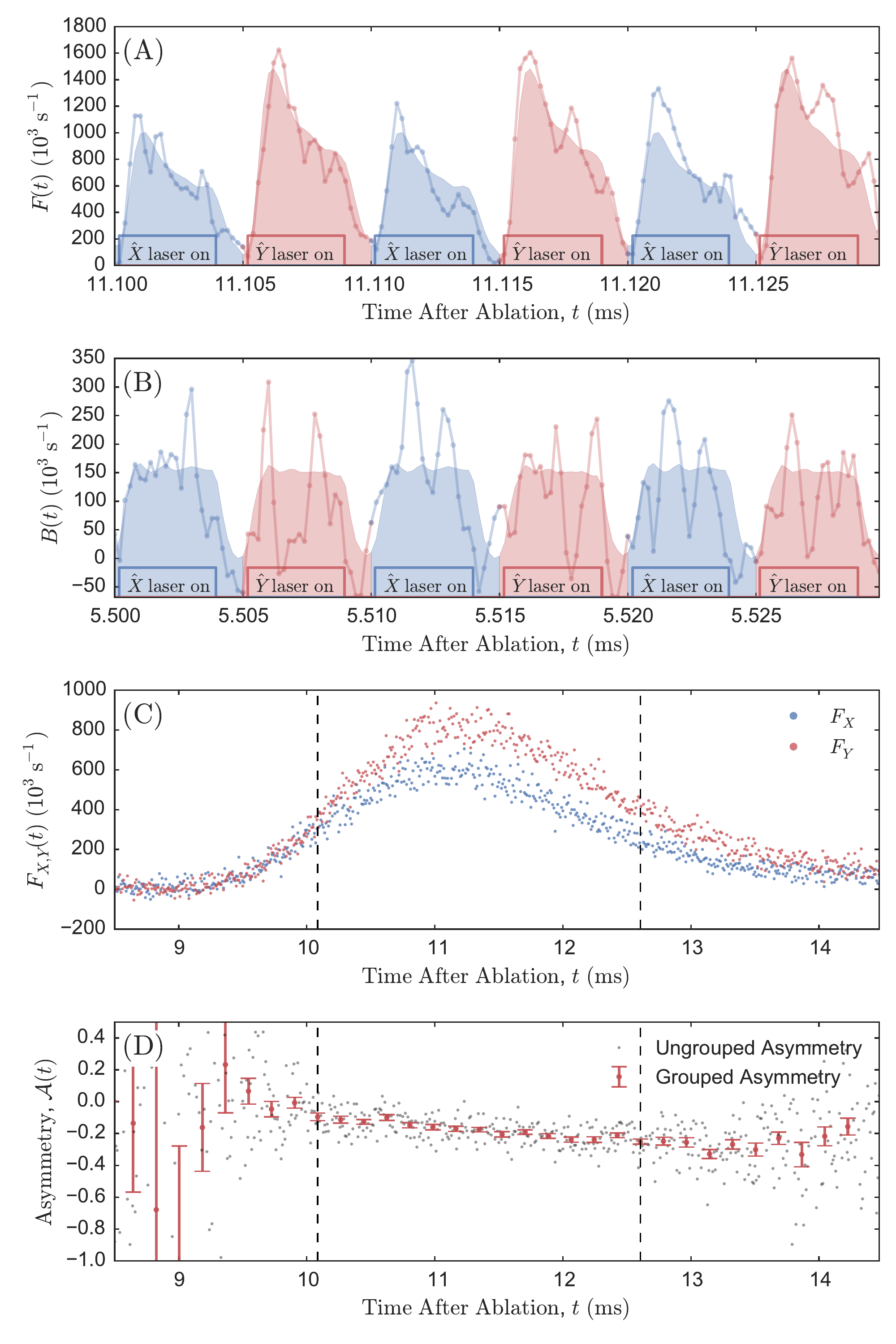}
\caption{(A) Molecule fluorescence signal $F(t)$ in photoelectrons/s induced by $\hat{X}$ (blue) and $\hat{Y}$ (red) readout laser polarisations. Lines show the raw data for a single trace consisting of an average of 25 molecule pulses. Shaded regions show the waveform averaged over 16 traces. (B) Background signal $B(t^{\prime})$ in photoelectrons/s obtained before the arrival of molecules in the state readout region. (C) Integrated fluorescence signals $F_X$ and $F_Y$ throughout the molecule pulse. Dashed lines denote the region with $F=(F_X+F_Y)/2>3\times10^5~\rm{s}^{-1}$, used as a typical cut for inclusion in eEDM data. Points are spaced by 5~$\upmu$s. (D) Computed asymmetry throughout the molecule pulse. In this example, 18 of the ungrouped asymmetry points are grouped together to compute the mean and uncertainty shown as the grouped asymmetry.}\label{fig:modulation}
\end{figure}

The rapid switching of the laser polarisation resulted in a modulated PMT signal, $S(t)$, as shown in figure~\ref{fig:modulation}. For the following discussion we consider the polarisation state to switch at a time $t=0$. Immediately after, there is a rapid increase in fluorescence as the molecules in the laser beam are quickly excited; while $\Omega_rt\ll1$, where $\Omega_r\sim2\pi\times1$~MHz is the Rabi frequency on the $H$ to $C$ transition, the fluorescence increases as $S(t)\propto\Omega_r^2\times t^2$. At later times, when $\Omega_rt\gtrsim1$, population is about evenly mixed between the $H$ and $C$ states (since $\Omega_r\gtrsim\gamma_C$); hence, $S(t)$ decays exponentially with a time constant of roughly $1/(2\gamma_C)\approx1~\upmu$s. Molecules that were not present at $t=0$ continue to enter the laser beam, causing $S(t)$ to approach a steady state. The laser is then turned off and the signal decays exponentially with time constant $1/\gamma_C\approx0.5~\upmu$s. The next laser pulse, with orthogonal polarisation, is turned on 1.2~$\upmu$s $\approx2.5/\gamma_C$ after the end of the previous one to prevent significant overlap of contributions to $S(t)$ induced by different polarisations. A low-pass filter in the PMT voltage amplifier with a cut-off frequency of $2\pi\times2$~MHz removed any short timescale dynamics from $S(t)$, and prevented aliasing of high frequency components in the signal given our fixed digitization rate of 5 MSa/s.

To determine the fluorescence $F(t)$ produced by each polarisation state, we subtracted a time-dependent background, $B(t^{\prime})$, taken from data with no molecule fluorescence present, i.e. $F(t)=S(t)-B(t^{\prime})$. Examples of the extracted $F(t)$ and $B(t^{\prime})$ time series are shown in figure~\ref{fig:modulation}A and B, respectively. $B(t^{\prime})$ was modulated in time due to scattered light from the state readout laser beam and has a DC electronic offset intrinsic to the PMTs. The first millisecond of data, which contains no fluorescence, was used to determine $B(t^{\prime})$. We assumed that $B(t^{\prime})$ was periodic with the switching of the laser polarisation but did not depend on the polarisation; we inferred its value by averaging together the recorded PMT signal across all polarisation bins for ${\approx}1$~ms of data taken before the arrival of the molecule pulse. Since molecule beam velocity variations caused jitter in the temporal position of the molecule pulse within the trace, 9~ms of data were collected per pulse, despite the fact that only the ${\approx}$2~ms of strong signal with $F(t)\gg B(t^{\prime})$ and ${\approx}1$~ms of background contained useful information for the spin precession measurement. 

Integrating $F(t)$ over times associated with pairs of orthogonally polarised laser pulses resulted in signals $F_X, F_Y$. The integration was performed over a specified time window that we denoted as a `polarisation bin'. Figure~\ref{fig:cuts}B shows two typical choices of polarisation bin and illustrates that the extracted eEDM is not significantly affected by this choice. Figure \ref{fig:pixel_plot} shows that most of the extracted quantities did not vary linearly within the polarisation bin (Pol.\ Cycle Time Dependence column).

After polarisation binning, the data displayed a fluorescence signal modulated by the envelope of the molecule pulse, as in figure~\ref{fig:modulation}C. Figure~\ref{fig:modulation}D shows the asymmetry, $\mathcal{A}$, computed from these data. 
The asymmetry is computed for each 10~$\upmu$s polarisation cycle, so that for the $i^{\rm{th}}$ cycle we have
\begin{equation}
\A_i=\frac{F_{X,i}-F_{Y,i}}{F_{X,i}+F_{Y,i}}.
\label{eq:asym_bins}
\end{equation}
The molecule phase, and hence asymmetry (see equation~\ref{eq:asymmetry}), had a linear dependence on the time after ablation because the molecules precessed in a magnetic field over a fixed distance; the slower molecules, which arrive later, precessed more than the faster molecules, which arrived earlier. We applied a fluorescence signal threshold cut of around $F=(F_X+F_Y)/2\ge3\times10^5~{\rm s}^{-1}$, indicated by dashed lines in figure~\ref{fig:modulation}C,D. Section \ref{sec:data_cuts} describes the threshold choice in detail.

To determine the statistical uncertainty in $\A$, $n\approx$ 20--30 adjacent asymmetry points were grouped together. For each group, $j$, centred around a time after ablation $t_j$, we calculated the mean, $\bar{\mathcal{A}}_j$, and the uncertainty in the mean, $\delta\bar{\mathcal{A}}_j$, depicted as red points and error bars in figure~\ref{fig:modulation}D. For smaller $n$, the variance in the sample variance in the mean grows, in which case, error propagation that utilises a weighted mean of data ultimately leads to an understimate of the final statistical uncertainty \cite{Kenny1951}. For larger $n$, the mean significantly varies within the group due to velocity dispersion, and the variance in the mean grows in a manner not determined by random statistical fluctuations. For the range $n=$ 20--30 we observed no significant change in any quantities which were deduced from the measured asymmetry. 

As described earlier in this section, the background, $B(t^{\prime})$, which we subtracted from the PMT signal, $S(t)$, was observed to be correlated with the fast switching of the readout laser beam polarisation. This can arise, for example, if the two polarisations have different laser beam intensities or pointings. 
We chose to use a polarisation independent $B(t^{\prime})$ by averaging over the two polarisation states. This produced an asymmetry offset as per equation~\ref{eq:asym_bins} and hence a significant $\Phi^{\rm nr}$ associated with the polarisation-dependent background. However, we did not consider $\Phi^{\rm nr}$ to be a crucial physical or diagnostic quantity.
We found that this methodology produced accurate estimates of the uncertainties of quantities computed from the measured asymmetry, as verified by $\chi^2$ analysis of measurements of $\Phi^{\N\E}$. We also found that none of the phase channels of interest changed significantly dependent on whether a polarisation-dependent $B(t^{\prime})$ was used.

\subsection{Computing Contrast and Phase}
\begin{figure}[!ht]
\centering
\includegraphics[width=12cm]{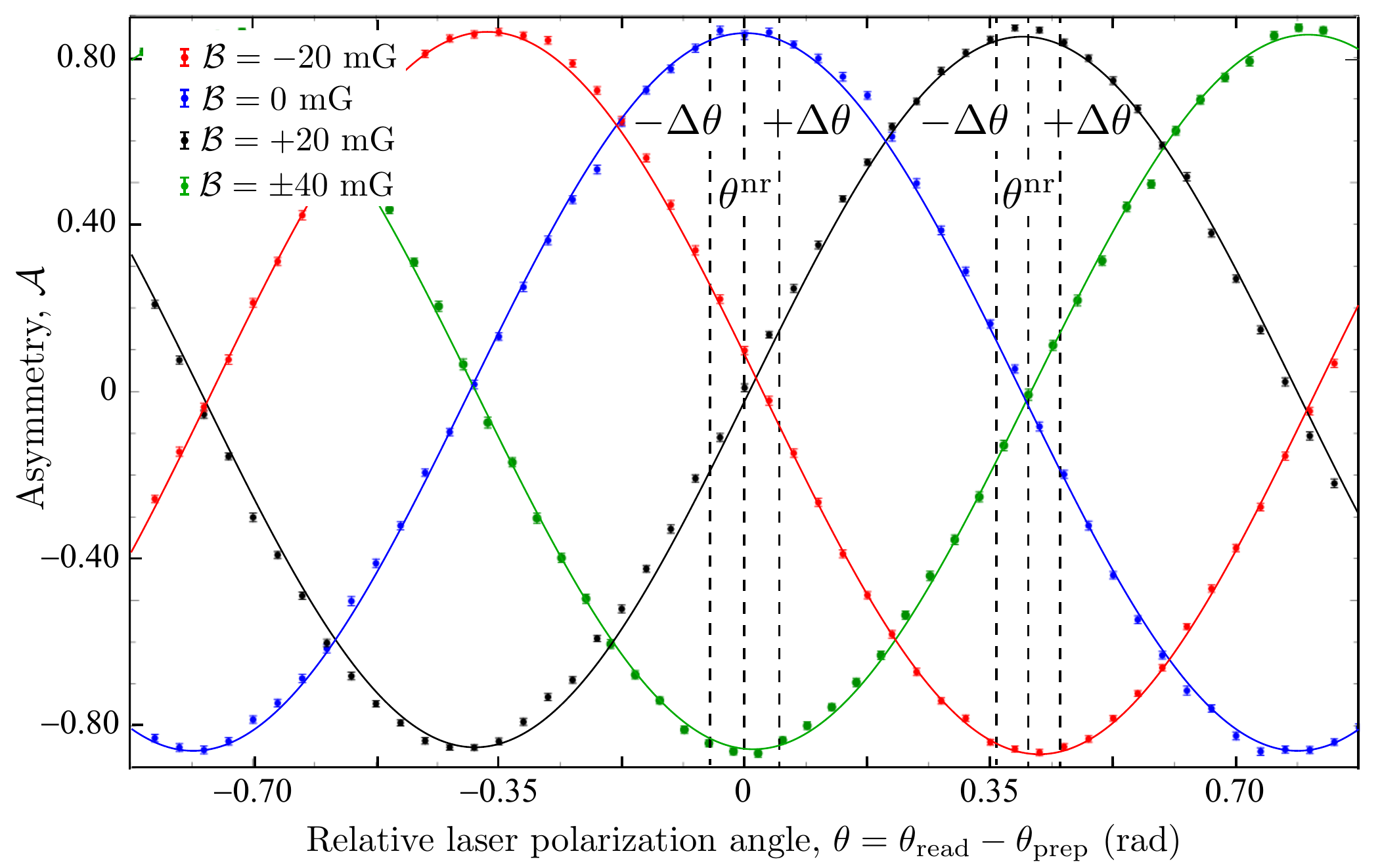}
\caption{Asymmetry vs.\ relative laser polarisation angle $\theta=\theta_{\rm read}-\theta_{\rm prep}$ for several magnetic field values. The value of $\theta$ was dithered about the value $\theta^{\rm nr}$ by $\pm\Delta\theta=\pm0.05$~rad to measure fringe contrast, $\mathcal{C}$. To stay on the steepest part of the fringe, we chose $\theta^{\rm nr}=0$ rad for $\B=\pm20$~mG and $\theta^{\rm nr}=\pi/4$ rad for $\B=\pm1,\pm40$~mG. For these data $|\mathcal{C}|<90\%$ due to low preparation laser power; typically, however, $|\mathcal{C}|\approx 95\%$. Solid lines represent the expected behaviour for a given magnetic field and contrast.}
\label{fig:fringe}
\end{figure}

To compute the measured phase $\Phi$ we must also measure the fringe contrast $\mathcal{C}$ and relative laser polarisation angle $\theta=\theta_{\rm read}-\theta_{\rm prep}$, as described in section~\ref{sec:Measurement_scheme}. The $\hat{X}$ and $\hat{Y}$ laser polarisations were set by a $\lambda/2$ waveplate and were determined absolutely by auxiliary polarimetry measurements \cite{Hess2014}. The contrast, defined as either $2\mathcal{C}=-\partial\A/\partial\theta$ or $2\mathcal{C}=\partial\A/\partial\phi$\footnote{Recall that in practice we consider $\mathcal{C}$ as an unsigned quantity for the purposes of data analysis.}, can be determined by dithering either the accumulated phase $\phi$ (by varying $\mathcal{B}_z$) or the relative laser polarisation angle $\theta$. We chose the latter as it could be changed quickly ($<1$~s) by rotating a half-wave plate with a stepper-motor-driven rotation stage. Figure~\ref{fig:fringe} shows the asymmetry as a function of $\theta$, for a range of values of applied magnetic field. We ran the experiment at the steepest part of the asymmetry fringe (where $\theta=\theta^{\rm nr}$) and measured the contrast, $\mathcal{C}_j$, for each asymmetry group, $\bar{\A}_j$, by switching $\theta$ between two angles, $\theta=\theta^{\rm nr}+\Delta\theta\Thsw$, for $\Thsw=\pm1$ and $\Delta\theta=0.05$~rad:
\begin{equation}
\label{eq:contrast_1}
\mathcal{C}_j=-\frac{\bar{\A}_{j}(\tilde{\theta}=+1)-\bar{\A}_{j}(\tilde{\theta}=-1)}{4\Delta\theta}.
\end{equation}

\begin{figure}[!htbp]
\centering
\includegraphics[trim=7mm 0mm 0mm 0mm,scale=0.6]{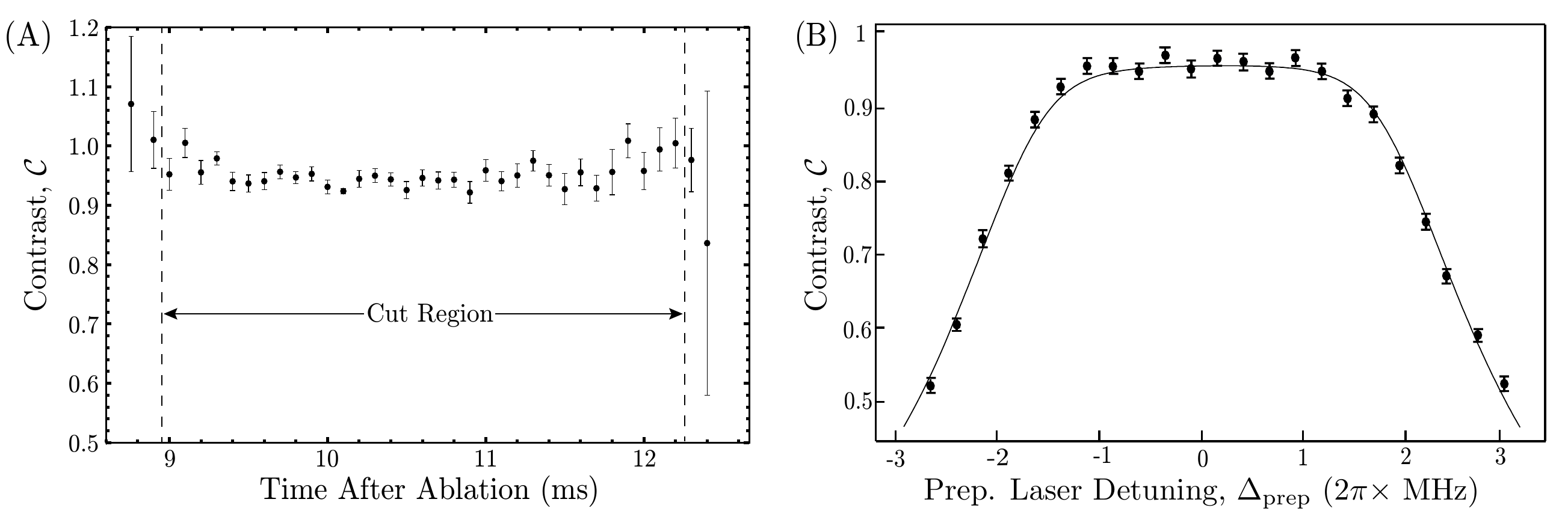}
\caption{(A) Contrast vs time after ablation, averaged over 64~traces. The signal threshold window is indicated by dashed lines (cf.\ figure~\ref{fig:modulation}). (B) Contrast vs preparation laser detuning. Error bars were computed as the standard error associated with 64 averaged traces. The solid line is a fit of the form $\mathcal{C}=a\times{\rm tanh}(b\gamma_C^2/(4\Delta_{\rm prep}^2+\gamma_C^2))$, motivated by solution of a classical rate equation.}
\label{fig:contrast}
\end{figure}

Because the fringe contrast was fairly constant over the duration of the molecule pulse (figure~\ref{fig:contrast}A), we used a weighted average\footnote{Each $\mathcal{C}_j$ measurement is weighted by its computed uncertainty.} of all $\mathcal{C}_j$ measurements within the cut region for that trace to extract the accumulated phase. We also performed the analysis by fitting $\mathcal{C}_j$ to a 2nd-order polynomial as a function of time during the ablation pulse; this led a better fit to the data, but had no significant effect on the results. We typically found $|\mathcal{C}|\approx95$\%. We believe that this was limited by a number of effects including: imperfect state preparation/readout, decay from the $C$ state back to the $H$ state and dispersion in the spin precession. We also observed that this value was constant over a $\pm2\pi\times1$~MHz detuning range of the state preparation laser (figure~\ref{fig:contrast}B), indicating complete optical pumping over this frequency range. Recall that, as defined, $\C$ can be positive or negative, depending on the sign of the asymmetry fringe slope (see figure~\ref{fig:fringe}, or equation~\ref{eq:contrast_1}). Given that we worked near zero asymmetry where the fringe slope was steepest, and that $\theta^{\rm nr}$ was always chosen to be 0 or $\pi/4$, we computed the total accumulated phase as
\begin{equation}
\label{eq:phase_1}
\Phi_j=\frac{\bar{\A}_j(\tilde{\theta}=+1)+\bar{\A}_j(\tilde{\theta}=-1)}{4\C}+q\frac{\pi}{4}.
\end{equation}
Here, $q=0,\pm1$ or $\pm2$, corresponds to applied magnetic fields of $\pm1$, $\pm20$, and $\pm40$~mG, respectively. We chose to apply a small magnetic field, $\B=1$ mG when operating at $q=0$ rather than turning off the magnetic field completely so that we would not need to change the experimental switch sequence or data analysis routine for data taken under this condition. Figure~\ref{fig:fringe} illustrates the correspondence between $\theta^{\rm nr}$ and applied magnetic field needed to remain on the steepest part of the asymmetry fringe.

\subsubsection{Accounting for Correlated Contrast}
\label{sec:corr_contrast}
\hspace*{\fill} \\
It was possible for the magnitude of the contrast $|\C|$ to vary between different experimental states. For example, if the state preparation laser detuning or fluorescence signal background were correlated with any of the block switches $\Nsw$, $\Esw$, or $\Bsw$, then contrast would also be correlated with those switches. As described in section~\ref{sec:systematics}, we observed both $\Nsw$- and $\Nsw\Esw$-correlated contrast. The latter was particularly troubling since it could lead to a systematic offset in the measured eEDM if not properly accounted for: since $\A=-\C\cos[2(\phi-\theta)]$, a nonzero $\A^{\mathcal{NE}}$ could occur due to either $\C^{\N\E}$ or $\phi^{\N\E}$. We accounted for contrast correlations by calculating $\C$ separately for each combination of $\Nsw$, $\Esw$, and $\Bsw$ experimental states (`state-averaged' contrast\footnote{Since there were $2^3=8$ different $\Nsw$, $\Esw$, and $\Bsw$ states in each 64-trace block, 64/8 = 8 traces were averaged together to determine the contrast for each experimental state.}):
\begin{equation}
\label{eq:contrast_2}
\C_j(\Nsw,\Esw,\Bsw)=-\frac{\bar{\A}_j(\tilde{\theta}=+1,\Nsw,\Esw,\Bsw)-\bar{\A}_j(-\tilde{\theta}=-1,\Nsw,\Esw,\Bsw)}{4\Delta \theta}.
\end{equation}
As previously discussed, we averaged or applied a quadratic fit to all $\C_j(\Nsw,\Esw,\Bsw)$ within a molecule pulse to compute $\bar{\C}_j(\Nsw,\Esw,\Bsw)$. The precession phase was calculated from each state-specific asymmetry and contrast measurement (cf.\ equation~\ref{eq:contrast_1}):
\begin{equation}
\label{eq:phase_2}
\Phi_j(\Nsw,\Esw,\Bsw)=\frac{\bar{\A}_j(\Nsw,\Esw,\Bsw)}{2\bar{\C}_j(\Nsw,\Esw,\Bsw)}+q \frac{\pi}{4},
\end{equation}
where
\begin{equation}
\label{eq:asymmetry_avg}
\bar{\A}_j(\Nsw,\Esw,\Bsw)=\frac{\bar{\A}_j(\tilde{\theta}=+1,\Nsw,\Esw,\Bsw)+\bar{\A}_j(\tilde{\theta}=-1,\Nsw,\Esw,\Bsw)}{2}
\end{equation}
is the average asymmetry over the two $\Thsw$ states in a data block that share identical values of $\Nsw$, $\Esw$ and $\Bsw$. By construction, phases computed from state-averaged contrast are immune to contrast correlations. We also computed phases by ignoring contrast correlations (i.e. treating contrast as independent of $\Nsw,\, \Esw,\, \Bsw$) and the result did not change significantly.

\subsubsection{Computing Phase and Frequency Correlations}
\label{sec:compute_phase}
\hspace*{\fill} \\
After extracting the measured phase $\Phi_j(\Nsw,\Esw,\Bsw)$, we performed the basis change described in equation \ref{eq:general_parity}, from this experiment switch \emph{state} basis to the experiment switch \emph{parity} basis, denoted by $\Phi^p_j$, where $p$ is a placeholder for a given experiment switch parity. 

We observed that the molecule beam forward velocity, and hence the spin precession time $\tau_j$, fluctuated by up to 10\% over a 10 minute time period. Since $\B_z$ and $g_1$ are known from auxiliary measurements to a precision of around 1\%, we were able to extract $\tau_j$ from each block from the Zeeman precession phase measurement, $\Phi^\B_j=-\mu_{\rm B}g_1\B_z\tau_j$ (see section~\ref{sec:Measurement_scheme_more_detail}). Velocity dispersion caused $\tau_j$ to vary across the molecule pulse with a nominally linear dependence on time after ablation, $t$, however we observed significant deviations from linearity. Thus, we fit $\tau_j$ to a 3rd order polynomial in $t$ in order to evaluate $\bar{\tau}_j$. Then, we evaluated the measured spin precession frequencies defined as
\begin{equation}
\omega^p_j=\Phi^p_j/\bar{\tau}_j,
\label{eq:omega_def}
\end{equation}
for all phase channels $p$ (see equation \ref{eq:phase_parity} for definition). We extracted the eEDM from $\omega_j^{\N\E}$, which in the absence of systematic errors would be given by $\omega^{\N\E}=-d_e\Eeff$ independent of $j$.

From here on we will drop the $j$ subscript that denotes a grouping of $n$ adjacent asymmetry points about a particular time after ablation $t_j$; it is implicit that independent phase measurements were computed from many separate groups of data, each with different values of $t_j$ across the duration of the molecule pulse. At the end of the analysis, and whenever it was convenient to do so, we implicitly performed weighted averaging across the $j$ subscript.
\begin{figure}[htbp]
\centering
\includegraphics[scale=0.9]{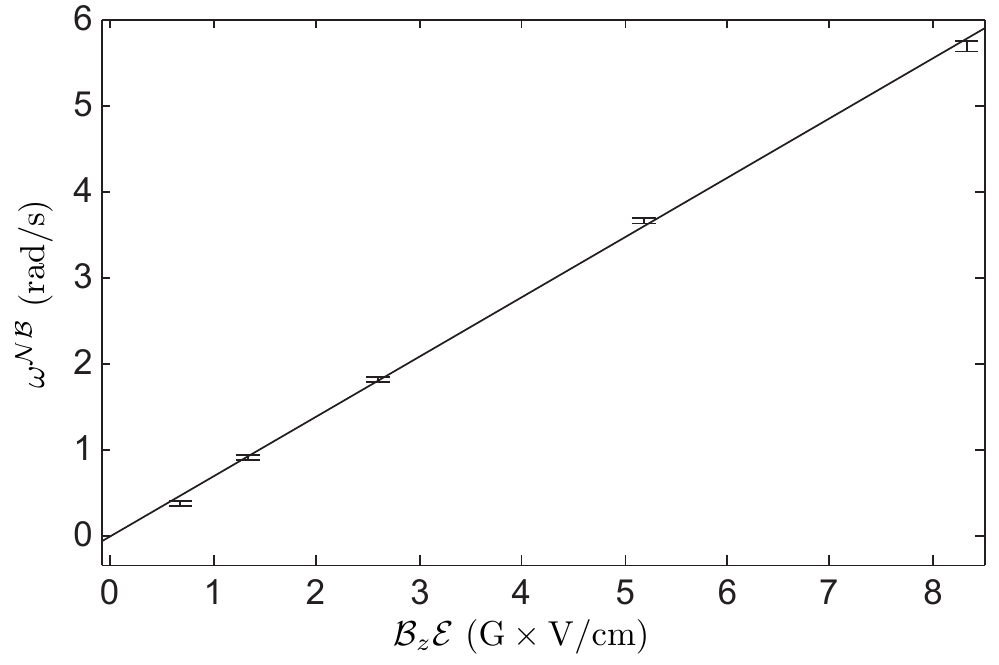}
\caption{The difference between magnetic moments of the two $\Omega$-doublet levels as measured by $\omega^{\N\B}$. As expected, this phase component scales linearly with $\E$ and $\B_z$. The constant of proportionality is $\eta \mu_{\rm B}$. Reproduced with permission from \cite{Petrov2014}}
\label{fig:delta_g}
\end{figure}

Other phase channels couuld be used to search for and monitor systematic errors, discussed in detail in section~\ref{sec:systematics}, or to measure properties of ThO, as is the case with $\omega^{\N\B}$. We discuss the latter case here. This channel provided a measure of $\Delta g$, the magnetic moment difference between upper and lower $\Nsw$-levels, arising from perturbations due to other electronic and rotational states \cite{Bickman2009,Petrov2014}. Because this difference limits the extent to which the $\Nsw$ reversal can suppress certain systematic errors \cite{Vutha2010}, it is an important quantity both in our experiment and in other experiments measuring eEDMs in molecules with $\Omega$-doublet structure \cite{JILAEDM}. Figure~\ref{fig:delta_g} illustrates an observed linear dependence $\Delta g/2=\eta \E$, as predicted \cite{Bickman2009,Petrov2014}. Since $\E$ and $\B_z$ are precisely known from auxiliary measurements, the constant $\eta$ can be directly calculated from our angular frequency measurements:
\begin{equation}
\label{eq:eta_2}
\eta=-\frac{\omega^{\N\B}}{\mu_{\rm B}\E\B_z}.
\end{equation}
Our measured value of $\eta=-0.79\pm0.01~\rm{nm}/\rm{V}$ was approximately half of what one would compute using the methods developed to understand the effect in the PbO molecule \cite{Hamilton2010,Bickman2009}. This discrepancy was subsequently understood as being primarily due to coupling to other fine-structure components in the $^3\Delta$ manifold \cite{Petrov2014,HutzlerThesis}. The $\omega^{\N\B}$ channel illustrates the importance of understanding phase channels besides that corresponding to the eEDM.

\subsection{Data Cuts}
\label{sec:data_cuts}
Three data cuts were applied as part of the analysis: fluorescence rate threshold (see section~\ref{sec:signal_asymmetry}), polarisation bin (see below), and contrast threshold (see below). These cuts made sure that we only used data taken under appropriate experimental conditions (e.g. only when lasers remained locked etc.) and thus ensured a high signal to noise ratio for the data used to extract the eEDM value. We thoroughly investigated how each of these cuts affected the calculated eEDM mean and uncertainty.

As previously mentioned, a fluorescence threshold cut of about $F_{\rm{cut}}=3\times10^5$~s$^{-1}$, was applied to each trace (average of 25 molecule pulses) to ensure that the fluorescence rate would always be larger than the background rate. This threshold was chosen to include the maximum number of asymmetry points in our measurement while also excluding low signal-to-noise asymmetry measurements that would increase the overall eEDM uncertainty, as described below. We also removed entire blocks (complete sets of $\Nsw$,$\Esw$,$\Bsw$,$\tilde{\theta}$) of data from the analysis if 
any of the block's experiment states had $\lesssim0.5$~ms of fluorescence data above $F_{\rm{cut}}$. 

\begin{figure}[!htbp]
\centering
\includegraphics[width=14cm]{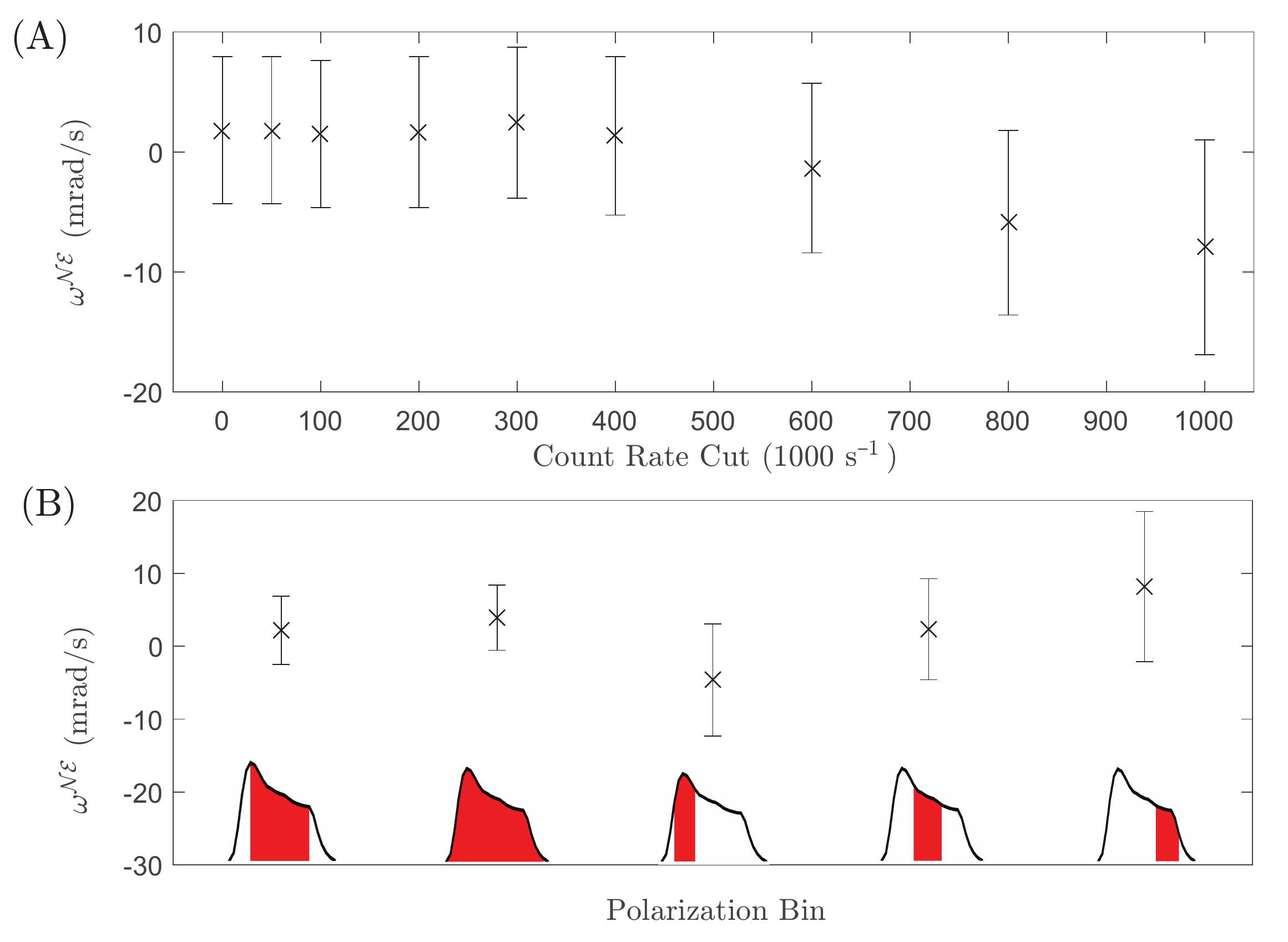}
\caption{Measured eEDM mean and uncertainty as a function of (A) fluorescence signal threshold and (B) polarisation bin size and position. For the former, a value of $3\times10^5$~s$^{-1}$ was used for the final result. For the latter, the two leftmost data points correspond to the polarisation bins used.}
\label{fig:cuts}
\end{figure}

The count rates of uncorrelated fluorescence photoelectrons exhibit Poissonian statistics. In each block we averaged together four traces with the same experimental configuration. After such averaging, the number of detected photoelectrons within a pair of laser polarisation bins was ${\gtrsim}50$, which was large enough that the photoelectron number distribution closely resembled a normal distribution. Because the asymmetry was defined as a ratio of two approximately normally distributed random variables ($F_X-F_Y$ and $F_X+F_Y$), its distribution was not necessarily normal. Rather, it approached a normal distribution in the limit of large $F_X+F_Y$ 
\cite{HutzlerThesis}. The same followed for all quantities computed from the asymmetry, including the eEDM. The fluorescence threshold cut therefore ensured that the distribution of eEDM measurements was very nearly normally distributed. Including low-signal data would have caused the distribution to deviate from normal and increase the overall uncertainty. To check that this signal size cut did not lead to a systematic error in our determination of $d_e$, the eEDM mean and uncertainty were calculated for multiple $F_{\rm{cut}}$ values, as shown in figure~\ref{fig:cuts}. If the cut was increased above $6\times10^5$~s$^{-1}$ the mean value was seen to move slightly (but within the computed uncertainties), and the uncertainty to increase. However, for all plausible values of the cuts the resulting value of $d_e$ was consistent, within uncertainties, with our final stated value.

As described in section~\ref{sec:signal_asymmetry}, data points within a polarisation bin were averaged together when calculating the asymmetry (cf. figure~\ref{fig:modulation}A). These data points were separated by 200~ns. Numbering these points from when the readout laser beam polarisation is switched, we binned points 5--20 or points 0--25, depending on the analysis routine (see section~\ref{ssec:differences_between_data_analysis_routines} below) when reporting our final result. The former choice was made to cut out background signal and overlapping fluorescence between polarisation states while retaining as much of the fluorescence signal as possible whereas the latter was chosen to minimize the statistical uncertainty given the lack of evidence for systematic errors that depended on time within the polarisation switching cycle. As shown in figure~\ref{fig:cuts}, we checked for systematic errors associated with this choice by also using several different polarisation bins to compute the eEDM. The eEDM uncertainty increased, as expected, for polarisation bins that cut out data with significant fluorescence levels, but the mean values were all consistent with each other within their respective
uncertainties.

In order for a block of data to be included in our final measurement, we also required that 
each of the 8 $(\Nsw,\Esw,\Bsw)$ experiment states
had a measured fringe contrast above 80\%. The primary cause of blocks failing to meet this requirement was the state preparation laser becoming unlocked. This cut resulted in less than 1\% of blocks being discarded. If the contrast cut was lowered, or not applied at all, the eEDM mean and uncertainty change by less than 3\% of our statistical uncertainty. As with the signal threshold, if this cut threshold was increased to 90\%, close to the average value of contrast, $\mathcal{C}$, then a larger fraction of data was neglected and the eEDM uncertainty was seen to increase.

For all the cuts discussed, we significantly varied the associated cut and in some cases removed it entirely. The eEDM mean and uncertainty were very robust against significant variation of each of these cuts, and the cuts were chosen before the blind offset applied to the eEDM channel was removed.

\subsection{Differences Between Data Analysis Routines}
\label{ssec:differences_between_data_analysis_routines}
As a systematic error check, we performed three independent analyses of the data. Each routine followed the general analysis method described above, but varied in many small details such as background subtraction method, cut thresholds, numbers of points grouped together to compute asymmetry, polarisation bin choice, etc. The analyses differed in the polynomial order of the fits applied to both the contrast $\mathcal{C}$ and the precession time $\tau$ vs.\ time after ablation $t$. The analyses also differed in the inclusion of a subset of the eEDM data that featured a particularly large unexplained signal in the $\omega^\N$ channel.

Each of the three analyses independently computed the eEDM channel and the systematic error in the eEDM channel. The uncertainties for all three routines were nearly identical, and the means agreed to within $\Delta\omega^{\N\E}<3~\rm{mrad/s}$, which is within the statistical uncertainty of the measurement $\delta\wNE=\pm4.8$~mrad/s. The eEDM mean and uncertainty were averaged over the three analyses to produce the final result.

\subsection{EDM Mean and Statistical Uncertainty}
\begin{figure}[!htbp]
\centering
\includegraphics[width=15.5cm]{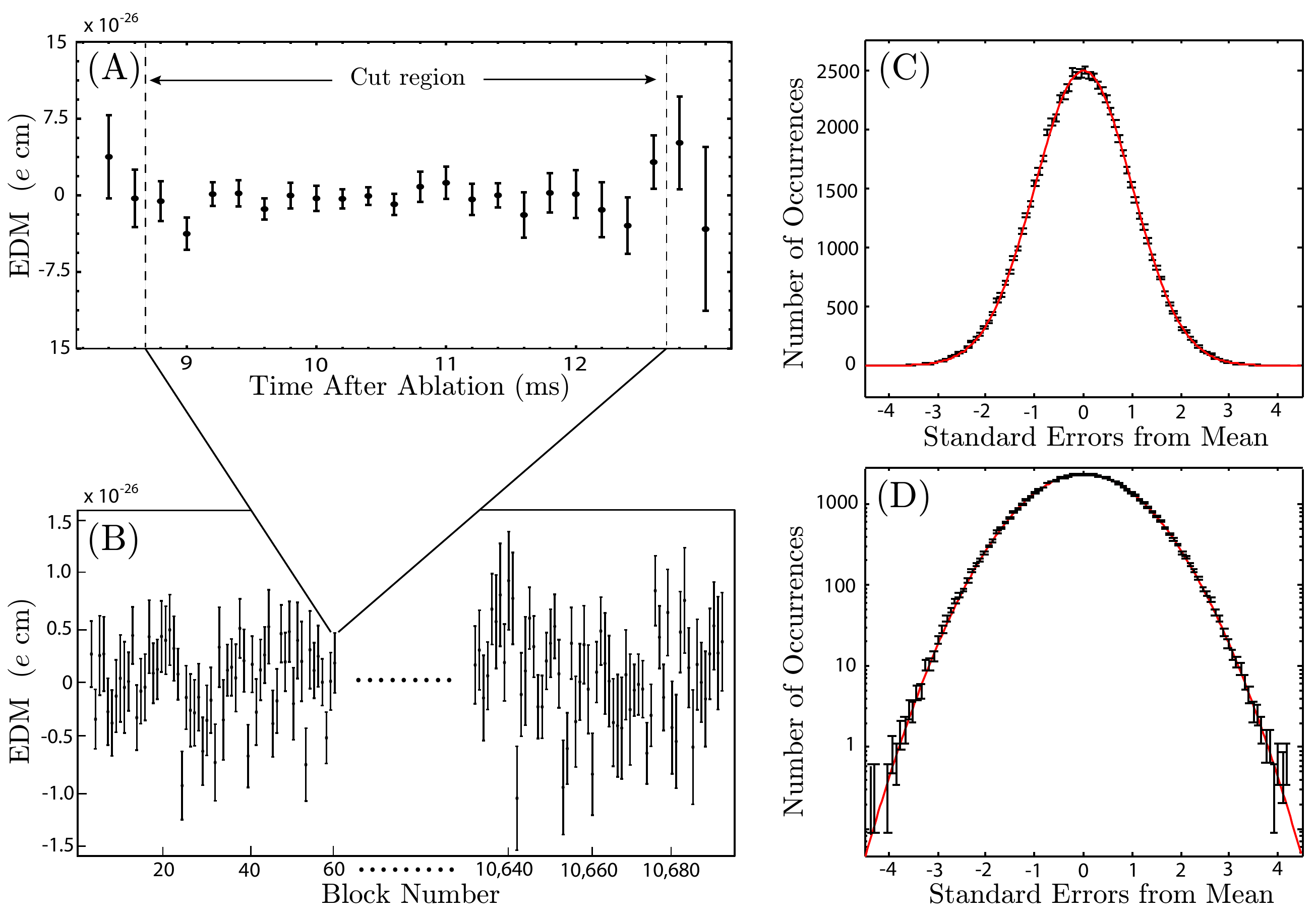}
\caption{The data set associated with our reported eEDM limit. \textbf{(A)} Variations in the extracted eEDM as a function of position within the molecular pulse. \textbf{(B)} Over 10,000 blocks of data were taken over a combined period of about two weeks. \textbf{(C)}-\textbf{(D)} The distribution of $\sim$200,000 separate eEDM measurements (black) matches very well with a Gaussian fit (red). The same data is plotted with both a linear and a log scale. In these histograms the mean of each individual measurement was normalized to its corresponding error bar.}
\label{fig:statistics}
\end{figure}
The final data set used to report our result is shown in figure~\ref{fig:statistics}. It consisted of ${\sim}10^4$ blocks of data taken over the course of $\sim$2~weeks (figure~\ref{fig:statistics}B); each block contains ${\approx}20$ separate eEDM measurements distributed over the duration of the molecule pulse (Figure~\ref{fig:statistics}A). All ${\approx}2\times10^5$~measurements were combined with standard Gaussian error propagation to obtain the reported mean and uncertainty. Figure~\ref{fig:statistics}C,D shows histograms of all measurements on a linear (C) and log (D) scale, showing the distribution agrees extremely well with a Gaussian fit. The resulting uncertainty was about 1.2 times that expected from the photoelectron shot-noise limit, taking into account the photoelectron rate from molecule fluorescence, background light, and PMT dark current. When the eEDM measurements were fit to a constant value, the reduced $\chi^2$ was $0.996\pm0.006$ where this uncertainty represents the $1\sigma$ width of the $\chi^2$ distribution for the appropriate number of degrees of freedom.
\begin{figure}[htbp]
\centering
\includegraphics[width=10cm]{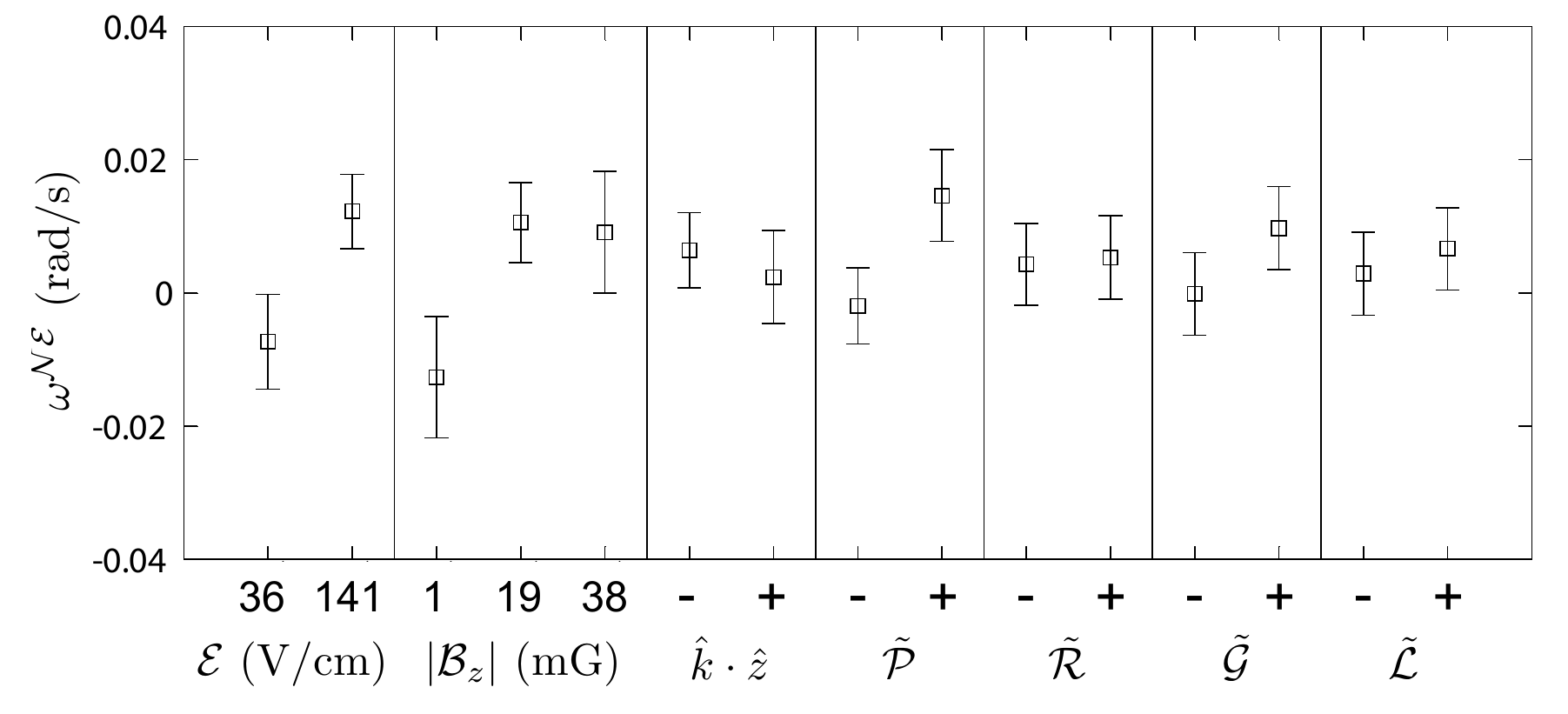}
\caption{Measured $\wNE$ values grouped by the states of $\left|\mathcal{B}_{z}\right|,$~$\left|\mathcal{E}_{z}\right|$, $\hat{k}\cdot\hat{z}$, and each superblock switch, before systematic error corrections. Reproduced with permission from \cite{Baron2014}.}
\label{fig:edm_vs_superblock}
\end{figure}

When computing the eEDM result, data from superblocks were averaged together. The mean could be either weighted or unweighted by the statistical uncertainty in each superblock state. Weighted averaging minimized the resulting statistical uncertainty, but unweighted averaging could  suppress systematic errors that have well-defined superblock parity from entering into the extracted value for $\omega^{\N\E}$.

Due to molecule number fluctuations, each block of data had a different associated uncertainty. However, roughly equal amounts of data were gathered for the $2^4$ superblock states defined by the state readout parity $\Psw$, field plate lead configuration $\Lsw$, state readout laser polarisation $\Rsw$, and global laser polarisation $\Gsw$. For the reported eEDM value, unweighted averaging (or to be precise, performing the basis change prescribed by equation \ref{eq:general_parity}) was used to combine data from the different $\Psw$, $\Rsw$, $\Lsw$, $\Gsw$ experiment states, since there were known systematic errors with well-defined superblock parity that were suppressed by these switches (see, for example, sections \ref{sssec:stark_interference_between_E1_and_M1_transition_amplitudes} and \ref{ssec:asymmetry_effects}). Note, however, that figure~\ref{fig:edm_vs_superblock} shows that these systematic errors produced no significant eEDM shift, and that the overall uncertainty was comparable (within 10\%) when the data was combined with weighted or unweighted averaging.

Unequal amounts of data were collected for the $\B_z$, $\E$, and $\hat{k}\cdot\hat{z}$ experimental states. For example, 40\% (60\%) of data were gathered with the state preparation and readout laser beams pointing east (west), $\hat{k}\cdot\hat{z}=-1 (+1)$. To account for this, we performed state-by-state analysis of the systematic errors: the primary systematic errors (described in section \ref{sssec:correlated_laser_parameters}) were allowed to depend on the magnitude of the magnetic field (though $\B_z=1,40$ mG were grouped together), and the pointing direction, and separate systematic error subtractions were performed for each ($\B_z$, $\hat{k}\cdot\hat{z}$) state. After this subtraction, the systematic uncertainties were added in quadrature with the statistical uncertainties for each state, and the data from each state was averaged together weighted by the resulting combined statistical and systematic uncertainties.

The reported statistical uncertainty was obtained via the method above assuming no systematic uncertainty. The reported systematic uncertainty was defined such that the quadrature sum of the reported statistical and systematic uncertainties gives the same value as when incorporating the state-by-state analysis. A description of the methods used to evaluate the systematic error and the systematic uncertainty in the measurement is provided in section \ref{ssec:total_systematic_error_budget}.
To prevent experimental bias we performed a blind analysis by adding an unknown offset to the mean of the eEDM channel, $\wNE$. The offset was randomly generated in software from a Gaussian distribution with standard deviation $\sigma=150$~mrad/s and mean zero. The mean, statistical error, procedure for calculating the systematic error, and procedure for computing the reported confidence interval were all determined before revealing and subtracting the blind offset.

\section{Systematic Errors}
 \label{sec:systematics}
According to Sozzi \cite{sozzi2008}, `The best way to handle systematic effects is not to have any...'. We approached our experimental design in a way to be very resilient to the systematic errors that impacted eEDM experiments in the past. We performed searches for unforeseen systematic errors, observed some, developed models to understand them, and carefully quantified them in auxiliary measurements, as described in this section. 

A true eEDM should contribute to the $\Nsw\Esw$-correlated spin precession frequency, $\wNE$, with a signal $\wNEt$ that does not vary with any experimental parameter. To discriminate between a systematic error in $\wNE$ and $\wNEt$, we pursued a strategy to vary a large number of experimental parameters and imperfections while closely monitoring $\wNE$. If $\wNE$ changes then there must be a systematic error correlated with that varied parameter. During our search for systematic errors we varied parameters including: applied electric and magnetic fields; magnetic field gradients; molecule beam pointing; and laser beam shape, pointing, detuning, and polarisation. In addition to monitoring $\wNE$, we monitored the spin precession frequency, contrast, fluorescence signal, and a number of additional experimental conditions such as molecule beam velocity, vacuum pressure and room temperature. We examined the correlations of these quantities with the experiment switches to determine whether there are any spurious signals that might point to unforeseen systematic errors, or a gap in our understanding of the experiment \cite{Hess2014}.

\subsection{Determining Systematic Errors and Uncertainties}
\label{ssec:determining_systematic_uncertainty}

\begin{table}[tbp]
\caption{\label{tbl:syst_check}Parameters varied during our systematic error search. Left: Category I Parameters --- These were ideally zero under normal experimental running conditions and we were able to vary them significantly from zero. For each of these parameters direct measurements or limits were placed on possible systematic errors. Right: Category II Parameters --- These had no single ideal value. Although direct limits on these systematic errors could not be derived, they served as checks for the presence of unanticipated systematic errors. See the main text for more details on all the systematic errors referenced.}
\begin{minipage}[t]{0.5\textwidth}
\begin{tabular}[t]{l}
\br
Category I Parameters\\
\mr
\textbf{Magnetic Fields}\\
- Non-reversing $\B$-field: $\B_{z}^{\nr}$\\
- Transverse $\B$-fields: $\B_{x},\B_{y}$\\
(both even and odd under $\Bsw$)\\
- $\B$-field gradients: \\
$\frac{\partial\B_{x}}{\partial x},\frac{\partial\B_{y}}{\partial x},\frac{\partial\B_{y}}{\partial y},\frac{\partial\B_{y}}{\partial z},\frac{\partial\B_{z}}{\partial x},\frac{\partial\B_{z}}{\partial z}$\\
(both even and odd under $\Bsw$)\\
- $\Esw$ correlated $\B$-field: $\B^{\E}$
(to simulate\\
$\vec{v}\times\vecE$/geometric phase/leakage current)\\
\textbf{Electric Fields}\\
- Non-reversing $\E$-field: $\E^{\nr}$\\
- $\E$-field ground offset\\
\textbf{Laser Detunings}\\
- State preparation/readout lasers: $\Delta_{\p}^{\rm nr}$, $\Delta_{\ro}^{\rm nr}$\\
- $\Psw$ correlated detuning, $\Delta^{\mathcal{P}}$\\
- $\Nsw$ correlated detunings: $\Delta^{\N}$\\
\textbf{Laser Pointings}\\
- Change in pointing of prep./read lasers\\
- State readout laser $\hat{X}/\hat{Y}$ dependent pointing\\
- $\Nsw$ correlated laser pointing\\
- $\Nsw$ and $\hat{X}/\hat{Y}$ dependent laser pointing\\
\textbf{Laser Powers}\\
- $\Nsw\Esw$ correlated power $\rabi^{\N\E}$\\
- $\Nsw$ correlated power $\rabi^{\N}$\\
- $\hat{X}/\hat{Y}$ dependent state readout laser power, $\rabi^{XY}$\\
\textbf{Laser Polarisation}\\
- Preparation laser ellipticity, $S_{\p}$\\
\textbf{Molecular Beam Clipping}\\
- Molecule beam clipping along $\hat{y}$ and $\hat{z}$\\
(changes $\left\langle v_{y}\right\rangle $,$\left\langle v_{z}\right\rangle $,$\left\langle y\right\rangle $,$\left\langle z\right\rangle $
of molecule beam)\\
\br
\end{tabular}
\end{minipage}
\begin{minipage}[t]{0.5\textwidth}
\begin{tabular}[t]{l}
\br
Category II Parameters \\
\mr
\textbf{Laser Powers}\\
- Power of prep./read lasers \\
\textbf{Experiment Timing}\\
- $\hat{X}$/$\hat{Y}$ polarisation switching rate\\
- Number of molecule pulses averaged \\
per experiment trace\\
\textbf{Analysis}\\
- Signal size cuts, asymmetry size cuts,\\
contrast cuts\\
- Difference between two PMT detectors\\
- Variation with time within molecule pulse\\
(serves to check $v_{x}$ dependence)\\
- Variation with time within polarisation \\
switching cycle\\
- Variation with time throughout the \\
full data set (autocorrelation)\\
- Search for correlations between all channels \\
of phase, contrast and fluorescence signal \\
- Correlations with auxiliary measurements\\
of $\B$-fields, laser powers, vacuum pressure\\
and temperature\\
- 3 independent data analysis routines\\
\br
\end{tabular}
\end{minipage}
\end{table}

In total, we varied more than 40 separate parameters during our search for systematic errors (see Table \ref{tbl:syst_check}). These fall into two categories.
Category I contains parameters $P$ which are optimally zero; $P\neq0$ represents an experimental imperfection. We were able to use experimental data to put a direct limit on the size of possible systematic errors proportional to these parameters. Category II contains parameters that have no optimum value and which we could vary significantly without affecting the nature of the spin precession measurement. The variation of these parameters could reveal systematic errors and serve as a check that we understood the response of our system to those parameters, but no quantitative bounds on the associated systematic errors were derived. 

For each Category I parameter $P$, we exaggerated the size of the imperfection by a factor greater than $10$, if possible, relative to the maximum size of the imperfection under normal operating conditions, $\bar{P}$, which was obtained from auxiliary measurements. Following previous work \cite{Regan2002,Griffith2009,Hudson2011}, we assumed a linear relationship between $\wNE$ and $P$, and extracted the sensitivity of the $\wNE$ to parameter $P$, $\partial\wNE/\partial P$. The systematic error under normal operating conditions was computed as $\omega_P^{\N\E}=(\partial\omega^{\N\E}/\partial P)\bar{P}$. The statistical uncertainty in the systematic error (henceforth referred to as the systematic uncertainty) $\delta\omega_P^{\N\E}$ was obtained from linear error propagation of uncorrelated random variables,
\begin{equation}
\delta\omega_P^{\N\E}= \sqrt{\left(\frac{\partial\omega^{\N\E}}{\partial P}^{\:}\delta\bar{P}\right)^{2}+\left(\bar{P}^{\:}\delta\frac{\partial\omega^{\N\E}}{\partial P}\right)^{2}},
\end{equation}
where $\delta\bar{P}$ is the uncertainty in $\bar{P}$ and $\delta\partial\omega^{\N\E}/\partial P$ is the uncertainty in $\partial\omega^{\N\E}/\partial P$. 

For parameters that had been observed to produce statistically significant shifts in $\wNE$, such as the non-reversing electric field, $\E^{\nr}$, we monitored the size of the systematic error throughout the reported data set during \textit{Intentional Parameter Variations} (described in section \ref{sec:Measurement_scheme_more_detail}) and deducted this quantity from $\wNE$ to give a value of the spin precession frequency due to T-odd interactions in the H state of ThO, $\wNEt=\omega^{\N\E}-\sum_{P}\omega_P^{\N\E}$. Most Category I parameters did not cause a statistically significant $\omega^{\N\E}_P$ and were not monitored. For these parameters, we did not subtract $\omega_P^{\N\E}$ from $\wNE$, but rather included an upper limit of $\left[(\omega_P^{\N\E})^{2}+(\delta\omega_P^{\N\E})^{2}\right]^{1/2}$ in the systematic uncertainty on $\wNEt$, or chose to omit this parameter from the systematic error budget altogether based on the criteria described in section \ref{ssec:total_systematic_error_budget}.

Where applicable, we also fit higher-order polynomial functions to $\wNE$ with respect to $P$ during the systematic error searches. No significant increase in the systematic uncertainty was observed using such fits and hence the contributions to the systematic error budget in Table \ref{tbl:syst_error} were all estimated from linear fits. We note, however, that certain non-linear dependences of $\wNE$ on $P$ could lead to underestimates of the systematic uncertainty, for example if $\wNE$ has a small (large) nonzero value for large (small) values of $P$. In efforts to avoid this, data were taken over as wide a range as possible, it is, however, always possible that such non-linear dependence is present between the parameter values for which we took data. We had no models by which non-linear dependence could manifest by variation of the parameters investigated, so we believe the procedure outlined above produced accurate estimates of the systematic errors.


\subsection{Systematic Errors Due to Imperfect Laser Polarisations}
\label{ssec:systematic_errors_due_to_imperfect_laser_polarizations}

The dominant systematic errors in our experiment were due to imperfections in the laser beams used to prepare the molecular and read out the molecular state. Non-ideal laser polarisations combined with laser parameters correlated with the expected eEDM signal resulted in three distinct systematic errors which we refer to as the $\E^{\nr}$, $\rabi^{\N\E}$, and Stark Interference (S.I.) systematic errors. In this section, we model the effects of several types of polarisation imperfections on the measured phase $\Phi$ (sections~\ref{sssec:stark_interference_between_E1_and_M1_transition_amplitudes} and \ref{sssec:AC_stark_shift_phases}) and discuss the correlated laser parameters that couple to these polarisation imperfections to result in systematic errors (section~\ref{sssec:correlated_laser_parameters}). We then discuss how we were able to suppress and quantify the residual systematic errors in the eEDM experiment (sections~\ref{sssec:suppression_of_the_AC_stark_shift_phases} and \ref{sssec:correlated_laser_parameters}).

\subsubsection{Idealized Measurement Scheme with Polarisation Offsets}
\label{sssec:idealized_measurement_scheme_with_polarization_offsets}

As described in section \ref{sec:Measurement_scheme}, the molecules initially enter the state preparation laser beam in an incoherent mixture of the two states $|\pm,\Nsw\rangle$. The bright state $|B(\hat{\epsilon}_{\p},\Nsw,\Psw_{\p})\rangle$ is then optically pumped away through $|C,\Psw_{\p}\rangle$ leaving behind the dark state $|D(\hat{\epsilon}_{\p},\Nsw,\Psw_{\p})\rangle$ as the initial state for the spin precession. The molecules then undergo spin precession by angle $\phi$ evolving to a final state $|\psi_{f}\rangle=U(\phi)|D(\hat{\epsilon}_{\p},\Nsw,\Psw_{\p})\rangle$ where $U(\phi)=\sum_{\pm}e^{\mp i\phi}|\pm,\Nsw\rangle\langle\pm,\Nsw|$ is the spin precession operator. The molecules then enter the state readout laser that optically pumps the molecules with alternating polarisations $\hat{\epsilon}_{\rX}$ and $\hat{\epsilon}_{\rY}$
(which are nominally linearly polarised and orthogonal) between $|\pm,\Nsw\rangle$ and $|C,\Psw_{\ro}\rangle$. For each polarisation, the optical pumping results in a fluorescence count rate proportional to the projection of the state onto the bright state, $F_{\rX,\rY}=fN_{0}|\langle B(\hat{\epsilon}_{\rX,\rY},\Nsw,\Psw_{\ro})|\psi_{f}\rangle|^{2}$ where $f$ is the photon detection efficiency, and $N_{0}$ is the number of molecules in the addressed $\Nsw$ level. We then compute the asymmetry, $\mathcal{A}=(F_{\rX}-F_{\rY})/(F_{\rX}+F_{\rY})$, dither the linear polarisation angles in the state readout laser beams to evaluate the fringe contrast, $\mathcal{C}=(\partial\mathcal{A}/\partial\phi)/2\approx-(\partial\mathcal{A}/\partial\theta_{\ro})/2$, and extract the measured phase, $\Phi=\mathcal{A}/(2\mathcal{C})+q\pi/4$.\footnote{Recall $q$ is chosen to be an integer which depends on the size of the applied magnetic field.} We then report the result of the measurement in terms of an equivalent phase precession frequency $\omega=\Phi/\tau$ where $\tau\approx1^{\:}\mathrm{ms}$ is the spin precession time, which was measured for each block as described in section~\ref{sec:compute_phase}.

Let us first consider the idealized case in which all laser polarisations are exactly linear, $\Theta_{i}=\pi/4$ for each laser $i\in\left\{\p,\rX,\rY \right\}$, the angle between the state preparation laser polarisation ($\p$) and state readout basis ($\rX,\rY$) is $\pi/4$, $\theta_{\ro}-\theta_{\p}=-\pi/4$, and the accumulated phase is small, $\left|\phi\right|\ll1$ (i.e. no magnetic field is applied). Under these conditions, the measured phase $\Phi$ is equal to the accumulated phase $\phi$. Now consider the effect of adding polarisation offsets $d\vec{\epsilon}_{i}$ to each of the three laser beams such that $\hat{\epsilon}_{i}\rightarrow\hat{\epsilon}_{i}+\kappa d\vec{\epsilon}_{i}$, where $\kappa = 1$ is a perturbation parameter. It is useful to cast the polarisation imperfections in terms of linear angle imperfections, $\theta_{i}\rightarrow\theta_{i}+\kappa d\theta_{i}$ and ellipticity imperfections, $\Theta_{i}\rightarrow\Theta_{i}+\kappa d\Theta_{i}$ where $S_{i}=-2d\Theta_{i}$ is the laser ellipticity Stokes parameter; these are related by
\begin{equation}
\frac{\hat{z}\cdot(\hat{\epsilon}_{i}\times d\vec{\epsilon}_{i})}{\hat{\epsilon}_{i}\cdot\hat{\epsilon}_{i}}=d\theta_{i}-id\Theta_{i}.\label{eq:extracting_polarization_imperfection_components}
\end{equation}
Note that laser polarisations can have a nonzero projection in the $\hat{z}$ direction, but we assume in the discussion above that $\hat{\epsilon}_{i}$ represents a normalized projection of the laser polarisation onto the $xy$ plane.\footnote{The $z$-component of the polarisation can only drive $\Delta M=0$ transitions, which are far off resonance from the state preparation/readout lasers.} With these polarisation imperfections in place, the measured phase $\Phi$ gains additional terms:
\begin{equation}
\Phi=\phi+\kappa(d\theta_{\p}-\frac{1}{2}(d\theta_{\rX}+d\theta_{\rY}))-\kappa^{2}\Psw_{\p}\Psw_{\ro}d\Theta_{\p}(d\Theta_{\rX}-d\Theta_{\rY})+O\left(\kappa^{3}\right),\label{eq:Measured_Phase_with_Polarization_Imperfections}
\end{equation}
up to second order in $\kappa$. In the eEDM measurement, we switch between two values of $\Psw\equiv\Psw_{\ro}$, the parity of the excited state addressed during state readout, and we set $\Psw_{\p}=+1$, the parity of the excited state addressed during state preparation. It is worth dwelling on equation \ref{eq:Measured_Phase_with_Polarization_Imperfections} for a moment. A rotation of all polarisations by the same angle leaves the measured phase unchanged: $d\theta_{i}\rightarrow d\theta_{i}+d\theta\implies\Phi\rightarrow\Phi$, as expected. A deviation in the relative angle between the state preparation and readout beams, $d\theta_{\p}\rightarrow d\theta_{\p}+d\theta$ and $d\theta_{\rX,\rY}\rightarrow d\theta_{\rX,\rY}-d\theta$, enters into the phase measurement as $\Phi\rightarrow\Phi+2d\theta$, but is benign so long as $d\theta$ is uncorrelated with the expected eEDM signal. The laser ellipticities affect the
phase measurement only when the state readout beams differ in ellipticity, and this contribution to the phase can be distinguished from the others by switching the excited state parity, $\Psw$. This last term is particularly interesting because it allows for multiplicative couplings between polarisation imperfections in the state preparation and state readout beams to contribute to the measured phase.

Although the polarisation imperfection terms in equation \ref{eq:Measured_Phase_with_Polarization_Imperfections} are uncorrelated with the $\Nsw\Esw$ and hence do not contribute to the systematic error, we will see in later sections that additional imperfections can lead to changes in the molecule state that is prepared or read-out that are equivalent to correlations $d\theta_{i}^{\N\E}$ and $d\Theta_{i}^{\N\E}$. The framework of equation \ref{eq:Measured_Phase_with_Polarization_Imperfections} is useful for understanding how these correlations result in systematic errors in the eEDM measurement extracted from $\Phi^{\N\E}$.

\subsubsection{Stark interference between E1 and M1 transition amplitudes}
\label{sssec:stark_interference_between_E1_and_M1_transition_amplitudes}

In this section we describe in detail how interference between multipole transition amplitudes can lead to a measured phase that mimicks an eEDM spin precession phase. We develop a general framework illustrating how such phases depend on laser polarisation and pointing.

In an applied electric field, opposite parity levels are mixed, allowing both odd parity (E1, M2,...) and even parity (M1, E2,...) electromagnetic multipole amplitudes to contribute when driving an optical transition. These amplitudes depend on the orientation of the electric field relative to the light polarisation $\hat{\epsilon}$ and the laser pointing direction $\hat{k}$. This Stark interference (S.I.) effect forms the basis of precise measurements of weak interactions through parity non-conserving amplitudes in atoms and molecules \cite{Bouchiat1974,Demille2008,Wood1997}. However, it can also generate a systematic error in searches for permanent electric dipole moments which look for spin precession correlated with the orientation of an applied electric field. These Stark interference amplitudes have been calculated and measured for optical transitions in Rb \cite{Chen1994,Hodgdon1991} and Hg \cite{Lamoreaux1992,Loftus2011}, and have been included in the systematic error analysis in the Hg EDM experiment \cite{Griffith2009,Swallows2013}.

In this section, we consider Stark interference as a source of systematic errors in the ACME experiment. There are two important differences between molecular and atomic systems. First, molecular states such as the $H^{3}\Delta_{1}$ state in ThO can be highly polarisable and opposite parity states can be completely mixed by the application of a modest laboratory electric field. Second, molecular selection rules can be much weaker than atomic selection rules: the $H^{3}\Delta_{1}\rightarrow C^{3}\Pi_{1}$ transition that we drive is nominally an E1 forbidden spin-flip transition ($\Delta\Sigma=1$, where $\Sigma$ is the projection of the total electron spin $S=1$ onto the intermolecular axis), 
but these states have significant subdominant contributions from other spin-orbit terms \cite{Paulovic2003}, between some of which the E1 transition is allowed. Both of these effects significantly amplify the effect of Stark interference in molecules relative to atoms. In this section we will derive the effect of Stark interference on the measured phase $\Phi$.

Consider a plane wave vector potential $\vec{A}$ with real amplitude $A_{0}$, oscillating at frequency $\omega$, that is resonant with a molecular optical transition $\left|g\right\rangle \rightarrow\left|e\right\rangle$,with wave vector $\vec{k}=\left(\omega/c\right)\hat{k}$, and complex polarisation $\hat{\epsilon}$:
\begin{align}
\vec{A}\left(\vec{r},t\right)= & A_{0}\hat{\epsilon}e^{i\vec{k}\cdot\vec{r}-i\omega t}+\mathrm{c.c}.
\end{align}
The interaction Hamiltonian $H_{\mathrm{int}}$ between this classical light field and the molecular system is given by:
\begin{align}
H_{\mathrm{int}}\left(t\right)= & -\sum_{a}\frac{e^{a}}{m^{a}}\vec{A}\left(\vec{r}^{\,a},t\right)\cdot\vec{p}^{a}
\end{align}
where $a$ indexes a sum over all of the particles in the system with charge $e^{a}$, mass $m^{a}$, position $\vec{r}^{\,a}$ and momentum $\vec{p}^{a}$. Typically we apply the multipole expansion on the transition matrix element between states $\left|g\right\rangle $ and $\left|e\right\rangle$; the matrix element can then be written as 
\begin{equation}
\mathcal{M}\equiv\langle e|H_{\mathrm{int}}|g\rangle=iA_{0}\omega_{eg}\sum_{\lambda=1}^{\infty}\langle e|\hat{\epsilon}\cdot\vec{E}_{\lambda}+(\hat{k}\times\hat{\epsilon})\cdot\vec{M}_{\lambda}|g\rangle,
\end{equation}
where $\vec{E}_{\lambda}$ describes the electric interaction of order $O((\vec{k}\cdot\vec{r})^{\lambda-1})$ and $\vec{M}_{\lambda}$ describes the magnetic interaction of order $O(\alpha(\vec{k}\cdot\vec{r})^{\lambda-1})$
(where $\alpha$ is the fine structure constant) such that
\begin{align}
\vec{E}_{\lambda}= & \frac{\left(i\right)^{\lambda-1}}{\lambda!}\sum_{a}e^{a}\vec{r}^{\,a}\left(\vec{k}\cdot\vec{r}^{\,a}\right)^{\lambda-1},\\
\vec{M}_{\lambda}= & \frac{\left(i\right)^{\lambda-1}}{\left(\lambda-1\right)!}\sum_{a}\left(\frac{e^{a}}{2m^{a}}\right)\left[\left(\vec{k}\cdot\vec{r}^{\,a}\right)^{\lambda-1}\left(\frac{1}{\lambda+1}\vec{L}^{a}+\frac{1}{2}g^{a}\vec{S}^{a}\right)+\left(\frac{1}{\lambda+1}\vec{L}^{a}+\frac{1}{2}g^{a}\vec{S}^{a}\right)\left(\vec{k}\cdot\vec{r}^{\,a}\right)^{\lambda-1}\right],\nonumber 
\end{align}
where $L^{a}$ is the orbital angular momentum, $S^{a}$ is the spin angular momentum, and $g^{a}$ is the spin g-factor for particle of index $a$ (see e.g.\ \cite{Sachs1987}). For typical atomic or molecular optical transitions, if all moments are allowed, we expect the dominant
corrections to the leading order E1 transition moment to be on the order of M1/E1 $\sim\alpha\sim10^{-2}$--$10^{-3}$ and E2/E1$\sim ka_{0}\sim10^{-3}$--$10^{-4}$, where $a_0$ is the Bohr radius. In this work we neglect the higher order contributions beyond E2, though the effects may by evaluated by using the expansion
above.


\begin{center}
\begin{table}
\centering
\begin{tabular}{cccc}
\hline 
\multicolumn{4}{c}{}\tabularnewline
\multicolumn{4}{c}{$\left\langle e\left|H_{\mathrm{int}}\left(O^{\lambda}\right)\right|g\right\rangle =iA_{0}\omega_{eg}\left[\hat{\epsilon}_{+1}^{*}\left\langle e\left|T_{+1}^{\lambda}\left(O^{\lambda}\right)\right|g\right\rangle +\hat{\epsilon}_{-1}^{*}\left\langle e\left|T_{-1}^{\lambda}\left(O^{\lambda}\right)\right|g\right\rangle \right]\cdot\vec{V}\left(O^{\lambda}\right)+\dots$}\tabularnewline
\tabularnewline
\hline 
\multirow{2}{*}{Term} & Tensor & \multirow{2}{*}{Molecular Operator, $O^{\lambda}$} & \multirow{2}{*}{Light Vector, $\vec{V}\left(O^{\lambda}\right)$}\tabularnewline
 & rank, $\lambda$ &  & \tabularnewline
\hline 
E1 & 1 & $\Sigma_{a}e^{a}r_{i}^{a}$ & $\hat{\epsilon}$\tabularnewline
M1 & 1 & $\Sigma_{a}\frac{e^{a}}{2m^{a}}\left(L_{i}^{a}+g^{a}S_{i}^{a}\right)$ & $\hat{k}\times\hat{\epsilon}$\tabularnewline
E2 & 2 & $\frac{\omega}{2c}\sum_{a}e^{a}r_{i}^{a}r_{j}^{a}$ & $\frac{i}{\sqrt{2}}\left[\hat{\epsilon}(\hat{k}\cdot\hat{z})+\hat{k}(\hat{\epsilon}\cdot\hat{z})\right]$\tabularnewline
M2 & 2 & $\frac{\omega}{c}\Sigma_{a}\frac{e^{a}}{2m^{a}}\left\{ r_{i}^{a},\frac{1}{3}L_{j}^{a}+\frac{1}{2}g^{a}S_{j}\right\} $ & $\frac{i}{\sqrt{2}}\left[\hat{k}((\hat{k}\times\hat{\epsilon})\cdot\hat{z})+(\hat{k}\times\hat{\epsilon})(\hat{k}\cdot\hat{z})\right]$\tabularnewline
\hline 
\end{tabular}
\par
\protect\caption{Only spherical tensor operators $T_{q}^{\lambda}$ with projection $q=\pm1$ contribute to the $\left|H\right\rangle \rightarrow\left|C\right\rangle$ transition amplitude. With this simplifying assumption, we can write the matrix element for each multipole operator in the form shown at the top of this table, which factors the molecule properties and the light properties (where $\hat{\epsilon}_{\pm}=\mp\left(\hat{x}\pm i\hat{y}\right)/\sqrt{2}$ are the spherical basis vectors, and $\hat{z}$ is the direction of the electric field). Here, the molecular operators $O^{\lambda}$ and the corresponding light vectors $\vec{V}\left(O^{\lambda}\right)$ are listed for the E1, M1, E2, and M2 operators.}
\label{tab:spher_tens}
\end{table}

\par\end{center}

During the state preparation and readout of the molecule state, transitions are driven between the state $|g\rangle=\sum_{\pm}d_{\pm}|\pm,\Nsw\rangle$ and $|e\rangle=|C,\Psw\rangle$, where $d_{\pm}$ are state amplitudes that denote the particular superposition in $\left|H\right\rangle $ that is being interrogated. The particular $d_{\pm}$ combination that results in $\mathcal{M}=0$ describes the state that is dark, and the orthogonal state is bright and is optically pumped away.

It is convenient to expand the Hamiltonian $H_{\mathrm{int}}$ in terms of spherical tensor operators. Furthermore, the laser is only resonant with $\Delta M=\pm1$ transitions, so the spherical tensor
operators with angular momentum projections other than $\pm1$ can be reasonably omitted. In table \ref{tab:spher_tens}, we factor the first 4 multipole operators into products of molecule and light field operators and express the molecular operators in terms of spherical tensors $T_{\pm1}^{\lambda}$ of rank $\lambda=1,2$. The E1 and M1 terms consist of vector operators with $\lambda=1$. The E2 and M2 operators are rank 2 cartesian operators which can have spherical tensor operator contributions for $\lambda=0,1,2$. The rank $\lambda=0$ components of the E2 and M2 operators, and the $\lambda=1$ component of the E2 operator, vanish. The rank $\lambda=1$ component of the M2 operator does not vanish, but the light field angular dependence of this operator is equivalent to E1, so we may treat it as such.

Using well-known properties of angular momentum matrix elements \cite{Brown2003}, we may write the transition matrix element in the following form,
\begin{align}
\mathcal{M}= & iA_{0}\omega_{eg}c_{\rm E1}\frac{1}{\sqrt{2}}\left[\left(-1\right)^{J+1}\Psw\right]^{(1-\mathcal{\tilde{N}}\Esw)/2}\left(\hat{\epsilon}_{-1}^{*}d_{+}+\Psw\left(-1\right)^{J'}\hat{\epsilon}_{+1}^{*}d_{-}\right)\cdot\vec{\varepsilon}_{\mathrm{eff}},
\end{align}
such that $\vec{\varepsilon}_{\mathrm{eff}}$ is the `effective E1 polarisation' (i.e. including the effects of interference between multipole transition matrix elements is equivalent to an E1 transition with this polarisation) with the form
\begin{align}
\vec{\varepsilon}_{\mathrm{eff}}=& \hat{\epsilon}-a_{\rm M1}i\hat{n}\times(\hat{k}\times\hat{\epsilon})+a_{\rm E2}(\Psw)i(\hat{k}(\hat{\epsilon}\cdot\hat{n})+\hat{\epsilon}(\hat{k}\cdot\hat{n}))+\dots\label{eq:effective E1 polarization}
\end{align}
where $\hat{n}=\Nsw\Esw\hat{z}$ is the orientation of the internuclear axis in the laboratory frame,  $a_{\rm E2}(\Psw)=c_{\rm E2}(\Psw)/(\sqrt{2}c_{\rm E1})$ and $a_{\rm M1}=c_{\rm M1}/c_{\rm E1}$ are real dimensionless ratios describing the strength of the M1 and E2 matrix elements relative to E1, and the $c$ coefficients are matrix elements,
\begin{align}
c_{\rm E1}= & \left\langle C,J,0,1\left|\mathrm{E1}\right|H,J',1,1\right\rangle \\
c_{\rm M1}= & \left\langle C,J,0,1\left|\mathrm{M1}\right|H,J',1,1\right\rangle \\
c_{\rm E2}(\Psw)= & \left\langle C,J,0,1\left|\mathrm{E2}\right|H,J',1,1\right\rangle +\nonumber \\
 & \Psw\left(-1\right)^{J}\left\langle C,J,0,1\left|\mathrm{\rm E2}\right|H,J',1,-1\right\rangle ,
\end{align}
which are defined using the state notation $\left|A,J,M,\Omega\right\rangle $ for electronic state $A$, and `E1, M1, E2' refer to the corresponding molecular operators in table \ref{tab:spher_tens}. It is useful to define the Rabi frequency $\rabi=|\mathcal{M}|$ as the magnitude of the amplitude connecting to the bright state, and the unit vector $\hat{\varepsilon}_{\mathrm{eff}}$ corresponding to the projection of $\vec{\varepsilon}_{\mathrm{eff}}$ onto the $xy$ plane,
\begin{align}
\hat{\varepsilon}_{\mathrm{eff}}= & \frac{\vec{\varepsilon}_{\mathrm{eff}}-(\vec{\varepsilon}_{\mathrm{eff}}\cdot\hat{z})\hat{z}}{\sqrt{|\vec{\varepsilon}_{\mathrm{eff}}|^{2}-|\vec{\varepsilon}_{\mathrm{eff}}\cdot\hat{z}|^{2}}}.\label{eq:unit vector}
\end{align}
This completely determines the bright and dark states, which have been previously defined in equations \ref{eq:bright_state} and \ref{eq:dark_state} for solely E1 transition matrix elements.

The odd parity E1 and even parity M1 and E2 contributions to the effective polarisation differ by a factor of $\Nsw\Esw$, which is correlated with the expected eEDM signal. Expanding the effective E1 polarisation in terms of switch parity components, $\hat{\varepsilon}_{\mathrm{eff}}=\hat{\varepsilon}_{\mathrm{eff}}^{\nr}+\Nsw\Esw d\vec{\varepsilon}_{\mathrm{eff}}^{\N\E}$, and evaluating the effective $\Nsw\Esw$ correlated polarisation imperfections using equation \ref{eq:extracting_polarization_imperfection_components}, we find that the bright and dark states have effective polarisation correlations given by:
\begin{align}
\frac{\hat{z}\cdot(\hat{\varepsilon}_{\mathrm{eff}}^{\nr}\times d\vec{\varepsilon}_{\mathrm{eff}}^{\N\E})}{\hat{\varepsilon}_{\mathrm{eff}}^{\nr}\cdot\hat{\varepsilon}_{\mathrm{eff}}^{\nr}}\approx & \, d\theta_{\mathrm{eff}}^{\N\E}-id\Theta_{\mathrm{eff}}^{\N\E}\\
\approx & -i(a_{M1}-a_{E2}(\Psw))(\hat{\epsilon}\cdot\hat{z})((\hat{k}\times\hat{\epsilon})\cdot\hat{z}).
\label{eq:state_correlations}
\end{align}
It is useful to use a particular parameterization of the laser pointing $\hat{k}$ and polarisation $\hat{\epsilon}$ to expand the expression in equation \ref{eq:state_correlations} in terms of pointing and polarisation imperfections. The state preparation laser $\hat{k}$-vector is aligned along (or against) the $\hat{z}$ direction in the laboratory, so it is convenient to parameterize the pointing deviation from normal by spherical angle $\vartheta_{k}$, and the direction of this pointing imperfection by polar angle $\varphi_{k}$ in the $xy$ plane, such that:
\begin{align}
\hat{k}= & \cos\varphi_{k}\sin\vartheta_{k}\hat{x}+\sin\varphi_{k}\sin\vartheta_{k}\hat{y}+\cos\vartheta_{k}\hat{z}.
\label{eq:pointing_imperfection}
\end{align}
We may use a parameterization for the polarisation $\hat{\epsilon}$ that is similar to that in equation \ref{eq:polarization_parametrization}, but a slight modification is required to ensure that $\hat{k}\cdot\hat{\epsilon}=0$:
\begin{align}
\hat{\epsilon}= & N_{\epsilon}\left(-e^{-i\theta}\cos\Theta\hat{\epsilon}_{+1}+e^{i\theta}\sin\Theta\hat{\epsilon}_{-1}+\epsilon_{z}\hat{z}\right)\\
\epsilon_{z}= & -\frac{1}{\sqrt{2}}\tan\theta_{k}\left(e^{-i\left(\theta-\varphi_{k}\right)}\cos\Theta+e^{i\left(\theta-\varphi_{k}\right)}\sin\Theta\right)
\end{align}
where $N_{\epsilon}$ is a normalization constant that ensures that $\hat{\epsilon}^{*}\cdot\hat{\epsilon}=1$. With these parameterizations in place, and expanding about small ellipticities $d\Theta$ such that $\Theta=\pi/4+d\Theta$, and small laser pointing deviation, $\vartheta_{k}\ll1$, we find that the $\Nsw\Esw$-correlated effective laser polarisation imperfections are given by:
\begin{align}
d\theta_{\mathrm{eff}}^{\N\E}\approx & -\frac{1}{2}(a_{M1}-a_{E2}(\Psw))\vartheta_{k}^{2\:}S^{\:}\cos(2(\theta-\varphi_{k}))\\
d\Theta_{\mathrm{eff}}^{\N\E}\approx & -\frac{1}{2}(a_{M1}-a_{E2}(\Psw))\vartheta_{k}^{2\:}\sin(2(\theta-\varphi_{k}))
\end{align}
where $S_{i}=-2d\Theta_{i}$ describe the laser ellipticities. Hence, following equation \ref{eq:Measured_Phase_with_Polarization_Imperfections}, there is a systematic error in $\wNE$:
\begin{align}
\omega_{\mathrm{S.I.}}^{\N\E}=&\frac{1}{\tau}\frac{1}{4}\left(a_{M1}-a_{E2}\left(\Psw\right)\right)\times\\&\left[\vartheta_{k,\p}^{2}\left(-2S_{\p}c_{\p}+\Psw s_{\p}\left(S_{\rX}-S_{\rY}\right)\right)+\right.\\&\left.\vartheta_{k,\rX}^{2}\left(S_{\rX}c_{\rX}+\Psw S_{\p}s_{\rX}\right)+\vartheta_{k,\rY}^{2}\left(S_{\rY}c_{\rY}-\Psw S_{\p}s_{\rY}\right)\right]
\end{align}
where $c_{i}\equiv\cos\left(2(\theta_{i}-\varphi_{i,k})\right)$ and $s_{i}\equiv\sin\left(2(\theta_{i}-\varphi_{i,k})\right)$ describe the dependence of the systematic error on the difference between the linear polarisation angle $\theta_{i}$ and the pointing angle $\varphi_{i,k}$ in the $xy$ plane. 

There is another contribution to this systematic error that arises when the coupling to the off-resonant
opposite parity excited state $|C,-\Psw\rangle$ is also taken into account. This additional contribution becomes significant when the ellipticities are comparable to or smaller than $\gamma_{C}/\Delta_{\Omega,C,J=1}\approx0.5\%$.

The eEDM channel, $\wNE$, was defined to be even under the superblock switches (including $\Psw$), hence those terms proportional to $\Psw$ in the equation above do not contribute to our reported result. Additionally, the $\Gsw$ and $\Rsw$ switches rotate the polarisation angles for each laser by roughly $\theta_{i}\rightarrow\theta_{i}+\pi/2$ periodically and the resulting $\wNE$ signal is averaged over these states. Provided that the pointing drift is much slower than the timescale of these switches, and to the extent that the laser polarisations constituting the $\Rsw$ and $\Gsw$ states are orthogonal, then these systematic errors should dominantly contribute to the $\omega^{\mathcal{NEG}}$ and $\omega^{\mathcal{NER}}$ channels which were found to be consistent with zero (see Figure~\ref{fig:pixel_plot}). 

An indirect limit on the size of the systematic error due to Stark interference, $\wNE_{\mathrm{S.I.}}$, may be estimated by assuming a reasonable suppression factor by which the effects in $\omega^{\N\E\R}$ and $\omega^{\N\E\G}$ may `leak' into $\wNE$. We monitored the pointing drift on a beam profiler and observed pointing drifts up to $d\vartheta_k\sim 50~\upmu\rm{rad}$  throughout a full set of superblock states. The absolute pointing misalignment angle was not well known but was estimated to be larger than $\vartheta_k\gtrsim0.5~\rm{mrad}$. Hence we may estimate a conservative suppression factor $d\vartheta_k/\vartheta_k\lesssim1/10$ by which pointing drift may contaminate $\wNE$ from $\omega^{\N\E\R}$ and $\omega^{\N\E\G}$. The two $\mathcal{\tilde{R}}$ states are very nearly orthogonal, but the $\Gsw$ states deviate sufficiently from orthogonal (see section~\ref{sssec:suppression_of_the_AC_stark_shift_phases}) such that the leakage from $\omega^{\N\E\G}\rightarrow\omega^{\N\E}$ will dominate the systematic error; we estimate a suppression factor of about $c_p^{\rm{nr}}/c_p^{\mathcal{G}}\sim s_p^{\rm{nr}}/s_p^{\mathcal{G}}\sim1/5$. Based on the upper limits on the measured values for $\omega^{\N\E\R}$ and $\omega^{\N\E\G}$ combined with leakage from $\omega^{\N\E\R}$ and $\omega^{\N\E\G}$ into $\omega^{\N\E}$ due to pointing drift, and leakage from $\omega^{\N\E\G}$ into $\omega^{\N\E}$ due to non-orthogonality of the two $\Gsw$ states, we estimate the possible size of the systematic error to be $\omega_{\mathrm{S.I.}}^{\N\E}\lesssim 1^{\:}\mathrm{mrad}/\mathrm{s}$.

Note that the mechanism for this systematic error was not discovered until after the publication of our result \cite{Baron2014} and hence was not included in our systematic error analysis there. Furthermore, since we did not observe this effect, this systematic error does not match any of the inclusion criteria outlined in section \ref{ssec:total_systematic_error_budget} and hence is not included in the systematic error budget in this paper. Since we did not understand the mechanism for this systematic error while running the apparatus, we were not able to place direct limits on the size of this systematic error. We estimate that the absolute pointing deviation from ideal was at most $5^{\:}\mathrm{mrad}$ and the ellipticity of each laser was no more than $S_{i}\approx5\%$. The E1/M1 interference coefficient is $a_{M1}\approx0.1$ for the $H\rightarrow C$ transition. This gives an estimate of $\omega_{\mathrm{S.I.}}^{\N\E}\sim0.1^{\:}\mathrm{mrad}/\mathrm{s}$ before suppression due to the $\Rsw$ and $\Gsw$ switches. Hence, we do not believe that this systematic error significantly shifted the result of our measurement.

\subsubsection{AC Stark shift phases}
\label{sssec:AC_stark_shift_phases}
In this section we describe contributions to the measured phase $\Phi$ that depend on the AC Stark shifts induced by the state preparation and readout lasers. We describe mechanisms by which such phase contributions may arise, and we describe mechanisms by which $\Nsw\Esw$ correlated experimental imperfections may couple to these phases to result in eEDM-mimicking phases. Concise descriptions of some of the effects described here can be found in \cite{Hess2014,SpaunThesis,HutzlerThesis}.


During our search for systematic errors as described in section \ref{ssec:determining_systematic_uncertainty}, we empirically found that there was a contribution to the measured phase $d\Phi(\Delta,\rabi)$ that had an unexpected linear dependence on the laser detuning, $\Delta$, a quadratic dependence on laser detuning $\Delta$ in the presence of a nonzero magnetic field, and a linear dependence on small changes to the magnitude of the Rabi frequency, $d\rabi/\rabi$, in the presence of a nonzero magnetic field,
\begin{align}
d\Phi\left(\Delta,\rabi\right)=&\sum_{i}\left[\alpha_{\Delta,i}\Delta_{i}+\alpha_{\Delta^{2},i}\Delta_{i}^{2}+\beta_{d\rabii}(d\rabii/\rabii)+\dots\right].
\label{eq:Empirical_AC_Stark_Shift_Phase_Result}
\end{align}
where $i\in\left\{ \p,\rX,\rY \right\}$ indexes the state preparation and readout lasers. The coefficients we measured were $\alpha_{\Delta}\sim1^{\:}\mathrm{mrad}/(2\pi\times\mathrm{MHz})$, $\alpha_{\Delta^{2}}\sim1^{\:}\mathrm{mrad}/\left(2\pi\times\mathrm{MHz}\right)^{2}$ and $\beta_{d\rabi}\sim10^{-3}$. We performed these measurements by independently varying the laser detunings $\Delta_i$ across resonance using AOMs or modulating the laser power using AOMs with the set-up depicted in figure \ref{fig:HC_transitions_setup} and extracting the measured phase $\Phi$. Examples of such measurements are given in figure~\ref{fig:phase_vs_detuning}.

We determined that this behaviour can be caused by mixing between bright and dark states, due to a small non-adiabatic laser polarisation rotation or Zeeman interaction present during the optical pumping used to prepare and read out the spin state. The mixed bright and dark states differ in energy by the AC Stark shift, which leads to a relative phase accumulation between the bright and dark state components that depends on the laser parameters $\Delta$ and $\rabi$. We shall now derive the AC Stark shift phase that results in equation \ref{eq:Empirical_AC_Stark_Shift_Phase_Result}, under simplifying assumptions amenable to analytic calculations.

\begin{figure}[htbp]
\centering
\includegraphics[width=15cm]{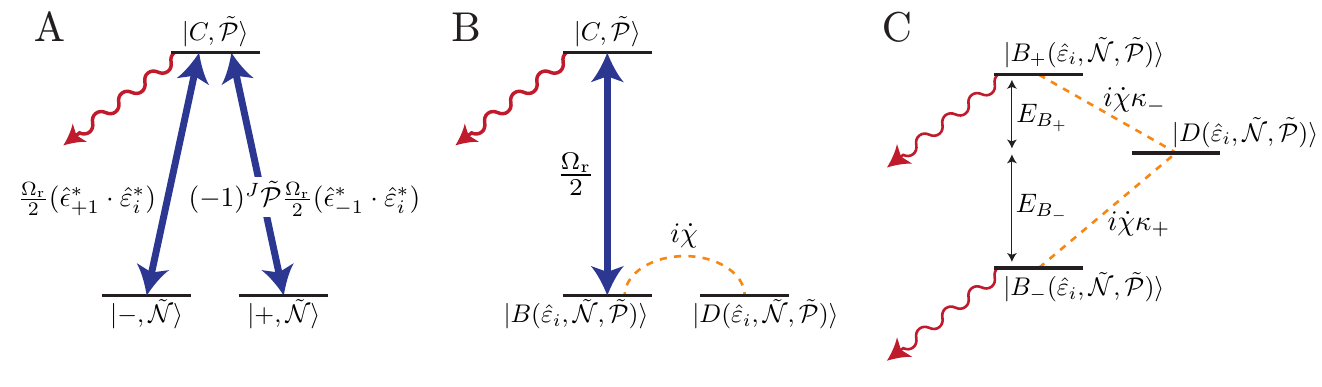}
\caption{Energy level diagrams depicting the Hamiltonian when the three-level $H\rightarrow C$ transition is addressed by the state preparation or readout lasers in three different bases. Solid double-sided blue arrows denote strong laser couplings between $H$ and $C$. Wiggly red arrows denote spontaneous emission from $C$ to states outside of the three level system. Dashed orange lines denote weak couplings induced by laser polarisation rotation. Basis A is useful for describing the spin precession phase induced by the Zeeman and eEDM Hamiltonians. Basis B is useful for describing the states that are prepared and read-out in the spin precession measurement. Basis C is useful for evaluating the AC Stark Shift phases induced by laser polarisation rotations.}
\label{fig:bases}
\end{figure}

Consider a three level system consisting of the bright $|B(\hat{\varepsilon},\Nsw,\Psw)\rangle$ and dark $|D(\hat{\varepsilon},\Nsw,\Psw)\rangle$ states and the lossy excited state $|C,\Psw\rangle$ with decay rate $\gamma_{C}$. For simplicity, assume that there is no applied
magnetic field for the time being. In this system, the instantaneous eigenvectors (depicted in figure \ref{fig:bases}C) are
\begin{align}
|B_{\pm}\rangle\equiv & \pm\kappa_{\pm}|C,\Psw\rangle+\kappa_{\mp}|B(\hat{\varepsilon},\Nsw,\Psw)\rangle,^{\:\:}|D\rangle\equiv|D(\hat{\varepsilon},\Nsw,\Psw)\rangle,
\label{eq:inst_eigv}
\end{align}
and the instantaneous eigenvalues are
\begin{align}
E_{B\pm}= & \frac{1}{2}\left(\Delta\pm\sqrt{\Delta^{2}+\rabi^{2}}\right),^{\:}E_{D}=0,
\label{eq:inst_eig}
\end{align}
such that the mixing amplitudes $\kappa_{\pm}$ are given by 
\begin{align}
\kappa_{\pm}= &\frac{1}{\sqrt{2}} \sqrt{1\pm\frac{\Delta}{\sqrt{\Delta^{2}+\rabi^{2}}}}.
\end{align}

The effect of the decay of the excited state (which occurs almost entirely to states outside of the three level system) may be taken into account by adding an anti-Hermitian operator term in the Schrodinger equation, $|\dot{\psi}\rangle=-i(H-i\frac{1}{2}\Gamma)|\psi\rangle$, where $\Gamma=\gamma_{C}|C,\Psw\rangle\langle C,\Psw|$ is the decay operator. This formulation is equivalent to the Lindblad master equation,
\begin{align}
\dot{\rho}= & -i\left[H,\rho\right]-\frac{1}{2}\left\{ \Gamma,\rho\right\} ,
\end{align}
that governs the time evolution of the density matrix $\rho=|\psi\rangle\langle\psi|$. In practice, we implement this decay term by calculating the time evolution of the system according to $H$, and then making the substitution $\Delta\rightarrow\Delta-i\gamma_C/2$ before calculating squares of amplitudes. 

It is useful to work in the dressed state basis, $|D\rangle$, $|B_\pm\rangle$, (basis C in figure \ref{fig:bases}) because these are nearly stationary states and have simple time evolution in the case that laser polarisation and Rabi frequency are stationary. If we allow the laser polarisation to vary in time, then the dressed state basis varies in time, and the system evolves according to the Hamiltonian,
\begin{align}
\tilde{H}= & UHU^{\dagger}-iU\dot{U}^{\dagger},
\end{align}
where $U$ is the transformation from time independent basis A to time dependent basis C (from figure \ref{fig:bases}), $UHU^{\dagger}$ is diagonal, and $-iU\dot{U}^{\dagger}$ is a fictitious force term that arises because we are working in a non-inertial frame when the laser polarisation is time dependent \cite{Budker2008}.

Assuming that the polarisation is nearly linear, $\Theta\approx\pi/4$, but allowing the polarisation to rotate slightly, and allowing for a nonzero two photon detuning due to the Zeeman shift $\delta=-\g\mu_{\rm B}\B_z\Bsw$, the Hamiltonian in the dressed state picture is:
\begin{align}
\tilde{H}= & \left(\begin{array}{ccc}
0 & -i\dot{\chi}^{*}\kappa_{+} & -i\dot{\chi}^{*}\kappa_{-}\\
i\dot{\chi}\kappa_{+} & E_{B-} & -\frac{i}{2}\frac{\dot{\rabi}\Delta-\rabi\dot{\Delta}}{\Delta^{2}+\Omega^{2}}\\
i\dot{\chi}\kappa_{-} & \frac{i}{2}\frac{\dot{\rabi}\Delta-\rabi\dot{\Delta}}{\Delta^{2}+\rabi^{2}} & E_{B+}
\end{array}\right)\begin{array}{c}
\ket{D} \\
\ket{B_{-}} \\
\ket{B_{+}} 
\end{array}
\end{align}
where $\dot{\chi}=\dot{\Theta}-i(\dot{\theta}+\delta)$ can be considered to be a complex polarisation rotation rate, $\dot{\Omega}_{\rm r}$ is the rate of change of the Rabi frequency, and $\dot{\Delta}$ is
the rate of change of the detuning. Note that this Hamiltonian implies that the effect of a two photon detuning arising from the Zeeman shift is equivalent to that of a linear polarisation rotating at a constant rate.

We may then apply first order time-dependent perturbation theory in this picture to determine the extent of bright/dark state mixing due to $\dot{\chi}$ in the time evolution of the system. If we parameterize the time-dependent state as
\begin{align}
\left|\psi\left(t\right)\right\rangle = & c_{D}\left(t\right)\left|D\right\rangle +c_{B+}\left(t\right)\left|B_{+}\right\rangle +c_{B-}\left(t\right)\left|B_{-}\right\rangle,
\end{align}
then in the case of a uniform laser field $\dot{\Omega}_{\rm r}=0$, of duration $t$ and with a constant detuning $\dot{\Delta}=0$, the time evolution of the coefficients is given at first order by:
\begin{align}
c_{D}\left(t\right)= & c_{D}\left(0\right)-\sum_{\pm}\int_{0}^{t}\dot{\chi}^{*}\left(t'\right)\kappa_{\mp}\left(t'\right)e^{-iE_{B\pm}t'}c_{B\pm}\left(0\right)\enspace dt'\label{eq:first order perturbation theory dark state}\\
c_{B\pm}\left(t\right)= & e^{-iE_{B\pm}t}c_{B\pm}\left(0\right)+e^{-iE_{B\pm}t}\int_{0}^{t}\dot{\chi}\left(t'\right)\kappa_{\mp}\left(t'\right)e^{iE_{B\pm}t'}c_{D}\left(0\right)\enspace dt'.\label{eq:first order perturbation theory bright state}
\end{align}

In the state preparation region, the molecules begin in an incoherent mixture of the states $|B(\hat{\varepsilon}_{\p},\Nsw,\Psw)\rangle$ and $|D(\hat{\varepsilon}_{\p},\Nsw,\Psw)\rangle$ and then enter the state preparation laser beam. In the ideal case of uniform laser polarisation, molecules that were in the bright state are optically pumped out of the three level
system, and molecules that are in the dark state remain there; this results in a pure state, $|D(\hat{\varepsilon}_{\p},\Nsw,\Psw)\rangle$. However, if there is a small polarisation rotation by amount $d\chi\equiv\int_{0}^{t}\dot{\chi}\left(t'\right)dt'\equiv d\Theta-i(d\theta-\g\mu_{\rm B}\B_z\Bsw t)$, such that $\left|d\chi\right|\ll1$, then the dark state obtains a bright state admixture that may not be completely optically pumped away before leaving the laser beam.\footnote{This is most liable to occur just before a molecule leaves the laser beam, such that complete optical pumping does not occur.} In this case, the final state can be written as
\begin{align}
|D(\hat{\varepsilon}'_{\p},\Nsw,\Psw)\rangle= & |D(\hat{\varepsilon}_{\p},\Nsw,\Psw)\rangle+d\chi\Pi|B(\hat{\varepsilon}_{\p},\Nsw,\Psw)\rangle
\end{align}
where $\hat{\varepsilon}'_{\p}$ is the effective polarisation that parameterizes the initial state in the spin precession region 
\begin{align}
\hat{\varepsilon}'_{\p}= & \hat{\varepsilon}_{\p}+d\chi\Pi i\hat{z}\times\hat{\varepsilon}_{\p}^{*},
\end{align}
and $\Pi$ is an amplitude that accounts for the AC Stark shift phase and the time dependent dynamics of the non-adiabatic mixing due to the polarisation rotation,
\begin{equation}
\Pi=\sum_{\pm}(\kappa_{\mp})^{2}e^{-iE_{B\pm}t}\int_{0}^{t}dt'^{\:}\frac{\dot{\chi}\left(t'\right)}{d\chi}e^{iE_{B\pm}t'}.\label{eq:Pi_def}
\end{equation}
The deviations between the effective polarisation and the actual laser polarisation can be viewed as effective polarisation imperfections,
\begin{align}
d\theta_{\p,\mathrm{eff}}= & -d\Theta_{\p}\text{Im}\Pi+(d\theta_{\p}-\g\mu_{\rm B}\B_z\Bsw t)\text{Re}\Pi,\\
d\Theta_{\p,^{\:}\mathrm{eff}}= & -d\Theta_{\p}\text{Re}\Pi-(d\theta_{\p}-\g\mu_{\rm B}\B_z\Bsw t)\text{Im}\Pi,
\end{align}
that lead to shifts in the measured phase $\Phi$ as described in equation \ref{eq:Measured_Phase_with_Polarization_Imperfections}.
For definiteness, consider the case in which the polarisation rotation rate $\dot{\chi}(t')=d\chi/t$ is a constant for the duration of the optical pumping pulse $t$. In this case,
\begin{align}
\Pi= & \sum_{\pm}(\kappa_{\mp})^{2}e^{-iE_{B\pm}t/2}\mathrm{sinc}(E_{B\pm}t/2).
\end{align}

This function has the property that $\text{Im}\Pi$ is an odd function in $\Delta$ that can take on values up to order unity across resonance (a frequency range on the order of $\gamma_{C}$) and is exactly zero on resonance. $\text{Re}\Pi$ is an even function quadratic in $\Delta$ about resonance, and depends on Rabi frequency on resonance. If the laser beam intensity reduces quickly as the molecule leaves it then most of the AC Stark shift phase arises from the last Rabi flopping period before the molecule exits the laser beam (provided $\dot{\chi}$ is nonzero during that time). If the intensity reduces slowly, the AC Stark shift phase can be exacerbated since the bright state amplitude is not as effectively optically pumped away while $\rabi<\gamma_C$. Nevertheless, beamshaping tests shown in figure \ref{fig:phase_vs_detuning} and numerical simulations indicate that $\Pi$ is not very sensitive to the shape of the spatial intensity profile of the laser beam or the shape of the spatial variation of the polarisation.

If we consider only the first order contribution to the shift in the measured phase in equation \ref{eq:Measured_Phase_with_Polarization_Imperfections}, $d\theta_{\p,\mathrm{eff}}$, and neglect the second order shift that arises due to $d\Theta_{\p,\mathrm{eff}}$, then we can relate the parameters in equation \ref{eq:Empirical_AC_Stark_Shift_Phase_Result} to the amplitude $\Pi$ accounting for the AC Stark shift phase and the complex polarisation rotation $d\chi$, by
\begin{align}
\alpha_{\Delta,\p}\approx&-\frac{\partial\text{Im}\Pi}{\partial\Delta_{\p}}d\Theta_{\p}\\
\alpha_{\Delta^{2},\p}\approx&\frac{\partial^{2}\text{Re}\Pi}{\partial\Delta_{\p}^{2}}\left(d\theta_{\p}-g\mu_{\rm B}\B_{z}\tilde{\B}t\right)\\
\beta_{d\rabi,\p}\approx&\rabi\frac{\partial\text{Re}\Pi}{\partial\rabi}\left(d\theta_{\p}-g\mu_{\rm B}\B_{z}\tilde{\B}t\right).\label{eq:bdOmega}
\end{align}
We can interpret these results as follows. The linear dependence of the measured phase on detuning, $\alpha_{\Delta,\p}$, comes from a spatially varying ellipticity in the $x$ direction coupling to the AC Stark shift phase. Similarly, the quadratic dependence of $\Phi$ on $\Delta$, $\alpha_{\Delta^{2},\p}$, and the dependence of $\Phi$ on a relative change in $\rabi$, $\beta_{d\rabi, \p}$, come from either a spatially varying linear polarisation in the $x$ direction or a Zeeman shift, each coupling to the AC Stark shift phase. Here, we only analyzed the phase shift that results from AC Stark shift effects in the state preparation laser beam, but there is an analogous phase shift in the state readout beam.

There are several other subdominant effects that also contribute to the AC Stark shift phase behavior described in equation \ref{eq:Measured_Phase_with_Polarization_Imperfections} in the presence of polarisation imperfections. The opposite parity excited state $|C,-\Psw\rangle$ couples strongly to the dark state, but the mixing between these two states is weak because the transition frequency is off-resonant by a detuning $\Delta_{\Omega,C,J=1}\approx2\pi\times51~\mathrm{MHz}\gg\gamma_{C}$. In the case that an optical pumping laser has nonzero ellipticity, the bright state gains a weak coupling to the opposite-parity excited state proportional to this ellipticity. Then, two-photon bright-dark state mixing ensues in such a way that the mixing amplitude, and hence the measured phase, depends on the laser detuning.

The rapid polarisation switching of the state readout beam can also introduce AC Stark shift-induced phases in the absence of a polarisation gradient, if the average ellipticity between the two polarisations is nonzero. Suppose a particular molecule is first excited by the $\hat{\epsilon}_{\rX}$ polarised beam. The two bright eigenstates $\ket{B_{\pm}}$ are mostly optically pumped away, resulting in a fluorescence signal $F_{\rX}$. The population remaining in the bright eigenstates acquires a phase relative to the dark state, due to the AC Stark shift. Then the molecules are optically pumped by the $\hat{\epsilon}_{\rY}$ polarised beam. If there is a nonzero average ellipticity, $\hat{\epsilon}_{\rY}$ is not quite orthogonal to $\hat{\epsilon}_{\rX}$ and the new bright eigenstates that give rise to the fluorescence signal $F_Y$ are superpositions of the former bright and dark states that acquired a relative AC Stark shift phase. This results in a fluorescence signal, and hence measured phase component, that depends linearly on laser detuning $\Delta$. 

\subsubsection{Polarisation Gradients from Thermal Stress-Induced Birefringence}
\label{sssec:polarization_gradients_from_thermal_stress_induced_birefringence}

\hspace*{\fill} \\
The AC Stark shift phases described in the previous section can be induced by polarisation gradients in $\hat{x}$ across the state preparation and readout laser beams. In this section we describe a known mechanism by which these arose. Recall that these laser beams passed through transparent, ITO-coated electric field plates. For an absorbance $\alpha$ and laser intensity $I$, the rate of heat deposition into the plates is $\dot{Q}\left(x,y\right)=\alpha\, I\left(x,y\right)$. The laser beam profile is stretched in the $y$ direction to ensure that all molecules are addressed. For simplicity we assume that the heating distribution, $\dot{Q}\left(x,y\right)=\dot{Q}\left(x\right)$, is completely uniform in the $y$ direction. We also assume that there are no shear stresses, i.e.\ local expansion of the glass is isotropic. Under these assumptions, the relationship between the heating rate, $\dot{Q}$, and the internal stress tensor $\sigma_{ij}$ (where $i,j$ are Cartesian indices) is
\begin{equation}
\frac{\partial^{2}\sigma_{yy}}{\partial x^{2}}= \frac{E\alpha_{V}}{\kappa}\dot{Q}\left(x\right),
\end{equation}
where $E$, $\alpha_{V}$ and $\kappa$ are the Young's modulus, coefficient of thermal expansion, and thermal conductivity, respectively \cite{Barber2010}. Unit vectors $\hat{x}$ and $\hat{y}$ correspond to the principal axes of the stress tensor due to the symmetry of the heating function, hence the off-diagonal (shear) elements are zero, $\sigma_{xy}=0$. The other diagonal component, $\sigma_{xx}$, is uniform across the plates, and equal to $\sigma_{yy}$ far away from the laser. The stress-optical law states that the birefringence and stress are linearly proportional along the principal axes of the stress tensor \cite{Dally1991}. The difference between the indices of refraction in the $x$ and $y$ directions is then $\Delta n=K\left(\sigma_{xx}-\sigma_{yy}\right)$, where $K\approx4\times10^{-6}\,\mbox{MPa}^{-1}$ is the stress-optical coefficient for Borofloat glass \cite{Schott2013b}. The retardance of an incident laser beam of index $i$ is $\Gamma_i=2\pi\Delta n\left(t/\lambda\right)$, where $t$ is the thickness of the field plates (in the $z$ direction), and $\lambda$ is the wavelength of light. Hence, in this limit, the retardance due to thermal stress-induced birefringence is related to the laser intensity by:
\begin{equation}
\frac{\partial^{2}\Gamma}{\partial x^{2}}=\eta\frac{t}{\lambda}I\left(x\right),
\label{eq:retarddiff}
\end{equation}
where $\eta=2\pi KE\alpha_{V}\alpha/\kappa\approx26\times10^{-6}$~W$^{-1}$ is a material constant of Borofloat glass \cite{Schott2013b}.
The ellipticity imprinted on the nominally linearly polarised laser beam is given by
\begin{equation}
S_i=\Gamma_i(x)\sin\left(2(\theta_i-\phi_{\Gamma,i})\right),
\label{eq:s3}
\end{equation}
where $\theta_i$ is the linear polarisation angle and $\phi_{\Gamma,i}$ is the orientation of the fast axis of the birefringent material (nominally $\hat{x}$ in our case).

Assuming the laser has total power $P$, a Gaussian profile in $x$ with standard deviation $w_{x}$, and a top-hat profile in $y$ with half width $w_{y}$, the intensity is given by
\begin{equation}
I\left(x\right)= \frac{P}{\sqrt{8\pi}w_{x}w_{y}}e^{-\frac{x^{2}}{2w_{x}^{2}}}
\end{equation}
where $2w_y \gg w_x$. There is then an analytic solution to equation~\ref{eq:retarddiff} from which we extracted a retardance gradient in the laser tail, $x=w_{x}$, of
\begin{equation}
\frac{\partial\Gamma}{\partial x}\approx\frac{{\rm erf}(1/\sqrt{2})P\kappa t}{4w_y\lambda}\approx0.03~\mathrm{rad}/{\rm mm}
\label{eq:retardance_gradient}
\end{equation}
for a nominal laser power of ${\approx}2$~W. Similar results were obtained from numerical finite element analysis. Thermal stress-induced birefringence has been observed in similar systems such as in UHV vacuum windows \cite{Solmeyer2011}, laser output windows \cite{Eisenbach1992}, and Nd:YAG rods \cite{Koechner1970}.
\begin{figure}
\begin{centering}
\includegraphics[width=10cm]{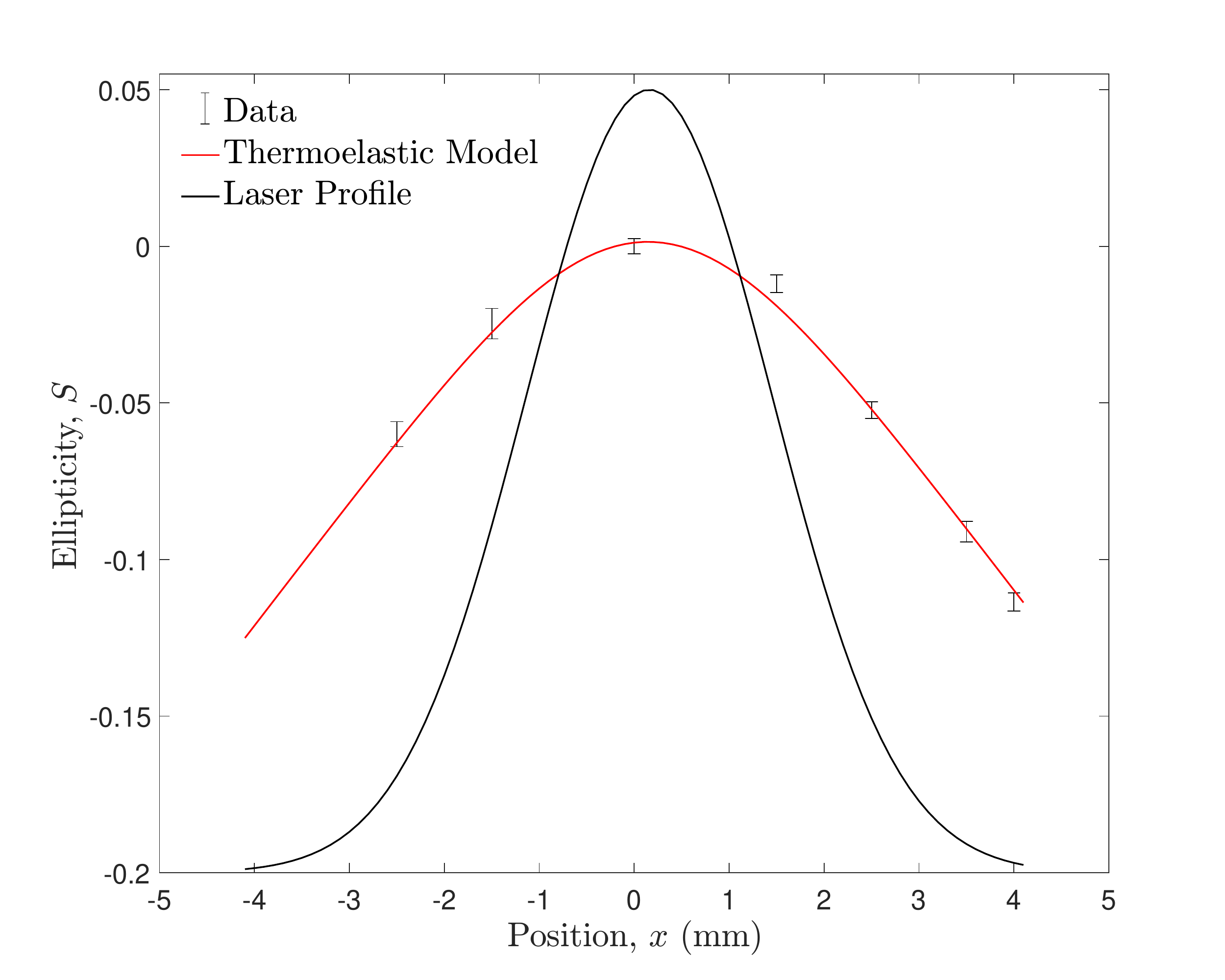}
\par\end{centering}
\caption{Measurement of the ellipticity, $S$, as a function of position along $x$ within the state readout laser beam. A fit to the thermo-elastic model, which assumes a Gaussian laser profile and has the amplitude and offset in $S$ as free parameters, is overlaid.}
\label{ite:birefringence}
\end{figure}

The estimates of the ellipticity gradient agree well with measurements of the polarisation of the beam, as shown in figure~\ref{ite:birefringence}. These polarimetry measurements were adapted from the procedure described in \cite{Berry1977}; a polarimeter was constructed consisting of a rotating quarter-wave plate, fixed polariser, and fast photodiode.
The use of a fast photodetector allows for polarimetry of the probe beam during the 100~kHz polarisation switching. The resolution of the system was such that we could quickly measure the normalized circular Stokes parameter, $S$, to a few percent, which is sufficient to measure typical birefringence gradients of ${\sim}10\%$ across the beam.

\subsubsection{Suppression of AC Stark Shift Phases}
\label{sssec:suppression_of_the_AC_stark_shift_phases}

\begin{figure}[htbp]
\centering
\includegraphics[width=15cm]{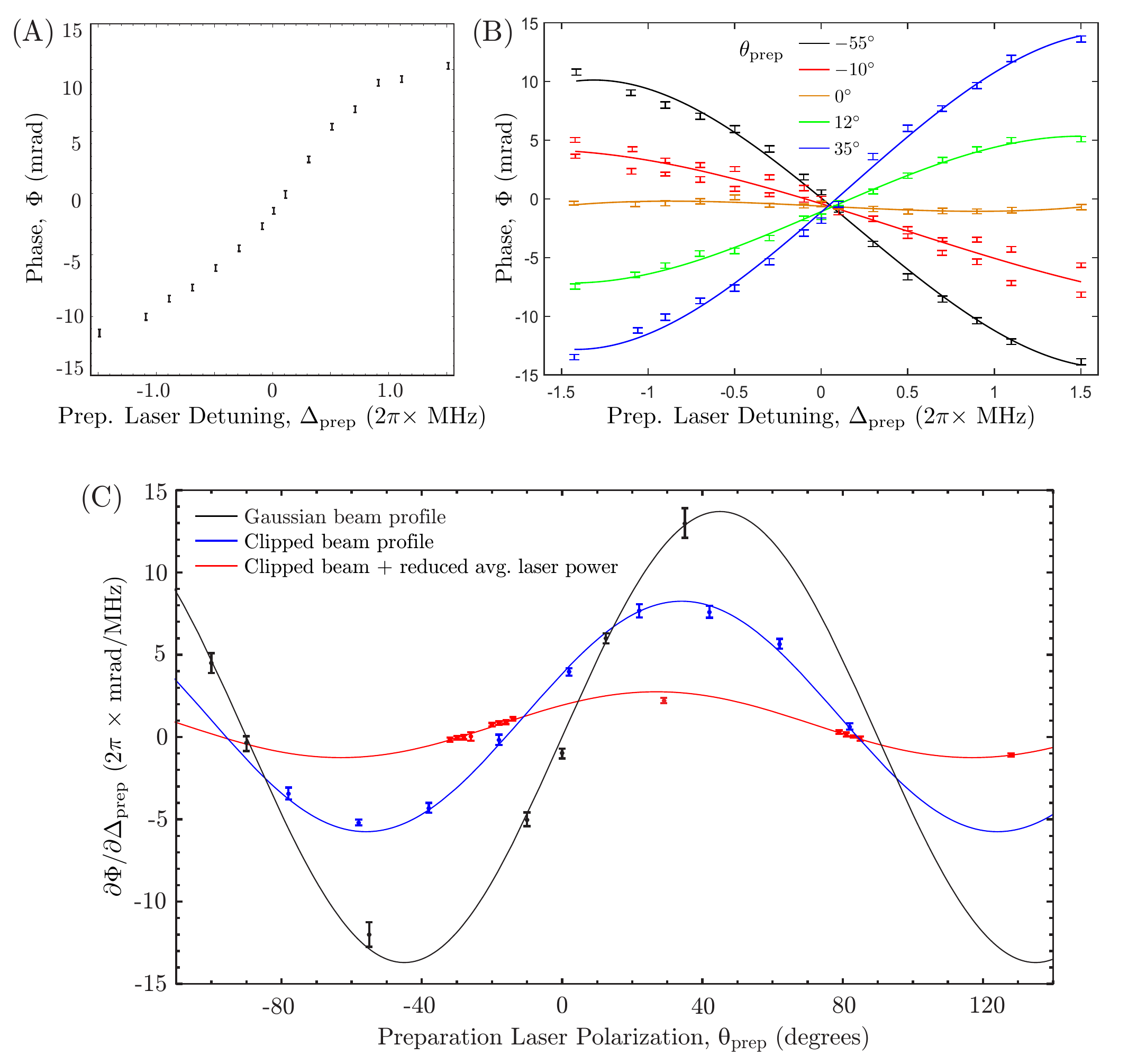}
\caption{(A) Measured molecule phase as a function of preparation laser detuning. The slope agrees with originally observed $\Phi^{\N\E}$ dependence on $\Delta^{\N\E}$. (B) Phase dependence on detuning for multiple preparation laser polarisation angles. (C) $\partial \Phi/\partial \Delta^{\nr}$ shows clear sinusoidal dependence on preparation laser polarisation. The magnitude of $\partial \Phi/\partial \Delta^{\nr}$ decreases for all polarisation angles when the Gaussian beam tails are clipped (blue) and when the average laser power is reduced with a chopper wheel (red).}
\label{fig:phase_vs_detuning}
\end{figure}

\hspace*{\fill} \\
We were able to suppress the magnitude of the AC Stark shift phases in several different ways that are illustrated in figure~\ref{fig:phase_vs_detuning}. The ellipticity gradient across the state preparation laser beam was suppressed by tuning the linear polarisation angle: as per equation~\ref{eq:s3}, the ellipticity gradient is proportional to $\sin(2\theta_{\p}-2\phi_{\Gamma,\p})$, which vanishes when the polarisation is aligned along a birefringence axis, i.e. $\theta_{\p}=\phi_{\Gamma,\p},\phi_{\Gamma,\p}+\pi/2$. To determine $\phi_{\Gamma,\p}$ we measured the total accumulated phase as a function of laser detuning for various $\theta_{\p}$ and then extracted the slope $\alpha_{\Delta,\p}^{\nr}=\partial\Phi^{\nr}/\partial\Delta_{\p}$ for small detuning values. Note that when fitting the phase vs.\ detuning data we found that cubic functions provided significantly better fits over the detuning ranges used (see Figure~\ref{fig:phase_vs_detuning}(B)). We then selected $\theta_{\p}$ to minimize $\alpha_{\Delta,\p}^{\nr}$. This suppressed $\alpha_{\Delta,\p}^{\nr}$ by about a factor of 50 relative to its original value, to $\alpha_{\Delta,\p}^{\nr}\lesssim0.1$~mrad/(2$\pi~\times$~MHz). 

Another method implemented to suppress AC Stark shift phases was to reduce the time-averaged power of the state preparation laser incident on the field plates. We used a chopper wheel to modulate the laser at 50~Hz, synchronous with the molecular beam pulses, with a 50\% duty cycle. We estimated the time scale for thermal changes to be on the order of $Q/\dot{Q}\sim2\rho Cw_{x}^{2}/\kappa\sim10^{\:}\mathrm{s}$, where $\rho$ and $C$ are the density and heat capacity of Borofloat respectively, so did not anticipate any significant transient effects
to be introduced. This modification reduced the retardance gradient, and hence the value of $\alpha_{\Delta,\p}^{\nr}$, by about a factor of two, as shown in Figure~\ref{fig:phase_vs_detuning}(C).

Finally, $\alpha_{\Delta,\p}^{\nr}$ was suppressed by shaping of the laser beam intensity profile. AC Stark shift phases were most significant at the downstream edge of the state preparation laser beam. Here, the intensity is such that bright-dark state mixing is still occurring but the bright state is not efficiently optically pumped away. By making the spatial intensity profile drop off more rapidly, we reduced the time that molecules spent in this intermediate intensity regime. This was achieved by taking advantage of the aspherical distortion introduced by misaligning a telescope immediately before the laser beam entered the spin-precession region. This suppressed $\alpha_{\Delta,\p}$ and $\beta_{\Delta^{2},\p}$ by ${\approx}2$, as shown in Figure~\ref{fig:phase_vs_detuning}(C). In addition to a phase suppression, we noticed that the optimal laser polarisation angle changed after implementing the steps described, as can be seen in Figure~\ref{fig:phase_vs_detuning}(C). The reason for this change is not definitively known, but we suspect that as we suppressed the birefringent contribution to the AC Stark shift phase, the non-birefringent contributions (i.e. the phase due to nonzero ellipticity causing bright-dark state mixing via the off-resonant opposite parity excited state) became fractionally larger, and we needed to tune the polarisation angle to obtain cancellation between these two classes of effects.

We observed much smaller AC Stark shift phases in the state readout laser beam than in the state preparation laser beam. This is not surprising since the effect is largely birefringent; the contributions to the effective polarisation imperfections for the $\hat{\epsilon}_{\rX}$ and $\hat{\epsilon}_{\rY}$ polarised lasers should be opposite in sign, $d\theta_{\rX}\propto\sin(2(\theta_{\ro}-\phi_{\Gamma,\ro}))$, $d\theta_{\rY}\propto\sin(2(\theta_{\ro}-\phi_{\Gamma,\ro}+\pi/2))$, such that they cancel each other in the measured phase (cf.\ equation~\ref{eq:Measured_Phase_with_Polarization_Imperfections}). The residual AC Stark shift phases measured in the state readout beam gave $\alpha_{\Delta,\ro}^{\nr}\approx0.5^{\:}\mathrm{mrad}/(2\pi\times\mathrm{MHz})$. This was sufficiently small that the methods of suppression described above were only implemented in the state preparation region.

\subsubsection{Systematic Errors due to Correlated Laser Parameters}
\label{sssec:correlated_laser_parameters}

In the discussion above, we described how polarisation imperfections can lead to contributions to the measured phase that depend on the AC Stark shifts and hence on the laser detunings $\Delta_{i}$ and Rabi frequencies $\rabii$. However, these phases only produce a systematic error in $\wNE$ if there is a nonzero correlation $\Delta_{i}^{\N\E}$ or $\rabii^{\N\E}$ of the laser detuning or Rabi frequency. We observed such correlations and discuss them in this section. We will also describe how we evaluated the associated systematic errors.

In section~\ref{sec:state_prep_read} (see figure~\ref{fig:Enr_wNE}) we discussed how a non-reversing component of the applied electric field, $\Enr$, could produce a $\Delta^{\N\E}$. In an entirely analogous manner, the Rabi frequency magnitude $\rabi$ of the $H\rightarrow C$ transition can exhibit the following correlations:
\begin{align}
\rabii=&\rabii^{\nr}+\Nsw\rabii^{\N}+\Nsw\Psw\rabii^{\N\P}+\Nsw\Esw\rabii^{\N\E}+\dots
\end{align}
Here, $\rabii^{\nr}$ is the dominant component of the Rabi frequency for laser $i\in\left\{ \p,\rX,\rY\right\} $, which could fluctuate in time on the order of 5\% due to laser power instability. $\rabii^{\N}$ is generated by a laser power difference between the $\tilde{\N}$ states. This arose because we routed the laser light along different paths through a series of AOMs for each state. We measured this effect with photodiodes and found that the largest fractional power correlation was $\rabi^{\N}/\rabi^{\nr}\approx2.5\times10^{-3}$. An additional contribution to $\rabii^{\N}$ and a contribribution to $\rabii^{\N\P}$ on the same order arises due to Stark mixing between rotational levels in $H$ and $C$, leading to $\Nsw$- and $\Nsw\Psw$-correlated transition amplitudes on the $H\rightarrow C$ transition.

\begin{figure}[htbp]
\centering
\includegraphics[width=12cm]{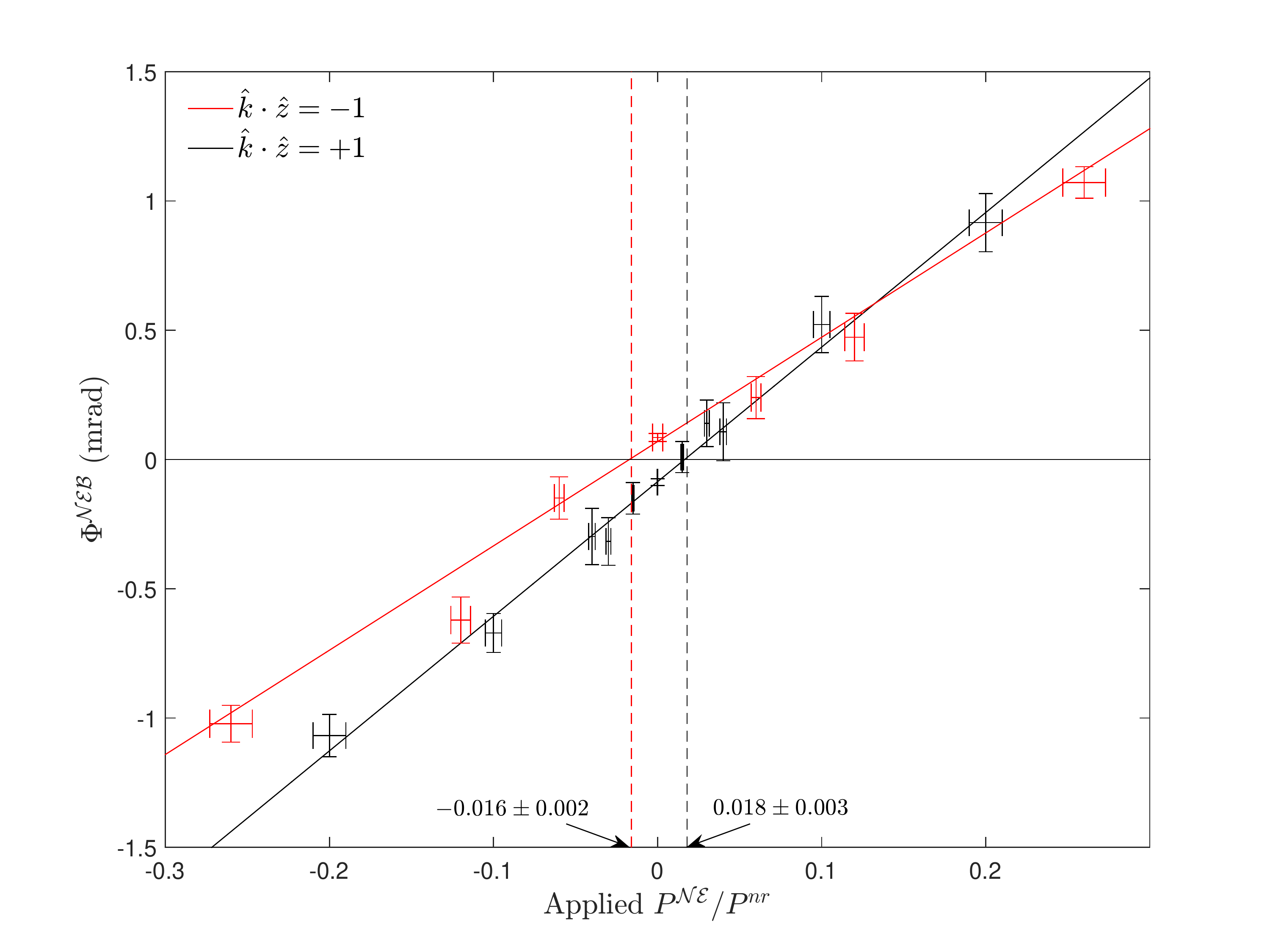}
\caption{$\Phi^{\N\E\B}$ as a function of applied $\Nsw\Esw$-correlated laser power, $P^{\N\E}$, for both directions of laser pointing, $\hat{k}\cdot\hat{z}$. The artificial $\rabi^{\N\E}$ resulting from correlated power $P^{\N\E}$ systematically shifts $\omega^{\N\E\B}$ in accordance with equation~\ref{eq:Empirical_AC_Stark_Shift_Phase_Result}. $\Phi^{\N\E\B}$ is zero when the applied $P^{\N\E}$ is such that there is no net $\Nsw\Esw$-correlated Rabi frequency. The intrinsic $\rabi^{\N\E}$ (i.e. that inferred when $P^{\N\E}=0$) changed sign with $\kz$ within the resolution of the measurement. The slopes between the two measurements differ due to differences in the AC Stark shift phase, believed to be due to differences in the spatial intensity profile and polarisation structure between the two measurements.}
\label{fig:Omega_NE}
\end{figure}

Although we did not observe a laser power correlation with $\Nsw\Esw$ we did observe signals consistent with a Rabi frequency correlation, $\rabi^{\N\E}$. A nonzero $\Nsw\Esw$-correlated fluorescence signal (as defined in section \ref{sec:signal_asymmetry}) that also reversed with the laser propagation direction $\kz$, $F^{\N\E}/F^{\nr}\approx-(2.4\times10^{-3})(\kz)$, together with a nonzero $\omega^{\N\E\B}\approx(2.5{}^{\:}\mathrm{mrad}/\mbox{s})(\B_{z}/\mathrm{mG})(\kz)$, provided the first evidence that a nonzero $\rabi^{\N\E}$ existed in our system. We believe that this fluorescence correlation arises from a linear dependence of the fluorescence signal size on Rabi frequency, $F^{\N\E}=(\partial F/\partial\rabi^{\nr})\rabi^{\N\E}$, which is nonzero since the state readout transitions were not fully saturated. We believe that the signal in $\omega^{\N\E\B}$ was caused by a coupling between the Rabi-frequency correlation and the $\B$-odd AC Stark shift phase, $\omega^{\N\E\B}=\frac{1}{\tau}\beta_{d\rabi}^{\B}\B_{z}(\rabi^{\N\E}/\rabi^{\nr})$. We were able to verify a linear dependence of both of these channels on $\rabi^{\N\E}$ by intentionally correlating the laser intensity with $\Nsw\Esw$ using AOMs; this is shown for the $\Phi^{\N\E\B}$ channel in Figure~\ref{fig:Omega_NE}. Varying the size of this artificial $\rabi^{\N\E}$ allowed us to measure the value present in the experiment under normal operating conditions, $\rabi^{\N\E}/\rabi^{\nr}=(-8.0\pm0.8)\times10^{-3}(\kz)$. $\rabi^{\N\E}$ can couple to $\beta_{d\rabi,i}^{\nr}$ as per equations~\ref{eq:Empirical_AC_Stark_Shift_Phase_Result} and \ref{eq:bdOmega} to result in a systematic error in $\wNE$. A nonzero $\beta_{d\rabi,i}^{\nr}$ can be produced by a linear polarisation angle gradient (not observed in the experiment) or by a non-reversing Zeeman shift component $\g\mu_{\rm B}\B_{z}^{\nr}$.

While searching for a model to explain the intrinsic $\rabi^{\N\E}$, we developed the Stark interference model presented in section \ref{sssec:stark_interference_between_E1_and_M1_transition_amplitudes}. For unnormalized effective polarisation $\vec{\varepsilon}_{\mathrm{eff}}=\vec{\varepsilon}_{\mathrm{eff}}^{\nr}+\Nsw\Esw d\vec{\varepsilon}_{\mathrm{eff}}^{\N\E}$,
this model predicts $\rabi^{\N\E}/\rabi^{\nr}\approx\text{Re}(\vec{\varepsilon}_{\mathrm{eff}}^{\nr *}\cdot d\vec{\varepsilon}_{\mathrm{eff}})\approx-\text{Im}\left[(a_{M1}+a_{E2})\right](\kz)$, which correctly predicts the dependence of $\rabi^{\N\E}$ on the laser propagation direction $\kz$. However, the factors $a_{\rm M1}$ and $a_{\rm E2}$, which correspond to the ratio of M1 and E2 amplitudes to the E1 amplitude, must be real for a plane wave, so $\text{Im}\left[(a_{M1}+a_{E2})\right]=0$. Hence this model fails to explain this Rabi frequency correlation unless there is some additional effect that introduces a phase shift between the E1 and M1 amplitudes. For example, interference between the E1 amplitude due to the incident laser beam, and a phase shifted M1 amplitude due to a (low intensity) reflected beam can lead to a nonzero $\rabi^{\N\E}$ by this model. However, this phase factor oscillates spatially on the scale of the light wavelength, which is very small compared to the size of the molecule cloud and hence should average out over the entire molecular beam cloud. The origin of the intrinsic $\rabi^{\N\E}$ is still not fully understood, and we are continuing to explore models to understand this effect.

\begin{figure}[!ht]
\centering
\includegraphics[width=12cm]{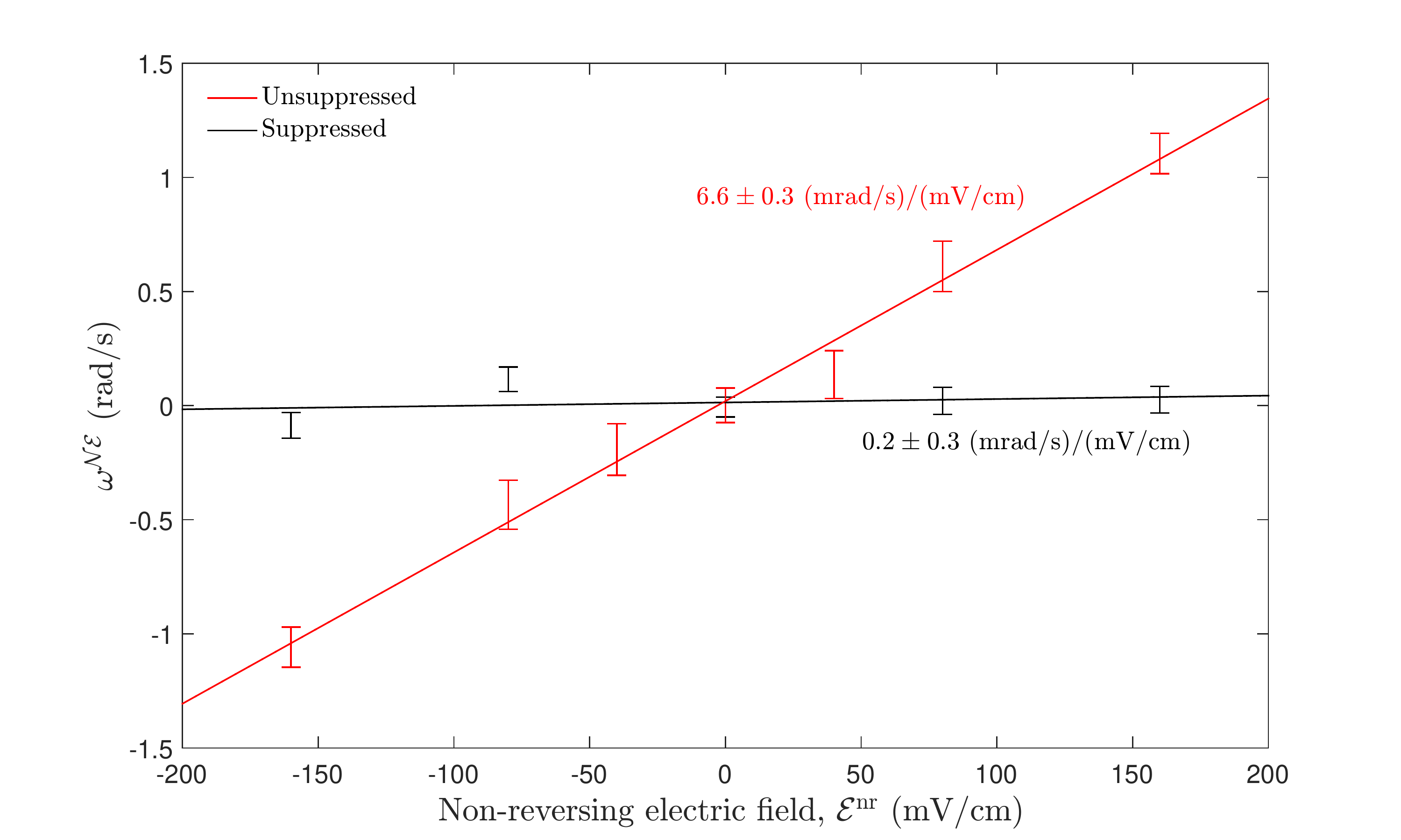}
\caption{Linear dependence of the $\wNE$ channel on an applied non-reversing electric field observed before (red) and after (black) we suppressed the known AC Stark shift phase by optimizing the preparation laser beam shape, time-averaged power and polarisation.}
\label{fig:Enr_slope}
\end{figure}

Given the empirical AC Stark shift phase model in equation~\ref{eq:Empirical_AC_Stark_Shift_Phase_Result}, the resulting systematic errors in the frequency measurement are given by
\begin{align}
\omega^{\N\E}_{\E^{\nr}} & =\frac{1}{\tau}\sum_{i\in\left\{ \p,\rX,\rY\right\} }\alpha_{\Delta,i}^{\nr}D_1\E^{\nr}(x_{i})\\
\omega^{\N\E}_{\rabi^{\N\E}} & =\frac{1}{\tau}\sum_{i\in\left\{ \p,\rX,\rY\right\} }\beta_{d\rabi,i}^{\nr}(\rabi^{\N\E}/\rabi^{\nr}).
\end{align}
Early in the experiment, we observed a nonzero systematic shift $\omega^{\N\E}_{\E^{\nr}}$ and took the steps outlined in section~\ref{sssec:suppression_of_the_AC_stark_shift_phases} to suppress it. To verify that the steps taken were effective, we examined $\wNE$ as a function of an intentionally applied non-reversing electric field. The resulting data are shown in figure~\ref{fig:Enr_slope}. The original slope, $\partial\omega^{\N\E}/\partial\E^{\nr}=(6.7\pm0.4)(\mathrm{rad/s})/(\mbox{V/cm})$,
corresponded to a systematic shift of $\omega^{\N\E}_{\E^{\nr}}\approx-34~\mathrm{mrad}/\mathrm{s}$ when combined with the measured $\E^{\nr}\approx-5^{\:}\mathrm{mV/\mathrm{cm}}$. Following the modifications described above, the $\partial\omega^{\N\E}/\partial\E^{\nr}$ slope was greatly suppressed, reducing the systematic error to $\omega^{\N\E}_{\E^{\nr}}<1~\mathrm{mrad}/\mathrm{s}$, well below the statistical uncertainty in the measurement of $\wNE$.

Because we observed that the parameters $\E^{\nr}$ and $\rabi^{\N\E}$ caused systematic errors in $\wNE$, we intermittently measured the size of the associated systematic errors throughout the datasets that were used for our reported result. We measured $\partial\omega^{\N\E}/\partial\E^{\nr}$ by applying a range of large non-reversing electric fields, up to around 70 times that present under normal running conditions. The value of $\partial\omega^{\N\E}/\partial\rabi^{\N\E}$ was measured by applying a correlated laser power $P^{\N\E}$ in the state preparation and state readout beams with a magnitude corresponding to an applied $\rabi^{\N\E}$ that was up to 20 times that measured under normal operating conditions. These parameters were measured for multiple values of the magnetic field magnitude, $\B_{z}$, for which different state readout laser beam polarisations were required. Due to known birefringent behavior of the AC stark shift phases, we allowed for this possibility for all AC stark shift phase systematic errors. We measured  $\partial\omega^{\N\E}/\partial\E^{\nr}$ for both $\hat{k}\cdot\hat{z}=\pm1$, but the $\rabi^{\N\E}$ systematic error was only discovered after the $\hat{k}\cdot\hat{z}=+1$ dataset and hence $\partial\omega^{\N\E}/\partial\rabi^{\N\E}$
was only monitored during the $\hat{k}\cdot\hat{z}=-1$ dataset. The $\rabi^{\N\E}$ systematic error during the $\hat{k}\cdot\hat{z}=+1$ dataset was determined from auxiliary measurements of the AC Stark
shift phase. As described in section \ref{ssec:efields}, $\E^{\nr}(x)$ exhibits significant spatial variation along the beam-line axis, $x$. However the $\E^{\nr}$ that was intentionally applied to determine $\partial\omega^{\N\E}/\partial\E^{\nr}$ was spatially uniform, and hence these measurements were insensitive to the difference $(\E^{\nr}(x_{\p})-\E^{\nr}(x_{\ro}))$ between the state preparation laser beam at $x_{\p}$ and the state readout beam at $x_{\ro}$. For this reason, we deduced the systematic error proportional to the difference $(\E^{\nr}(x_{\p})-\E^{\nr}(x_{\ro}))$ from auxiliary measurements of the AC Stark shift phase parameters, $\alpha_{\Delta,i}^{\nr}$.

In summary, the systematic errors proportional to $\E^{\nr}$ and $\rabi^{\N\E}$ that were evaluated and subtracted from $\omega^{\N\E}$ to report a measured value of $\wNEt$ can be expressed as

\begin{align}
\omega_{\E^{\nr}}^{\N\E}= & \left(\frac{\partial\omega^{\N\E}}{\partial\E^{\nr}}\right)\frac{1}{2}(\E^{\nr}(x_{\p})+\E^{\nr}(x_{\ro}))\nonumber\\
&+\frac{1}{\tau}(\alpha_{\Delta,\p}^{\nr}-\alpha_{\Delta,\rX}^{\nr}-\alpha_{\Delta,\rY}^{\nr})\frac{1}{2}(\E^{\nr}(x_{\p})-\E^{\nr}(x_{\ro}))\label{eq:Enr_systematic_error_value}\\
\omega_{\rabi^{\N\E}}^{\N\E}= & \begin{cases}
\frac{1}{\tau}\sum_{i\in\left\{ \p,\rX,\rY\right\} }\beta_{d\rabi,i}^{\nr}\left(\frac{\rabi^{\N\E}}{\rabi^{\nr}}\right) & (\hat{k}\cdot\hat{z})=+1\\
\left(\frac{\partial\omega^{\N\E}}{\partial\rabi^{\N\E}}\right)\rabi^{\N\E} & (\hat{k}\cdot\hat{z})=-1
\end{cases}
\end{align}
where $(\partial\omega^{\N\E}/\partial\E^{\nr})$ and $(\partial\omega^{\N\E}/\partial\rabi^{\N\E})$ were monitored by \emph{Intentional Parameter Variations} (see section~\ref{sec:Measurement_scheme_more_detail}) throughout the dataset used for our reported result, and $\E^{\nr}(x_{\p})$, $\E^{\nr}(x_{\ro})$,
$\rabi^{\N\E}$, $\alpha_{\Delta,i}^{\nr}$, and $\beta_{d\rabi,i}^{\nr}$ were obtained from auxiliary measurements. These two systematic errors account for almost all of the systematic offset that was subtracted from $\wNE$ to obtain $\wNEt$ as described in section~\ref{ssec:total_systematic_error_budget}.


\subsection{\texorpdfstring{$\A^{\N\E}$}{ANE} asymmetry effects}
\label{ssec:asymmetry_effects}
In addition to the dependence of the measured phase on laser detuning and Rabi frequency, we observed dependence of the asymmetry $\A$ (as defined in section~\ref{sec:signal_asymmetry}) on the laser parameters $\Delta_{\ro}$ and $\Omega_{{\rm r},\ro}$, due to differences between the properties of the $X$ and $Y$ readout laser beams. The laser-induced fluorescence signal $F(\Delta,\rabi)$ varies quadratically with detuning (for small detuning) and linearly with Rabi frequency. Under normal conditions, the signal sizes from $X$ and $Y$ are comparable, $F_X \approx F_Y \approx F$. If the $X$ and $Y$ beams have different wavevectors, $\vec{k}_{X,Y} = \vec{k}^{\mathrm{nr}} \pm \vec{k}^{XY}$, and $\vec{k}^{XY}$ has some component along $\hat{x}$, then the two beams will acquire different Doppler shifts. This leads to a linear dependence of the asymmetry on detuning, which in turn can couple to $\Delta^{\N\E}$ to result in a contribution to $\A^{\N\E}$,
\begin{equation}
\A^{\N\E} \approx\frac{1}{F} \frac{\partial^2 F}{\partial \Delta_{\ro}^2} (\vec{k}^{XY}\cdot\langle\vec{v}\rangle)\Delta^{\N\E}.
\end{equation}
Similarly, if the two readout beams differ in Rabi frequency, $\Omega_{r, X/Y} \approx \rabi^{\mathrm{nr}} \pm \rabi^{XY}$, the asymmetry becomes linearly dependent on Rabi frequency, which in turn can couple to $\rabi^{\N\E}$ to result in a contribution to $\A^{\N\E}$,
\begin{equation}
\A^{\N\E} \approx -\left(\frac{1}{F} \frac{\partial F}{\partial \rabi}\right)^2 \rabi^{XY} \rabi^{\N\E}.
\end{equation}

However, these asymmetry effects are very distinguishable from spin precession phases and polarisation misalignments. Since the $\Psw$ and $\Rsw$ switches effectively swap the role of the $X$ and $Y$ readout beams, the $\A^{\N\E}$ effects described above do not contribute to $\wNE$ when summed over these switches. Additionally, asymmetry effects, once converted to an equivalent frequency or phase, depend on the sign of the contrast, $\C$, unlike true phases. In the $\B_z\approx 20$~mG configuration, ${\rm sgn}(\C)={\rm sgn}(\B_z)$, but ${\rm sgn}(\C)$ has no dependence on ${\rm sgn}(\B_z)$ for $\B_z\approx 1,~40$~mG. Hence asymmetry correlations $\A^{\N\E}$ are mapped onto frequency correlations $\omega^{\N\E\P\R}$ or $\omega^{\N\E\B\P\R}$ depending on the magnetic field magnitude. 

If the pointing or Rabi frequency differences between the $X$ and $Y$ beams drift on timescales comparable to or shorter than the $\Psw$ or $\Rsw$ switches, these effects can occasionally `leak' into the `adjacent' channels $\omega^{\N\E\P}$, $\omega^{\N\E\R}$, $\omega^{\N\E\B\P}$, $\omega^{\N\E\B\R}$; however, we have not seen any evidence of these effects contributing to the $\omega^{\N\E}$ channel itself, and hence did not include systematic error contributions due to these effects in our systematic error budget.

\subsection{\texorpdfstring{$\Esw$}{E}-Correlated Phase}
\label{ssec:E_correlated_phase}

Previous eEDM measurements have often been limited by a variety of systematic errors that would have produced an $\Esw$-correlated phase precession frequency in our experiment, $\omega^{\E}$ \cite{Khriplovich1997,Murthy1989,Regan2002,Eckel2013}, such as $\Esw$-correlated leakage currents, geometric phases, and motional magnetic fields. Our ability to spectroscopically reverse the molecular orientation through a choice of $\Nsw$ distinguished these effects from an eEDM-generated phase. In addition, the aforementioned effects scale with the magnitude of the applied electric field, which was orders of magnitude smaller in our experiment than previous similar eEDM experiments due to the high polarisability of ThO \cite{Regan2002}. Also, because the molecular polarisation was saturated, the eEDM phase should have been independent of the magnitude of the applied field. We also note that any shifts from leakage currents and motional magnetic fields coupled through the magnetic dipole moment, which is near-zero in the $H$-state of ThO. Thus we expected $\omega^{\E}$ to be substantially suppressed, and that it should not enter $\omega^{\N\E}$ at any significant level.

\begin{figure}[htbp]
\centering
\includegraphics[width=12cm]{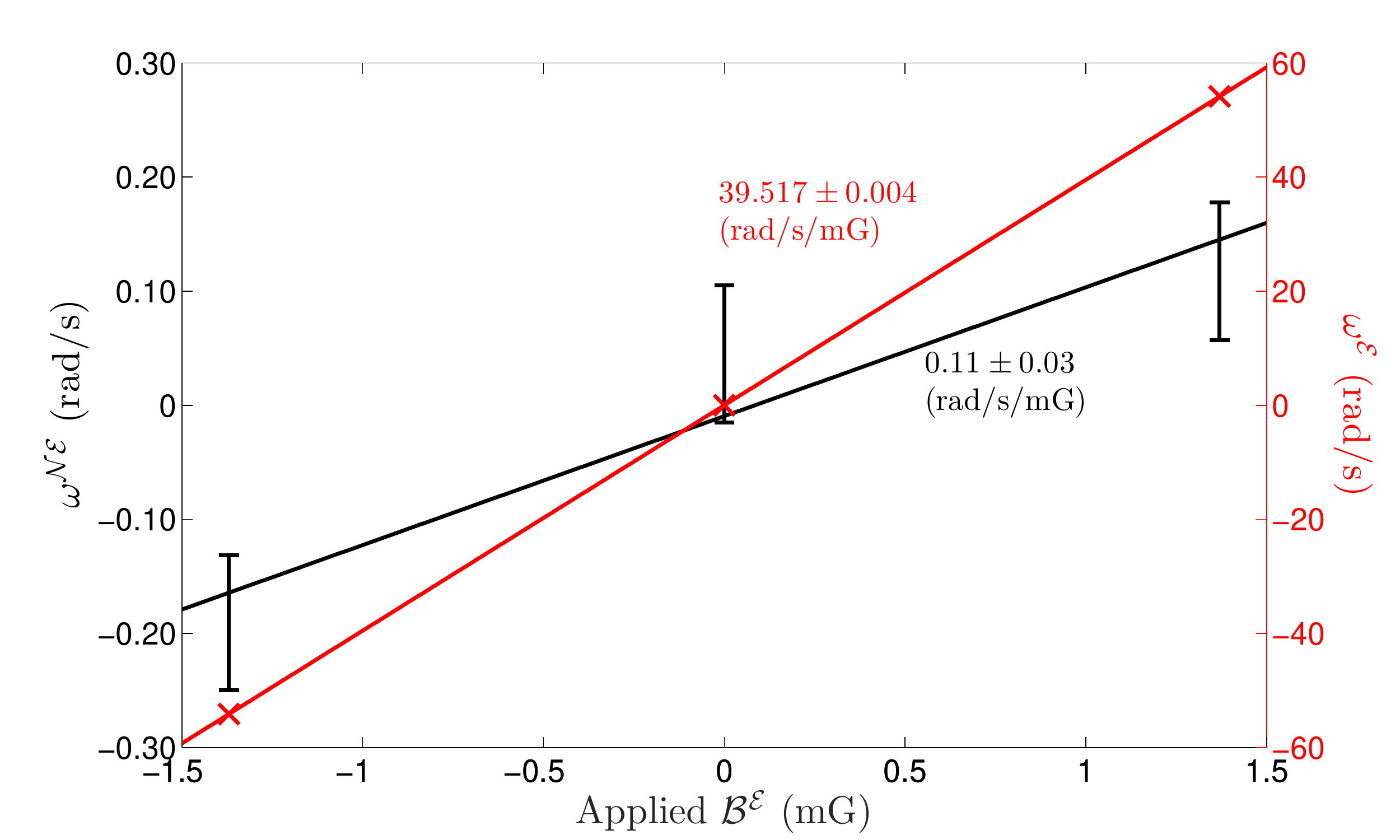}
\caption{Illustration of the $\sim$1000-fold suppression of systematic errors associated with $\omega^{\E}$ provided by the $\Nsw$ switch. Large values of $\omega^{\E}$ occur when there is a component of $\B_z$ correlated with $\Esw$, $\B^\E$. In previous eEDM experiments, this would have corresponded to a systematic error. In our experiment a much smaller shift in $\wNE$ results from the small difference in magnetic moments between the two $\Nsw$ levels. Error bars for the $\omega^{\E}$ data are significantly smaller than the data points. Data were taken with $\E=142~\mathrm{V}/\mathrm{cm}$ and the measured ratio of the slopes, $(\partial\wNE/\partial\B^\E)/(\partial\omega^{\E}/\partial\B^\E)=(2.8\pm0.8)\times 10^{-3}$ is consistent with the expected value $\eta \E/\g=(2.5\pm0.1)\times 10^{-3}$.}
\label{fig:phi_E}
\end{figure}

The reversal of $\Nsw$ did not, however, entirely eliminate an eEDM-like phase due to $\omega^{\E}$. As discussed in section~\ref{sec:compute_phase}, there was a small and $\E$-field dependent difference between the $g$-factors of the two $\Nsw$ levels \cite{Bickman2009,Petrov2014}, which meant that a systematic error in the $\omega^\E$ channel showing up in $\omega^{\N\E}$ at a level given by $\omega^{\N\E}_{\omega^\E}=(\eta\E/\g)\omega^{\E}$. 
We verified this relation by intentionally correlating a 1.4~mG component of our applied magnetic field with $\Esw$. This deliberate $\B^{\E}$ resulted in a large shift in the value of $\omega^\E$ and a $\sim$1000-times smaller offset of $\wNE$, as illustrated in figure~\ref{fig:phi_E}.

The intentionally applied $\B^\E$ was the only experimental parameter that was observed to produce a measurable shift in $\omega^{\E}$. Even large ($\sim$20~mG) magnetic fields components along $\hat{x}$ and $\hat{y}$, which exaggerate the effect of motional magnetic fields, did not shift $\omega^{\E}$ (this is expected, since the large tensor Stark shift in $\ket{H,J=1}$ dramatically suppresses the effect of motional magnetic fields \cite{Player1970}). For our eEDM data set, $\omega^{\E}$ was consistent with zero. We included a contribution from $\omega^{\E}$ in our error budget for $\wNE$ by multiplying the mean and uncertainty of the extracted $\omega^{\E}$ by our measured $|\E|$-dependent suppression factors $\eta\E/\g$.

\subsection{\texorpdfstring{$\Nsw$}{N}-Correlated Laser Pointing}
\label{ssec:N_correlated_pointing}
We discovered a nonzero, time-dependent signal in $\omega^{\N}$ which was associated with an $\Nsw$-correlated laser pointing, $\hat{k}^{\N}\approx5~\upmu$rad. An investigation into the mechanism behind this effect was inconclusive. We found that the pointing correlation appeared downstream of the AOMs that created the rapid polarisation switching and improved alignment was able to reduce the effect. We also found that the observed pointing was in some way correlated with the seed power and input angle of incidence into the high-power fiber amplifier immediately upstream of the polarisation switching, despite the fact that pointing out of the amplifier did not fluctuate. Since we used four different sets of AOMs to perform the $\Nsw$ and $\Psw$ switches before the amplifier, we observed laser pointing correlated with both of these switches. By matching the characteristics of these four beam paths we were able to suppress $\hat{k}^{\N}$ to $<1~\upmu$rad.

The effect of $\hat{k}^{\N}$ on $\omega^{\N}$ was studied by exaggerating the former with piezoelectrically actuated mirrors. Examining $\partial\omega^{\N}/\partial\hat{k}^{\N}$ showed significant fluctuations in its value. We were unable to identify the mechanism by which $\hat{k}^\N$ affected $\omega^{\N}$.

We had no evidence that the effect causing the observed variation in $\omega^{\N}$ also caused a systematic error in $\wNE$, but to be cautious we included an associated systematic uncertainty in our systematic error budget (section \ref{ssec:total_systematic_error_budget}). Assuming a linear relationship between $\omega^{\N\E}$ and $\omega^{\N}$, we extracted $\partial\omega^{\N\E}/\partial\omega^{N}$ from a combination of data taken under normal conditions and with an exaggerated $\omega^{\N}$ induced by an exaggerated $\hat{k}^{\N}$. We then placed an upper limit on a possible systematic error $\wNE_{\omega^{\N}}$ based on the value of $\omega^{\N}$ obtained under normal running conditions.
The resulting systematic uncertainty was four times smaller than our statistical uncertainty.

\subsection{Laser Imperfections}
\label{ssec:laser_imperfections}
Of the lasers used in our experiment, only the state preparation and readout lasers were known to produce possible systematic errors; imperfections in the rotational cooling, optical pumping or target ablation lasers simply resulted in a reduction in usable molecule flux. As part of our search for systematic errors, we intentionally exaggerated all known state preparation and readout laser imperfections possible without dismantling the apparatus (cf.\ table~\ref{tbl:syst_check}). In this section we describe this procedure and the resulting contributions to our systematic error budget.

\subsubsection{Laser Detuning}
\label{sssec:laser_detuning}
\hspace*{\fill} \\
The correlated components of the state preparation and readout laser beam detunings are described in detail in section~\ref{sec:state_prep_read}. Each detuning component was separately exaggerated and in some cases multiple components were simultaneously exaggerated. Most of the detuning terms in equation~\ref{eq:detuningcorrelations} were exaggerated to $\pm2\pi\times1$--2~MHz. No detuning or detuning correlation produced a significant shift in $\wNE$ other than $\Delta^{\N\E}$ caused by $\E^{\nr}$, discussed in section~\ref{sssec:correlated_laser_parameters}. In some cases, shifts in other phase channels were induced, but all shifts were consistent with well-understood AC Stark shift and asymmetry models described in sections \ref{sssec:AC_stark_shift_phases} and \ref{ssec:asymmetry_effects}.
For example, the combination of nonzero $\Delta^{\N}$ and $\Delta^{\nr}$ coupled to the $\B$-dependent component of the AC stark shift phase (equation~\ref{eq:bdOmega}) induces a significant shift in $\omega^{\N\B}$ (cf.\ equation~\ref{eq:Empirical_AC_Stark_Shift_Phase_Result}). 
Asymmetry correlations also resulted from these detuning correlations, but these were only manifested in channels odd with respect to $\Psw$ and $\Rsw$, and hence had no plausible effect on $\wNE$. Because the YbF eEDM experiment \cite{Kara2012} observed unexplained dependence of the measured eEDM value on state preparation microwave detuning, we included a systematic error contribution from all detuning imperfections in our systematic error budget.

\subsubsection{Laser Pointing and Intensity}
\label{sssec:laser_pointing_and_intensity}
\hspace*{\fill} \\
Similar to detuning imperfections, the state preparation and readout lasers could have imperfect pointing and correlated intensities. Ideally the laser propagation direction, $\hat{k}$, would have been parallel to the laboratory electric field. This would have diminished the amount of $\hat{z}$ polarised light experienced by the molecules, which could drive unwanted off-resonant transitions, and prevented stray retroflection from the ITO field plate surfaces. Using this ITO retroflection as a guide, we aligned $\hat{k}$ perpendicular to the field plate surface, and therefore parallel to $\hat{\E}$, to within $\sim{3}$~mrad. To test for errors related to imperfect pointing, both the state preparation and readout pointing misalignments were exaggerated in the $x$-direction to $\pm$10~mrad, as was the relative pointing of the $\hat{X}$ and $\hat{Y}$ state readout beams. The vacuum windows and $\sim$3.8~cm wide holes in the magnetic shields prevented us from further misaligning the beams. To decouple pointing imperfections from detuning imperfections, the state preparation and readout laser frequencies were tuned to resonance after each pointing adjustment. No shift in $\wNE$ was observed and no systematic error contribution from pointing imperfections was included. Pointing imperfections were only observed to affect the signal asymmetry, as previously discussed in section \ref{ssec:asymmetry_effects}.

Unlike laser pointing and detuning, there was no `ideal' value for laser intensity. The state preparation and readout laser intensities were chosen such that we were driving optical pumping to completion on the $H\rightarrow C$ transition without producing unnecessary thermal stress on the field plates. We decreased each laser intensity by a factor of four to check that there was no variation in $\wNE$. We observed a nonzero $\rabi^{\N}$ 
caused by the $\Nsw$-correlated seed power into the high-power fiber amplifiers and by Stark mixing between rotational levels in $H$ and $C$ as discussed in section \ref{sssec:correlated_laser_parameters}. We exaggerated this imperfection by a factor of 20. Only $\omega^{\N\B}$ was shifted, consistent with our understanding of the $\B$-correlated AC Stark shift phase. These intensity systematic error checks were not included in the systematic error budget.

\subsection{Magnetic Field Imperfections}
\label{ssec:magnetic_field_imperfections}
The $H$ state is very insensitive to a magnetic field $\B_z$ due to its small $g$-factor, as discussed in section~\ref{sec:tho_molecule}. Sensitivity to the transverse fields is even further suppressed by the large size of the tensor Stark shift relative to the Zeeman interaction. Nevertheless, there are known mechanisms by which magnetic field imperfections can contribute to systematic errors: $\B_z^{\nr}$ can contribute to the $\wNE_{\rabi^{\N\E}}$ systematic error discussed in section~\ref{sssec:correlated_laser_parameters}, and transverse fields $\B_x^{\nr}$ and $\B_y^{\nr}$ can lead to the geometric phase systematic errors \cite{Vutha2010} discussed in section~\ref{ssec:E_correlated_phase}. We designed the experiment to allow a wide variety of magnetic field tilts and gradients to be applied as described in section~\ref{sec:bfields} and we directly looked for systematic errors resulting from these magnetic field imperfections.

Both $\B$-correlated and uncorrelated imperfections were applied. We did not precisely measure the residual values of each of these parameters along the molecule beam line until we had studied all systematic errors and collected our published data set. Based on the projected ${\sim}10^5$ magnetic shielding factor, we expected all stray magnetic fields and gradients to be on the order of 10~$\upmu$G and 1 $\upmu$G/cm, respectively. For this reason we only exaggerated these imperfections to $\sim$2~mG and $\sim$0.5~mG/cm. When we mapped out the magnetic field with a magnetometer inserted between the electric field plates as described in section \ref{sec:bfields}, we discovered that several imperfections were much larger than we expected (e.g.\ $\B_y \approx 0.5$~mG). This was caused by poor magnetic shielding due to insufficient shield degaussing. For this reason we gathered additional eEDM data with some magnetic field parameters exaggerated by an additional factor of five. $\wNE$ and nearly all other frequency channels, apart from $\omega^{\nr}$ and $\omega^{\B}$ were not observed to be affected by any of these magnetic field parameters. 
Because uncorrelated stray magnetic fields and magnetic field gradients caused unexpected eEDM offsets in the PbO eEDM experiment \cite{Eckel2013}, we included contributions from all uncorrelated magnetic field imperfections in our systematic error budget described in section \ref{ssec:total_systematic_error_budget}.

\subsection{Electric Field Imperfections}
\label{ssec:electric_field_imperfections}
Unlike the magnetic field, we do not have the ability to control electric field gradients and stray electric fields, aside from the average value of $\Enr$. The field plates were located at the center of the experiment, inside the vacuum chamber and magnetic shields and coils, with no direct access available. To search for systematic errors related to the electric field, equal amounts of eEDM data were gathered with two different electric field magnitudes. The $\wNE$ values from both field magnitudes were consistent with each other. The YbF eEDM experiment observed unexplained eEDM dependence on the voltage offset common to both field plates. For this reason we exaggerated this offset by a factor of 1000 (relative to its residual value of ${\approx}5$~mV) and, even though it did not shift our eEDM measurement, included it in our systematic error budget. 

\subsubsection{Molecule Beam} \label{ssec:molecular_beam}
\hspace*{\fill} \\
The molecule beam should have ideally travelled parallel to the electric field plates and well-centred between the plates. This minimizes Doppler shifts, protects the plates from being coated with ThO, and ensures that the molecules experience the most uniform electric field. The entire beam source vacuum chamber sat on a two axis ($yz$) translation stage. The exit aperture of the buffer gas cell was aligned to within 1~mm of the centre of the fixed collimators and electric field plates, using a theodolite. Geometric constraints only allowed us to exaggerate the cell misalignment by roughly a factor of three (up to 3~mm) before the molecules would have hit the sides of the field plates. We also 
varied the transverse spatial and velocity distributions
by using adjustable collimators between the beam source and spin-precession region to block half of the beam from the $\pm\hat{x},\pm\hat{z}$ directions. The value of $\wNE$ was not observed to shift with any molecule beam parameter adjustment.


\subsection{Searching for Correlations in the eEDM Data Set} \label{ssec:correlations_in_the_eEDM_data_set}

\begin{figure}[htbp]
\centering
\includegraphics[width=\textwidth]{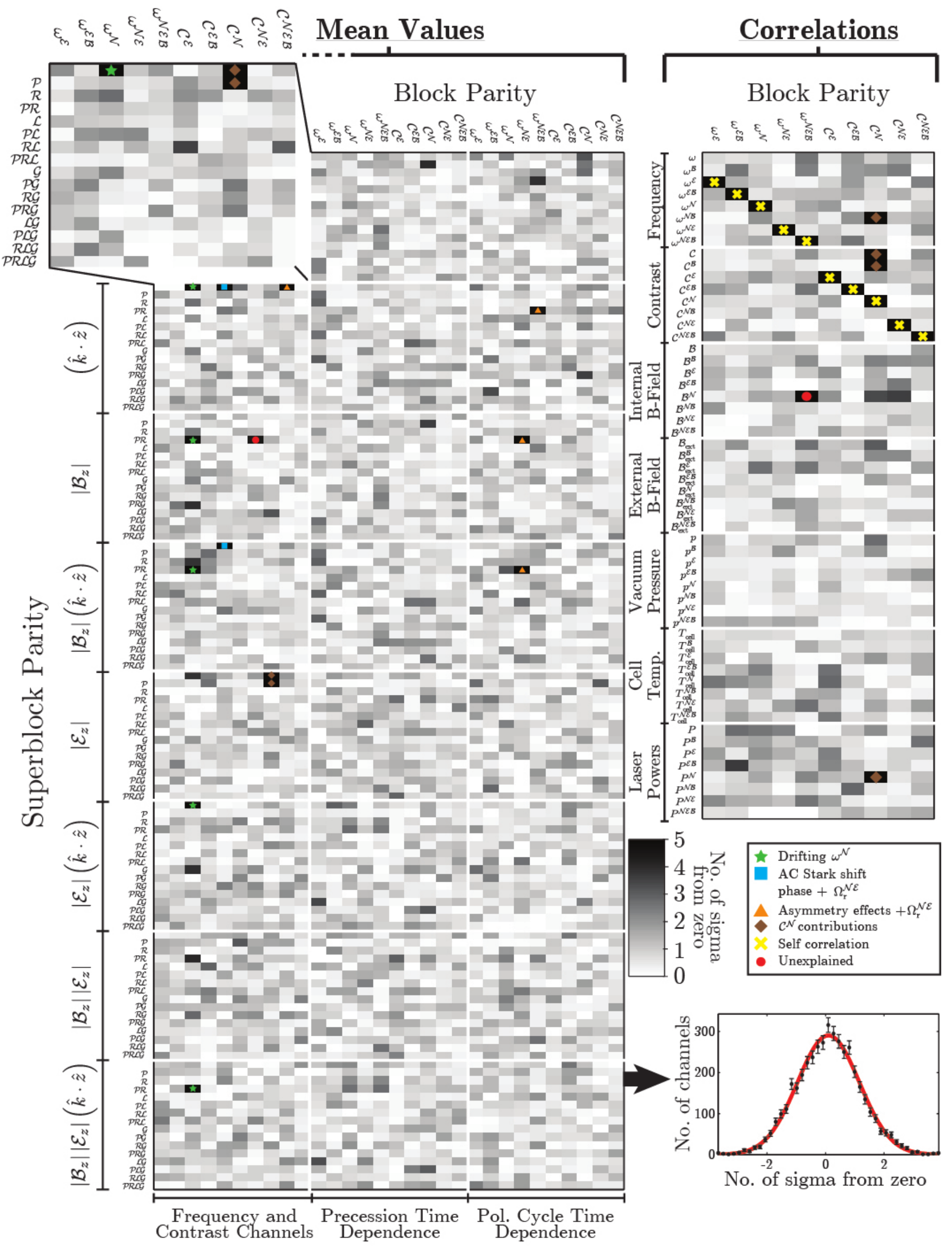}
\caption{Over 4,000 switch-parity channels (left) and correlations between switch parity channels (upper right) computed from the eEDM data set. The deviation of each quantity from zero in units of the statistical uncertainty is indicated by the grayscale shading. We set a significance threshold of $4\sigma$ above which there is a probability of $p=25\%$ of finding at least 1 false positive. We mark each significant channel/correlation with a symbol corresponding to a model known to produce a signal in that channel. The quantities below this threshold exhibit a normal distribution, shown in the lower right.
}
\label{fig:pixel_plot}
\end{figure}

In addition to performing systematic error checks for possible variations of $\omega^{\N\E}$ with various experimental parameters, we searched for statistically nonzero values within the set of 1536 possible correlations with the block and superblock switches. This analysis was performed for our primary measured quantities $\omega$, $\C$, and $\F$ and for a wide range of auxiliary measurements such as laser powers, magnetic field, room temperature, etc. We also examined the switch-parity channels of $\omega$, $\C$, and $\F$ as a function of time within the molecule beam pulse, and as a function of time within the polarisation switching cycle. We used the Pearson correlation coefficient to look for correlations between the aforementioned switch-parity channels and used the autocorrelation function to look for signs of time variation of the mean within those channels. Figure~\ref{fig:pixel_plot} illustrates data from such a search with a subset of the previously described quantities. 
In this search, we looked at 4390 quantities and we set the significance threshold at $4\sigma$ which correponds to a probability of $p\approx0.25$ that there will be one or more false positives above that threshold. We represented the significance of each of these quantities with a grayscale pixel. Each pixel that was significant at the $4\sigma$ level is marked with a symbol corresponding to a known explanatory physical model, or a red dot if the signal is not yet explained. The fact that we understand most of the significant signals present in our experiment, combined with the fact that the statistical distribution of the remaining signals below the significance threshold is consistent with a normal distribution, gives us added confidence in our models of the experiment and our reported eEDM result.

Channels/correlations marked with symbols are significantly nonzero due to known mechanisms as follows:
\begin{itemize}
    \item Green stars: Correlations due to the nonzero and drifting signal in the $\omega^{\N}$ channel described in section~\ref{ssec:N_correlated_pointing}.
    \item Light blue squares: Signals in $\omega^{\N\E\B}$ channels due to the $\Bsw$-odd AC stark shift phase coupling to $\rabi^{\N\E}$ as described in section~\ref{sssec:correlated_laser_parameters}.
    \item Orange triangles: Correlations due to contrast or asymmetry coupling to $\rabi^{\N\E}$. Contrast correlations arise simply because there is a linear dependence of total contrast on Rabi frequency, and the asymmetry correlation is described in section~\ref{ssec:asymmetry_effects}.
    \item Brown diamond: Correlations in $\C^{\N}$ and related contrast channels due to nonzero Rabi frequency correlations $\rabi^{\N}$ and $\rabi^{\N\P}$. These arise due to laser power correlations with the $\Nsw$ and $\Psw$ switches and due to Stark mixing between rotational levels in $H$ and $C$, which create $\Nsw$- and $\Psw$-correlated transition amplitudes on the $H\rightarrow C$ transition as described in section~\ref{sssec:laser_pointing_and_intensity}.
    \item Red dot: Signals above our significance threshold for which we have been unable to find a plausible explanation. Even if these quantities arise from real physical effects, they would need to couple to other correlated quantities to contribute to $\omega^{\N\E}$ and there is no evidence for this in the eEDM dataset. 
\end{itemize}


\subsection{Systematic Error Budget}
\label{ssec:total_systematic_error_budget}

The method used for construction of a systematic uncertainty varies from experiment to experiment (see for example \cite{sinervo2003,barlow2002}), and it is ultimately a subjective quantity. Even if individual contributions are derived from objective measurements, their inclusion or exclusion in the systematic uncertainty is subjective. Furthermore, the systematic uncertainty cannot possibly be a measure of the uncertainty in all systematic errors in the experiment, but rather only those which were identified and searched for. Although we work hard to identify all significant systematic errors in the measurement, we cannot rule out the possibility that some were missed.

Our criteria for including a given quantity in the systematic uncertainty consist of three classes of systematic errors in order of decreasing importance of inclusion:

\begin{enumerate}[(A)]
\item If we measured a nonzero correlation between $\wNE$ and some parameter which had an ideal value in the experiment, we performed auxiliary measurements to evaluate the corresponding systematic error and subtract that error from $\wNE$ to obtain $\wNEt$. The statistical uncertainty in the shift made to $\wNE$ contributed to the systematic uncertainty.
\item If we observed a signal in a channel that we deemed important to understand, and it was not understood, but was not observed to be correlated with $\wNE$, we set an upper limit on the shift in $\wNE$ due to a possible correlation between the two channels. Since such a signal represented a gap in our understanding of the experiment, we added this upper limit as a contribution to the systematic uncertainty.
\item If a similar experiment saw a nonzero, not understood correlation between their measurement channel and some parameter with an ideal experimental value, but we did not observe an analogous correlation, we set an upper limit on the shift in $\wNE$ due to this imperfection. Since this signal may have signified a gap in our understanding of our experiment, we added this upper limit as a contribution to the systematic uncertainty.
\end{enumerate}

\begin{table}[tbp]
\centering
\caption{Systematic error shifts and uncertainties for $\wNE$, in units of mrad/s grouped by inclusion class (defined in the text). Total uncertainties are calculated by summing the individual contributions in quadrature. Note that $\wNE\approx1.3$~mrad/s corresponds roughly to $1\times10^{-29}~\ecm$ for our experiment.}
\begin{tabular}{llcc}
\br
Class & Parameter & Shift (mrad/s) & Uncertainty (mrad/s)\\
\mr
A & $\E^{\nr}$ correction & $-0.81$ & $0.66$\\
A & $\rabi^{\N\E}$ correction & $-0.03$ & $1.58$\\
A & $\omega^{\E}$ correlated effects & $-0.01$ & $0.01$\\
B & $\omega^{\N}$ correlation &  & $1.25$\\
C & Non-reversing $\B$-field $\left(\B_{z}^{\nr}\right)$ &  & $0.86$\\
C & Transverse $\B$-fields $\left(\B_{x}^{\nr},\B_{y}^{\nr}\right)$ &  & $0.85$\\
C & $\B$-field gradients &  & $1.24$\\
C & Prep./readout laser detunings &  & $1.31$\\
C & $\Nsw$ correlated detuning &  & $0.90$\\
C & $\E$-field ground offset &  & $0.16$\\
\mr
  & Total Systematic & $-0.85$ & $3.24$\\
\mr
  & Statistical Uncertainty &  & $4.80$\\
\mr
  & Total Uncertainty &  & $5.79$\\
\br
\end{tabular}
\label{tbl:syst_error}
\end{table}

Table \ref{tbl:syst_error} contains a list of the contributions to our systematic error, grouped by inclusion class, with the corresponding shifts and/or uncertainties. Accounting for class A systematic errors was obligatory, and the removal of these errors from $\wNE$ can be viewed as a redefinition of the measurement channel to $\wNEt$ which does not contain those unwanted effects. These systematic errors consisted of those that depended on the parameters $\E^{\nr}$, $\rabi^{\N\E}$, and $\omega^{\E}$ as described in sections \ref{sssec:correlated_laser_parameters} and \ref{ssec:E_correlated_phase}, and as such our reported measurement of the $T$-odd spin precession frequency is defined as $\wNEt=\wNE-\wNE_{\E^{\nr}}-\wNE_{\rabi^{\N\E}}-\wNE_{\omega^\E}$. The class B and class C systematic errors were included in the systematic uncertainty to lend credance to our result despite unexplained signals and unexplained systematic errors in experiments similar to ours.
All uncertainties in the contributions to the systematic error were added in quadrature to obtain the systematic uncertainty.

With reference to the class B criterion, we deemed the following channels as important to understand: $\omega^{\N},$ $\omega^{\E},$ $\omega^{\E\B}$, and $\omega^{\N\E\B}$. Signals were initially not expected in any of these channels and could be measured with the same precision as $\wNE$. The $\omega^{\nr}$, $\omega^{\B}$ and $\omega^{\N\B}$ channels were not included in our systematic error since the Zeeman spin precession signals present in these channels had non-stationary means and additional noise due to drift in the molecule beam velocity. Only one of these channels, $\omega^{\N}$, described in section \ref{ssec:N_correlated_pointing}, met the class B inclusion criterion. 

With reference to the class C criterion, we defined the set of experiments similar to ours to include other eEDM experiments performed in molecules: the YbF experiment \cite{Hudson2011} and the PbO experiment \cite{Eckel2013}. The PbO experiment observed unexplained systematic errors coupling to stray magnetic fields and magnetic field gradients (cf.\  section~\ref{ssec:magnetic_field_imperfections}), and the YbF experiment observed unexplained systematic errors proportional to detunings (cf.\ section~\ref{sssec:laser_detuning}) and a field plate ground voltage offset (cf.\ section~\ref{ssec:electric_field_imperfections}). Thus we included the systematic uncertainty associated with the aforementioned effects in our budget.

After having accounted for the systematic errors and systematic uncertainty, we reported $\wNEt$, the contribution to the channel $\wNE$ induced by $T$-odd interactions present in the $H$ state of ThO, as
\begin{align}
\label{eq:wNEt_num}
\wNEt=&2.6 \pm 4.8_{\rm{stat}}\pm 3.2_{\rm{syst}}~\rm{mrad}/\rm{s}\\
     =&2.6 \pm 5.8~\rm{mrad}/\rm{s}, \label{eq:wNEt_num_err_comb}
\end{align}
where the combined uncertainty is defined as the quadrature sum of the statistical and systematic uncertainties, $\sigma^2=\sigma_{\rm{stat}}^2+\sigma_{\rm{syst}}^2$. This result is consistent with zero within $1\sigma$. Since $\sigma_{\rm{syst}}$ is to some extent a subjective quantity, its inclusion should be borne in mind when interpreting confidence intervals based on $\sigma$. Nevertheless, this inclusion decision does not have a large impact on the meaning of the resulting confidence intervals since $\sigma$ is only about 20\% larger than $\sigma_{\rm{stat}}$.


\section{Interpretation}
 \label{sec:interpretation}

\subsection{Confidence Intervals}
\label{ssec:confidence_intervals}
A classical (i.e.\ frequentist) confidence interval \cite{Riley2006} is a natural choice for reporting the result of an eEDM measurement.
For repeated and possibly different experiments measuring the eEDM, the frequency with which the confidence intervals include or exclude the value $\de=0$ suggests whether the results are consistent or inconsistent, respectively, with the Standard Model.
Furthermore, the confidence level (C.L.) represents an objective measure of the \emph{a priori} probability that the confidence interval assigned to any one of these measurements, selected at random, includes the unknown true value of the eEDM $d_{e,{\rm true}}$. 
Since no statistically significant eEDM has yet been observed, the recent custom has been for electron eEDM experiments to report an upper limit at the 90\% C.L. \cite{Regan2002,Hudson2011}. 
The proper interpretation of such limits is that if the experiment were performed a large number of times, and the confidence interval were \emph{computed in the same way} for each experimental trial, $d_{e,\rm true}$ would fall within the interval 90\% of the time.

Feldman and Cousins pointed out that in order for this interpretation to be valid, the confidence interval construction must be independent of the result of the measurement \cite{Feldman1998}. If the procedure for constructing 90\% confidence intervals is chosen contingent upon the measurement outcome, the resulting intervals may `undercover', i.e. fail to include the true value more than 10\% of the time. This happens, for example, if an upper bound is reported whenever the measured result falls within a few standard deviations of zero, and a two-sided confidence interval is reported whenever the measured result is significant at more than a few-sigma level. Feldman and Cousins termed this inconsistent approach `flip-flopping'.


In order to avoid flip-flopping, we chose a confidence interval construction, the Feldman-Cousins method described in reference~\cite{Feldman1998}, that consistently unifies these two limits. We applied this method to a model with Gaussian statistics, in which the measured magnitude of the eEDM channel, $x=|\omega^{\N\E}_{T,{\rm meas}}|$, is sampled from a folded Gaussian distribution
\begin{equation}\label{eq:foldednormal}
P(x|\mu)=\frac{1}{\sigma\sqrt{2\pi}}\left(\exp\left[-\frac{(x-\mu)^2}{2\sigma^2}\right]+\exp\left[-\frac{(x+\mu)^2}{2\sigma^2}\right]\right),
\end{equation}
where the location parameter is the unknown true magnitude of the eEDM channel, $\mu=|\omega^{\N\E}_{T,{\rm true}}|$, and the scale parameter $\sigma$ is equal to the quadrature sum of the statistical and systematic uncertainties given in equation~\ref{eq:wNEt_num_err_comb} and at the bottom of table~\ref{tbl:syst_error}. 

The central idea of the Feldman-Cousins approach is to use an ordering principle which, for each possible value of the parameter of interest $\mu$, ranks each possible measurement outcome $x$ by the `strength' of the evidence it provides that $\mu$ is the true value. The values of $x$ that provide the strongest evidence for each value of $\mu$ are included in the confidence band for that value. In the Feldman-Cousins method, the metric for the strength of evidence is the likelihood of $\mu$ given that $x$ is measured [i.e. $\mathcal{L}(\mu|x) = P(x|\mu)$], divided by the largest probability $x$ can possibly achieve for any value of $\mu$. The denominator in this prescription takes into account the fact that an experimental result that is somewhat improbable under a particular hypothesis can still provide good evidence for that hypothesis if the result is similarly improbable under even the most favorable hypothesis. This approach has its theoretical roots in likelihood ratio testing \cite{Stuart1999}.

\begin{figure}[!ht]
\centering
\includegraphics[width=0.49\textwidth]{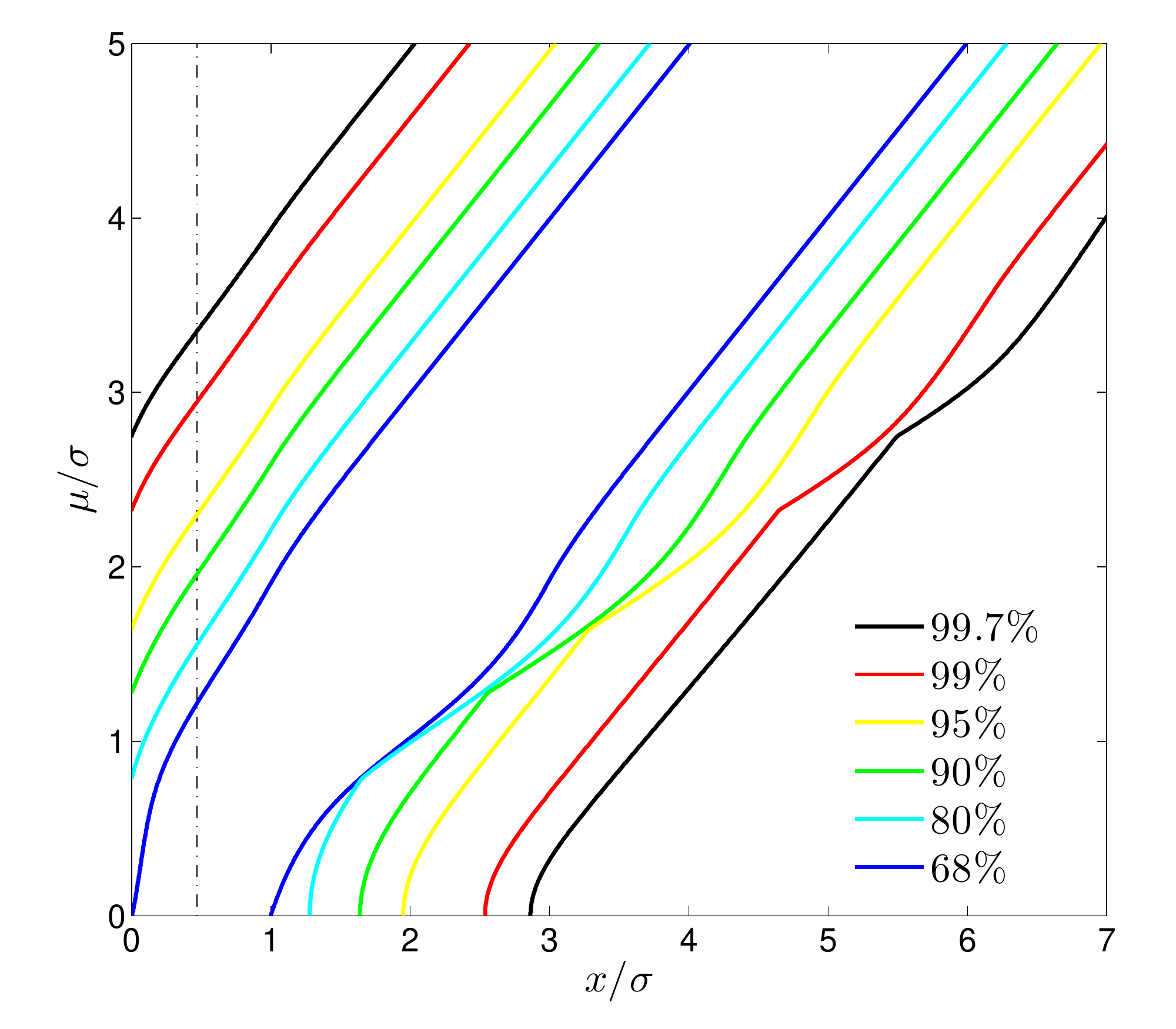}
\includegraphics[width=0.49\textwidth]{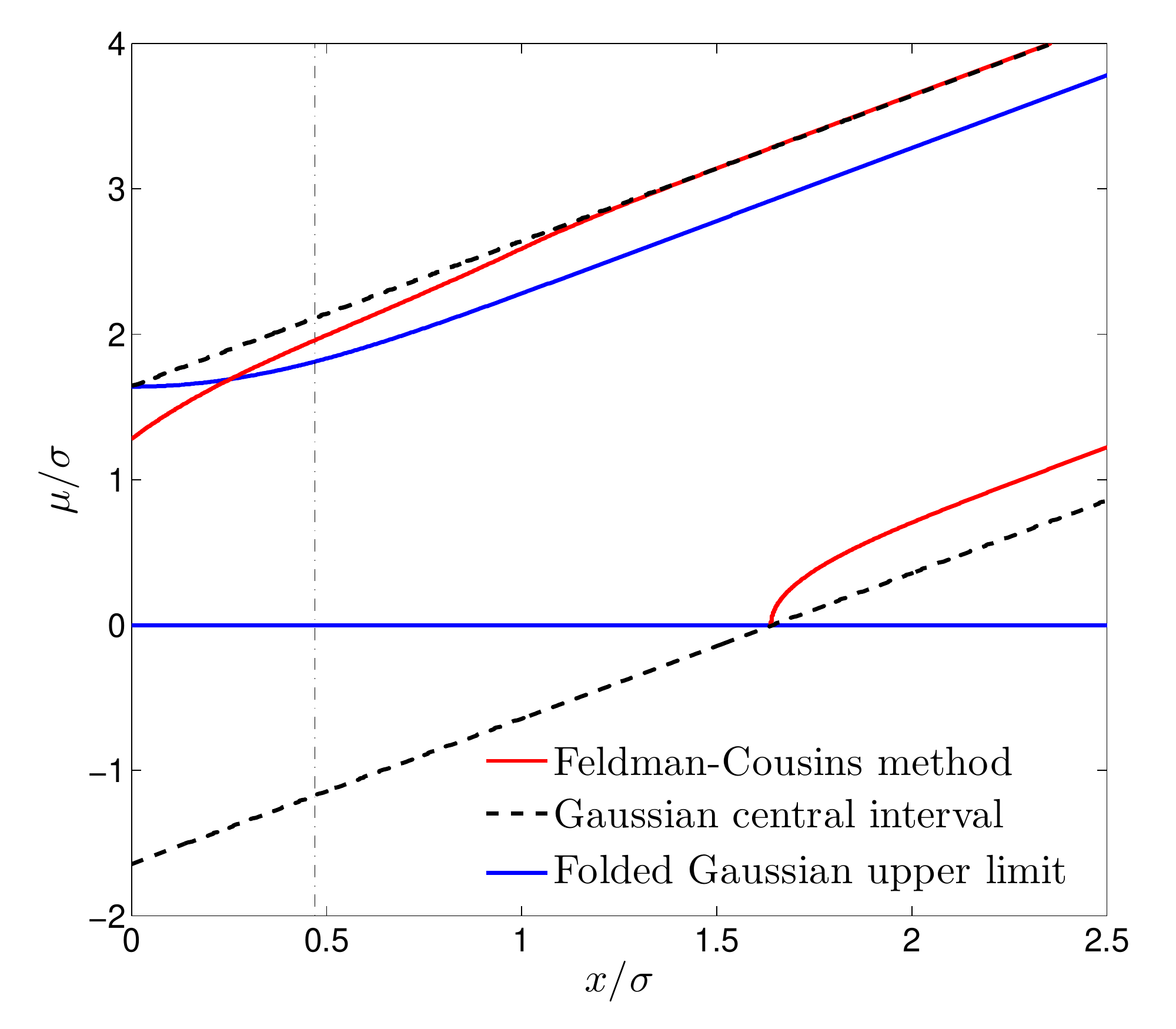}
\caption{Left: Feldman-Cousins confidence bands for a folded Gaussian distribution, constructed as described in the text, for a variety of confidence levels. Each pair of lines indicates the upper and lower bounds of the confidence band associated with each C.L. To the left of the $x$-intercepts, the lower bounds are zero. Confidence bands are plotted as a function of the possible measured central values $x$ scaled by the standard deviation $\sigma$, and our result is plotted as a vertical dot-dashed line. The $\mu$-value of the point at which our result line intersects with each of the colored lines gives the upper limit of our measurement at different C.L.'s. Right: Comparison between 90\% confidence intervals computed using three different methods, described in the text. Confidence bands are plotted as a function of the possible measured central values of a quantity $x$ scaled by the standard deviation $\sigma$. Our result, $|\omega^{\N\E}_{T,{\rm meas}}|/\sigma=0.46$, is plotted as a vertical dot-dashed line. The $\mu$-values of the points at which our result line intersects the upper and lower line for each method give the upper and lower bounds of three possible 90\% confidence intervals for our measurement. To avoid invalidating the confidence interval by flip-flopping, our result should be interpreted using the Feldman-Cousins method, which we chose before unblinding.}
\label{fig:fc_conf_int}
\end{figure}

Our specific  procedure for computing confidence intervals was a numerical calculation performed using the following recipe (cf.\ figure~\ref{fig:fc_conf_int}):
\begin{enumerate}
\item{Construct the confidence bands on a Cartesian plane, of which the horizontal axis represents the possible values of $x$ and the vertical axis the possible values of $\mu$. Divide the plane into a fine grid with $x$-intervals of width $\Delta_x$ and $\mu$-intervals of height $\Delta_{\mu}$. We will consider only the discrete possible values $x_i = i \Delta_x$ and $\mu_j = j \Delta_{\mu}$, where the index $i$($j$) runs from $0$ to $n_x$($n_{\mu}$).}\label{it:setup}
\item{For all values of $i$, maximize $P(x_i|\mu_j)$ with respect to $\mu_j$. Label the maximum points $\mu^{\mathrm{max},i}$.}\label{it:ymax}
\item{For some value of $j$, say $j=0$, compute the likelihood ratio $R(x_i) = P(x_i|\mu_j)/P(x_i|\mu^{\mathrm{max},i})$ for every value of $i$.}\label{it:R}
\item{Construct the `horizontal acceptance band' at $\mu_j$ by including values of $x_i$ in descending order of $R(x_i)$. Stop adding values when the cumulative probability reaches the desired C.L. of 90\%, i.e., $\displaystyle \sum_{x_i}P(x_i|\mu_j)\Delta_x = 0.9$.}\label{it:horiz}
\item{Repeat steps (\ref{it:R})--(\ref{it:horiz}) for all values of $j$.}
\item{To determine the reported confidence interval, draw a vertical line on the plot at $x = |\omega^{\N\E}_{T,{\rm meas}}|$. The 90\% confidence interval is the region where the line intersects the constructed confidence band.}
\end{enumerate}


The left-hand plot in figure~\ref{fig:fc_conf_int} was generated using the prescription above at several different C.L.'s. 
Note that the 90\% confidence intervals switch from upper bounds to two-sided confidence intervals when the value of $|\omega^{\N\E}_{T,{\rm meas}}|$ becomes larger than $1.64 \sigma$. This is the level of statistical significance required to exclude the value $\de = 0$ from a 90\% C.L.\ central Gaussian confidence band.

From equation~(\ref{eq:wNEt_num_err_comb}), we find $|\omega^{\N\E}_{T,{\rm meas}}|=0.46\sigma$ with $\sigma=5.79~\rm{mrad}/\rm{s}$. In our confidence interval construction, this corresponds to an upper bound of $|\wNEt|<1.9\sigma=11~\rm{mrad}/\rm{s}$ (90\% C.L.). 
A comparison between three different 90\% confidence interval constructions for small values of $\mu$ is shown in the right-hand plot of figure~\ref{fig:fc_conf_int}. The black dashed lines represent the central confidence band for the signed values (rather than the magnitude) of $\mu$ and $x$, where $\mu$ is the mean of a Gaussian probability distribution in $x$. The blue lines give an upper bound constructed by computing the the value of $\mu$ such that the cumulative distribution function for the folded Gaussian in equation~\ref{eq:foldednormal} is equal to $0.9$ for each value of $x$. It should be noted that this upper bound is more conservative than a true classical 90\% confidence band, as it overcovers for small values of $\mu$ (e.g., if the true value were $\mu_{\rm true} < 1.64 \sigma$, the confidence intervals of 100\% of experimental results would include $\mu_{\rm true}$). We nevertheless include this construction for comparison because we believe that previous experiments have reported EDM upper bounds using this method \cite{Hudson2011,Griffith2009,Regan2002}. These intervals have a valid interpretation as Bayesian `credible intervals' conditioned on a uniform prior for $\mu$ \cite{Feldman1998}. Finally, the red lines represent the Feldman-Cousins approach described here, which unifies upper limits and two-sided intervals. For our measurement outcome, indicated by the vertical dot-dashed line, the Feldman-Cousins intervals yield a $7\%$ larger eEDM limit than the folded Gaussian upper bound would have.

\subsection{Physical Quantities}
\label{ssec:physical_quantities}


Under the most general interpretation, our experiment is sensitive to any $P$- and $T$-violating interaction that produces an energy shift $\wNEt$. The eEDM is not the only such predicted interaction for diatomic molecules \cite{Kozlov1995}, and in particular a $P$- and $T$-odd nucleon-electron scalar-pseudoscalar interaction would also manifest as a $\Nsw\Esw$-odd phase in our experiment. Thus, we write
\begin{equation}
\wNEt=-d_e\Eeff + W_{\rm S}C_{\rm S},\footnote{Note that the sign of the $C_{\rm S}$ term is opposite to that used, incorrectly, in our original paper \cite{Baron2014}. In addition, here $W_{\rm S}$ differs in magnitude from the related quantity $W_{\rm T,P}$ given explictly in \cite{Denis2016,Skripnikov2016}. A detailed discussion of the sign and notational conventions for this Hamiltonian is provided in \ref{sec:sign_conventions}.}
\end{equation}
where $W_{\rm S}$ is a (calculated) energy scale specific to the species of study \cite{Skripnikov2013,Dzuba2011a,DzubaErratum2012,Denis2016,Skripnikov2016} and $C_{\rm S}$ is a dimensionless constant characterizing the strength of the $T$-violating nucleon-electron scalar-pseudoscalar coupling relative to the ordinary weak interaction. 

We can use our measurement to set an upper limit on $\de$ by assuming that $C_{\rm S}=0$ and that $\wNEt$ is therefore entirely attributable to the eEDM. Taking the effective electric field to be the unweighted mean of the two most recent calculations of this quantity \cite{Denis2016,Skripnikov2016}, $\Eeff=78~{\rm GV/cm}$, we can interpret our result in equation~(\ref{eq:wNEt_num_err_comb}) as:
\begin{align}
\de&=(-2.2\pm4.8)\times10^{-29}~e\cdot{\rm cm}\\
\Rightarrow|\de|&<9.3\times10^{-29}~e\cdot{\rm cm}~(90\%\,{\rm C.L.}),
\end{align}
where the second line is obtained by appropriately scaling the upper bound on $\wNEt$ derived in section \ref{ssec:confidence_intervals}.

If, instead, we assume that $d_e=0$, our measurement of $\wNEt$ in ThO can be restated as a measurement of $C_{\rm S}$. Using an unweighted mean of the most recent calculations of the interaction coefficient, $W_{\rm S} = -2\pi\times 282~{\rm kHz}$ \cite{Denis2016,Skripnikov2016}, we obtain:
\begin{align}
C_{\rm S}&=(-1.5\pm3.2)\times10^{-9}\\
\Rightarrow|C_{\rm S}|&<6.2\times10^{-9} \:(90\%\,{\rm C.L.}),
\end{align}
which, at the time, was an order of magnitude smaller than the existing best limit set by the ${}^{199}$Hg EDM experiment \cite{Swallows2013}, and is still a factor of 2 smaller than the recently improved limit from the same group \cite{Heckel2016}.

\section{Summary and Outlook}
 \label{sec:conclusion}
Our new limit on the size of the electron's electric dipole moment \cite{Baron2014}:
\begin{equation}
d_e\le9.3\times10^{-29}~e\cdot{\rm cm}.
\end{equation}
represented an order of magnitude improvement on previous bounds \cite{Hudson2011,Regan2002} and more strongly constrained the viable parameter space for many extensions to the Standard Model, while probing one-loop effects of new physics at a mass scale of ${\sim}10$~TeV.


We have presented our experimental method for measuring an eEDM-induced precession phase in the dipolar molecule ThO, 
detailing the way we utilise several experimental switches to isolate the component of accumulated phase with the correct symmetry properties. We described the apparatus that we used to carry out our measurement and have presented
a thorough analysis of the systematic errors present in the experiment, showing in detail the approach to finding and quantifying shifts of the eEDM-associated phase and their corresponding uncertainties. 

Despite the success of the experiment in reducing the limit on the value of the eEDM, there are several aspects of the experimental procedure that we are improving on which will significantly further enhance our statistical sensitivity. These upgrades include:
\begin{itemize}
\item\emph{Thermochemical Source}: Instead of relying on ablation to generate ThO molecules from a ThO$_2$ target, a relatively uncontrolled process, we are developing a new method using a thermochemical reaction-based beam source. This relies on the specific reaction \cite{Ackermann1963,OECD2008}
\begin{equation}
{\rm Th (s)}+{\rm ThO}_2 {\rm (s)} \rightarrow {\rm 2ThO (g)}
\end{equation}
occurring in a precursor target made of a Th/ThO$_2$ mixture. Preliminary tests have demonstrated a roughly factor of 10 increase in the time-averaged molecular flux produced via this method.

\item\emph{Beam Geometry}: In the current experiment, the molecules in the spin-precession region subtend a solid angle of ${\sim}60~\upmu$sr relative to the beam source, meaning only ${\sim}10^{-5}$--$10^{-4}$ of molecules produced reach the state readout region. This useful fraction of molecules can be increased in two ways: by shortening the distance between beam source and spin-precession region, and by increasing the spacing between the electric field plates so as to accomodate a beam with a larger transverse size. By making both of these changes to the apparatus we can increase the usable molecule number by a factor of ${\approx}8$.

\item\emph{State Preparation}: In the current experiment we transfer molecules into the $H$ state by optically pumping via $\ket{A,J=0}$ (see section~\ref{sec:state_prep_read}). This procedure is inefficient; only ${\sim}35\%$ of molecules addressed by the excitation laser are transferred into the $H$-state manifold, within which 1/6 of the population is in the desired superposition state $\ket{B(\hat{\epsilon}_{\rm prep},\Nsw,\Psw)}$. We can significantly increase the number of molecules prepared by using stimulated Raman adiabatic passage (STIRAP) to perform coherent population transfer from $X$ to $H$ via $C$. We have demonstrated an estimated efficiency of 75\% which will increase the usable molecule number by a factor of ${\approx}12$ \cite{Panda2016}.

\item\emph{State Readout}: We will be changing the transition which we perform our state readout on. The $I$ state of ThO is another $\Omega=1$ state which has a number of advantages over the $C$ state, namely a ${\sim}10$ times higher transition dipole moment from the $H$ state, a larger branching ratio to the $X$ state and a shorter fluorescence wavelength to $X$ \cite{Steimle2014,Hess2014,SpaunThesis}. The latter allows for higher quantum efficiency detection of photons. In addition, the efficiency with which we collect the light is improved by using light pipes instead of fiber bundles. Together we anticipate a factor of ${\approx}6$ improvement in signal \cite{Panda2016}. 

\item\emph{Other Improvements}: The suppression of known systematic effects was limited only by statistics. To the best of our knowledge, the limit on $\partial\wNE/\partial\Enr$ (see section~\ref{ssec:systematic_errors_due_to_imperfect_laser_polarizations}) could have been 10 times smaller if we had collected the data required to tune out that slope with such precision. Therefore, there is no reason to believe that the systematic effects we have discovered in this first generation measurement will limit the next generation of the experiment. However, we are taking additional measures to suppress such systematics, such as new electric field plates designed to minimise the absorbed laser power and hence the birefringence.
\end{itemize}

The ACME search for the electric dipole moment of the electron is now entering its second generation and we anticipate a new measurement that will either find a nonzero value of $d_e$, or constrain it to be ${\lesssim}10^{-29}~e\cdot{\rm cm}$, thus probing one-loop interactions at an energy scale of ${\sim}30$~TeV.

\section*{Acknowledgements}
This work was supported by the National Science Foundation.

\appendix
\section{\texorpdfstring{Conventions Used in the $T$-odd Hamiltonian}{Conventions Used in the T-odd Hamiltonian}}

\label{sec:sign_conventions}
Since there is not a consistent notation used throughout the literature on eEDM measurements in molecules, we describe here the conventions used throughout this paper and our other recent work.  We suggest that future work in the field consider adopting the same conventions if possible.

We begin with notation to describe the relevant molecular structure.  Following standard conventions from Ref. \cite{Brown2003}, we define the operators $\Lambda = \vec{L}\cdot \hat{n}$, $\Sigma = \vec{S}\cdot \hat{n}$, and $\Omega = \vec{J}_e \cdot \hat{n}  = \Lambda + \Sigma$, where $\hat{n}$ is a unit vector along the internuclear axis, $\vec{L}$ and $\vec{S}$ are the total electronic orbital and spin angular momentum, respectively, and $\vec{J}_e \equiv \vec{L} + \vec{S}$ is the total electronic angular momentum.  It is also useful to define the quantities $\vec{\Lambda}$, $\vec{\Sigma}$, and $\vec{\Omega}$, which are the vector components along $\hat{n}$ of $\vec{L}$, $\vec{S}$, and $\vec{J}_e$, respectively. For example, $\vec{\Omega} = \Omega \hat{n} = (\vec{J}_e\cdot\hat{n}) \hat{n}$.  We draw attention to our convention for the direction of the internuclear axis, $\hat{n}$: we choose it to point from the negative oxygen ion to the positive thorium ion, i.e. such that $\hat{n}$ is aligned with the molecule's electric dipole moment, $\vec{D}$. This choice (for which there appears to be no consensus in the literature) impacts the physical meaning associated with the sign of the quantum numbers $\Lambda$, $\Sigma$, and $\Omega$ and their vector analogues.

In the molecule-fixed frame, the Hund's case (a) basis consists of eigenstates of $S$, $\Lambda$, and $\Sigma$, and hence also of $\Omega$. There is a degeneracy between states with opposite signs of all these quantum numbers, i.e. between $|\Lambda; S, \Sigma; \Omega \rangle $ and $| -\Lambda; S, -\Sigma; -\Omega \rangle$.  In the laboratory frame, with no external fields applied, the eigenstates of energy, parity $P$, and total angular momentum $\vec{J} = \vec{J}_e + \vec{R}$ (where $\vec{R}$ is the pure rotational angular momentum) and its projection along the laboratory quantization axis $\hat{z}$, $J_z$ (with quantum number $M$), correspond to even and odd superpositions of these molecule-frame states.  The associated eigenstates can be written as 
\begin{equation}
||\Lambda |; S, |\Sigma |;  J, |\Omega |, M; P \rangle = 
\left[ 
|\Lambda; S, \Sigma; \Omega\rangle |J, \Omega, M \rangle 
+ P(-1)^{(J-S)} |-\Lambda; S, -\Sigma; -\Omega\rangle |J, -\Omega, M \rangle  
\right]/\sqrt{2}.
\end{equation}

The opposite-parity states with otherwise equal quantum numbers have a small energy splitting $\Delta_\Omega$ (due to Coriolis coupling). (In the $\ket{H,J=1}$  state of ThO, we refer to this splitting as $\Delta_{\Omega,1}$.)  In a sufficiently strong polarizing electric field $\vec{\mathcal{E}} = \mathcal{E} \hat{z}$, such that $|D \mathcal{E}| \gg \Delta_\Omega$, these states fully mix. If in addition $|D \mathcal{E}| \ll BJ$ (where $B$ is the rotational constant), this results in energy eigenstates where $J$, $M$, and (signed) $\Omega$ are all good quantum numbers, as described for ThO in the $\ket{H,J=1}$ manifold in sections \ref{sec:tho_molecule} and \ref{sec:Measurement_scheme}. In this limit, $\tilde{\mathcal{N}}\equiv \mathrm{sgn}\left(\langle \vec{\mathcal{E}}\cdot\hat{n} \rangle\right) = \mathrm{sgn}(\Omega) \mathrm{sgn}(M) \mathrm{sgn}(\mathcal{E})= \pm 1$ is a good quantum number.

The $H$ state of ThO can be described, to a fair approximation \cite{Paulovic2003}, as a pure $^3\Delta_1$ state in the Hund's case (a) basis, i.e. with $|\Lambda | = 2$, $|\Sigma | = 1$, and $|\Omega | = |\Lambda + \Sigma | = 1$. Hence in this approximation, in the ThO $H$ state $\Sigma= -\Omega$ and $\vec{\Sigma}$ is antiparallel to $\vec{\Lambda}$ and $\vec{\Omega}$. While $\Lambda$, $\Sigma$, and $\Omega$ are good quantum quantum numbers in the Hund's case (a) basis, in the more general case of Hund's case (c) coupling---which very accurately describes the $H$ state of ThO \cite{Paulovic2003}---only $\Omega$ is well-defined. Hence it is common in the literature of the field to express relevant molecule-frame matrix elements in terms of their dependence only on the value of $\Omega$. We follow this convention as well.  However, in the $H$ state of ThO, the expectation values of the operators  $\vec{\Lambda}$ and $\vec{\Sigma}$ (evaluated in a state with a given value of $\Omega$) are not far from their values in the Hund's case (a) basis.  Since these expectation values have signs that are linked to the sign of $\Omega$, it is useful to write them in terms of the 
good quantum number $\Omega$; for example, in the $H$ state of ThO, $\left\langle \Sigma \right\rangle = - |\!\left\langle \Sigma \right\rangle\! | \Omega \approx -\Omega$. This approximation is often used elsewhere in the literature.  

In our experiment we apply a magnetic field $\vec{\mathcal{B}} = \mathcal{B}\hat{z}$; hence we are also concerned with the molecular magnetic dipole moment, $\vec{\mu}$.  In the laboratory frame, we write $\vec{\mu} \equiv g_J \mu_B \vec{J}$, so that under the Zeeman Hamiltonian $H_{\rm Z} = -\vec{\mu}\cdot\vec{\mathcal{B}}$, a lab-frame eigenstate with quantum numbers $J,M$ has energy shift $\Delta E_{\rm Z} = -g_J \mu_B M \mathcal{B}$.  Since $g_1$, the value of $g$ in the $\ket{H,J=1}$ state of ThO, is negative, $\vec{\mu}$ is antiparallel to $\vec{J}$.

\subsection{eEDM Interaction}

To make contact with common language in the literature about the eEDM in molecules, we first write the effective, nonrelativistic eEDM interaction in terms of an internal electric field $\vec{\mathcal{E}}_{\rm int}$.  (As we will see, this is closely related, but not identical, to the effective field $\vec{\mathcal{E}}_{\rm eff}$.)  We choose a convention where $\vec{\mathcal{E}}_{\rm int} = -\mathcal{E}_{\rm int} \hat{n}$. This means that the internal field vector is defined to be directed \textit{opposite} to $\hat{n}$, i.e., along the average direction of the electric field \textit{inside} the molecule (here, from positive Th ion to negative O ion) when 
$\mathcal{E}_{\rm int}$ is positive.   We also adopt the convention that, in the $H$ state of ThO, there is an effective eEDM $\vec{d}_e^{\rm eff} = d_e\vec{S}$ (where again $S=1$ to a fair approximation). This choice appears, at first glance, to contradict the discussion in section~\ref{sec:theory}, where for a single electron we wrote $\vec{d}_e=2d_e\vec{s}$ (where $s=1/2$). However, these two definitions are in fact consistent when taking into account that in the $H~^3\Delta_1$ state of ThO only one of the two valence electrons (the one in the $\sigma$ orbital) contributes significantly to the EDM energy shift, while both electrons contribute to the total spin $S=1$. Hence, in our formulation, the molecule-frame projection $\vec{d}_e^{\rm eff}\cdot\hat{n}$ can take extreme values $\pm d_e$, as expected for a single contributing electron. (This `single contibuting electron' approximation is valid for all molecules used to date in searches for the eEDM.)

We then write the effective eEDM Hamiltonian $H_{\rm EDM}^{\rm eff}$ in the standard form for interaction of an electric dipole moment with the internal electric field:
\begin{equation}
H_{\rm EDM}^{\rm eff}=-\vec{d}_e^{\rm eff}\cdot\vec{\mathcal{E}}_{\rm int}= +d_e \mathcal{E}_{\rm int} \vec{S}\cdot\hat{n},
\label{eq:HEDMapp1a}
\end{equation}
where the + sign in the final expression arises from the sign convention for $\vec{\mathcal{E}}_{\rm int}$.
In eigenstates of $\Omega$, the expectation value of $H_{\rm EDM}^{\rm eff}$---that is, the energy shift $\Delta E_{\rm EDM}$ due to the eEDM---can be written as
\begin{equation}
\Delta E_{\rm EDM} = +d_e \mathcal{E}_{\rm int} \left\langle\Sigma\right\rangle
 = +d_e \left( \mathcal{E}_{\rm int} |\!\left\langle\Sigma\right\rangle\! | \right) \mathrm{sgn}\left(\langle\Sigma\rangle\right).
\label{eq:HEDMapp1b}
\end{equation}

\begin{figure}[!ht]
\centering
\includegraphics[width=\linewidth]{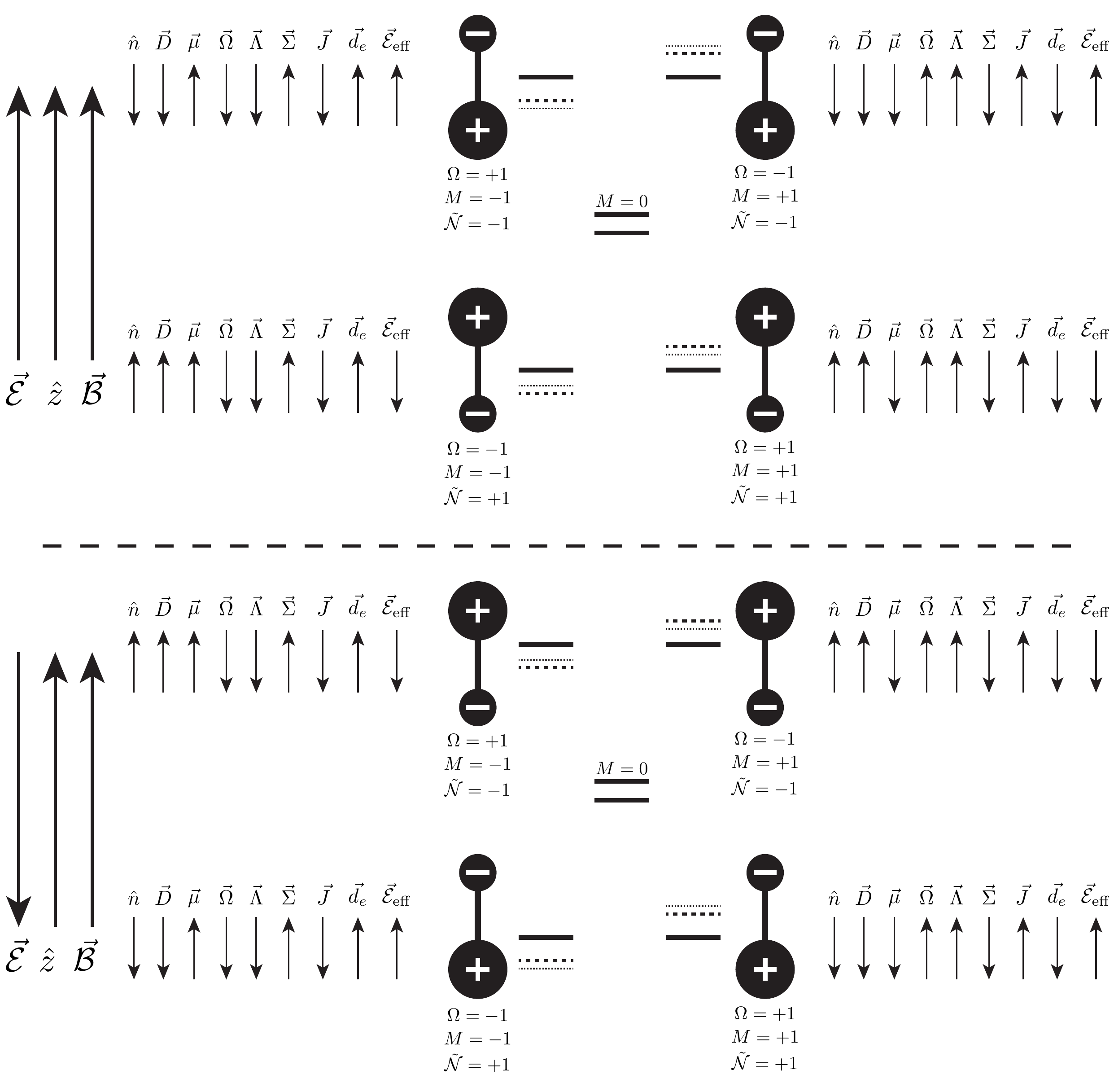}
\caption{Summary of sign conventions used in the ACME experiment. All vectors depict expectation values of operators defined in the text, in the states $| H, J=1, \tilde{\N},M\rangle$. Note the difference between
scalar $\Omega$ and vector $\vec{\Omega}$. The figure is drawn with a negative $g$-factor, i.e. the magnetic moment $\vec{\mu}$ opposes $\vec{J}$, and with positive values of $d_e$ and $\Eeff$. Energy levels are shown in the centre of the figure --- solid lines show the Stark-shifted levels ($M=0$ levels are unaffected), dashed lines include Zeeman shifts and dotted lines include a non-zero eEDM interaction. Figure inspired by \cite{Lee2009}.}
\label{fig:ACME_signs}
\end{figure}

Now, we finally re-introduce the effective electric field $\Eeff$ used throughout the main text of this paper. This is related to the internal field introduced above, via 
\begin{equation}
   \Eeff \equiv |\!\left\langle\Sigma\right\rangle\! | \mathcal{E}_{\rm int},
\end{equation}
We can then use this notation to describe the effective nonrelativistic eEDM interaction, within a given electronic state and eigenstate of $\Omega$ (and otherwise independent of molecular structure), as follows:
\begin{eqnarray}
\vec{\mathcal{E}}_{\rm eff} \equiv -\Eeff \hat{n}; \\
\vec{d}_e \equiv d_e \vec{S}/ | \left\langle \Sigma \right\rangle |; \\
H_{\rm EDM}^{\rm eff} = -\vec{d}_e \cdot \vec{\mathcal{E}}_{\rm eff} = +d_e \Eeff \Sigma/ | \left\langle \Sigma \right\rangle |; \\
\Delta E_{\rm EDM} = \mathrm{sgn}(\left\langle \Sigma \right\rangle) d_e \Eeff,
\label{eq:HEDMapp1d}
\end{eqnarray}
where the sign in the last expressions arises from the defined definitions of $\Sigma$ (component of $\vec{S}$ along $\hat{n}$) and $\vec{\mathcal{E}}_{\rm eff}$ (antiparallel to $\hat{n}$). All relevant quantities are summarised pictorially in figure~\ref{fig:ACME_signs}.

In most of the theoretical literature on this subject, this energy shift is written in the unambiguous form $\Delta E_{\rm EDM} = +d_e W_d \Omega$.  However, there has been no consistent definition in the literature for the relation between $W_d$ and $\Eeff$. In particular, both their relative signs and the dependence of their relative magnitude on the value of $|\Omega |$ (encompassing both the case of one- and two-electron systems) are often defined differently, or imprecisely.  
In our notation, the expressions above imply a general relationship between $\Eeff$ and $W_d$:
\begin{equation}
\Eeff = W_d \Omega \mathrm{sgn}(\left\langle \Sigma \right\rangle).
\label{eq:genHEDMapp1}
\end{equation}
This relation is valid for systems with one or two valence electrons (in the `single contributing electron' approximation for the latter case), and regardless of the relative directions of $\vec{\Sigma}$ and $\vec{\Omega}$.

Now we apply these general considerations to the specific case of the $H$ state of ThO. Here, since $\left\langle \Sigma \right\rangle \approx -\Omega$, we find that $\Eeff = -W_d$ with our conventions.  Thus, the energy shifts can be written for ThO as
\begin{equation}
\Delta E_{\rm EDM} = -d_e\Eeff \Omega.
\label{eq:HEDMapp1c}
\end{equation}
In our experiment, this gives rise to energy shifts, for a given direction of the laboratory electric field $\E$, given by
\begin{equation}
\langle H, J=1, \tilde{\N},M|H_{\rm eEDM}^{\rm eff}| H, J=1, \tilde{\N},M\rangle=-d_e\E_{\rm eff}M\tilde{\N}\tilde{\E},
\label{eq:HEDMapp2}
\end{equation}
since in our notation $\Omega=M\tilde{\N}\tilde{\E}$. 
Then, finally, the experimentally determined energy shift arising from the eEDM is 
\begin{eqnarray}
\omega^{\N\E}_{\rm EDM} = \frac{1}{2}\frac{1}{\tilde{\N}\tilde{\E}} 
\left[ \langle  H, J=1, \tilde{\N},M=+1|H_{\rm eEDM}^{\rm eff}| H, J=1, \tilde{\N},M=+1 \rangle \right. \nonumber \\
\hphantom{H,J=1,\N,} -\left. \langle H, J=1, \tilde{\N},M=-1|H_{\rm eEDM}^{\rm eff}| H, J=1, \tilde{\N},M=-1 \rangle \right] \nonumber \\
\hphantom{\omega^{\N\E}_{\rm EDM}} = -d_e\E_{\rm eff} .
\label{eq:HEDMapp2a}
\end{eqnarray}


\subsection{Scalar-Pseudoscalar Nucleon-Electron Interaction}

We next turn to notation describing the $T$-violating scalar-pseudoscalar (SP) interaction between a nucleon and an electron.
The relativistic Hamiltonian for this interaction can be written as
\begin{equation}
H_{\rm SP}=i\frac{G_{\rm F}}{\sqrt{2}}(ZC_{\rm S,p}+NC_{\rm S,n})\gamma_0\gamma_5\rho_{\rm N}(\vec{r}),
\end{equation}
where $G_F$ is the Fermi coupling constant, $\gamma_i$ are Dirac matrices, $\rho_{\rm N}(\vec{r})$ is the normalised nuclear density, $Z(N)$ is the proton (neutron) number, and $C_{\rm S,p}$ and $C_{\rm S,n}$ are dimensionless constants which describe the interaction strength (relative to that of the ordinary weak interaction) specifically for protons and neutrons, respectively. 
Using the definition
\begin{equation}
C_{\rm S}=\frac{Z}{A} C_{\rm S,p} + \frac{N}{A} C_{\rm S,n} = \frac{Z}{A} C_{\rm S,p} + \left( 1-\frac{Z}{A}\right) C_{\rm S,n},
\end{equation}
where $A=Z+N$, $C_\mathrm{S}$ represents a weighted average of the couplings to protons and neutrons, and is different for every nuclear species. However, since the ratio $Z/A$ is nearly the same for all heavy nuclei used in molecular and atomic EDM experiments (ranging only from $Z/A=0.41$ for $^{133}$Cs to $Z/A=0.39$ for $^{232}$Th), typically a common value for $C_{\rm S}$ is assumed for all experiments of this type.
Thus we can write
\begin{equation}
H_{\rm SP}=i\frac{G_{\rm F}}{\sqrt{2}} AC_{\rm S} \gamma_0\gamma_5\rho_{\rm N}(\vec{r}).
\label{eq:HPSapp1}
\end{equation}
In a given molecular electronic state, this gives rise to a non-relativistic, single-electron effective Hamiltonian of the form
$ H_{\rm SP}^{\rm eff} = 2\vec{s}\cdot\hat{n} C_{\rm S} Y_{\rm S}$; the factor of 2 is included so that the maximal energy shifts due to this term have the simple form $\Delta E_\mathrm{SP}^\mathrm{max} = \pm C_\mathrm{S} Y_\mathrm{S}$. 
By analogy with our discussion of the eEDM Hamiltonian, in a molecular state with $S = 1$ and a `single contributing electron', as in the $H~^3\Delta_1$ state of ThO, we rewrite this in the form
\begin{equation}
H_{\rm SP}^{\rm eff} = \vec{S}\cdot\hat{n} C_{\rm S} Y_{\rm S}.
\end{equation}
Hence, the energy shift due to this interaction can be written as
\begin{equation}
\Delta E_{\rm SP} = \langle \vec{S}\cdot\hat{n} \rangle C_{\rm S} Y_{\rm S} = Y_{\rm S} \left[ \langle \Sigma \rangle / \Omega \right] \Omega,
\end{equation}
where the term in square brackets is a constant of the molecular state, determined by the fixed relative size and orientation of $\vec{\Sigma}$ and $\vec{\Omega}$, with value $\approx -1$ in the $H~^3\Delta_1$ state of ThO. In the literature on molecular eEDM systems, this energy shift is typically written in the simpler form
\begin{equation}
\Delta E_{\rm SP} = C_{\rm S} W_{\rm S}\Omega.
\end{equation}
Here, in our notation, $W_{\rm S}\equiv Y_{\rm S}\left[\langle\Sigma\rangle/\Omega\right]$ ($\approx -Y_{\rm S}$ in ThO).
However, quantities analogous to $Y_{\rm S}$ (in terms of which the energy shifts depend explicitly on the spin direction) are rarely introduced in the literature; instead, only forms analogous to $W_{\rm S}$ (where the energies depend only on $\Omega$) are used.


Our definition for $C_{\rm S}$ was historically a standard notation used in the literature. However, in some recent papers (e.g. references \cite{Skripnikov2015,Denis2016}) it is implicitly assumed that the neutron coupling $C_{\rm S,n}$ vanishes.  In these papers, the factor $AC_{\rm S}$ in equation \ref{eq:HPSapp1} is replaced by $ZC_{\rm S,p}$ (or its equivalent in a different notation),\footnote{In reference \cite{Skripnikov2015} our $C_{\rm S,p}$ is denoted as $k_{\rm T,P}$ and our $W_{\rm S,p}$ as $W_{\rm T,P}$; in Ref.\ \cite{Denis2016}, our $W_{\rm S,p}$ is denoted simply as $W_{\rm S}$. References \cite{Dzuba2011,DzubaErratum2012} denote our $C_{\rm S}$ as $C^{\rm SP}$ and our $W_{\rm S}$ as $W_{\rm c}$.}
and the energy shift is written in the analogous form 
$\Delta E_{\rm SP} = C_{\rm S,p} W_{\rm S,p}\Omega$.  
These papers report values of $W_{\rm S,p}$ in the $H$ state of ThO, based on sophisticated calculations of the molecular wavefunctions.  However, since there is no particular reason to expect this interaction to couple more strongly to protons than to neutrons, we prefer to report our results in terms of $C_{\rm S}$. To do so, we use the relation $W_S = (A/Z)W_{\rm S,p}$ to determine $W_S$ from the reported values for $W_{\rm S,p}$.  

Finally, the experimentally determined energy shift arising from the nucleon-electron SP interaction is 
\begin{equation}
\omega^{\N\E}_{\rm SP} = C_{\rm S} W_{\rm S},
\label{eq:HPSapp3}
\end{equation}
and the total T-violating energy shift is
\begin{equation}
\omega^{\N\E}_{\rm T}= -d_e\E_{\rm eff} + C_{\rm S} W_{\rm S} = d_e W_d + C_{\rm S} W_{\rm S}.
\label{eq:HSP}
\end{equation}
Note that the sign of the $C_{\rm S}$ term is opposite to that used, incorrectly, in our original paper \cite{Baron2014}.

\subsection{Relation to other notations in the literature}

Table~\ref{tab:sign_convs} shows some of the conventions used in the literature to describe the $T$-violating electron-nucleon interaction in molecular systems, and how they relate to our conventions. We note in particular three key differences between the (shared) conventions of references \cite{Skripnikov2015,Denis2016}---which currently provide the most accurate values for $W_d$ and $W_S$---and ours.  First: these references define $\hat{n}$ in the direction opposite to $\vec{D}$, and hence opposite to ours.  This in turn means that their definition of $\Omega$ has opposite sign to ours.  Hence, the same physical energy shifts (defined as $\Delta E_{\rm EDM} = W_d\Omega$ both there and here) are obtained only if we take $W_d$ to have sign opposite to that of the reported $W_d$ in these papers. Second: these references define the eEDM energy shift as $\Delta E_{\rm EDM} = +d_e\Eeff \Omega$, while we have shown that in our notation $\Delta E_{\rm EDM} = -d_e\Eeff \Omega$.  Here there are two sign differences (one from the overall sign, one from the definition of $\Omega$).  Hence, the same physical energy shifts are obtained when taking $\Eeff$ to have the same sign as reported in these papers.  Third: these references formulate the scalar-pseudoscalar nucleon-electron interaction in terms of a quantity equivalent to our $W_{\rm S,p}$ rather than our $W_{\rm S}$.  Hence we must rescale these values as described above, using $W_S = (A/Z)W_{\rm S,p}$. In addition, the same physical energy shifts $\Delta E_{\rm SP} = W_S\Omega$ are obtained only if we take $W_{\rm S,p}$ to have sign opposite to that of the reported $W_{\rm S,p}$ in these papers.

\begin{center}
\begin{threeparttable}
\centering
\begin{tabular}{C{2.9cm}C{1.1cm}C{0.6cm}C{4cm}C{4cm}}
\hline 
& $\hat{n}$ & $\vec{\mathcal{E}}_{\rm eff}$ & $\Delta E_{\rm EDM}$ & $\Delta E_{\rm SP}$ \tabularnewline
\hline
ACME & \includegraphics[width=10pt]{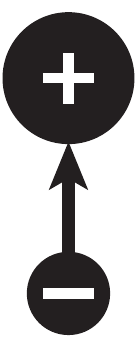} & \includegraphics[width=10pt]{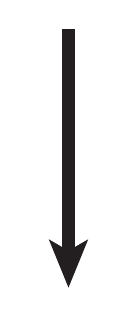} & $d_e W_d\Omega=-d_e \E_{\rm eff} \Omega = -\vec{d}_e\cdot\vec{\mathcal{E}}_{\rm eff}$ & $C_{\rm S}W_{\rm S}\Omega$ \tabularnewline
Lee et al. \cite{Lee2009} & \includegraphics[width=10pt]{nup.pdf} & \includegraphics[width=10pt]{adown.pdf} & $-\vec{d}_e\cdot\vec{\mathcal{E}}_{\rm eff}$\tnote{a} & \tabularnewline
YbF \cite{Kara2012} & \includegraphics[width=10pt]{nup.pdf} & \includegraphics[width=10pt]{adown.pdf} & $-\vec{d}_e\cdot\vec{\mathcal{E}}_{\rm eff}$\tnote{b} & \tabularnewline
Kozlov et al. \cite{Kozlov1995,Kozlov2002} & \includegraphics[width=10pt]{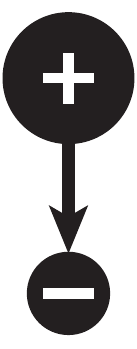} & & $+W_dd_e\Omega$\tnote{c} & \tabularnewline
Skripnikov et al. \cite{Skripnikov2013,Skripnikov2015,Skripnikov2016} & \includegraphics[width=10pt]{ndown.pdf} & & $+W_dd_e\Omega=+d_e\Eeff{\rm sgn}(\Omega)$\tnote{d} & $+W_{T,P}k_{T,P}\Omega$\tnote{e}, where $k_{T,P}=AC_{\rm S}/Z$\tnote{f} \tabularnewline
Fleig et al. \cite{Fleig2014,Fleig2013,Denis2016} & \includegraphics[width=17.2pt]{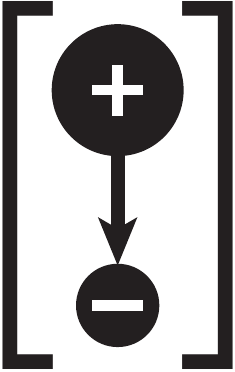} & & $+W_dd_e\Omega=+d_e\Eeff[{\rm sgn}(\Omega)]$\tnote{g} & $+W_{P,T}k_S\Omega$\tnote{h}, where $k_S=AC_S/Z$ \tabularnewline
Dzuba et al. \cite{Dzuba2011a,DzubaErratum2012} & \includegraphics[width=17.2pt]{bndown.pdf} & & $+W_dd_e[{\rm sgn}(\Omega)]=-d_e\Eeff[{\rm sgn}(\Omega)]$\tnote{i} & $+W_cC^{\rm SP}[{\rm sgn}(\Omega)]$\tnote{j} \tabularnewline
\hline
\end{tabular}
\begin{tablenotes}
    \item[a] Reference \cite{Lee2009}, p.\ 2007
    \item[b] Reference \cite{Kara2012}, p.\ 3
    \item[c] Reference \cite{Kozlov1995}, above equation 6.27
    \item[d] Reference \cite{Skripnikov2015}, equation 1 and following
    \item[e] Reference \cite{Skripnikov2015}, equation 4
    \item[f] Reference \cite{Skripnikov2015}, equation 4 and \cite{Dzuba2011a}, equation 25 and following
    \item[g] Reference \cite{Fleig2014}, equation 1 and Reference \cite{Fleig2013} equations 2--4
    \item[h] Reference \cite{Denis2015}, equations 3 and 4
    \item[i] Reference \cite{Dzuba2011a}, equation 24 and table IV
    \item[j] Reference \cite{Dzuba2011a}, equation 25
\end{tablenotes}
\par
\protect\caption{Summary of the different conventions used in some of the literature relating to eEDM measurements/theory. Where entries are left blank the convention is not stated in the reference provided. Quantities in square brackets are not explicitly stated in the references but are implied. In some cases, nomenclature has been modified for consistency. Footnotes provide specific references for the equations shown.}

\label{tab:sign_convs}
\end{threeparttable}
\end{center}

\section{Glossary of Abbreviations and Symbols}
\label{sec:glossary}
\subsection{Experiment Switches}
During the course of the experiment, we performed many parameter switches. Most of these switch parameter symbols are denoted by a superscript tilde $\tilde{\mathcal{X}}$, which indicates that that parameter takes on two
values, $\tilde{\mathcal{X}}=\pm1$.
\begin{description}
\item [{$\Nsw$}] Used as a quantum number, $\Nsw\approx\Esw\rm{sgn}\left(M\Omega\right)$,
for states of $\left|M\right|>0$, $\left|\Omega\right|>0$, that
refers to states with opposite molecular alignment with respect to
the applied electric field. It is also used to refer to the experiment
switch between spectroscopically addressing states in $\left|H,J=1\right\rangle $
with opposite values of $\Nsw$.
\item [{$\Esw$}] Denotes the alignment of the applied electric
field with respect to the laboratory $\hat{z}$ axis, $\Esw=\rm{sgn}\left(\vec{\mathcal{E}}\cdot\hat{z}\right)$ where $\vec{\mathcal{E}}$ is the applied electric field.
\item [{$\Bsw$}] Denotes the alignment of the applied magnetic
field with respect to the laboratory $\hat{z}$ axis, $\Bsw=\rm{sgn}\left(\vec{\mathcal{B}}\cdot\hat{z}\right)$ where $\vec{\mathcal{B}}$ is the applied magnetic field.
\item [{$\tilde{\theta}$}] Denotes the state of the polarisation dither
that is used to extract the contrast in the spin precession measurement.
It refers to the direction of the offset angle in the $xy$ plane
of the state readout polarisation basis $\hat{X},\hat{Y}$, relative
to the average polarisation of these lasers.
\item [{$\Psw$}] Used as a quantum number to denote the
parity (eigenvalue of the parity operator $P$) of a given molecular
state of well-defined parity. It is also used to refer to the experiment
switch between spectroscopically addressing states in $\left|C,J=1\right\rangle $
with opposite values of $\Psw$ with the state readout
lasers.
\item [{$\Lsw$}] Denotes the state of the mapping between
the two output channels of the electric field voltage supply, and
the two electric field plates which can be either connected normally
(+1), or inverted relative to normal (-1).
\item [{$\Rsw$}] Denotes the state of an experimental switch
of the state readout polarisation basis offset angle with respect
to the $x$-axis by either 0 $\left(+1\right)$ or $\pi/2$ $\left(-1\right)$.
\item [{$\Gsw$}] Denotes the state of an experimental switch
of the global polarisation; the state preparation and state readout lasers are rotated synchronously by a common angle. This can be thought of as a redefinition
of the $\hat{x}$ and $\hat{y}$ axes in the $xy$ plane.
\item [{$\B_{z}$}] Denotes the magnitude of the magnetic field
along the $\hat{z}$ direction in the laboratory, $\B_{z}=|\vec{\B}\cdot\hat{z}|$.
This parameter is switched between three values differing by about
$20^{\:}\mathrm{mG}$. In figure~\ref{fig:pixel_plot}, channels $X$ that are `odd'
with respect to this parameter refer to the linear variation $\partial X/\partial\mathcal{B}_{z}$.
\item [{$\E$}] Denotes the magnitude of the electric field, $\E=|\vec{\E}|$.
This parameter is switched between two values.
\item [{$\hat{k}\cdot\hat{z}$}] Denotes the orientation of both the state
preparation and the state readout laser pointing directions with
respect to the laboratory $\hat{z}$ axis. This is a binary switch,
$\hat{k}\cdot\hat{z}=\pm1$, but we do not denote this switch with
a tilde as we do with the other binary switch parameters.
\end{description}

\subsection{Laser Parameters}
There are a variety of laser parameters which are used to describe
the state preparation laser that is denoted with a subscript `prep',
or the state readout lasers that are denoted with a subscript `read'
if the property applies to both state readout lasers, or with subscripts
$X$ and $Y$, if the parameter can vary between the two readout lasers.
\begin{description}
\item [{$\hat{k}$}] Laser pointing direction. In this paper, the pointing
direction is always nearly aligned or antialigned with respect to
the laboratory $\hat{z}$ axis such that $\hat{k}\cdot\hat{z}\approx\pm1$.
\item [{$\vartheta_{k}$}] Defined in equation~\ref{eq:pointing_imperfection}. Polar angle of deviation
of the pointing $\hat{k}$ from aligned or anti-aligned with the $\hat{z}$
axis.
\item [{$\varphi_{k}$}] Defined in equation~\ref{eq:pointing_imperfection}. Azimuthal angle denoting
the direction in the $xy$ plane, relative to the $x$-axis, of the deviation of the pointing
$\hat{k}$ from the $\hat{z}$ axis.
\item [{$\hat{\epsilon}$}] Complex laser polarisation. The readout laser
polarisations are also referred to as $\hat{X}$ and $\hat{Y}$ as
an alternative to $\hat{\epsilon}_{X}$ and $\hat{\epsilon}_{Y}$
at some points.
\item [{$\hat{\varepsilon}$}] Effective polarisation. Used to parameterize
the effect of experiment imperfections on the molecule state as the
polarisation vector that would be required to obtain the same molecule
state in the absence of those experiment imperfections.
\item [{$\theta$}] Defined in section~\ref{sec:Measurement_scheme} and equation~\ref{eq:polarization_parametrization} as the linear polarisation angle of the complex polarisation vector.
\item [{$\Theta$}] Defined in section~\ref{sec:Measurement_scheme} and equation~\ref{eq:polarization_parametrization} as encoding the ellipticity of the complex polarisation vector.
\item [{$S$}] Defined in section~\ref{sec:Measurement_scheme_more_detail} as the relative circular Stokes parameter,
$S\equiv S_{3}/I=\cos2\Theta$.
\item [{$\omega_{\rm L}$}] Laser frequency.
\item [{$P$}] Laser power.
\item [{$\Omega_{\rm r}$}] Rabi frequency for a particular laser beam and transition. Defined as the transition dipole matrix element multiplied by the amplitude of the electric field associated with the laser beam.
\item [{$\Gamma$}] Optical retardance for some birefrigent element along
the laser beam path.
\item [{$\phi_{\Gamma}$}] Angle in the $xy$ plane of the fast axis associated
with an optical retardance $\Gamma$.
\end{description}

\subsection{Molecular States and Parameters}
These symbols are all used to describe the molecular energy level structure and the manner in which our laser light interacts with the molecules, in particular for the state preparation and readout processes.
\begin{description}
\item [{$J$}] Total angular momentum.
\item [{$M$}] Projection of $J$ onto the laboratory $\hat{z}$-axis.
\item [{$\Omega$}] Projection of $J$ onto the internuclear axis, $\hat{n}$.
\item [{$B_H$}] Rotational constant of the $H$ state.
\item [{$\Eeff$}] `Effective electric field' to which we consider the eEDM to be subjected.
\item [{$\Delta_{\Omega,1}$}] The $\Omega$-doublet splitting of the $\ket{H,J=1}$ state.
\item [{$D_1$}] Expectation value of the molecular electric dipole moment of the $\ket{H,J=1}$ state.
\item [{$g_1$}] The $g$-factor of the $\ket{H,J=1}$ state.
\item [{$\eta$}] Defined in equation~\ref{eq:eta_2}, it is proportional to the $g$-factor difference between the two $\Nsw$ states.
\item [{$\left|\pm,\Nsw\right\rangle $}] Sublevels within the $\ket{H,J=1}$ (eEDM sensitive) manifold, labelled by their values of $M$ and $\Nsw$.
\item [{$\left|C,\Psw\right\rangle $}] Sublevel to which molecules are excited during state preparation and readout. One of two sublevels in the $\ket{C,J=1}$ manifold, with $M=0$ and parity $\Psw=\pm1$.
\item [{$\left|B(\hat{\epsilon}),\Nsw,\Psw\right\rangle $}] Superposition of $M$ sublevels within the $\ket{H,J=1,\Nsw}$ manifold that is depleted during state preparation with a laser beam of polarisation $\hat{\epsilon}$, as defined in equation~\ref{eq:bright_state}.
\item [{$\left|D(\hat{\epsilon}),\Nsw,\Psw\right\rangle $}] Superposition of $M$ sublevels within the $\ket{H,J=1,\Nsw}$ manifold that remains after state preparation with a laser beam of polarisation $\hat{\epsilon}$, as defined in equation~\ref{eq:dark_state}.
\item [{$\left|B_{\pm}(\hat{\epsilon}),\Nsw,\Psw\right\rangle$}] Instantaneous eigenvectors of the three-level system formed by $\left|B(\hat{\epsilon}),\Nsw,\Psw\right\rangle$, $\left|D(\hat{\epsilon}),\Nsw,\Psw\right\rangle$ and $\left|C,\Psw\right\rangle$, as defined in equation~\ref{eq:inst_eigv}.
\item [{$\Delta$}] One-photon detuning from resonance, discussed in section~\ref{sec:state_prep_read} and defined in equation~\ref{eq:detuningcorrelations}.
\item [{$\gamma$}] Decay rate of the a given electronic state. The electronic state label is given in the subscript. In most of the paper, only $\gamma_C$, the decay rate of the $C$ state, is relevant.
\item [{$\rabi$}] Transition Rabi frequency, which is proportional
to the square root of the laser intensity.
\item [{$E_{B\pm},^{\:}E_{D}$}] Instantaneous eigenenergies of the dressed three-level system, defined in equation~\ref{eq:inst_eig}.
\item [{$\dot{\chi}$}] Complex polarisation rotation rate defined in section~\ref{sssec:AC_stark_shift_phases}.
\item [{$\Pi$}] Defined and discussed in section~\ref{sssec:AC_stark_shift_phases} and equation~\ref{eq:Pi_def}. This is a factor in the AC Stark shift phase that is independent of laser polarisation but depends on the laser detuning and Rabi frequency.
\item [{$v_{\parallel}$}] The mean longitudinal velocity of the molecular beam.
\end{description}

\subsection{Measurement Quantities}
These symbols represent quantities related to the measurement of the accumulated phase and the way in which it is extracted during data analysis, as well as some related quantities pertaining to systematic studies.
\begin{description}
\item [{$N$}] Total number of measurments performed, equivalent to the number of detected photoelectrons.
\item [{$N_0$}] Number of molecules in the state readout region in the particular $\Nsw$ level being addressed.
\item [{$f$}] Fraction of fluorescence photons emitted in the state readout region that are detected.
\item [{$S$}] Recorded photoelectron count rate measured on the photodetectors.
\item [{$F$}] Photoelectron count rate due to the molecule fluorescence.
$F_{X,Y}$ is used to denote the molecular fluorescence induced by
the $X$ and $Y$ state readout lasers, respectively. $F_{\mathrm{cut}}$
is used to denote the fluorescence threshold above which data was
included in the analysis.
\item [{$B$}] Background count rate primarily due to scattered
light from the state readout lasers. This background signal is subtracted
from the raw photoelectron signal $S$ to obtain the fluorescence
photoelectron count rate, $F=S-B$.
\item [{$\mathcal{A}$}] Signal asymmetry as defined in equation~\ref{eq:Asymmetry}.
\item [{$\mathcal{C}$}] Spin precession fringe contrast, as defined in
equation~\ref{eq:Contrast_Definition}, is the sensitivity of the asymmetry to molecular
spin precession.
\item [{$\phi$}] Actual spin precession phase of the molecules as defined in equation~\ref{eq:total_phase}.
\item [{$\Phi$}] Measured spin precession phase as described in section~\ref{sec:Measurement_scheme_more_detail}, $\Phi=\mathcal{A}/(2\mathcal{C})$.
\item [{$\tau$}] Measured spin precession time as described in sections~\ref{sec:Measurement_scheme} and \ref{sec:compute_phase}.
\item [{$\omega$}] Measured spin precession frequency, as defined in equation~\ref{eq:omega_def}, $\omega=\Phi/\tau$.
\item [{$\chi^{2}$}] Reduced chi-squared statistic, $\chi^2=\frac{1}{N_{\rm dof}}\sum_i\left(\frac{x_i - f_i(\{x\})}{dx_i}\right)^2$,
where $N_{\rm dof}$ is the number of degrees of freedom, $x_i$ are the data points, $dx_i$ are the uncertainties, and $f_i(\{x\})$ is a fit function that can depend on $i$ and the ensemble of all of the data, $\{x\}$. 
For normally distributed data that fits well to the applied fit function, $\chi^2$ should be consistent with 1.
\item [{$\wNE$}] The measurement channel of interest, the spin precession frequency channel that is correlated with $\Nsw$ and $\Esw$. The expected eEDM signal should contribute to this channel.
\item [{$\wNE_T$}] The contribution to spin precession frequency $\wNE$ induced by $T$-odd spin precession effects in the $H$ state in ThO.
\item [{$\wNE_P$}] A systematic error in the $\wNE$ channel that is proportional to some parameter $P$.
\end{description}

\clearpage

\bibliography{library}

\providecommand{\newblock}{}
\begin{thebibliography}{100}
\expandafter\ifx\csname url\endcsname\relax
  \def\url#1{{\tt #1}}\fi
\expandafter\ifx\csname urlprefix\endcsname\relax\def\urlprefix{URL }\fi
\providecommand{\eprint}[2][]{\url{#2}}

\bibitem{Michelson1887}
Michelson A~A and Morley E~W 1887 {\em Am. J. Phys.\/} {\bf 34} 33

\bibitem{Wu1957}
Wu C~S, Ambler E, Hayward R~W, Hoppes D~D and Hudson R~P 1957 {\em Phys.
  Rev.\/} {\bf 105}(4) 1413--1415
  \urlprefix\url{http://link.aps.org/doi/10.1103/PhysRev.105.1413}

\bibitem{Christenson1964}
Christenson J~H, Cronin J~W, Fitch V~L and Turlay R 1964 {\em Phys. Rev.
  Lett.\/} {\bf 13} 138--140

\bibitem{Lee1957}
Lee T~D and Yang C~N 1957 {\em Phys. Rev.\/} {\bf 105}(5) 1671--1675
  \urlprefix\url{http://link.aps.org/doi/10.1103/PhysRev.105.1671}

\bibitem{Kobayashi1973}
Kobayashi M and Maskawa T 1973 {\em Prog. Theor. Phys.\/} {\bf 49} 652

\bibitem{Aad2012short}
{ATLAS Collaboration} 2012 {\em Physics Letters B\/} {\bf 716} 1--29 ISSN
  0370-2693
  \urlprefix\url{http://www.sciencedirect.com/science/article/pii/S037026931200857X}

\bibitem{Englert1964}
Englert F and Brout R 1964 {\em Phys. Rev. Lett.\/} {\bf 13}(9) 321--323
  \urlprefix\url{http://link.aps.org/doi/10.1103/PhysRevLett.13.321}

\bibitem{Ambrosino2006}
Ambrosino F, Antonelli A, Antonelli M, Bacci C, Beltrame P, Bencivenni G,
  Bertolucci S, Bini C, Bloise C, Bocchetta S, Bocci V, Bossi F, Bowring D,
  Branchini P, Caloi R, Campana P, Capon G, Capussela T, Ceradini F, Chi S,
  Chiefari G, Ciambrone P, Conetti S, Lucia E~D, Santis A~D, Simone P~D, Zorzi
  G~D, Dell'Agnello S, Denig A, Domenico A~D, Donato C~D, Falco S~D, Micco B~D,
  Doria A, Dreucci M, Felici G, Ferrari A, Ferrer M, Finocchiaro G, Fiore S,
  Forti C, Franzini P, Gatti C, Gauzzi P, Giovannella S, Gorini E, Graziani E,
  Incagli M, Kluge W, Kulikov V, Lacava F, Lanfranchi G, Lee-Franzini J, Leone
  D, Martini M, Massarotti P, Mei W, Meola S, Miscetti S, Moulson M, Müller S,
  Murtas F, Napolitano M, Nguyen F, Palutan M, Pasqualucci E, Passeri A, Patera
  V, Perfetto F, Pontecorvo L, Primavera M, Santangelo P, Santovetti E,
  Saracino G, Sciascia B, Sciubba A, Scuri F, Sfiligoi I, Sibidanov A, Spadaro
  T, Testa M, Tortora L, Valente P, Valeriani B, Venanzoni G, Veneziano S,
  Ventura A, Versaci R and Xu G 2006 {\em Physics Letters B\/} {\bf 642}
  315--321 ISSN 0370-2693
  \urlprefix\url{http://www.sciencedirect.com/science/article/pii/S0370269306012251}

\bibitem{Kostelecky2010}
Kosteleck\'y V~A and Van~Kooten R~J 2010 {\em Phys. Rev. D\/} {\bf 82}(10)
  101702 \urlprefix\url{http://link.aps.org/doi/10.1103/PhysRevD.82.101702}

\bibitem{Dehmelt1999}
Dehmelt H, Mittleman R, Van~Dyck R~S and Schwinberg P 1999 {\em Phys. Rev.
  Lett.\/} {\bf 83}(23) 4694--4696
  \urlprefix\url{http://link.aps.org/doi/10.1103/PhysRevLett.83.4694}

\bibitem{DiSciacca2013}
DiSciacca J, Marshall M, Marable K, Gabrielse G, Ettenauer S, Tardiff E, Kalra
  R, Fitzakerley D~W, George M~C, Hessels E~A, Storry C~H, Weel M, Grzonka D,
  Oelert W and Sefzick T (ATRAP Collaboration) 2013 {\em Phys. Rev. Lett.\/}
  {\bf 110}(13) 130801

\bibitem{Bennett2004}
Bennett G~W, Bousquet B, Brown H~N, Bunce G, Carey R~M, Cushman P, Danby G~T,
  Debevec P~T, Deile M, Deng H, Dhawan S~K, Druzhinin V~P, Duong L, Farley
  F~J~M, Fedotovich G~V, Gray F~E, Grigoriev D, Grosse-Perdekamp M, Grossmann
  A, Hare M~F, Hertzog D~W, Huang X, Hughes V~W, Iwasaki M, Jungmann K, Kawall
  D, Khazin B~I, Krienen F, Kronkvist I, Lam A, Larsen R, Lee Y~Y, Logashenko
  I, McNabb R, Meng W, Miller J~P, Morse W~M, Nikas D, Onderwater C~J~G, Orlov
  Y, \"Ozben C~S, Paley J~M, Peng Q, Polly C~C, Pretz J, Prigl R, zu~Putlitz G,
  Qian T, Redin S~I, Rind O, Roberts B~L, Ryskulov N, Semertzidis Y~K, Shagin
  P, Shatunov Y~M, Sichtermann E~P, Solodov E, Sossong M, Sulak L~R, Trofimov
  A, von Walter P and Yamamoto A 2004 {\em Phys. Rev. Lett.\/} {\bf 92}(16)
  161802 \urlprefix\url{http://link.aps.org/doi/10.1103/PhysRevLett.92.161802}

\bibitem{Michimura2013}
Michimura Y, Matsumoto N, Ohmae N, Kokuyama W, Aso Y, Ando M and Tsubono K 2013
  {\em Phys. Rev. Lett.\/} {\bf 110}(20) 200401
  \urlprefix\url{http://link.aps.org/doi/10.1103/PhysRevLett.110.200401}

\bibitem{Hohensee2010}
Hohensee M~A, Stanwix P~L, Tobar M~E, Parker S~R, Phillips D~F and Walsworth
  R~L 2010 {\em Phys. Rev. D\/} {\bf 82}(7) 076001
  \urlprefix\url{http://link.aps.org/doi/10.1103/PhysRevD.82.076001}

\bibitem{Herrmann2009}
Hohensee M~A, Stanwix P~L, Tobar M~E, Parker S~R, Phillips D~F and Walsworth
  R~L 2010 {\em Phys. Rev. D\/} {\bf 82}(7) 076001
  \urlprefix\url{http://link.aps.org/doi/10.1103/PhysRevD.82.076001}

\bibitem{Gagnon2004}
Gagnon O and Moore G~D 2004 {\em Phys. Rev. D\/} {\bf 70}(6) 065002
  \urlprefix\url{http://link.aps.org/doi/10.1103/PhysRevD.70.065002}

\bibitem{Khriplovich1997}
Khriplovich I~B and Lamoreaux S~K 1997 {\em {CP Violation Without
  Strangeness}\/} (Springer)

\bibitem{Bernreuther1991}
Bernreuther W and Suzuki M 1991 {\em Reviews of Modern Physics\/} {\bf 63} 313

\bibitem{Pendlebury2015}
Pendlebury J~M, Afach S, Ayres N~J, Baker C~A, Ban G, Bison G, Bodek K,
  Burghoff M, Geltenbort P, Green K, Griffith W~C, van~der Grinten M,
  Gruji\ifmmode~\acute{c}\else \'{c}\fi{} Z~D, Harris P~G, H\'elaine V,
  Iaydjiev P, Ivanov S~N, Kasprzak M, Kermaidic Y, Kirch K, Koch H~C, Komposch
  S, Kozela A, Krempel J, Lauss B, Lefort T, Lemi\`ere Y, May D~J~R, Musgrave
  M, Naviliat-Cuncic O, Piegsa F~M, Pignol G, Prashanth P~N, Qu\'em\'ener G,
  Rawlik M, Rebreyend D, Richardson J~D, Ries D, Roccia S, Rozpedzik D,
  Schnabel A, Schmidt-Wellenburg P, Severijns N, Shiers D, Thorne J~A, Weis A,
  Winston O~J, Wursten E, Zejma J and Zsigmond G 2015 {\em Phys. Rev. D\/} {\bf
  92}(9) 092003
  \urlprefix\url{http://link.aps.org/doi/10.1103/PhysRevD.92.092003}

\bibitem{Graner2016}
Graner B, Chen Y, Lindahl E~G and Heckel B~R 2016 {\em Phys. Rev. Lett.\/} {\bf
  116}(16) 161601
  \urlprefix\url{http://link.aps.org/doi/10.1103/PhysRevLett.116.161601}

\bibitem{Pospelov2005}
Pospelov M and Ritz A 2005 {\em Ann. Phys.\/} {\bf 318} 119--169 ISSN 00034916
  \urlprefix\url{http://linkinghub.elsevier.com/retrieve/pii/S0003491605000539}

\bibitem{Streater2000}
Streater R~F and Wightman A~S 2000 {\em {PCT, Spin and Statistics, and All
  That}\/} (Princeton University Press)

\bibitem{Fukuyama2012}
Fukuyama T 2012 {\em International Journal of Modern Physics A\/} {\bf 27}
  1230015 ISSN 0217-751X
  \urlprefix\url{http://www.worldscientific.com/doi/abs/10.1142/S0217751X12300153}

\bibitem{Engel2013}
Engel J, Ramsey-Musolf M~J and van Kolck U 2013 {\em Progress in Particle and
  Nuclear Physics\/} {\bf 71} 21--74 ISSN 01466410
  \urlprefix\url{http://linkinghub.elsevier.com/retrieve/pii/S0146641013000227}

\bibitem{Sakharov1967}
Sakharov A~D 1967 {\em JETP Lett.\/} {\bf 5} 27--30

\bibitem{Gavela1994}
Gavela M~B, Hern\'{a}ndez P, Orloff J and P\`{e}ne O 1994 {\em Modern Physics
  Letters A\/} {\bf 09} 795--809 (\textit{Preprint}
  \eprint{http://www.worldscientific.com/doi/pdf/10.1142/S0217732394000629})
  \urlprefix\url{http://www.worldscientific.com/doi/abs/10.1142/S0217732394000629}

\bibitem{Baron2014}
{The ACME Collaboration}, Baron J, Campbell W~C, DeMille D, Doyle J~M,
  Gabrielse G, Gurevich Y~V, Hess P~W, Hutzler N~R, Kirilov E, Kozyryev I,
  O’Leary B~R, Panda C~D, Parsons M~F, Petrik E~S, Spaun B, Vutha A~C and
  West A~D 2014 {\em Science\/} {\bf 343} 269--272
  \urlprefix\url{http://www.sciencemag.org/content/343/6168/269.abstract}

\bibitem{Denis2016}
Denis M and Fleig T 2016 {\em J. Chem. Phys.\/} {\bf 145} 214307
  \urlprefix\url{http://dx.doi.org/10.1063/1.4968597}

\bibitem{Skripnikov2016}
Skripnikov L~V 2016 {\em J. Chem. Phys.\/} {\bf 145} 214301
  \urlprefix\url{http://dx.doi.org/10.1063/1.4968229}

\bibitem{Pospelov1991}
Pospelov M~E and Khriplovich I~B 1991 {\em Yad. Fiz.\/} {\bf 53} 1030--1033

\bibitem{Abel2001}
Abel S, Khalil S and Lebedev O 2001 {\em Nuclear Physics B\/} {\bf 606} 151 --
  182 ISSN 0550-3213
  \urlprefix\url{http://www.sciencedirect.com/science/article/pii/S0550321301002334}

\bibitem{Nir1999}
Nir Y 1999 {CP} violation in and beyond the standard model Lectures given in
  the XXVII SLAC Summer Institute on Particle Physics

\bibitem{Sandars1964}
Sandars P~G~H and Lipworth E 1964 {\em Physical Review Letters\/} {\bf 13} 718

\bibitem{Stein1967}
Stein T~S, Carrico J~P, Lipworth E and Weisskopf M~C 1967 {\em Phys. Rev.
  Lett.\/} {\bf 19} 741

\bibitem{Weisskopf1968}
Weisskopf M~C, Carrico J~P, Gould H, Lipworth E and Stein T~S 1968 {\em Phys.
  Rev. Lett.\/} {\bf 21} 1645

\bibitem{Player1970}
Player M~A and Sandars P~G~H 1970 {\em J. Phys. B\/} {\bf 3} 1620--1635
  \urlprefix\url{http://stacks.iop.org/0022-3700/3/1620}

\bibitem{Lamoreaux1987}
Lamoreaux S~K, Jacobs J~P, Heckel B~R, Raab F~J and Fortson N 1987 {\em Phys.
  Rev. Lett.\/} {\bf 59} 2275

\bibitem{Murthy1989}
Murthy S~A, Krause D, Li Z~L and Hunter L~R 1989 {\em Phys. Rev. Lett.\/} {\bf
  63}(9) 965--968
  \urlprefix\url{http://link.aps.org/doi/10.1103/PhysRevLett.63.965}

\bibitem{Abdullah1990}
Abdullah K, Carlberg C, Commins E~D, Gould H and Ross S~B 1990 {\em Phys. Rev.
  Lett.\/} {\bf 65} 2347

\bibitem{Commins1994}
Commins E, Ross S, DeMille D and Regan B 1994 {\em Physical Review A\/} {\bf
  50} 2960--2977 ISSN 1050-2947
  \urlprefix\url{http://link.aps.org/doi/10.1103/PhysRevA.50.2960}

\bibitem{Regan2002}
Regan B, Commins E, Schmidt C and DeMille D 2002 {\em Physical Review
  Letters\/} {\bf 88} 18--21 ISSN 0031-9007
  \urlprefix\url{http://link.aps.org/doi/10.1103/PhysRevLett.88.071805}

\bibitem{Hudson2011}
Hudson J~J, Kara D~M, Smallman I~J, Sauer B~E, Tarbutt M~R and Hinds E~A 2011
  {\em Nature\/} {\bf 473} 493--6 ISSN 1476-4687
  \urlprefix\url{http://www.ncbi.nlm.nih.gov/pubmed/21614077}

\bibitem{Schiff1963}
Schiff L~I 1963 {\em Physical Review\/} {\bf 132} 2194

\bibitem{Salpeter1958}
Salpeter E~E 1958 {\em Phys. Rev.\/} {\bf 112}(5) 1642--1648

\bibitem{Sandars1965}
Sandars P~G~H 1965 {\em Physics Letters\/} {\bf 14} 194 ISSN 00319163
  \urlprefix\url{http://dx.doi.org/doi:10.1016/0031-9163(65)90583-4}

\bibitem{Commins2007}
Commins E~D, Jackson J~D and DeMille D~P 2007 {\em American Journal of
  Physics\/} {\bf 75} 532 ISSN 00029505
  \urlprefix\url{http://link.aip.org/link/AJPIAS/v75/i6/p532/s1&Agg=doi}

\bibitem{Commins2010}
Commins E~D and DeMille D 2010 {The Electric Dipole Moment of the Electron}
  {\em Lepton Dipole Moments\/} ed Roberts B~L and Marciano W~J (World
  Scientific) chap~14, pp 519--581

\bibitem{DeMille2000}
DeMille D, Bay F, Bickman S, Kawall D, Krause D, Maxwell S and Hunter L 2000
  {\em Physical Review A\/} {\bf 61} 52507 ISSN 1050-2947
  \urlprefix\url{http://link.aps.org/doi/10.1103/PhysRevA.61.052507}

\bibitem{Hinds1997}
Hinds E~A 1997 {\em Phys. Scr.\/} {\bf T70} 34--41

\bibitem{Eckel2013}
Eckel S, Hamilton P, Kirilov E, Smith H~W and DeMille D 2013 {\em Physical
  Review A\/} {\bf 87} 052130 ISSN 1050-2947
  \urlprefix\url{http://link.aps.org/doi/10.1103/PhysRevA.87.052130}

\bibitem{Leanhardt2011}
Leanhardt A~E, Bohn J~L, Loh H, Maletinsky P, Meyer E~R, Sinclair L~C, Stutz
  R~P and Cornell E~A 2011 {\em Journal of Molecular Spectroscopy\/} {\bf 270}
  1--25 ISSN 0022-2852
  \urlprefix\url{http://dx.doi.org/10.1016/j.jms.2011.06.007}

\bibitem{Meyer2006}
Meyer E~R, Bohn J~L and Deskevich M~P 2006 {\em Phys. Rev. A\/} {\bf 73} 62108
  ISSN 1050-2947
  \urlprefix\url{http://link.aps.org/doi/10.1103/PhysRevA.73.062108}

\bibitem{Mosyagin1998}
Mosyagin N~S, Kozlov M~G and Titov A~V 1998 {\em Journal of Physics B: Atomic,
  Molecular and Optical Physics\/} {\bf 31} L763--L767 ISSN 0953-4075
  \urlprefix\url{http://stacks.iop.org/0953-4075/31/i=19/a=002?key=crossref.fc40d2593364d311c6a64905cd54e3ef}

\bibitem{Sauer2011}
Sauer B, Hudson J, Kara D, Smallman I, Tarbutt M and Hinds E 2011 {\em Physics
  Procedia\/} {\bf 17} 175--180 ISSN 18753892
  \urlprefix\url{http://linkinghub.elsevier.com/retrieve/pii/S1875389211003658}

\bibitem{Abe2014}
Abe M, Gopakumar G, Hada M, Das B~P, Tatewaki H and Mukherjee D 2014 {\em Phys.
  Rev. A\/} {\bf 90}(2) 022501
  \urlprefix\url{http://link.aps.org/doi/10.1103/PhysRevA.90.022501}

\bibitem{Herzberg1989}
Herzberg G 1989 {\em {Molecular Spectra and Molecular Structure: Spectra of
  Diatomic Molecules , 2nd ed.}\/} (Krieger)

\bibitem{Petrov2014}
Petrov A~N, Skripnikov L~V, Titov A~V, Hutzler N~R, Hess P~W, O'Leary B~R,
  Spaun B, DeMille D, Gabrielse G and Doyle J~M 2014 {\em Phys. Rev. A\/} {\bf
  89}(6) 062505
  \urlprefix\url{http://link.aps.org/doi/10.1103/PhysRevA.89.062505}

\bibitem{Vutha2011}
Vutha A~C, Spaun B, Gurevich Y~V, Hutzler N~R, Kirilov E, Doyle J~M, Gabrielse
  G and DeMille D 2011 {\em Physical Review A\/} {\bf 84} 034502 ISSN 1050-2947
  \urlprefix\url{http://link.aps.org/doi/10.1103/PhysRevA.84.034502}

\bibitem{Herzberg1971}
Herzberg G 1971 {\em {The Spectra and Structures of Simple Free Radicals}\/}
  (Ithaca: Cornell University Perss)

\bibitem{Brown2003}
Brown J~M and Carrington A 2003 {\em {Rotational Spectroscopy of Diatomic
  Molecules}\/} (Cambridge University Press) ISBN 0521530784
  \urlprefix\url{http://books.google.com/books/about/Rotational\_Spectroscopy\_of\_Diatomic\_Mole.html?id=veUq07zAoxMC\&pgis=1}

\bibitem{Landau1981}
Landau L~D and Lifshitz E~M 1981 {\em Quantum Mechanics: Non-relativistic
  Theory\/} (Butterworth-Heinemann)

\bibitem{Kirilov2013}
Kirilov E, Campbell W~C, Doyle J~M, Gabrielse G, Gurevich Y~V, Hess P~W,
  Hutzler N~R, O’Leary B~R, Petrik E, Spaun B, Vutha A~C and DeMille D 2013
  {\em Physical Review A\/} {\bf 88} 013844 ISSN 1050-2947
  \urlprefix\url{http://arxiv.org/abs/1305.2179
  http://link.aps.org/doi/10.1103/PhysRevA.88.013844}

\bibitem{DeMille2001}
DeMille D, Bay F, Bickman S, Kawall D, Hunter L, Krause D, Maxwell S and Ulmer
  K 2001 {Search for the electric dipole moment of the electron using
  metastable PbO} {\em AIP Conference Proceedings\/} vol 596 (AIP) pp 72--83
  ISSN 0094243X \urlprefix\url{http://link.aip.org/link/?APC/596/72/1&Agg=doi}

\bibitem{HutzlerThesis}
Hutzler N 2014 {\em A New Limit on the Electron Electric Dipole Moment: Beam
  Production, Data Interpretation, and Systematics\/} Ph.D. thesis Harvard
  University (advisor: J. Doyle)

\bibitem{Hess2014}
Hess P 2014 {\em Improving the Limit on the Electron EDM: Data Acquisition and
  Systematics Studies\/} Ph.D. thesis Harvard Univ. (advisor: G. Gabrielse)

\bibitem{Kawall2004}
Kawall D, Bay F, Bickman S, Jiang Y and DeMille D 2004 {Progress towards
  measuring the electric dipole moment of the electron in metastable PbO} {\em
  AIP Conference Proceedings\/} vol 698 (AIP) pp 192--195 ISSN 0094243X
  \urlprefix\url{http://link.aip.org/link/?APC/698/192/1&Agg=doi}

\bibitem{Vutha2010}
Vutha A~C, Campbell W~C, Gurevich Y~V, Hutzler N~R, Parsons M, Patterson D,
  Petrik E, Spaun B, Doyle J~M, Gabrielse G and DeMille D 2010 {\em Journal of
  Physics B\/} {\bf 43} 74007 ISSN 0953-4075
  \urlprefix\url{http://stacks.iop.org/0953-4075/43/i=7/a=074007?key=crossref.114fcc8fa117ce235d58bc66e7812c5f}

\bibitem{Regan2001}
Regan B 2001 {\em {A Search for Violation of Time-Reversal Symmetry in Atomic
  Thallium}\/} Ph.D. thesis Berkeley

\bibitem{Hutzler2012}
Hutzler N~R, Lu H~I and Doyle J~M 2012 {\em Chemical Reviews\/} {\bf 112}
  4803--27 ISSN 1520-6890
  \urlprefix\url{http://pubs.acs.org/doi/abs/10.1021/cr200362u
  http://www.ncbi.nlm.nih.gov/pubmed/22571401}

\bibitem{Patterson2007}
Patterson D and Doyle J~M 2007 {\em The Journal of Chemical Physics\/} {\bf
  126} 154307 ISSN 0021-9606
  \urlprefix\url{http://www.ncbi.nlm.nih.gov/pubmed/17461626}

\bibitem{Maxwell2005}
Maxwell S~E, Brahms N, DeCarvalho R, Glenn D~R, Helton J~S, Nguyen S~V,
  Patterson D, Petricka J, DeMille D and Doyle J~M 2005 {\em Physical Review
  Letters\/} {\bf 95} 173201 ISSN 0031-9007
  \urlprefix\url{http://link.aps.org/doi/10.1103/PhysRevLett.95.173201}

\bibitem{Hutzler2011}
Hutzler N~R, Parsons M~F, Gurevich Y~V, Hess P~W, Petrik E, Spaun B, Vutha A~C,
  DeMille D, Gabrielse G and Doyle J~M 2011 {\em Physical Chemistry Chemical
  Physics : PCCP\/} {\bf 13} 18976--85 ISSN 1463-9084
  \urlprefix\url{http://www.ncbi.nlm.nih.gov/pubmed/21698321}

\bibitem{Edvinsson1985}
Edvinsson G and Lagerqvist A 1985 {\em J. Mol. Spectrosc.\/} {\bf 113} 93

\bibitem{Paulovic2003}
Paulovic J, Nakajima T, Hirao K, Lindh R and Malmqvist P~A 2003 {\em J. Chem.
  Phys.\/} {\bf 119} 798--805

\bibitem{Budker2008}
Budker D, Kimball D and DeMille D 2008 {\em {Atomic physics: An exploration
  through problems and solutions}\/} (Oxford University Press, USA) ISBN
  0199532419
  \urlprefix\url{http://www.amazon.com/Atomic-physics-exploration-problems-solutions/dp/0199532419}

\bibitem{Bickman2009}
Bickman S, Hamilton P, Jiang Y and DeMille D 2009 {\em Physical Review A\/}
  {\bf 80} 023418 ISSN 1050-2947
  \urlprefix\url{http://link.aps.org/doi/10.1103/PhysRevA.80.023418}

\bibitem{VuthaThesis}
Vutha A~C 2011 {\em {A search for the electric dipole moment of the electron
  using thorium monoxide}\/} Ph.D. thesis Yale University

\bibitem{Petricka2007}
Petricka J 2007 {\em {A New Cold Molecule Source: The Buffer Gas Cooled
  Molecular Beam}\/} Ph.D. thesis Yale University

\bibitem{Sushkov2008}
Sushkov A and Budker D 2008 {\em Phys. Rev. A\/} {\bf 77} 42707 ISSN 1050-2947
  \urlprefix\url{http://link.aps.org/doi/10.1103/PhysRevA.77.042707}

\bibitem{Patterson2009}
Patterson D, Rasmussen J and Doyle J~M 2009 {\em New Journal of Physics\/} {\bf
  11} 055018 ISSN 1367-2630
  \urlprefix\url{http://stacks.iop.org/1367-2630/11/i=5/a=055018?key=crossref.ada50c0bc5df340d553560bba982f04c}

\bibitem{Campbell2009Review}
Campbell W~C and Doyle J~M 2009 {Cooling, Trap Loading, and Beam Production
  Using a Cryogenic Helium Buffer Gas} {\em Cold molecules: theory, experiment,
  applications\/} ed Krems R~V, Stwalley W~C and Friedrich B (CRC Press)
  chap~13, pp 473--508

\bibitem{Tu2009}
Tu M~F, Ho J~J, Hsieh C~C and Chen Y~C 2009 {\em The Review of scientific
  instruments\/} {\bf 80} 113111 ISSN 1089-7623
  \urlprefix\url{http://www.ncbi.nlm.nih.gov/pubmed/19947721}

\bibitem{Patterson2010}
Patterson D 2010 {\em {Buffer Gas Cooled Beams and Cold Molecular
  Collisions}\/} Ph.D. thesis Harvard University

\bibitem{Barry2011}
Barry J~F, Shuman E~S and Demille D 2011 {\em Phys. Chem. Chem. Phys.\/} {\bf
  13} 18936 ISSN 1463-9084
  \urlprefix\url{http://www.ncbi.nlm.nih.gov/pubmed/21706119}

\bibitem{Lu2011}
Lu H~I, Rasmussen J, Wright M~J, Patterson D and Doyle J~M 2011 {\em Physical
  chemistry chemical physics : PCCP\/} {\bf 13} 18986--90 ISSN 1463-9084
  \urlprefix\url{http://www.ncbi.nlm.nih.gov/pubmed/21796294}

\bibitem{Skoff2011PRA}
Skoff S~M, Hendricks R~J, Sinclair C~D~J, Hudson J~J, Segal D~M, Sauer B~E,
  Hinds E~A and Tarbutt M~R 2011 {\em Physical Review A\/} {\bf 83} 23418 ISSN
  1050-2947 \urlprefix\url{http://link.aps.org/doi/10.1103/PhysRevA.83.023418}

\bibitem{Skoff2011Thesis}
Skoff S~M 2011 {\em {Buffer gas cooling of YbF molecules}\/} Ph.D. thesis
  Imperial College London

\bibitem{Hummon2013}
Hummon M~T, Yeo M, Stuhl B~K, Collopy A~L, Xia Y and Ye J 2013 {\em Physical
  Review Letters\/} {\bf 110} 143001 ISSN 0031-9007
  \urlprefix\url{http://link.aps.org/doi/10.1103/PhysRevLett.110.143001}

\bibitem{Bulleid2013}
Bulleid N~E, Skoff S~M, Hendricks R~J, Sauer B~E, Hinds E~A and Tarbutt M~R
  2013 {\em Physical chemistry chemical physics : PCCP\/} {\bf 15} 12299--307
  ISSN 1463-9084 \urlprefix\url{http://www.ncbi.nlm.nih.gov/pubmed/23775176}

\bibitem{Balakrishna1988}
Balakrishna P, Varma B~P, Krishnan T~S, Mohan T~R~R and Ramakrishnan P 1988
  {\em Journal of Materials Science Letters\/} {\bf 7} 657--660 ISSN 0261-8028
  \urlprefix\url{http://www.springerlink.com/index/10.1007/BF01730326}

\bibitem{KiggansPrivate}
Kiggans J private communication

\bibitem{SpaunThesis}
Spaun B~N 2014 {\em A Ten-Fold Improvement to the Limit of the Electron
  Electric Dipole Moment\/} Ph.D. thesis Harvard University (advisor: G.
  Gabrielse)

\bibitem{Edvinsson1965}
Edvinsson G, Selin L~E and Aslund N 1965 {\em Ark. Phys.\/} {\bf 30} 283--319

\bibitem{YuliaThesis}
Gurevich Y~V 2012 {\em Preliminary Measurements for an Electron EDM Experiment
  in ThO\/} Ph.D. thesis Harvard University (advisor: G. Gabrielse)

\bibitem{FarkasThesis}
Farkas D 2006 {\em An Optical Reference and Frequency Comb for Improved
  Spectroscopy of Helium\/} Ph.D. thesis Harvard University (advisor: G.
  Gabrielse)

\bibitem{Patten1971}
Patten R~A 1971 {\em Applied Optics\/} {\bf 10} 2717--2721

\bibitem{Vutha2009}
Vutha A and DeMille D 2009 {\em arXiv\/} (\textit{Preprint} \eprint{0907.5116})
  \urlprefix\url{http://arxiv.org/abs/0907.5116}

\bibitem{Robertson2006}
Robertson N~A, Blackwood J~R, Buchman S, Byer R~L, Camp J, Gill D, Hanson J,
  Williams S and Zhou P 2006 {\em Classical and Quantum Gravity\/} {\bf 23}
  2665--2680 ISSN 0264-9381
  \urlprefix\url{http://stacks.iop.org/0264-9381/23/i=7/a=026?key=crossref.d6e94b4d247636bdb793339078a56486}

\bibitem{Kara2012}
Kara D~M, Smallman I~J, Hudson J~J, Sauer B~E, Tarbutt M~R and Hinds E~A 2012
  {\em New Journal of Physics\/} {\bf 14} 103051 ISSN 1367-2630
  \urlprefix\url{http://stacks.iop.org/1367-2630/14/i=10/a=103051?key=crossref.935bac343ba2f6bf853baf0bc5df44bc}

\bibitem{Mager1969}
Mager A~J 1969 {\em IEEE Trans. Magn.\/} {\bf MAG-6} 67--75

\bibitem{Sumner1987}
Sumner T~J, Pendlebury J~M and Smith K~F 1987 {\em J. Phys. D\/} {\bf 20}
  1095--1101

\bibitem{Kenny1951}
Kenny J~F and Keeping E~S 1951 {\em {Mathematics of Statistics}\/} 2nd ed (New
  York: Van Nostrand)

\bibitem{JILAEDM}
Loh H, Cossel K~C, Grau M~C, Ni K~K, Meyer E~R, Bohn J~L, Ye J and Cornell E~A
  2013 {\em Science\/} {\bf 342} 1220--1222 (\textit{Preprint}
  \eprint{http://www.sciencemag.org/content/342/6163/1220.full.pdf})
  \urlprefix\url{http://www.sciencemag.org/content/342/6163/1220.abstract}

\bibitem{Hamilton2010}
Hamilton P 2010 {\em {Preliminary results in the search for the electron
  electric dipole moment in PbO}\/} Ph.D. thesis Yale University

\bibitem{sozzi2008}
Sozzi M 2008 {\em Discrete Symmetries and CP Violation: From Experiment to
  Theory\/} (OUP Oxford)

\bibitem{Griffith2009}
Griffith W, Swallows M, Loftus T, Romalis M, Heckel B and Fortson E 2009 {\em
  Physical Review Letters\/} {\bf 102} 101601 ISSN 0031-9007
  \urlprefix\url{http://link.aps.org/doi/10.1103/PhysRevLett.102.101601}

\bibitem{Bouchiat1974}
Bouchiat M and Bouchiat C 1974 {\em Journal de Physique\/} {\bf 34} 899--927

\bibitem{Demille2008}
DeMille D, Cahn S~B, Murphree D, Rahmlow D~A and Kozlov M~G 2008 {\em Phys.
  Rev. Lett.\/} {\bf 100} 023003 ISSN 0031-9007
  \urlprefix\url{http://link.aps.org/doi/10.1103/PhysRevLett.100.023003}

\bibitem{Wood1997}
Wood C~S, Bennett S~C, Cho D, Masterson B~P, Roberts J~L, Tanner C~E and Wieman
  C~E 1997 {\em Science\/} {\bf 275} 1759--1763

\bibitem{Chen1994}
Chen X, Huang-Hellinger F, Heckel B and Fortson E 1994 {\em Physical Review
  A\/} {\bf 50} 4729--4732

\bibitem{Hodgdon1991}
Hodgdon J, Heckel B and Fortson E 1991 {\em Physical Review A\/} {\bf 43}
  3343--3347

\bibitem{Lamoreaux1992}
Lamoreaux S and Fortson E 1992 {\em Physical Review A\/} {\bf 46} 7053--7059

\bibitem{Loftus2011}
Loftus T~H, Swallows M~D, Griffith W~C, Romalis M~V, Heckel B~R and Fortson E~N
  2011 {\em Physical Review Letters\/} {\bf 106} 253002 ISSN 0031-9007

\bibitem{Swallows2013}
Swallows M~D, Loftus T~H, Griffith W~C, Heckel B~R, Fortson E~N and Romalis M~V
  2013 {\em Phys. Rev. A\/} {\bf 87}(1) 012102
  \urlprefix\url{http://link.aps.org/doi/10.1103/PhysRevA.87.012102}

\bibitem{Sachs1987}
Sachs R 1987 {\em {The Physics of Time Reversal}\/} (University of Chicago
  Press)

\bibitem{Barber2010}
Barber J~R 2010 {\em {Elasticity, Solid Mechanics and its Applications}\/} 3rd
  ed (Springer) ISBN 9789048138081
  \urlprefix\url{http://www.lavoisier.fr/livre/notice.asp?id=6OKWSRA2SOSOWP}

\bibitem{Dally1991}
Dally J~W and Riley W~F 1991 {\em {Experimental Stress Analysis}\/} third edit
  ed (McGraw-Hill) ISBN 0070152187

\bibitem{Schott2013b}
Schott 2013 {Schott Borofloat Glass Product Information} Tech. Rep. 502 Schott
  \urlprefix\url{http://www.markoptics.com/files/Schott Borofloat 33.pdf}

\bibitem{Solmeyer2011}
Solmeyer N, Zhu K and Weiss D~S 2011 {\em The Review of scientific
  instruments\/} {\bf 82} 066105 ISSN 1089-7623
  \urlprefix\url{http://www.ncbi.nlm.nih.gov/pubmed/21721740}

\bibitem{Eisenbach1992}
Eisenbach S and Lotem H 1992 {\em SPIE 8th Meeting on Optical Engineering in
  Israel\/} {\bf 1972}
  \urlprefix\url{http://proceedings.spiedigitallibrary.org/proceeding.aspx?articleid=1013612}

\bibitem{Koechner1970}
Koechner W 1970 {\em Applied optics\/} {\bf 9} 2548--2553 ISSN 0003-6935
  \urlprefix\url{http://www.ncbi.nlm.nih.gov/pubmed/20094304}

\bibitem{Berry1977}
Berry H~G, Gabrielse G and Livingston A~E 1977 {\em Applied Optics\/} {\bf 16}
  3200--5 ISSN 0003-6935
  \urlprefix\url{http://www.ncbi.nlm.nih.gov/pubmed/20174328}

\bibitem{sinervo2003}
Sinervo P~K 2003 Definition and treatment of systematic uncertainties in high
  energy physics and astrophysics {\em Statistical Problems in Particle
  Physics, Astrophysics, and Cosmology\/} vol~1 (Citeseer) p 122

\bibitem{barlow2002}
Barlow R 2002 {\em arXiv preprint hep-ex/0207026\/}

\bibitem{Riley2006}
Riley K~F, Hobson M~P and Bence S~J 2006 {\em {Mathematical methods for physics
  and engineering: a comprehensive guide}\/} (Cambridge: Cambridge Univ. Press)

\bibitem{Feldman1998}
Feldman G~J and Cousins R~D 1998 {\em Physical Review D\/} {\bf 57} 3873--3889
  ISSN 0556-2821
  \urlprefix\url{http://link.aps.org/doi/10.1103/PhysRevD.57.3873}

\bibitem{Stuart1999}
Stuart A, Ord J~K and Arnold S 1999 {\em Classical Inference and the Linear
  Model\/} 6th ed ({\em Kendall's Advanced Theory of Statistics\/} vol~2A)
  (London: Arnold)

\bibitem{Kozlov1995}
Kozlov M~G and Labzowsky L~N 1995 {\em Journal of Physics B: Atomic, Molecular
  and Optical Physics\/} {\bf 28} 1933--1961 ISSN 0953-4075
  \urlprefix\url{http://stacks.iop.org/0953-4075/28/i=10/a=008?key=crossref.2547c28c7d7ac4a7fa9f914888c84fc6}

\bibitem{Skripnikov2013}
Skripnikov L~V, Petrov A~N and Titov A~V 2013 {\em J. Chem. Phys.\/} {\bf 139}
  221103 ISSN 1089-7690

\bibitem{Dzuba2011a}
Dzuba V~A, Flambaum V~V and Harabati C 2011 {\em Phys. Rev. A\/} {\bf 84}(5)
  052108 \urlprefix\url{http://link.aps.org/doi/10.1103/PhysRevA.84.052108}

\bibitem{DzubaErratum2012}
Dzuba V~A, Flambaum V~V and Harabati C 2012 {\em Phys. Rev. A\/} {\bf 85}(2)
  029901 \urlprefix\url{http://link.aps.org/doi/10.1103/PhysRevA.85.029901}

\bibitem{Heckel2016}
Graner B, Chen Y, Lindahl E~G and Heckel B~R 2016 {\em Phys. Rev. Lett.\/} {\bf
  116}(16) 161601
  \urlprefix\url{http://link.aps.org/doi/10.1103/PhysRevLett.116.161601}

\bibitem{Ackermann1963}
Ackermann R~J, Rauh E~G, Thorn R~J and Cannon M~C 1963 {\em Journal of Physical
  Chemistry\/} {\bf 67} 762--769 ISSN 0022-3654
  \urlprefix\url{http://pubs.acs.org/cgi-bin/doilookup/?10.1021/j100798a010}

\bibitem{OECD2008}
OECD 2008 {\em Chemical Thermodynamics of Thorium\/} (OECD Publishing)

\bibitem{Panda2016}
Panda C~D, O'Leary B~R, West A~D, Baron J, Hess P~W, Hoffman C, Kirilov E,
  Overstreet C~B, West E~P, DeMille D, Doyle J~M and Gabrielse G 2016 {\em
  Phys. Rev. A\/} {\bf 93}(5) 052110
  \urlprefix\url{http://link.aps.org/doi/10.1103/PhysRevA.93.052110}

\bibitem{Steimle2014}
Kokkin D~L, Steimle T~C and DeMille D 2014 {\em Phys. Rev. A\/} {\bf 90} 062503

\bibitem{Lee2009}
Lee J, Meyer E~R, Paudel R, Bohn J~L and Leanhardt A~E 2009 {\em Journal of
  Modern Optics\/} {\bf 56} 2005--2012

\bibitem{Skripnikov2015}
Skripnikov L~V and Titov A~V 2015 {\em J. Chem. Phys.\/} {\bf 142} 024301

\bibitem{Dzuba2011}
Dzuba V and Flambaum V 2011 {\em Physical Review A\/} {\bf 83} ISSN 1050-2947
  \urlprefix\url{http://link.aps.org/doi/10.1103/PhysRevA.83.042514}

\bibitem{Kozlov2002}
Kozlov M and DeMille D 2002 {\em Physical Review Letters\/} {\bf 89} 133001
  ISSN 0031-9007
  \urlprefix\url{http://link.aps.org/doi/10.1103/PhysRevLett.89.133001}

\bibitem{Fleig2014}
Fleig T and Nayak M~K 2014 {\em J. Mol. Spectrosc.\/} {\bf 300} 16--21

\bibitem{Fleig2013}
Fleig T and Nayak M~K 2013 {\em Phys. Rev. A\/} {\bf 88}(3) 032514
  \urlprefix\url{http://link.aps.org/doi/10.1103/PhysRevA.88.032514}

\bibitem{Denis2015}
Denis M, N{\o}rby M~S, Jensen H~J~A, Gomes A~S~P, Nayak M~K, Knecht S and Fleig
  T 2015 {\em New. J. Phys.\/} {\bf 17} 043005
  \urlprefix\url{http://dx.doi.org/10.1088/1367-2630/17/4/043005}

\end{thebibliography}

\end{document}